\documentclass[a4paper,11pt]{article}
\usepackage{jheppub} 
\usepackage{lineno}

\usepackage[utf8]{inputenc}

\usepackage{amsmath}
\usepackage{amssymb}
\usepackage{mathtools}
\usepackage{bbold}
\usepackage{braket}
\usepackage{accents}
\usepackage{comment}
\usepackage{graphicx}
\usepackage{bmpsize}
\usepackage{feynmp}

\usepackage{booktabs}
\usepackage{subfig}
\usepackage{siunitx}

\DeclareMathOperator{\Tr}{Tr}

\newcommand\numberthis{\addtocounter{equation}{1}\tag{\theequation}}

\makeatletter
\newcommand{\bigplus}{%
  \DOTSB\mathop{\mathpalette\mattos@bigplus\relax}\slimits@
}
\newcommand\mattos@bigplus[2]{%
  \vcenter{\hbox{%
    \sbox\z@{$#1\sum$}%
    \resizebox{!}{0.9\dimexpr\ht\z@+\dp\z@}{\raisebox{\depth}{$\m@th#1+$}}%
  }}%
  \vphantom{\sum}%
}
\makeatother

\makeatletter
\newcommand*{\rom}[1]{\expandafter\@slowromancap\romannumeral #1@}
\makeatother

\arxivnumber{2607.27133} 

\title{\boldmath Quantum Field Theory Of Cosmological Perturbations Induced By Ultralight Dark Matter}

 \author{Tomislav Prokopec}
 \author[1]{and Marco Vecchioni\note{Corresponding author.}}
 \affiliation{Institute for Theoretical Physics, Spinoza Institute \& EMME$\Phi$, Utrecht University, Princetonplein 5, 3584 CC Utrecht, The Netherlands}

\emailAdd{mrcvecchioni@gmail.com}
\emailAdd{t.prokopec@uu.nl}

\abstract{The growth of primordial perturbations during the matter-dominated era is primarily driven by dark matter. Ultralight scalar fields (ULDM) are a promising candidate for this role, conventionally modeled as operating in a classical, high-occupation regime. In this work, we develop a first-principles field-theoretic framework to investigate the impact of ULDM on linear cosmological perturbations during matter domination, explicitly retaining its quantum nature. Deriving a closed equation for the graviton field dynamics, we compute and regularize its source terms for a generic Gaussian initial state of the ULDM field within the adiabatic (WKB) approximation, employing the middle-point working assumption for non-local terms. After gauge-fixing we find that, contrary to previous claims, the classical condensate of ULDM has no influence on gravitational wave propagation. However, the time-dependent graviton effective mass induced by quantum pressure of the squeezed state can drive parametric resonance in specific primordial gravitational wave modes. We demonstrate this growth is negligible for non-relativistic ULDM at matter-radiation equality under the assumption of a power-law squeezing spectrum for masses in the range $m \sim 10^{-21}{-}10^{-24}$ eV.}

\begin{document}
\begin{fmffile}{2PI}

\maketitle
\flushbottom

\section{Introduction}

Recent evidence for a nanohertz stochastic gravitational-wave background (SGWB) in pulsar-timing data 
\cite{Agazie_2023, Xu_2023, 2023,10.1093/mnras/stae2571}, together with the extended redshift reach of modern 
galaxy surveys \cite{Dawson_2025} is opening new observational windows into the primordial Universe. Constraining early-Universe physics through these observations requires increasingly precise theoretical modeling of linear cosmological perturbations and their evolution throughout cosmic history. Within the inflationary paradigm, quantum fluctuations of the metric and matter fields are stretched to cosmological 
scales, yielding nearly scale-invariant primordial spectra \cite{mukhanov1981quantum}. Such perturbations 
subsequently source the anisotropies of the cosmic microwave background \cite{aghanim2020planck, 2020} 
and the formation of large-scale structure \cite{Abbott_2023}, and include a background of primordial tensor 
modes \cite{Chongchitnan_2006}. Reliable predictions for the associated observables require a consistent treatment of their evolution from inflation through the radiation- and matter-dominated eras to the late-time dark-energy epoch. During the matter-dominated era, the growth of gravitational perturbations 
is governed mainly by dark matter, which makes up roughly 84\% of the cosmic matter density \cite{ppplank}. 
Among the many candidates proposed, two have emerged as subjects of particular interest in recent studies: 
ultralight bosonic fields (ULDM), with masses in the range 
$m\sim10^{-24}{-}10^{-21}\,$eV \cite{Ferreira_2021, Hui:2016ltb}, and primordial black holes (PBHs) 
\cite{Carr_2022}.

In the high-occupancy regime, ULDM is commonly assumed to be well described by a coherent state, reducing its dynamics to that of a classical condensate. In this description its oscillations can, in principle, induce resonant growth 
of gravitational-wave (GW) amplitudes \cite{Delgado:2023psl, Cai_2024, Jenks:2026jpo}. 
A complementary classical particle picture has also 
been proposed, treating virialized dark matter through its phase-space distribution 
\cite{Friedrich:2017glg, Friedrich:2018qjv, Friedrich:2019zic}. However, even when occupation numbers are large, the underlying quantum state 
of the ULDM field may leave observable imprints. In particular, mode squeezing, a generic outcome of inflation, 
can introduce features that remain unsuppressed in this regime. Recent analyses have shown that the quantum 
stress of ULDM leads to oscillatory corrections to the Newtonian potential 
\cite{hwang2021oscillatinggravitationalpotentialultralight}, and that a squeezed-state produces distinctive 
signatures in density statistics \cite{Kalia_2026}. These results motivate a more systematic investigation 
of quantum signatures in cosmological observables, as inflation generically drives light scalar fields into 
highly squeezed states \cite{Albrecht_1994,Polarski:1995jg, Brandenberger:1992sr,Brandenberger:1992jh}. 

Motivated by these considerations, we develop a first-principles quantum field-theoretic framework for deriving the evolution equations of linear cosmological perturbations in the presence of a general Gaussian ULDM state during matter domination. The framework accommodates a nonzero condensate, corresponding to a coherent field amplitude, as well as large occupation numbers and explicit mode squeezing. Our starting point is the two-particle-irreducible (2PI) effective action \cite{PhysRevD.10.2428} for a minimally coupled scalar field on a on a perturbed spatially flat FLRW spacetime. We formulate the theory within the Schwinger–Keldysh, or closed-time-path, formalism \cite{schwinger1961brownian, Keldysh:1964ud}, thereby ensuring causal evolution equations. Related first-principles derivations of quantum-corrected evolution equations have recently been developed for the radiation-dominated era \cite{Ota:2022xni,Ota:2023iyh,Frob:2025sfq,Sasaki:2025zao,Glavan:2026wwl,Liu:2026, BaghouzianFennemaProkopec:2026}.

In the following chapter, we construct the semiclassical effective action, treating ULDM as a quantum scalar field while keeping gravitational perturbations classical. Its variation yields coupled evolution equations, whose perturbative expansion gives a linearized graviton equation sourced by matter background quantities.
An appropriate manipulation of the equation casts it into a form better suited for treatment under the adiabatic (WKB) approximation, in which the condensate contribution to the four-point self-energy drops out.
In Chapter 3, we solve for the homogeneous ULDM condensate and Keldysh two-point function for an arbitrary Gaussian initial state, using the mid-point approximation for non-local terms. We then compute the expectation value of the energy-momentum tensor and the one-loop graviton self-energy, with details of the calculation and renormalization provided in Appendix B. After gauge fixing, Chapter 4 derives the evolution equations for the scalar metric potentials and tensor modes. Contrary to earlier claims, the coherent ULDM condensate does not directly modify GW propagation. Instead, ULDM mode squeezing induces a periodic effective graviton mass, producing a quantum parametric resonance in the GW amplitude. In Chapter 5, we study the GW equation under the 
assumption of non-relativistic dark-matter with occupation number and squeezing distributions that follow 
power-law spectra in momentum. For different classes of initial states, we derive the solution for the 
strain and estimate the resonant growth of the GW amplitude. Importantly, we find that this resonant 
amplification is limited by the ULDM pressure: numerically, the enhancement remains extremely small 
(of order $\lesssim10^{-12}$) for plausible ULDM parameters. In other words, although the quantum-induced 
resonance effect is a novel theoretical possibility, its impact on observable GWs is negligibly tiny under 
realistic conditions.

\section{Semiclassical graviton field equation}

\subsection{Classical framework}

This chapter derives the linearized equation of motion for the graviton field. 
We consider a minimally coupled, real, massive scalar field $\phi$ in $D$ dimensions 
described by the classical action,
\begin{equation} \label{classicalaction}
\begin{split}
    S[\phi,g_{\mu \nu}] &= \frac{1}{\kappa^2} \int d^D\! x \, \sqrt{-g(x)}\,R(x) 
    - \frac{1}{2}\int d^D\! x \, \sqrt{-g(x)} \left[ g^{\mu \nu} \partial_{\mu} \phi \, \partial_{\nu} \phi + m^2 \phi^2 \right] \\[3pt]
    &\equiv S_g[g_{\mu \nu}]+ S_m[\phi, g_{\mu\nu}],
\end{split}
\end{equation}
where $\kappa^2 = 16\pi G$ is the loop counting parameter of quantum gravity, $R$ is the Ricci scalar, $g$ denotes the metric determinant, and $g^{\mu \lambda}g_{\lambda \nu}=\delta_{\nu}^{\mu}$. Anticipating dimensional regularisation $D$ is kept general. 
Here \(S_g\) and \(S_m\) denote the gravitational and matter parts of the action, respectively.
The metric is expanded around a spatially flat Friedmann–Lemaître–Robertson–Walker (FLRW) 
background in conformal time $\eta$,
\begin{equation} \label{metricexpansion}
    g_{\mu \nu} = a^2(\eta) \left( \eta_{\mu \nu} + \epsilon h_{\mu \nu} \right), \qquad \epsilon \ll 1,
\end{equation}
with $a(\eta)$ the scale factor and $h_{\mu\nu}$ the metric perturbation. 
Here $\epsilon$ denotes a perturbative parameter of order $\kappa$, which tracks powers 
of the perturbation field $h_{\mu\nu}$. To make the perturbative analysis clear, we keep the powers of \(\epsilon\) explicit throughout the body of the paper; all quantities (action, vertices, operators, etc.) are consequently of order \(\mathcal{O}(1)\).
To study linearized perturbations, we maintain terms up to order $\epsilon^2$ in the action. 
Expanding the gravitational sector yields:
\begin{equation}
S_g[a, h_{\mu \nu}] =S_g^{(0)}[a] + \epsilon S_g^{(1)}[a, h_{\mu \nu}] + \epsilon^2 S_g^{(2)}[a, h_{\mu \nu}] + \mathcal{B.T.} + \mathcal{O}(\epsilon^3),
\label{expanding the total action}
\end{equation}
where the gravitational action comprises the following contributions,
\begin{equation}
\begin{split}
S_g^{(0)}[a] &= \int d^D\! x \, a^{D-2}\left[ -(D-2)(D-1) \mathcal{H}^2 \right]; \\[8pt]
S_g^{(1)}[a, h_{\mu \nu}] &= \int d^D\! x \,a^{D-2}\left\{ h^{\mu \nu} \delta^0_{\mu} \delta^0_{\nu} (D-2) (\mathcal{H}'- \mathcal{H}^2) + h \frac{D-2}{2} \left[ \mathcal{H}^2 (D-3) + 2 \mathcal{H}' \right] \right\}; \\[8pt]
S_g^{(2)}[a, h_{\mu \nu}] &=  \int d^D\! x \,a^{D-2} \left\{  \left[ \frac{D-2}{4}  \left( \mathcal{H}^2 (D-3) + 2 \mathcal{H}' \right) \right] \left( \frac{1}{2}h^2 - h_{\mu \nu} h^{\mu \nu} \right)  \right. \\[4pt]
& \left.  + \delta^0_{\mu} \delta^0_{\nu} (D-2) (\mathcal{H}^2- \mathcal{H}') ( h^{\mu}_{\alpha}h^{\alpha \nu} - \frac{1}{2} h h ^{\mu \nu}) -  \frac{D-2}{2} \delta^0_{\nu} \mathcal{H} \left( h^{\mu \nu} \partial_{\mu} h\right)   \right. \\[4pt]
& - \left. \right. \frac{1}{2} (\partial_{\alpha} h)(\partial_{\mu} h^{\mu \alpha} ) + \frac{1}{4} (\partial_{\alpha} h) (\partial^{\alpha} h) - \frac{1}{4} (\partial_{\alpha} h_{\mu \nu}) (\partial^{\alpha} h^{\mu \nu}) + \left.  \frac{1}{2}(\partial_{\alpha} h_{\mu \nu}) (\partial^{\nu} h^{\mu \alpha}) \right\}
\,.
\end{split}
\end{equation}
The matter action in~(\ref{classicalaction}) can be expanded in an analogous fashion:
\begin{equation}
\begin{split}
S_m^{(0)}[\phi, a] &= - \int d^D\! x \,\,\frac{a^{D-2}}{2} \big( \partial \phi \cdot \partial \phi + m^2 \phi^2\big); \\[8pt]
S_m^{(1)}[\phi, a, h_{\mu \nu}] &= -  \int d^D\! x \, \frac{a^{D-2}}{2}\left[ \frac{h}{2}\big( \partial \phi \cdot \partial \phi + m^2 \phi^2\big) - h^{\mu \nu} \, \partial_{\mu} \phi \, \partial_{\nu} \phi \right]; \\[8pt]
S_m^{(2)}[\phi, a, h_{\mu \nu}] &=- \int d^D\! x \,\frac{a^{D-2}}{2}  \left[ \left( \frac{h^2}{8} - \frac{1}{4} h^{\mu \nu} h_{\mu \nu} \right) \big( \partial \phi \cdot \partial \phi + m^2 \phi^2\big) \right.
\\[3pt]
& \left. \qquad \qquad \qquad \qquad  + \left( h^{\alpha \mu} h_{\alpha}^{\ \nu} - \frac{h}{2} h^{\mu \nu} \right) \partial_{\mu} \phi \, \partial_{\nu} \phi  \right],
\end{split}
\end{equation}
where indices on $h_{\mu\nu}$ and $\partial_\mu$ are raised/lowered by the Weyl-rescaled background metric $\eta^{\mu\nu}/\eta_{\mu\nu}$ and $h=\eta^{\mu\nu}h_{\mu\nu}$.
Finally, $\mathcal{B.T.}$ in~(\ref{expanding the total action}) denotes the boundary terms obtained by performing integrations by parts in the gravitational part of the action,
\begin{equation}
    \begin{split}
        \mathcal{B.T.} =& \int d^D\! x \, \partial^{\mu} \bigg\{ \epsilon^2\, a^{D-2} \left[  \mathcal{H} \delta_{\mu}^0 \bigg(\frac{1}{2} h_{\alpha \beta}h^{\alpha \beta} - \frac{1}{4} h^2\bigg) + \mathcal{H}\delta_{\nu}^0 \bigg(\frac{D}{2} h h_{\mu}^{\nu} - D h_{\mu \alpha} h^{\alpha \nu}\bigg) \right.  \\[2pt]
& +\left. \left. \frac{1}{2} h \partial_{\alpha} h^{\alpha}_{\mu} - \frac{1}{2} h \partial_{\mu} h + h^{\alpha \nu} \partial_{\mu}h_{\alpha \nu} + h_{\mu}^{\alpha} \partial_{\alpha}h - h^{\alpha \nu} \partial_{\alpha}h_{\mu \nu} -  h_{\mu \nu} \partial_{\alpha}h^{\alpha \nu} \right]   \right.  \\[3pt]
& +\left. \right. \epsilon \, a^{D-2} \big[  (\partial_{\alpha} h^{\alpha}_{\mu} - \partial_{\mu} h) + (D-2) \,\mathcal{H} \,\delta_{\alpha}^0 (h \delta^{\alpha}_{\mu} - h^{\alpha}_{\mu}) \\[3pt]
& +   (D-1)\,\mathcal{H}\, (2 h^{\alpha}_{\mu} \delta^0_{\alpha} + h \,\eta_{\mu \alpha} \,\delta^{\alpha}_0) \big] +  a^{D-2}\, \mathcal{H}\, 2 (D-1) \eta_{\mu \alpha} \delta^{\alpha}_0\bigg\}.
    \end{split}
\end{equation}
These contributions are neglected throughout this work, and -- as one can show -- do not influence the bulk form of the kinetic operator
for gravitational perturbations.
Here $\mathcal{H}\equiv a'/a$ denotes the conformal Hubble rate and primed variables 
are derivatives with respect to conformal time $\eta$. Functional derivatives of $S_m^{(1)}$ 
and $S_m^{(2)}$ with respect to the matter and perturbation fields give the three‑ 
and four‑point vertices of the theory:
\begin{align*}
[iV_{(3), \phi}^{\alpha \beta}](x;y;y') &:= \frac{i\,\delta^3 S_m^{(1)}}{ \delta h_{\alpha \beta}(x) \delta \phi(y) \delta \phi(y')} \\[3pt]
&\equiv [iV_{(3), \phi}^{\alpha \beta}](x|1;2) \, \delta^D(x-y|1) \delta^D(x-y'|2), \numberthis\\[6pt]
[{}^{\alpha \beta}iV_{(4), \phi}^{ \rho \sigma}](x;x';y;y') &:= \frac{i\, \delta^4 S_m^{(2)}}{ \delta h_{\alpha \beta}(x)\delta h_{\rho \sigma}(x')\delta \phi(y) \delta \phi(y') }\\[3pt]
&\equiv [{}^{\alpha \beta}iV_{(4), \phi}^{ \rho \sigma}](x;x'|1;2) \delta^D(x-y|1) \delta^D(x-y'|2) .\numberthis
\end{align*}
Here we have introduced the vertices with amputated matter legs~\footnote{We could have removed one more delta function in the four point vertex.}, i.e. with the matter delta functions removed. Numerical labels ($1,2$) are assigned to each matter‑field 
leg to keep track of which derivative acts on which external line, thereby preserving 
the correct momentum dependence of each interaction. The corresponding Feynman rules are:
\\
\begin{flalign*}
&\vcenter{\hbox{%
  \begin{fmfgraph*}(80,40)
    \fmfleft{ul,dl}
    \fmfright{r}
    \fmf{plain}{ul,d1}
    \fmf{plain}{dl,d1}
    \fmf{gluon}{d1,r}
    \fmfdot{d1}
    \fmfv{label=$x$}{d1}
    \fmfv{label=$1$}{ul}
    \fmfv{label=$2$}{dl}
    \fmfv{label=\hspace{-0.8cm}\vspace{0.8cm}$\alpha\beta$}{r}
  \end{fmfgraph*}
}} \; \propto \, [iV_{(3), \phi}^{\alpha \beta}](x|1;2) = -i \epsilon  a^{D-2} \left[ \frac{\eta^{\alpha \beta}}{2} \big( \partial_1 \!\cdot \! \partial_2 + m^2\big) - \partial^{(\alpha}_1 \partial^{\,\beta)}_2 \right]; &
\end{flalign*}
\\
\begin{flalign*}
&\vcenter{\hbox{%
  \begin{fmfgraph*}(80,40)
    \fmfleft{ur,dl}
    \fmfright{ul,dr}
    \fmf{plain}{ul,d1}
    \fmf{plain}{dl,d1}
    \fmf{gluon}{d1,ur}
    \fmf{gluon}{d1,dr}
    \fmfdot{d1}
    \fmfv{label=\vspace{0.5cm}\hspace{-0.35cm}$x$}{d1}
    \fmfv{label=$1$}{ul}
    \fmfv{label=$2$}{dl}
    \fmfv{label=$\alpha\beta$,label.side=left}{ur}
    \fmfv{label=$\rho\sigma$,label.side=right}{dr}
  \end{fmfgraph*}%
}}\; \propto \,[{}^{\alpha \beta}iV_{(4), \phi}^{ \rho \sigma}](x;x'|1;2)=-i \epsilon^2  a^{D-2} \left[ \bigg( \frac{1 }{8}\eta^{\alpha \beta} \eta^{\rho \sigma} -\frac{1}{4} \eta^{\alpha(\rho}\eta^{\sigma) \beta} \bigg) \big( \partial_1 \! \cdot \!\partial_2  + m^2\big)\right.  &\\
   &  \qquad \, \, \, \qquad \qquad \qquad \qquad \qquad  \quad \qquad \qquad\qquad \qquad \qquad + \frac12 \eta^{\alpha)(\rho} \partial_1^{\, \sigma)}\partial_2^{(\beta}+ \frac12 \eta^{\alpha)(\rho} \partial_2^{\, \sigma)}\partial_1^{(\beta} \\
   &\qquad \, \, \, \qquad \qquad \qquad \qquad \qquad \quad \qquad \qquad \qquad\qquad \qquad \left.- \frac{1}{4} \eta^{\alpha \beta} \partial_1^{(\rho} \partial_2^{\, \sigma)} - \frac{1}{4} \eta^{\rho \sigma} \partial_1^{(\alpha} \partial_2^{\,\beta)} \right] \delta^D(x-x').
\end{flalign*}
\\
These vertices encode the interactions between the ULDM field $\phi$ and the graviton $h_{\alpha\beta}$. 

The analysis presented in this work is aimed at an adiabatic treatment of the matter sector, 
which is justified during matter domination, but not necessarily during radiation era~\cite{Liu:2026}. The calculation is simplified by introducing 
the rescaled matter field $\chi = a^{(D-2)/2} \phi$. This removes the overall factor $a^{D-2}$ 
from the vertices at the cost of adding terms proportional to powers of the conformal Hubble rate $\mathcal{H}$ and its time derivative $\mathcal{H}'$. 
In terms of $\chi$, the matter action becomes,
\begin{align}
S_m^{(0)}[\chi, a] &= -\frac{1}{2} \int d^D\! x \, \mathcal{L}_0(\chi, a), \\[6pt]
S_m^{(1)}[\chi, a, h_{\mu \nu}] &= -\frac{1}{2} \int d^D\! x \, \left[ \frac{h}{2} \mathcal{L}_0(\chi, a) - h^{\mu \nu} \mathcal{L}_{\mu \nu}(\chi, a) \right], \\[6pt]
S_m^{(2)}[\chi, a, h_{\mu \nu}] &= -\frac{1}{2} \int d^D\! x \, \Bigg[ \left( \frac{h^2}{8} - \frac{1}{4} h^{\mu \nu} h_{\mu \nu} \right) \mathcal{L}_0(\chi, a) \\
&\qquad \qquad \qquad + \left( h^{\alpha \mu} h_{\alpha}^{\ \nu} - \frac{h}{2} h^{\mu \nu} \right) \mathcal{L}_{\mu \nu}(\chi, a) \Bigg],
\end{align}
where we have defined
\begin{align}
\mathcal{L}_{\mu \nu}(\chi, a) &= \partial_{\mu} \chi\, \partial_{\nu} \chi 
- (D-2) \mathcal{H}\, \delta^0_{\nu} \chi\, \partial_{\mu} \chi+ \delta^0_{\mu} \delta^0_{\nu} \mathcal{H}^2 \chi^2 \left( \frac{D-2}{2} \right)^2 
, \\[6pt]
\mathcal{L}_0(\chi, a) &= \eta^{\mu \nu} \mathcal{L}_{\mu \nu}(\chi, a) + m^2 a^2 \chi^2 \\
&= \partial \chi \! \cdot\! \partial \chi + (D-2) \mathcal{H} \, \chi \partial_{0} \chi -  \mathcal{H}^2 \chi^2 \left(\frac{D-2}{2} \right)^2 + m^2 a^2 \chi^2.
\end{align}
The amputated three‑ and four‑point vertices $[iV_{(3),\chi}^{\alpha \beta}]$ and 
$[{}^{\alpha \beta}iV_{(4),\chi}^{ \rho \sigma}]$ consequently differ from those obtained for $\phi$. 
Their explicit expressions are longer and are presented in Appendix \ref{appendixA}. 
Since we will be using only the field $\chi$ from now on, we omit the corresponding suffix. This concludes the classical setup of our framework.

Quantizing the matter field ($\chi \to \hat{\chi}$) while treating the graviton classically, 
we construct the semiclassical two-particle-irreducible (2PI) effective action. 
This procedure yields a coupled system of equations governing the condensate $\bar{\chi}$, 
the matter propagator $i\Delta$, the graviton field $h_{\mu\nu}$, and the background scale factor $a(\eta)$. 
In this approach, full quantum gravity effects are neglected; we assume the backreaction of graviton loops to be suppressed during matter domination.~\footnote{This can be argued as follows. Parametrically, cold dark matter contributes to the one-loop self-energy as $\sim \kappa^2 T^4\sim \Omega_{DM} H^2$, while gravitational waves and scalar gravitational perturbations contribute as $\sim \Omega_{GW} H^2$ and $\Omega_{\psi} H^2$, respectively.
Since $\Omega_{GW}\ll \Omega_{\psi}\sim \Omega_\kappa \leq 10^{-3}$ and $\Omega_{DM}\sim 10^{-1}$
($\Omega_\kappa$ here refers to the relative energy density in spatial curvature),
we infer that the contribution from gravitational loops in matter era is suppressed when compared to that of dark matter.}

\subsection{Semiclassical 2PI effective action}

We employ the in-in (Schwinger-Keldysh) formalism, first introduced in 
\cite{schwinger1961brownian, Keldysh:1964ud}, to maintain causal equations of motion~\cite{Berges:2000ur,Prokopec:2002uw,Berges_2004}. 
This introduces two copies of each field, labeled by a Keldysh index $c \in \{+,-\}$, 
corresponding to forward ($+$) and backward ($-$) time evolution. The gravitational dynamical 
variables are the scale factor $a$ and the graviton field $h_{\mu\nu,c}$. The matter sector 2PI variables are the condensate \(\bar{\chi}_c\) and the Keldysh propagator \(i\Delta_{cd}\), whose components correspond, respectively, to the Feynman (\(++\)) and Dyson (\(--\)) propagators, and to the positive (\(-+\)) and negative (\(+-\)) Wightman functions.  
Retaining only terms up to $\mathcal{O}(\epsilon^2)$ for linear dynamics, the semiclassical 
2PI effective action takes the form:
\\
\begin{equation} 
\begin{split} \label{semiclassicalaction}
    \Gamma[a, h_{\mu \nu,c}, \bar{\chi}_c, i\Delta_{cd} ] &=  S_{SK}^{(0)}[a,\bar{\chi}_c] + \epsilon \, S^{(1)}_{SK}[a, h_{\mu \nu,c},\bar{\chi}_c] + \epsilon^2 \, S^{(2)}_{SK}[a, h_{\mu \nu,c}, \bar{\chi}_c]\\[6pt]
    &  - \frac{i}{2}  \Tr \left[ \log(i\Delta)\right] + \frac{1}{2} \Tr \left[ \mathcal{D}\cdot i\Delta\right] \, + \mathcal{O}(\epsilon ^3).
    \end{split}
\end{equation}
We define the Schwinger–Keldysh version of the classical action evaluated on the closed time contour as:
\begin{equation}
    S_{SK}^{(i)}[a,h_{\mu\nu,c},\bar{\chi}_c] := \sum_{b = \pm} b \times S^{(i)}[a,h_{\mu\nu,b},\bar{\chi}_b], \qquad i = 0,1,2,
\end{equation}
where the sum runs over the two branches of the contour, with $b = +$ ($-$) denoting 
the forward (backward) branch. The operator $\mathcal{D}_{cd}(x;x')$, defined as the 
second functional derivative of the matter action,
\begin{equation}
    \mathcal{D}_{cd}(x;x'):= \sum_{i=0}^2\frac{\delta^2 S^{(i)}_{KG}}{\delta \bar{\chi}_c(x) \delta \bar{\chi}_d(x')},
\end{equation}
is given explicitly in equation \eqref{appendixA:matteroperator} in terms of the theory's vertices. 
In the semiclassical theory, the interaction term $\Gamma_{(2)}$ vanishes identically, 
as the matter sector contains no quantum interactions or self-interactions. From a quantum 
field theory perspective, the effective action describes a free quantum scalar field $\bar{\chi}_c$ 
coupled to a classical dynamical source $h_{\mu\nu,c}$. Nevertheless, the action \eqref{semiclassicalaction} 
captures the interaction between gravity and matter through the term $\Tr \left[ \mathcal{D} \cdot i\Delta \right]$, 
which can be expanded using the vertices of the theory, $[iV_{(3),c}^{\alpha \beta}]$ and 
$[{}^{\alpha \beta}i V_{(4),c}^{ \rho \sigma}]$, introduced in Appendix \ref{appendixA}, 
which now carry a Keldysh index:
\begin{equation} \label{interactionsemiclassical}
\begin{split}
   \frac{1}{2} &\Tr \left[ \mathcal{D}\cdot i\Delta\right] =- \frac{1}{2} \, \sum_c \, \int_x \biggl\{ c\bigg[ \partial_1\cdot \partial_2  - m^2 a^2 + \left( \frac{D-2}{2}\right)^2 \mathcal{H}^2+ \frac{D-2}{2}\mathcal{H} \, (\partial_0^1 + \partial_0^2 )\bigg] \\[6pt]
    &
 + i \, \epsilon  \, h_{\alpha \beta,c}(x) \,[iV_{(3),c}^{\alpha \beta}](x|1;2) \, +  i\, \epsilon^2 \int_{x'} \,  h_{\alpha \beta,c}(x) h_{\rho \sigma,c}(x') \, [{}^{\alpha \beta}i V_{(4),c}^{ \rho \sigma}](x;x'|1;2)\biggr\} \, i\Delta_{cc} (x;x|1;2).
\end{split}
\end{equation}
The second line of eq. \eqref{interactionsemiclassical} encodes the matter–gravity interaction and can be represented diagrammatically through the semiclassical vertices of the theory: 
\vspace{-0.5cm}
\begin{flalign*}
  \hspace{1.5cm} 
&\vcenter{\hbox{%
  \begin{fmfgraph*}(70,70)
  \hspace{3cm} 
    \fmfleft{l1,l2,l3}
    \fmfright{r1,r2,r3}
    \fmfbottom{b1,b2,b3,b4,b5}
    \fmf{phantom}{l2,d1}
    \fmf{phantom}{d1,r2}
    \fmfv{decor.shape=circle,decor.fill=shaded,decor.size=10,label=\vspace{1cm}$h^{\alpha \beta}$,label.dist=3mm}{b3}
    \fmfv{label=\vspace{0.5cm}\hspace{-0.5cm}$x,,c$}{d1}
    \fmfdot{d1}
    \fmf{plain, label=$1$}{d1,l2}
    \fmf{plain, label=$2$, label.side=left}{d1,r2}
    \fmffreeze
    \fmf{gluon}{b3,d1}
\end{fmfgraph*}%
}} \propto \epsilon\, h_{\alpha\beta,c}(x)\,[iV_{(3),c}^{\alpha \beta}](x|1;2);&
\end{flalign*} \vspace{-0.3cm}
\begin{flalign*}
  \hspace{1.5cm} 
&\vcenter{\hbox{%
  \begin{fmfgraph*}(70,70)
    \fmfleft{l1,l2,l3}
    \fmfright{r1,r2,r3}
    \fmfbottom{b1,b2,b3,b4,b5}
    \fmf{phantom}{l2,d1}
    \fmf{phantom}{d1,r2}
    \fmfv{label=\vspace{0.5cm}\hspace{-0.5cm}$x,,c$}{d1}
    \fmfdot{d1}
    \fmf{plain, label=$1$}{d1,l2}
    \fmf{plain, label=$2$, label.side=left}{d1,r2}
    \fmffreeze
    \fmf{gluon}{b1,d1}
    \fmf{gluon}{d1,b5}
    \fmfv{decor.shape=circle,decor.fill=shaded,decor.size=10,label=$h^{\rho \sigma}$,label.dist=3mm}{b5}
    \fmfv{decor.shape=circle,decor.fill=shaded,decor.size=10,label=$h^{\alpha \beta}$,label.dist=3mm}{b1}
\end{fmfgraph*}%
}} \propto \epsilon^2\, \int_{x'} h_{\alpha\beta,c}(x)h_{\rho\sigma,c}(x')\,[{}^{\alpha \beta}i V_{(4),c}^{ \rho \sigma}](x;x'|1;2),&
\end{flalign*}
\vspace{1.2em}
\\ 
where graviton propagator insertions are replaced by insertions of the classical graviton field. The interaction terms in eq.\eqref{interactionsemiclassical} can therefore be represented by the following Feynman diagrams:
\begin{equation*}
   \vcenter{\hbox{\parbox{40mm}{ 
        \begin{fmfgraph*}(80,80)
         \fmftop{t}
         \fmfbottom{b1,b2,b3,b4,b5}
         \fmf{phantom}{b3,c1,c2,c3,t}
         \fmffreeze
         \fmf{plain,left}{c1,t}
         \fmf{plain,right}{c1,t}
        \fmffreeze
        \fmf{gluon}{b3,c1}
         \fmfv{decor.shape=circle,decor.fill=shaded, decor.size=10, label=$h^{\alpha \beta}$, label.dist=4mm}{b3}  
         \fmfv{label=$x,, c$, label.dist= -4mm}{c1}
         \fmfdot{c1}
        \end{fmfgraph*}
    }}}
    \bigplus \quad \quad
    \vcenter{\hbox{\parbox{40mm}{ 
        \begin{fmfgraph*}(80,80)
             \fmftop{t}
         \fmfbottom{b1,b2,b3,b4,b5}
         \fmf{phantom}{b3,c1,c2,c3,t}
         \fmffreeze
         \fmf{plain,left}{c1,t}
         \fmf{plain,right}{c1,t}
        \fmffreeze
        \fmfv{decor.shape=circle,decor.fill=shaded, decor.size=10, label=$h^{\alpha \beta}$, label.dist=4mm}{b2} 
        \fmfv{decor.shape=circle,decor.fill=shaded, decor.size=10, label=$h^{\rho \sigma}$, label.dist=4mm}{b4} 
        \fmf{gluon}{b2,c1,b4}
         \fmfv{label=$x,, c$, label.dist= -4mm}{c1} 
         \fmfdot{c1}
        \end{fmfgraph*}
    }}}.
\end{equation*}
\\ \\
These diagrams are the only 2PI diagrams that can be constructed with the contracted vertices. This interpretation of the $\mathcal{O}(\epsilon)$ terms in (\ref{interactionsemiclassical}) underscores the semiclassical nature of the action and its distinction from the full quantum theory. We now proceed to compute the functional derivatives of this 2PI effective action to obtain the equations of motion for the dynamical variables.
\subsection{Equations of motion}
\subsubsection*{Matter Sector}
We start by deriving the action (\ref{semiclassicalaction}) with respect to the condensate $\bar{\chi}_c$, obtaining the following equation of motion: 
\begin{equation}
\begin{split}
   &\sum_{b}\int_{y} \mathcal{D}_{cb}(x;y) \, \bar{\chi}_b(y) \, = \,0.
    \end{split}
\end{equation}
This equation has differential operators acting at all orders in $\epsilon$. We introduce an abbreviated notation for each of these contributions:
\begin{equation} \label{differentialoperatorsmatter}
    \begin{split}
        &\mathcal{D}^{(b)}_{cb} (x;y) := \, \delta_{cb}\, c \, \, \bigg[ \partial^2 - a^2m^2 + \mathcal{H}^2 \left(\frac{D-2}{2}\right)^{\!2} + \mathcal{H}' \frac{D-2}{2} \bigg] \, \delta^D(x-y)  ;\\[3pt]
        &\mathcal{D}^{(1)}_{cb} (x;y) := \, i \, \delta_{cb} \, \int_{w} \,  h_{\alpha \beta,c}(w) \,  [iV_{(3),c}^{\alpha \beta}](w|1;2) \, \delta^D(w-y|1) \, \delta^D(w-x|2) \, ;\\[3pt]
        &\mathcal{D}^{(2)}_{cb} (x;y) := \, i \, \delta_{cb} \, \int_{w_1,w_2} \! \! \! \! \!\! \! \!  h_{\alpha \beta,c} (w_1)\, h_{\rho \sigma,c} (w_2) \,[{}^{\alpha \beta}i V_{(4),c}^{ \rho \sigma}](w_1;w_2|1;2) \, \delta^D(w_1-y|1) \, \delta^D(w_1-x|2) \,,
    \end{split}
\end{equation}
where we have chosen the signs for later convenience. The superscript \((b)\) indicates that the corresponding operator acts at the background level.
Using these expressions, the equation for the dynamic of the matter condensate takes the following form:
\begin{equation} \label{condensateequationsemiclassical}
    \sum_b \, \int_{y} \, \bigl[ \mathcal{D}^{(b)}_{cb} (x;y) \, - \, \epsilon \, \mathcal{D}^{(1)}_{cb} (x;y) \, - \epsilon^2 \,\mathcal{D}^{(2)}_{cb} (x;y) \bigr] \, \bar{\chi}_b (y) \, = \, 0.
\end{equation}
The equation that determines the dynamic of the matter Keldysh propagator is obtained by deriving the action (\ref{semiclassicalaction}) with respect to $i\Delta_{cd}(x;y)$. Using the differential operators introduced in (\ref{differentialoperatorsmatter}), it takes the following form:
\begin{equation} \label{propagatorequationsemiclassical}
    \sum_b \, \int_{y} \, \bigl[ \mathcal{D}^{(b)}_{cb} (x;y) \, - \, \epsilon \, \mathcal{D}^{(1)}_{cb} (x;y) \, - \epsilon^2 \,\mathcal{D}^{(2)}_{cb} (x;y) \bigr] \, i\Delta_{bd} (y;x') \, = \, i\, \delta_{cd} \, \delta^{D}(x-x').
\end{equation}
The presence of the delta function $\delta_{cd}$ ensures that we have a propagator equation for the time-ordered and anti-time-ordered components of the Keldysh propagator, respectively $i\Delta_{++}$ and $i\Delta_{--}$, while the Wightman functions $i\Delta_{+-}$  and $i\Delta_{-+}$ satisfy the relevant homogeneous equation.
\subsubsection*{Gravity Sector}
The equations for both the scale factor and the perturbations are obtained by varying the action (\ref{semiclassicalaction}) with respect to the graviton field $h_{\mu\nu,c}$. This unified treatment stems from the metric decomposition (\ref{metricexpansion}), which implies that the functional derivative with respect to the full metric $g_{\mu\nu}$ is proportional to the derivative with respect to the perturbation field:
\begin{equation}
   \frac{\delta \Gamma}{\delta h_{\mu \nu}} \, =\,  \frac{a^{-2}}{\epsilon} \frac{\delta \Gamma}{ \delta g_{\mu \nu}}  \, .
\end{equation}
We therefore compute the functional derivative with respect to \(h_{\mu\nu,c}\). It is convenient to express the resulting equation in terms of the in-in FLRW Einstein tensor,
\begin{equation} \label{ein-tens}
\begin{split}
    \mathcal{G}^{\alpha \beta}_{(b),c}(x)  &:=\kappa^2 a^{2-D}  \frac{\delta S_{SK,g}^{(1)}}{\epsilon \delta h_{\mu \nu,c}(x)}\\[3pt]
    &=  - c\, \frac{D-2}{2} \bigg\{\delta^{\alpha}_0 \delta^{\beta}_0 \, 2(\mathcal{H}'-\mathcal{H}^2) +\, \eta^{\alpha\beta} \left[(D-3)\mathcal{H}^2 + 2\mathcal{H}' \right]\bigg\},
    \end{split}
\end{equation}
and the background energy-momentum tensor, which naturally splits into a classical contribution and a one-loop correction:
\begin{equation} \label{en-mom}
\begin{split}
   &\big< \hat{\mathcal{T}}^{\, \alpha \beta}_{(b),c}\big>[\bar{\chi}_c,\, i \Delta_{cc}] =\mathcal{T}^{\, \alpha \beta}_{\mathrm{cl},c} \, [\bar{\chi}_c] + \big< \hat{\mathcal{T}}^{\,\alpha \beta}_{c}\big>_{\mathcal{C}}\, [i\Delta_{cc}], \\[5pt]
   \end{split}
   \end{equation}
   where the suffix \(\mathcal{C}\) denotes the connected part of the expectation value. The classical part, arising from the tree-level matter action, is given by
   \begin{equation}
   \begin{split}
     \mathcal{T}^{\, \alpha \beta}_{\mathrm{cl},c} \, [\bar{\chi}_c](x)&:= 2a^{2-D}  \frac{\delta S_{SK,m}^{(1)}}{\epsilon \delta h_{\alpha \beta,c}(x)} \\[6pt]
     &= a^{2-D} \, c\, \bigg\{ \delta^{\alpha}_{0} \delta^{\beta}_{0}  \mathcal{H}^2 \left(\frac{D-2}{2}\right)^{\!2} \bar{\chi}_{c}^2 - (D-2)\mathcal{H} \delta^{(\alpha}_{0} \, \bar{\chi}_c\, \partial^{\,\beta)}\bar{\chi}_c + \partial^{ \alpha} \bar{\chi}_c \partial^{\beta} \bar{\chi}_c \\[3pt]
   & \, \, \, \,  \, \, \, \, - \frac{1}{2}\eta^{\alpha \beta} \bigg[ m^2 a^2\bar{\chi}_c^2 -  \mathcal{H}^2 \bigg(\frac{D-2}{2}\bigg)^{\!2}\bar{\chi}_c^2  - \big(\partial_0\bar{\chi}_c\big)^2 + (D-2)\mathcal{H}  \bar{\chi}_c\partial_0 \bar{\chi}_c \bigg]\bigg\},\\[2pt],
   \end{split}
   \end{equation}
while the one-loop contribution, originating from the trace term in the effective action, takes the form \vspace{0.2cm}
   \begin{equation}
   \begin{split}
   \big< \hat{\mathcal{T}}^{\,\alpha \beta}_{c}\big>_{\mathcal{C}}\, [i\Delta_{cc}](x) &:= 2a^{2-D} \frac{\delta \Tr \left[ \mathcal{D}\cdot i\Delta\right]}{\epsilon \delta h_{\alpha \beta,c}(x)}\bigg|_{h=0}  \\[6pt]
   &= a^{2-D} \, c\, \left\{ \partial^{(\alpha} \partial'^{\beta)} + \delta^{\alpha}_{0} \delta^{\beta}_{0}\mathcal{H}^2 \left(\frac{D-2}{2}\right)^{\!2}  \right.  - \frac{D-2}{2}\mathcal{H} \delta^{(\alpha}_{0} \big(\partial^{\,\beta)}+ \partial'^{\beta)} \big) \\
& \left.  -\frac{1}{2}\eta^{\alpha \beta} \bigg[ \partial \cdot \!\partial' + m^2 a^2 - \mathcal{H}^2 \bigg(\frac{D-2}{2}\bigg)^{\!2} + \frac{D-2}{2} \mathcal{H} \big( \partial_0 + \partial_0'\big) \bigg] \right\} i\Delta_{cc}(x;x') \bigg|_{x'\to x} .
    \end{split}
\end{equation}
Employing these definitions together with the FLRW Lichnerowicz operator \([{}^{\alpha \beta}\mathcal{L}^{\,\rho \sigma}_{cd}]\) given in eq.~\eqref{appendixA:gravitonoperator}, the linearized gravitational field equation takes the form:
\\
\begin{equation} \label{gravitysectorequation}
    \begin{split}
      \frac{a^{D-2}}{\kappa^2} \mathcal{G}^{\alpha \beta}_{(b),c} &+ \, \epsilon \, \sum_d \, \int_{x'} \, [{}^{\alpha \beta}\mathcal{L}^{\, \rho \sigma}_{cd}](x;x') \, h_{\rho \sigma,d} (x') =   \frac{a^{D-2}}{2}\mathcal{T}^{\,\alpha \beta}_{\mathrm{cl},c} \, [\bar{\chi}_c] +  \frac{a^{D-2}}{2} \big< \hat{\mathcal{T}}^{\,\alpha \beta}_{c}\big>_{\mathcal{C}}\, [i\Delta_{cc}] \\[4pt]
     & + i \, \epsilon \,\int_{x'} h_{\rho \sigma,c}(x')\,  [^{\alpha \beta}i V_{(4),c}^{ \rho \sigma}](x;x'|1;2) \, \big( i\Delta_{cc}(x;x|1;2)+ \bar{\chi}_c(x|1)\bar{\chi}_c(x|2) \big).\\[3pt]
    \end{split}
\end{equation}
Equations \eqref{condensateequationsemiclassical}, \eqref{propagatorequationsemiclassical} and \eqref{gravitysectorequation} constitute the complete 2PI dynamical system of the semiclassical theory, with solutions accurate to \(\mathcal{O}(\epsilon^2)\) throughout the time domain of interest. A general analytical solution to this coupled system is not available. To make further analytic progress, we therefore introduce an approximation scheme that enforces the symmetries of the FLRW background. Implementing the adiabatic approximation within this symmetry-preserving framework yields a tractable equation for the graviton field dynamics, which becomes the central focus of the remainder of this work.
\subsection{Dynamics of perturbations}
\begin{figure}
\centering
\includegraphics[scale=0.19]{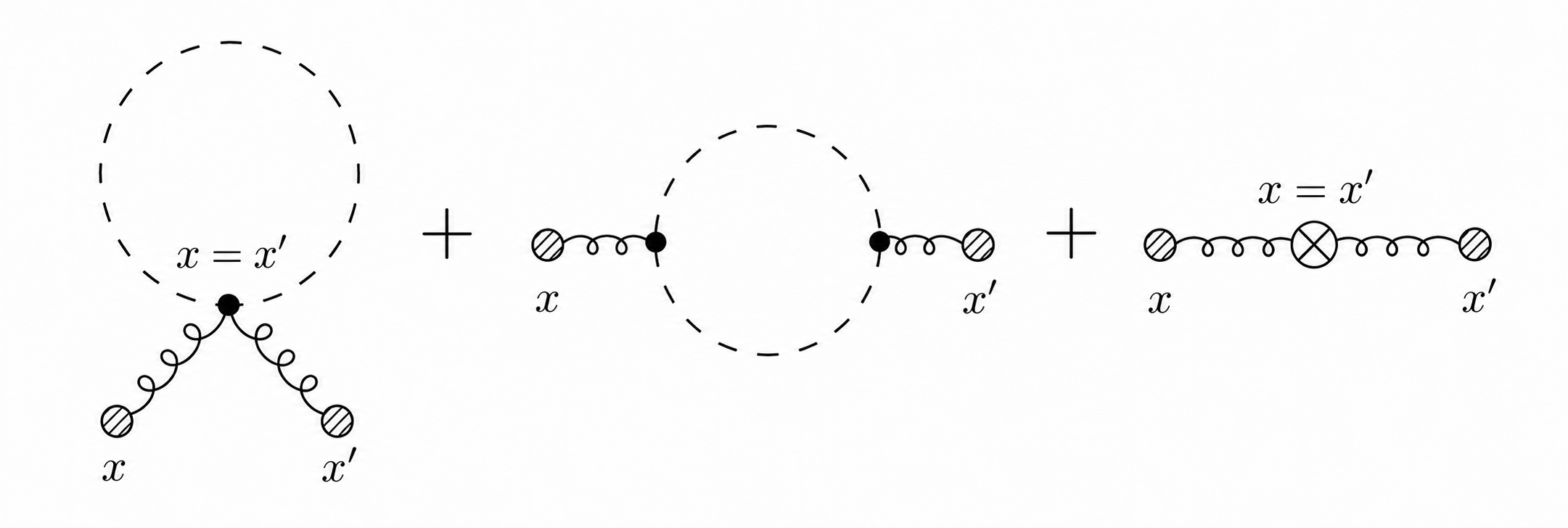}
\caption{Diagrammatical representation of the 4-point (left), 3-point (middle), and counterterm (right) contributions to the graviton self-energy $i[{}^{\alpha \beta}\Sigma^{\,\rho \sigma}]$. In this figure, one graviton leg needs to be truncated due to the functional derivative performed with respect to $h_{\alpha \beta}$. The dashed line indicates the background matter two-point function, $i\Delta^{(b)}(x;x')\, + \bar{\chi}^{(b)}(x) \bar{\chi}^{(b)}(x')$.}
\label{fig:esempio5}
\end{figure}

In this section, we leverage symmetry considerations and perturbative ordering to derive an analytically tractable equation for the graviton field $h_{\mu\nu,c}$. To preserve the symmetries of the background, a solution to equation (\ref{gravitysectorequation}) requires the Einstein tensor to be sourced exclusively by homogeneous and isotropic quantities ($h_{\mu\nu}=0$). However, the right-hand side of this equation breaks these symmetries through its dependence on $h_{\mu\nu,c}$. Furthermore, the dressed matter condensate $\bar{\chi}_c$ and propagator $i\Delta_{cd}$, being functionals of the metric perturbations, also acquire inhomogeneous components. To solve this issue, we isolate the perturbation-dependent contributions of \eqref{gravitysectorequation} from the background terms. This procedure decouples the background dynamics from the perturbations while retaining the influence of the background on their evolution. To this end, we introduce an ansatz that decomposes the matter propagator and condensate into background $(b)$ and perturbation-dependent $(h)$ parts:
\begin{equation} \label{ansatz1} 
    \begin{split}
        &\bar{\chi}_c = \bar{\chi}_c^{(b)} + \epsilon \, \bar{\chi}^{(h)}_c;\\[3pt]
        &\bar{\chi}_c^{(b)} \, : \, \bigg[ \partial_0^2 +a^2m^2 -\mathcal{H}^2 \left(\frac{D-2}{2}\right)^{\!2} - \mathcal{H}' \frac{D-2}{2} \bigg] \,\bar{\chi}_c^{(b)}(\eta)   = 0,
    \end{split}
\end{equation}
and 
\begin{equation} \label{ansatz2}
    \begin{split}
     &i\Delta_{cd} = i\Delta_{cd}^{(b)} + \epsilon \, i\Delta_{cd}^{(h)};\\[5pt]
     &i\Delta_{cd}^{(b)} \,:\, \bigg[ \partial^2 - a^2m^2 + \mathcal{H}^2 \left(\frac{D-2}{2}\right)^{\!2} + \mathcal{H}' \frac{D-2}{2} \bigg] \, i\Delta_{cd}^{(b)} (x;x')=  \, i \, c\,\delta_{cd} \, \delta^D(x-x').
    \end{split}
\end{equation} \vspace{0.2cm}
\\
Here, homogeneity and isotropy on constant-\(\eta\) hypersurfaces, as required by the FLRW spacetime, force the matter one-point function to depend only on conformal time: \(\bar{\chi}^{(b)}_c(\eta)\). 
The decomposition yields two coupled equations. The first is the semiclassical Einstein equation, which governs the expansion of the background spacetime:
\begin{equation} \label{mainsystem1}
 \mathcal{G}^{\alpha \beta}_{(b),c} =  \frac{\kappa^2}{2} \, \mathcal{T}^{\, \alpha \beta}_{\mathrm{cl},c} \, [\bar{\chi}_c^{(b)}] +  \frac{\kappa^2}{2}  \big< \hat{\mathcal{T}}^{\,\alpha \beta}_{c}\big>_{\mathcal{C}}\, [i\Delta_{cc}^{(b)}].
 \end{equation}
 \\
The second one governs the linearised graviton dynamics and reads as follows: \vspace{0.2cm}
 \begin{equation} \label{mainsystem2}
     \begin{split}
 &\, \epsilon \, \sum_d \, \int_{x'} \, [{}^{\alpha \beta}\mathcal{L}^{\, \rho \sigma}_{cd}](x;x') \, h_{\rho \sigma,d} (x') \, = \, \frac{a^{D-2}}{2}\, \delta \mathcal{T}^{\, \alpha \beta}_{\mathrm{cl},c} \, [\bar{\chi}_c] +  \frac{a^{D-2}}{2}\,  \delta\big< \hat{\mathcal{T}}^{\, \alpha \beta}_{c}\big>_{\mathcal{C}}\, [i\Delta_{cc}]\\[4pt]
 &\qquad \quad  + i \, \epsilon \,\int_{x'} h_{\rho \sigma,c}(x')\,  [^{\alpha \beta}i V_{(4),c}^{ \rho \sigma}](x;x'|1;2)  \, \big( i\Delta_{cc}(x;x'|1;2)+ \bar{\chi}_c(x|1)\bar{\chi}_c(x|2) \big) + \mathcal{O}\big(\epsilon^2 \big).\\[4pt]
     \end{split}
 \end{equation}
 \\
Here we have introduced the following notation: \vspace{0.2cm}
\begin{equation}
    \delta \mathcal{T}^{\,\alpha \beta}_{\mathrm{cl},c} \, [\bar{\chi}_c] =  \mathcal{T}^{\, \alpha \beta}_{\mathrm{cl},c} \, [\bar{\chi}_c] -  \mathcal{T}^{\,\alpha \beta}_{\mathrm{cl},c} \, [\bar{\chi}_c^{(b)}] , \quad \delta\big< \hat{\mathcal{T}}^{\, \alpha \beta}_{c}\big>_{\mathcal{C}}\, [i\Delta_{cc}] = \big< \hat{\mathcal{T}}^{\, \alpha \beta}_{c}\big>_{\mathcal{C}}\, [i\Delta_{cc}] - \big< \hat{\mathcal{T}}^{\, \alpha \beta}_{c}\big>_{\mathcal{C}}\, [i\Delta_{cc}^{(b)}].\vspace{0.2cm}
\end{equation}
The right-hand side of \eqref{mainsystem2} contains terms of order $\mathcal{O}(\epsilon^2)$. To reduce the complexity of the equation we adopt a perturbative approach, discarding these higher-order terms at the cost of losing the full non-perturbative power of the 2PI resummation. To implement this approach, we isolate the $\mathcal{O}(\epsilon)$ contributions from both the propagator $i\Delta^{(h)}_{cd}$ and condensate $\bar{\chi}_c^{(h)}$, yielding the following form for the matter condensate:
\begin{equation}
 \bar{\chi}_c(x) \, = \, \bar{\chi}_c^{(b)}(\eta) \, + \, \epsilon \, \bar{\chi}_c^{(1)}(x) \, + \mathcal{O}(\epsilon^2),
\end{equation}
and matter two-point function:
\begin{equation}
 i\Delta_{cd}(x;x') \, = \, i\Delta_{cd}^{(b)}(x;x') \, + \, \epsilon \, i\Delta_{cd}^{(1)}(x;x')\, + \mathcal{O}(\epsilon^2).
\end{equation}
We plug these expressions in the equation of motions \eqref{condensateequationsemiclassical}\eqref{propagatorequationsemiclassical}, obtaining the following: 
\begin{equation}
    \begin{split}
           & \sum_b \int_y \, \mathcal{D}^{(b)}_{cb}(x;y) \, \epsilon \, \bar{\chi}_b^{(1)}(y)=  \sum_b \int_y \, \epsilon \, \mathcal{D}^{(1)}_{cb} (x;y) \, \bar{\chi}_b^{(b)}(y) \, + \mathcal{O}(\epsilon^2); \\[7pt]
           &    \sum_b \int_y \, \mathcal{D}^{(b)}_{cb}(x;y) \, \epsilon \, i\Delta_{bd}^{(1)}(y;x' )=  \sum_b \int_y \, \epsilon \, \mathcal{D}^{(1)}_{cb} (x;y)\, i\Delta_{bd}^{(b)}(y;x')\, + \mathcal{O}(\epsilon^2).
    \end{split}
\end{equation}
Written as functionals of $h_{\mu \nu,c}$ and the background quantities, the solutions to such equations read as follow: 
\begin{equation} \label{matter1st}
\begin{split}
  & \epsilon \,  \bar{\chi}_c^{(1)}(x) =  \frac{1}{i}\sum_{b_1,b_2} \,  b_1 \,\int_{w_1, w_2}  \, i\Delta_{cb_1}^{(b)}(x;w_1)\, \epsilon \, \mathcal{D}_{b_1,b_2}^{(1)}(w_1;w_2) \,  \bar{\chi}_{b_2}^{(b)}(w_2);\\[7pt]
  & \epsilon \,  i\Delta^{(1)}_{cd} (x;x') =  \frac{1}{i}\sum_{b_1,b_2} \,  b_1 \,\int_{w_1, w_2}  \, i\Delta_{c b_1}^{(b)}(x;w_1)\, \epsilon \, \mathcal{D}_{b_1,b_2}^{(1)}(w_1;w_2) \,  i\Delta_{b_2 d}^{(b)}(w_2;x').
   \end{split}
\end{equation}
Inserting these solutions into the right-hand side of the graviton equation \eqref{mainsystem2} transforms the local term $\delta \mathcal{T}^{\, \alpha \beta}$, dependent on the fully dressed one- and two-point functions, into a non-local expression involving the background quantities $\bar{\chi}_c^{(b)}$ and $i\Delta_{cd}^{(b)}$. We identify this non-local contribution as the 3-point self-energy $i[{}^{\alpha \beta}\Sigma_{\text{3pt},cd}^{\,\rho \sigma}]$ of the graviton, since it reduces to the diagrammatic representation with two 3-vertices and two bare matter propagators, shown in figure \ref{fig:esempio5}\vspace{1mm}:
\begin{equation}
          \begin{split}
        & \frac{a^{D-2}}{2}\, \delta \mathcal{T}^{\,\alpha \beta}_{\, \text{cl},c} \, [\bar{\chi}_c] +  \frac{a^{D-2}}{2}\,  \delta\big< \hat{\mathcal{T}}^{\, \alpha \beta}_{c}\big>_{\mathcal{C}}\, [i\Delta_{cc}] \, = \, \epsilon \, \sum_d \, \int_y \, [{}^{\alpha \beta}\Sigma_{\text{3pt},cd}^{\,\rho \sigma}] (x;x')\, h_{\rho \sigma,d}(x') \, + \, \mathcal{O}(\epsilon^2),\\
        \end{split}
        \end{equation}
        where we defined \vspace{1mm}
        \begin{equation} \label{definition3ptse}
            \begin{split}
         & i\,[{}^{\alpha \beta}\Sigma_{\text{3pt},cd}^{\,\rho \sigma}](x;x')\, := \, -\frac{d}{2} \,  [i V^{\alpha \beta}_{(3),c}](x|1;2)\, [i V^{\rho \sigma}_{(3),d}](x'|3;4) \\[4pt]
        & \qquad \qquad \qquad \times \left(i\Delta^{(b)}_{cd}(x;x'|1;3) i\Delta^{(b)}_{cd}(x;x'|2;4)+ 2  i\Delta^{(b)}_{cd} (x;x'|1;3) \bar{\chi}^{(b)}_c (x|2)\bar{\chi}^{(b)}_d (x'|4) \right) .
         \end{split}
\end{equation}
Similarly, the remaining contribution in the graviton equation reduces to a local term corresponding to a loop diagram where a matter propagator connects the two non-contracted legs of the 4-vertex, shown in figure \ref{fig:esempio5}. This local contribution is therefore identified as the 4-point self-energy of the graviton:
\begin{equation}
\begin{split}
&i \, \epsilon  \int_{x'}\, h_{\rho \sigma,c}(x')\,  [{}^{\alpha \beta} i V_{(4),c}^{ \rho \sigma}](x;x'|1;2)  \, \big( i\Delta_{cc}(x;x|1;2)+ \bar{\chi}_c(x|1)\bar{\chi}_c(x|2) \big) \,\\[4pt]
&\equiv \, \epsilon \, \sum_d \, \int_{x'} \, [{}^{\alpha \beta}\Sigma_{\text{4pt},cd}^{\, \rho \sigma}] (x;x')\, h_{\rho \sigma,d}(x') \, + \, \mathcal{O}(\epsilon^2),
\end{split}
 \end{equation}
 where similarly we defined
 \\
 \begin{equation} \label{definition4ptse} \vspace{0.2cm}
          i\,[{}^{\alpha \beta}\Sigma_{\text{4pt},cd}^{\, \rho \sigma}] (x;x')\, := \, - \,\delta_{cd}\,[{}^{\alpha \beta} i V_{(4),c}^{ \rho \sigma}](x;x'|1;2) \,   \big( i\Delta^{(b)}_{cc}(x;x|1;2)+ \bar{\chi}^{(b)}_c(x|1)\bar{\chi}^{(b)}_c(x|2) \big) .
\end{equation} 
Through this treatment, we find that the primitive equation of motion for the semiclassical graviton one-point function in $D$ dimensions reads,
 \vspace{0.2cm}
\begin{equation} \label{selfen2} \vspace{0.2cm}
    \sum_d \int_{x'} \,  [{}^{\alpha \beta}\mathcal{L}_{cd}^{\, \rho \sigma}](x;x') \,  h_{\rho \sigma,d}(x') \,= \,  \sum_d \int_{x'}  \,[{}^{\alpha \beta}\Sigma_{cd}^{\, \rho \sigma}] (x;x')  \,  h_{\rho \sigma,d}(x') \, + \, \mathcal{O}(\epsilon) .
\end{equation}
Here we have added the two contributions to the graviton self-energy:
\begin{equation} \label{realselfenergy2}
    i\,[{}^{\alpha \beta}\Sigma_{cd}^{\,\rho \sigma}] (x;x') \,= \, i\,[{}^{\alpha \beta}\Sigma_{\text{3pt},cd}^{\,\rho \sigma}] (x;x') \,+ \, i\,[{}^{\alpha \beta}\Sigma_{\text{4pt},cd}^{\, \rho \sigma}] (x;x'), 
\end{equation}
which is represented in its diagrammatic form in Figure \ref{fig:esempio5}.
Equations~\eqref{mainsystem1} and \eqref{selfen2} constitute a coupled set of primitive equations in generic \(D\) dimensions. In \(D = 4\), these equations contain ultraviolet divergences. To obtain finite equations in four dimensions, we must introduce counterterms that regularize the connected part of the energy-momentum tensor \(\big\langle \hat{\mathcal{T}}^{\, \alpha \beta}_{c}\big\rangle_{\mathcal{C}}\) and the self-energy \(i[{}^{\alpha \beta}\Sigma_{cd}^{\,\rho \sigma}]\). The counterterm action that achieves this regularization is:
\begin{equation}\label{countertermact}
    S_{\mathrm{ct}}[a, h_{\mu \nu,c}] = \int d^Dx \, \sqrt{-g(x)} \left( c_1 C_{\mu \nu \rho \sigma}C^{\mu \nu \rho \sigma} + c_2 R^2 + c_3 m^2 R + c_4 m^4 \right),
\end{equation}
where the presence of terms beyond the Einstein–Hilbert action reflects the non-renormalizability of gravity. Here \(C_{\mu \nu \rho \sigma}\) denotes the Weyl tensor. The counterterm energy-momentum tensor and self-energy are obtained via functional derivatives:
\begin{equation} \label{ctselfen}
 \mathcal{T}^{\, \alpha \beta}_{\mathrm{ct},c}(x) := 2a^{2-D}\frac{\delta S_{\mathrm{ct}}}{\epsilon \delta h_{\alpha \beta,c}(x)} \bigg|_{h=0}, \qquad 
 i\, 
[{}^{\alpha \beta}\Sigma_{\text{ct},cd}^{\,\rho \sigma}](x;x') := \frac{i\, \delta^2 S_{\mathrm{ct}}}{\epsilon^2 \delta h_{\alpha \beta,c}(x) \delta h_{\rho \sigma,d}(x')} \bigg|_{h=0}.
\end{equation}
We then define the corresponding regularized quantities:
\begin{equation}
    \begin{split}
     &\big\langle \hat{\mathcal{T}}^{\, \alpha \beta}_{\mathrm{reg},c}\big\rangle_{\mathcal{C}} \, := \,  \big\langle \hat{\mathcal{T}}^{\, \alpha \beta}_{c}\big\rangle_{\mathcal{C}}  + \mathcal{T}^{\, \alpha \beta}_{\mathrm{ct},c}, \\[8pt] 
    & i\, [{}^{\alpha \beta}\Sigma_{\text{reg},cd}^{\,\rho \sigma}] := i\, 
[{}^{\alpha \beta}\Sigma_{cd}^{\,\rho \sigma}]+ i\,
[{}^{\alpha \beta}\Sigma_{\text{ct},cd}^{\,\rho \sigma}].
    \end{split}
\end{equation}
In Appendix~\ref{Appendix B} we show that, within the adiabatic regime, a suitable choice of the coefficients \(c_i\) $(i=1,2,3,4)$
renders these quantities finite.~\footnote{Because the regularization concerns ultraviolet divergences, it can be carried out independently of the adiabatic approximation.} From this point on, we replace the bare energy-momentum tensor and self-energy in eqs.~\eqref{mainsystem1} and \eqref{selfen2} with their regularized counterparts, dropping the superscript ``reg", as all quantities we will henceforth deal with are understood to be regularized unless otherwise stated.

Equation~\eqref{selfen2} depends solely on background variables and is formally solvable as a functional of the scale factor \(a(\eta)\), the initial graviton spectrum and the initial state of the ULDM field. However, a more convenient form can be obtained for studies that rely on the adiabatic approximation introduced below. Employing the semiclassical FLRW equation \eqref{mainsystem1}, we eliminate terms of order $\mathcal{O}(\mathcal{H}^2, \mathcal{H}')$ from the graviton sector, recasting them in terms of the matter condensate and propagator. To implement this, we rewrite the Lichnerowicz operator \eqref{appendixA:gravitonoperator} to explicitly display the components proportional to the average pressure and energy density of the matter fluid, distinguishing between classical contributions and 1-loop corrections. The energy density reads as follows:
\begin{equation}
    \begin{split}
               & \big<\rho_c\big>  = \rho_{\text{cl},c} \,+ \, \big< \rho_c \big>_{\mathcal{C}}  :=  a^{-2} \mathcal{T}^{\,00}_{\text{cl},c} \, + \, a^{-2}\big< \mathcal{T}^{\,00}_c\big>_{\mathcal{C}}= \frac{2a^{-2}}{\kappa^2} \mathcal{G}_{(b),c}^{00}= c\frac{a^{-2}}{\kappa^2}(D-1)(D-2) \mathcal{H}^2  ,
               \end{split}
               \end{equation}
               while the expression for the pressure yields:
               \begin{equation}
                   \begin{split}
                & \big<P_c\big>  = P_{\, \text{cl},c}\,+ \, \big< P_c \big>_{\mathcal{C}}  :=  \frac{a^{-2}}{D-1} \sum_i \, \bigl( \mathcal{T}^{\,ii}_{\text{cl},c} \, + \, \big< \mathcal{T}^{\,ii}_c\big>_{\mathcal{C}} \bigr) = \frac{2a^{-2}}{\kappa^2(D-1)} \sum_i \mathcal{G}_{(b),c}^{ii}\\
               & \qquad \qquad \qquad \qquad \qquad \qquad \qquad \qquad = - c\frac{a^{-2}}{\kappa^2} (D-2) \big[(D-3) \mathcal{H}^2 + 2 \mathcal{H}' \big] .
    \end{split}
\end{equation}
In terms of such quantities, the FLRW Lichnerowicz operator reads as follows:
\begin{equation}
    \begin{split}
      &[{}^{\alpha \beta}\mathcal{L}_{cd}^{\, \rho \sigma}](x;y) \, = \,   \delta_{cd} \, \, \delta^{D}(x-y) \,  \frac{a^{D-2}}{\kappa^2} \bigg\{ c\, [{}^{\alpha \beta}\mathcal{L}_{(\eta)}^{ \, \rho \sigma}](x) \,  \\[3pt]
&+ c\, \mathcal{H}\frac{D-2}{2} \big[ \eta^{\alpha \beta} \eta^{\rho \sigma} \partial_0 + \eta^{\rho \sigma} \delta^{(\alpha}_0 \partial^{\, \beta)} + \eta^{\alpha \beta} \delta^{(\rho}_0\partial^{\, \sigma)} - \eta^{\alpha ( \rho} \eta^{\sigma) \beta} \partial_0  - \delta^{(\alpha}_0 \eta^{\beta)(\rho} \partial^{\,\sigma)} -  \delta^{(\rho}_0 \eta^{\sigma)(\alpha} \partial^{\,\beta)} \big]  \\[3pt]
&+  \bigg( \frac{1}{8}\eta^{\alpha \beta} \eta^{\rho \sigma} - \frac{1}{4} \eta^{\alpha(\rho} \eta^{\sigma) \beta}  \bigg) \big( -2 a^2\kappa^2\braket{P_c} \big)+\frac{1}{8}\big(\eta^{\alpha \beta} \delta^{\rho}_0 \delta^{\sigma}_0+ \eta^{\rho \sigma} \delta^{\alpha}_0 \delta^{\beta}_0 \big) a^2\kappa^2\braket{\rho_c -P_c}\\[3pt]
& +\bigg( \eta^{\alpha)(\rho} \delta^{\sigma)}_0 \delta^{(\beta}_0 - \frac{1}{4} \eta^{\alpha \beta} \delta^{\rho}_0 \delta^{\sigma}_0  - \frac{1}{4} \eta^{\rho \sigma} \delta^{\alpha}_0 \delta^{\beta}_0 \bigg) \kappa^2 a^2 \braket{\rho_c + P_c} \bigg\},
    \end{split}
\end{equation}
where we introduced the Minkowski space amputated Lichnerowicz operator:
\begin{equation}
   [{}^{\alpha \beta}\mathcal{L}^{ \,\rho \sigma}_{(\eta)}] = \frac{1}{2} \big(- \eta^{\alpha \beta}\eta^{\rho \sigma} \partial^2 + \eta^{\rho \sigma} \partial^{\alpha}\partial^{\beta} + \eta^{\alpha \beta} \partial^{\rho} \partial^{\sigma} + \eta^{\alpha ( \rho} \eta^{\sigma) \beta} \partial^2 - 2 \partial^{(\alpha}\eta^{\beta) (\rho} \partial^{\, \sigma} \big).
\end{equation}
After moving the terms proportional to the components of the energy-momentum tensor to the right-hand side, the gravitational sector of the equation consists of the second-order differential operator \([{}^{\alpha \beta}\mathcal{L}_{(\eta)}^{\, \rho \sigma}]\) together with a dissipative term proportional to the conformal Hubble scale \(\mathcal{H}\):
\begin{equation} \label{equationwithlocalcondensate}
\begin{split}
& a^{D-2} \,c \, \bigg\{ \,[{}^{\alpha \beta}\mathcal{L}_{(\eta)}^{\, \rho \sigma} ]\, + \, \mathcal{H}\frac{D-2}{2} \big[ \eta^{\alpha \beta} \eta^{\rho \sigma} \partial_0 + \eta^{\rho \sigma} \delta_0^{(\alpha} \partial^{\beta} + \eta^{\alpha \beta} \delta_0^{(\rho}\partial^{\, \sigma)} \\
& \qquad \qquad \qquad \qquad \quad  \qquad - \eta^{\alpha ( \rho} \eta^{\sigma) \beta} \partial_0  - \delta_0^{(\alpha} \eta^{\beta)(\rho} \partial^{\,\sigma)} -  \delta_0^{(\rho} \eta^{\sigma)(\alpha} \partial^{\,\beta)} \big]   \bigg\} h_{\rho \sigma,c}(x) \\[4pt]
&= \kappa^2  h_{\rho \sigma,c} (x) \bigg\{-a^{D-2}  \bigg( \eta^{\alpha)(\rho} \delta_0^{\sigma)} \delta_0^{(\beta} - \frac{1}{4} \eta^{\alpha \beta} \delta_0^{\rho} \delta_0^{\sigma}  - \frac{1}{4} \eta^{\rho \sigma} \delta_0^{\alpha} \delta_0^{\beta} \bigg) \\
& \qquad \qquad \qquad \qquad \qquad \qquad \times \bigg[  \mathcal{T}^{\, 00}_{\text{cl},c} \, + \, \big< \mathcal{T}^{\,00}_c\big>_{\mathcal{C}} + \frac{1}{D-1} \sum_i  \bigl( \mathcal{T}^
{\,ii}_{\text{cl},c} \, + \, \big< \mathcal{T}^{\,ii}_c\big>_{\mathcal{C}} \bigr) \bigg]  \\
& + \bigg( \frac{1}{8}\eta^{\alpha \beta} \eta^{\rho \sigma} - \frac{1}{4} \eta^{\alpha(\rho} \eta^{\sigma) \beta}  \bigg) \, \frac{2a^{D-2}}{(D-1)} \sum_i  \bigl( \mathcal{T}^{\,ii}_{\text{cl},c} \, + \, \big< \mathcal{T}^{\, ii}_c\big>_{\mathcal{C}} \bigr) \\
&- \frac{a^{D-2}}{8} \big(\eta^{\alpha \beta}  \delta_0^{\rho} \delta_0^{\sigma} + \eta^{\rho \sigma} \delta_0^{\alpha} \delta_0^{\beta} \big) \bigg[  \mathcal{T}^{\,00}_{\text{cl},c} \, + \, \big< \mathcal{T}^{\, 00}_c\big>_{\mathcal{C}} + \frac{1}{D-1} \sum_i  \bigl( \mathcal{T}^{\,ii}_{\text{cl},c} \, + \, \big< \mathcal{T}^{\,ii}_c\big>_{\mathcal{C}} \bigr) \bigg] \, \bigg\}\\
&+  \sum_d \int_{x'} \,[{}^{\alpha \beta}\Sigma_{cd}^{\, \rho \sigma}(x;x')]   h_{\rho \sigma,d} (x').
\end{split} 
\end{equation}
Here, the energy-momentum tensor is evaluated at the background quantities $\bar{\chi}^{(b)}_c$ and $i\Delta_{cd}^{(b)}$. Additionally, we have included $\kappa^2$ in the self-energy expression to conform to standard notation.
In eq.~\eqref{equationwithlocalcondensate}, the homogeneity of the condensate causes its contribution to the first three lines of the right-hand side to cancel against the condensate contribution to the four-point self-energy, which can be written as:
\begin{equation}
\begin{split}
     &\sum_d \int_{x'} \, [{}^{\alpha \beta}\Sigma_{\text{4pt},cd}^{\,\rho \sigma}](x;x') \bigg|_{i\Delta^{(b)}=0} h_{\rho \sigma,d} (x') \,\\[4pt]
     & = \, \kappa^2  h_{\rho \sigma,c} \bigg[ a^{D-2}  \bigg( \eta^{\alpha)(\rho} \delta_0^{\sigma)} \delta_0^{(\beta} - \frac{1}{4} \eta^{\alpha \beta} \delta_0^{\rho} \delta_0^{\sigma}  - \frac{1}{4} \eta^{\rho \sigma} \delta_0^{\alpha} \delta_0^{\beta} \bigg)  \biggl(  \mathcal{T}^{\,00}_{\text{cl},c}  + \frac{1}{D-1} \sum_i  \, \mathcal{T}^{\,ii}_{\text{cl},c}\biggr)  \\[3pt]
& \qquad \quad  -\bigg( \frac{1}{8}\eta^{\alpha \beta} \eta^{\rho \sigma} - \frac{1}{4} \eta^{\alpha(\rho} \eta^{\sigma) \beta}  \bigg) \, \frac{2a^{D-2}}{(D-1)} \sum_i  \, \mathcal{T}^{\,ii}_{\text{cl},c}\, \bigg].
     \end{split}
\end{equation}
In Chapter~4 we show how this cancellation, which is a consequence of the symmetries of the background spacetime,  makes the propagation of gravitational waves independent of the ULDM condensate. 

The equation for the graviton field is obtained by setting the Keldysh index \(c = +\) and identifying the fields on the two branches as \(h_{\mu\nu,+} = h_{\mu\nu,-} = h_{\mu\nu}\) and \(\bar{\chi}^{(b)}_+ = \bar{\chi}^{(b)}_- = \bar{\chi}^{(b)}\). The components of the Keldysh propagator \(i\Delta_{cd}^{(b)}\) correspond, respectively, to the Feynman (\(++\)) and Dyson (\(--\)) propagators, and to the positive (\(-+\)) and negative (\(+-\)) Wightman functions. We set \(D = 4\), as the energy-momentum tensor and self-energy are understood to have been regularized via dimensional regularization. With these identifications, the semiclassical graviton equation of motion takes the following form:
\begin{equation} \label{endpoint}
\begin{split}
& a^{2} \,\bigg\{ \, [{}^{\alpha \beta}\mathcal{L}_{(\eta)}^{\, \rho \sigma}] \, + \, \mathcal{H} \big[ \eta^{\alpha \beta} \eta^{\rho \sigma} \partial_0 + \eta^{\rho \sigma} \delta_0^{(\alpha} \partial^{\,\beta)} + \eta^{\alpha \beta} \delta_0^{(\rho}\partial^{\,\sigma)} \\
& \qquad  \qquad \qquad \quad  \qquad - \eta^{\alpha ( \rho} \eta^{\sigma) \beta} \partial_0  - \delta_0^{(\alpha} \eta^{\beta)(\rho} \partial^{\,\sigma)} -  \delta_0^{(\rho} \eta^{\sigma)(\alpha} \partial^{\,\beta)} \big]   \bigg\} h_{\rho \sigma}(x) \\[4pt]
&= \kappa^2  h_{\rho \sigma} (x) \bigg\{-a^{2}  \bigg( \eta^{\alpha)(\rho} \delta_0^{\sigma)} \delta_0^{(\beta} - \frac{1}{4} \eta^{\alpha \beta} \delta_0^{\rho} \delta_0^{\sigma}  - \frac{1}{4} \eta^{\rho \sigma} \delta_0^{\alpha} \delta_0^{\beta} \bigg) \\
& \qquad \qquad \qquad \qquad \qquad \qquad \times \biggl( \, \big< \mathcal{T}^{\,00}\big>_{\mathcal{C}} + \frac{1}{3} \sum_i \, \big< \mathcal{T}^{\,ii}\big>_{\mathcal{C}} \biggr)  \\
& + \bigg( \frac{1}{8}\eta^{\alpha \beta} \eta^{\rho \sigma} - \frac{1}{4} \eta^{\alpha(\rho} \eta^{\sigma) \beta}  \bigg) \, \frac{2a^{2}}{3} \sum_i \, \big< \mathcal{T}^{\,ii}\big>_{\mathcal{C}} \\
&- \frac{a^{2}}{8} \big(\eta^{\alpha \beta}  \delta_0^{\rho} \delta_0^{\sigma} + \eta^{\rho \sigma} \delta_0^{\alpha} \delta_0^{\beta} \big) \bigg[  \mathcal{T}^{\,00}_{\text{cl}} \, + \, \big< \mathcal{T}^{\,00}\big>_{\mathcal{C}} + \frac{1}{3} \sum_i  \bigl( \mathcal{T}^{\,ii}_{\text{cl}} \, + \, \big< \mathcal{T}^{\, ii}\big>_{\mathcal{C}} \bigr) \bigg] \, \bigg\}\\
&+   \int_{x'} \,\big\{\,  [{}^{\alpha \beta}\Sigma_{\text{4pt},++}^{\, \rho \sigma}](x;x')\big|_{\bar{\chi}^{(b)}=0}+\,  [{}^{\alpha \beta}\Sigma_{\text{3pt},R}^{\,\rho \sigma}](x;x')  + [ {}^{\alpha \beta}\Sigma_{\mathrm{ct},++}^{\, \rho \sigma}](x;x')\big\} h_{\rho \sigma} (x').
\end{split} 
\end{equation}
The counterterm action contributes to the retarded self-energy only through its \(++\) component, as it originates from functional derivatives of an action that carries a single Keldysh index and therefore has a vanishing $+-$ component. The same holds for the four‑point contribution, as can be seen from eq.~\eqref{definition4ptse}.
In this expression we have introduced the retarded component of the 3-point self-energy:
\begin{equation} \label{retardedseproper3pt}
          \begin{split}
         & i\, [{}^{\alpha \beta}\Sigma_{\text{3pt},R}^{\,\rho \sigma}] (x;x')\, :=\, i\, [{}^{\alpha \beta}\Sigma_{\text{3pt},++}^{\,\rho \sigma}] (x;x')\, + \,  i\, [{}^{\alpha \beta}\Sigma_{\text{3pt},+-}^{\, \rho \sigma}] (x;x')\\[6pt]
         &= -\frac{\kappa^2}{2} \,  [i V_{(3)}^{\alpha \beta}](x|1;2)\, [i V_{(3)}^{\rho \sigma}](x'|3;4) \, \times \, \big[\, i\Delta^{(b)}_{++}(x;x'|1;3) i\Delta^{(b)}_{++}(x;x'|2;4)\, \\[4pt]
         &  \quad - \, i\Delta^{(b)}_{+-}(x;x'|1;3) i\Delta^{(b)}_{+-}(x;x'|2;4)   + 2  i\Delta^{(b)}_{R} (x;x'|1;3) \bar{\chi}^{(b)} (x|2)\bar{\chi}^{(b)}(x'|4) \big] ,\\[3pt]
        \end{split}
        \end{equation}  
where \(i\Delta^{(b)}_R = i\Delta^{(b)}_{++} - i\Delta^{(b)}_{+-}\) denotes the retarded propagator for the dark matter field on the FLRW background. Here, \(iV_{(3)}\) and \(iV_{(4)}\) are the standard in-out vertices, in which the dependence on the Keldysh contour index and its accompanying sign have been suppressed. In what follows, we will refer the sum of the self energies appearing in eq.~\eqref{endpoint} as the full retarded self energy
\begin{equation}
  i [{}^{\alpha \beta}\Sigma_{R}^{\, \rho \sigma}](x;x') :=  i [{}^{\alpha \beta}\Sigma_{\text{4pt},++}^{\, \rho \sigma}](x;x')\big|_{\bar{\chi}^{(b)}=0}+\, i [{}^{\alpha \beta}\Sigma_{\text{3pt},R}^{\, \rho \sigma}](x;x')  +i [ {}^{\alpha \beta}\Sigma_{\mathrm{ct},++}^{\,\rho \sigma}](x;x').
\end{equation}
Equation (\ref{endpoint}) represents the conclusion of the analysis carried out in this chapter. We will start from this expression in Chapter 4 to study the equations of tensor perturbations and gravitational potentials. 

A crucial feature of the in-in formalism is that it leads to an equation involving only the
retarded components of the self-energy. As we will make explicit in the next chapter, this ensures
that the value of the graviton field \(h_{\rho\sigma}(x)\) is determined solely by its chronology within the past
light-cone of the event \(x\) and is therefore manifestly causal~\cite{Berges:2000ur,Prokopec:2002uw,Baumgart2020ManifestlyCI}.

\section{Energy-momentum tensor and self-energy in the adiabatic regime}
\subsection{Adiabatic approximation in the matter sector}
The equation of motion~(\ref{endpoint}) resists analytical solution despite the implemented simplifications. However, for a specific range of matter field modes, the adiabatic (WKB) approximation substantially simplifies the evaluation of its right-hand side \cite{Parker:2012at}. This approximation neglects terms of order \(\mathcal{O}(\mathcal{H})\) in contributions proportional to the matter condensate and propagator, and is valid when the internal time scale of the matter Lagrangian (\(1/m\)) is much shorter than the external cosmological time scale (\(1/H\)). Under this condition, the dynamics approximate those of a time-independent system with a slowly varying parameter \(a(\eta)m\). The validity of this approximation requires the condition
\begin{equation}
  \epsilon_{m} \equiv \mathcal{O}\left(\frac{\mathcal{H}}{am}\right) \ll 1,
  \quad
  \mathcal{H} := \frac{1}{a}\frac{da}{d\eta} = a H,
\end{equation}
where \(H\) is the physical Hubble rate. For applications extending to matter-radiation equality (\(z_{\mathrm{eq}}\approx3400\)), where \(H(z_{\mathrm{eq}}) \sim 10^{-28}\,\mathrm{eV}\) and \(\mathcal{H}(z_{\mathrm{eq}}) \sim 10^{-31}\,\mathrm{eV}\), the adiabatic parameter is indeed small:
\begin{equation}\label{adiabatic}
  \epsilon_{m}(z_{\mathrm{eq}})
  \sim \frac{10^{-28}\,\mathrm{eV}}{10^{-21}\text{--}10^{-24}\,\mathrm{eV}}
  \sim 10^{-7}\text{--}10^{-4},
\end{equation}
and it decreases at lower redshifts.

Independently of the mass-based adiabatic condition, the terms in
the classical action $S_m$ that are controlled by spatial
gradients require the additional sub-horizon condition \(\|\partial_i\| \gg \mathcal{H}\), so that derivatives of the scale factor in gradient terms can be neglected. The subsequent analysis of the ULDM energy-momentum tensor and graviton self-energy is performed under this assumption and is therefore restricted to sub-horizon modes. It would be of interest to study the evolution of super‑Hubble modes, for which eq.~\eqref{endpoint} still holds; we leave this to future work. In the adiabatic regime, the amputated vertices simplify dramatically:
\begin{equation} \label{adiabatic vertices}
    \begin{split}
        &[iV_{(3),c}^{\alpha\beta}](x|1;2) \approx -i c\left[ \frac{\eta^{\alpha \beta}}{2} \big( \partial_1 \cdot \partial_2  + a^2m^2\big) - \partial^{(\alpha}_1 \partial^{\beta)}_2 \right]; \\[6pt]
      &[^{\alpha \beta}iV_{(4),c}^{\,\rho\sigma}](x|1;2) \approx -i  c \left[ \bigg( \frac{1 }{8}\eta^{\alpha \beta} \eta^{\rho \sigma} -\frac{1}{4} \eta^{\alpha(\rho}\eta^{\sigma) \beta} \bigg) \big( \partial_1 \cdot \partial_2  + a^2m^2\big)\right.  \\
   & \left. \qquad \qquad \quad \qquad \qquad + \frac12 \eta^{\alpha)(\rho} \partial_1^{\sigma)}\partial_2^{(\beta} + \frac12 \eta^{\alpha)(\rho} \partial_2^{\sigma)}\partial_1^{(\beta}  - \frac{1}{4} \eta^{\alpha \beta} \partial_1^{(\rho} \partial_2^{\sigma)} - \frac{1}{4} \eta^{\rho \sigma} \partial_1^{(\alpha} \partial_2^{\beta)} \right]. 
    \end{split}
\end{equation}
These simplified vertices are the cornerstone of the analytic treatment that follows. 
\subsection{Initial conditions}
Our analysis begins at matter-radiation equality ($z_{\mathrm{eq}}\approx 3400$), where we specify initial conditions for the ULDM field. We remain agnostic to specific production mechanisms, whether inflationary misalignment \cite{PhysRevD.101.083014, Paola_Arias_2012, Hui_2021}, gravitational production~\cite{PhysRevD.57.7120, Herring_2020, PhysRevD.35.2955}, or decay products \cite{Ghoshal_2024}, and consider the most general Gaussian state parameterized by occupation number, squeezing, and condensate components.~\footnote{ It is generally believed that the states of gravitational perturbations and dark matter perturbations are -- to a very good accuracy -- in Gaussian states at decoupling, as evidenced by the CMB temperature fluctuations.} Since in most production scenarios the relevant coupling constant responsible for ULDM production is weak, the Gaussian approximation is well justified and is expected to capture the leading-order effects on the graviton field $h_{\mu\nu}$. We therefore neglect non‑Gaussianities, leaving their inclusion for future work. From now on, we will omit the superscript $(b)$, since all the quantities we will be dealing with are evaluated at the background level. The evolution of the condensate is determined by eq.~\eqref{ansatz1} upon setting $D = 4$. This equation is more conveniently solved in cosmic time $t$, which turns out to be the natural variable for describing ULDM quantities during matter domination. Rescaling the field by a factor of $\sqrt{a}$, the equation of motion reads as follows:
\begin{equation} \label{fuffa2}
\left[ \partial_t^2 +m^2 - \frac{9}{4} H^2 - \frac{3}{2} \dot{H} \right] \left( \sqrt{a} \, \bar{\chi} \right) = 0,
\end{equation}
where $H = \dot{a}/a$ is the Hubble rate and dotted variables denote derivatives with respect to cosmic time $t$. During matter domination ($a(t) \propto t^{2/3}$), the terms proportional to $H^2$ and $\dot{H}$ cancel exactly in eq.~\eqref{fuffa2}, allowing us to solve for the condensate:
\begin{equation}
    \bar{\chi} = \frac{\phi_0}{\sqrt{a}} \cos(m t - \theta_0),
\end{equation}
where \(\phi_0\) and \(\theta_0\) are fixed by the initial conditions. For the two-point functions, we employ the WKB approximation, treating mode energies as slowly varying:
\begin{equation}
    \omega(k;t) = \sqrt{m^2 + \frac{k^2}{a^2}}, \quad |\partial_0\omega| \sim \bigg|\frac{aH k^2}{\omega} \bigg| \le |aH\omega| \ll \omega^2.
\end{equation}
Here and in what follows, we denote the magnitude of a spatial vector by \(v \equiv \|\vec{v} \, \|\).

The Wightman functions for a general Gaussian initial state in the adiabatic regime are obtained by adapting the initial-state construction of ref.~\cite{Anderson_2005} to account for evolution on a slowly varying background. The corresponding mode functions are evaluated at leading adiabatic order following ref.~\cite{Parker:2012at}. The initial state for fluctuations is characterized by mixing and squeezing parameters. The mixing parameter \(n_k \in \mathbb{R}^+\) is defined as the initial occupation number (i.e. before applying the squeezing Bogoliubov transformation) and determines the purity of the state: when $n_{k}=0$ for all $k$, the state is pure. Squeezing is introduced via a Bogoliubov transformation parametrized by a real amplitude \(r_k\) and a real phase \(\phi_k\).  Introducing the annihilation and creation operators of the ULDM field at initial time, \(\hat{a}_{\vec{k}}\) and \(\hat{a}^{\dagger}_{\vec{k}}\) respectively, their expectation values are expressed in terms of these parameters as follows:
\begin{equation}
\begin{split}
&\braket{\hat{a}^{\dagger}_{\vec{k}'}\hat{a}_{\vec{k}}}= \big[(2n_{k}+1)\sinh^2(r_{k}) + n_{k}\big](2\pi)^3\delta^3(\vec{k}-\vec{k}'\big)
;\\[6pt]
&\braket{\hat{a}_{\vec{k}}\hat{a}_{\vec{k}'}}= (2n_{k}+1)\sinh(r_{k})\cosh(r_{k}) e^{-i\phi_k}(2\pi)^3\delta^3(\vec{k}+\vec{k}'\big).
    \end{split}
\end{equation}
These identities define the physical occupation number and squeezing amplitude of the Gaussian state:
\begin{equation}
    N_{k} = (2n_{k}+1)\sinh^2(r_{k}) + n_{k}, \quad 
    S_{k} = (2n_{k}+1)\sinh(r_{k})\cosh(r_{k}),
\end{equation}
which satisfy the constraint $S_{k}^2 \leq N_{k}(N_{k}+1)$, with equality for pure states. 
The Wightman function consequently decomposes into three physically distinct contributions: \vspace{3mm}
\begin{equation} \label{Wightman function}
\begin{split}
    i\Delta_{-+}(x;x') = [i\Delta_{+-}(x;x')]^*=&\frac{1}{\sqrt{aa'}}\int_{\vec{k}} \frac{e^{i\vec{k}\cdot(\vec{x}-\vec{x}')}}{\sqrt{\omega(k;t)\omega(k;t')}} \bigg[ \frac{e^{-i (\Phi(k;t)- \Phi(k;t'))}}{2} \\[3pt]
    &  + N_{k}\cos[\Phi(k;t)- \Phi(k;t')]+ S_{k}\cos[\Phi(k;t)+\Phi(k;t') - \phi_{k}] \bigg]\\[3pt]
    \equiv i\Delta_{-+}^{V}(x;x') + i&\Delta_{-+}^{N}(x;x') +i\Delta_{-+}^{S}(x;x'),
\end{split}
\end{equation}
corresponding to vacuum fluctuations, occupation number effects, and squeezing contributions, respectively. Here we introduced the adiabatic phase, 
\begin{equation} \label{adiabatic phase og}
    \Phi(k;t) = \int^t_{t_{\text{eq}}} \sqrt{m^2 + \frac{k^2}{a^2(t')}} dt',
\end{equation}
and we employ the shorthand notation \(a \equiv a(t)\) and \(a' \equiv a(t')\).
During matter domination, this phase involves a hypergeometric function, \vspace{5mm}
\begin{equation} \label{phasesqueezing}
\begin{split}
    \Phi(k;t) &= mt \times {}_2F_1\left(-\frac{1}{2};\frac{1-D}{4}; \frac{5-D}{4}; -\frac{k^2}{a^2m^2}\right) \\
    &- mt_{\text{eq}} \times {}_2F_1\left(-\frac{1}{2};\frac{1-D}{4}; \frac{5-D}{4}; -\frac{k^2}{a^2_\text{eq} m^2}\right).
\end{split}
\end{equation}
Analytical tractability of non-local expressions is severely hindered by the exact adiabatic phase \(\Phi\). In the spirit of the adiabatic approximation, we adopt the \textit{mid-point working assumption}: for non-local source terms, we approximate
\begin{equation}
\Phi(k;t) - \Phi(k;t') \approx \omega(k;\bar{t})\,(t - t'), \qquad 
\Phi(k;t) + \Phi(k;t') \approx \omega(k;\bar{t})\,(t + t'),
\end{equation}
and similarly set \(\sqrt{\omega(k;t)\omega(k;t')} \approx \omega(k;\bar{t})\) and \(\sqrt{aa'} \approx a(\bar{t}) \equiv \bar{a}\). We assume that corrections to these approximations are adiabatically suppressed, a property supported by the construction of the effective action in \cite{canevarolo2024gradientcorrectionsquantumeffective}. By contrast, for local source terms we retain the exact phase \(\Phi\) throughout the calculations. Local terms therefore do not suffer from the uncertainties associated with this approximation.

The oscillatory nature of the squeezing term in \eqref{Wightman function} suggests possible resonant effects in graviton dynamics, particularly significant since squeezed states can support large occupation numbers while maintaining quantum purity. We therefore focus special attention on implications of non-zero $S_{k}$ for graviton evolution.

From (\ref{Wightman function}) we can obtain the time-ordered two-point function of the ULDM field which in turn can be broken down into contributions deriving from different properties of the initial state:
\begin{equation}
    \begin{split}
        & i\Delta_{++}(x;x') \,=\,  [i\Delta_{--}(x;x')]^{*} \, = \, i\Delta_{++}^{V}(x;x') \, + \, i\Delta_{++}^{N}(x;x')\, + \, i\Delta_{++}^{S}(x;x').
    \end{split}
\end{equation}
The state dependent contribution $i\Delta_{++}^{N}(x;x')\, + \, i\Delta_{++}^{S}(x;x')$ satisfies an homogeneous equation of motion and therefore does not differ from the positive (or negative) Wightman function:
\begin{equation} \label{blabla}
    \begin{split}
         & i\Delta_{++}^{N}(x;x') \, = \, i\Delta_{+-}^{N}(x;x')\equiv i\Delta_{N}(x;x') ; \\[5pt]
     & i\Delta_{++}^{S}(x;x')\, = \, i\Delta_{+-}^{S}(x;x')\equiv i\Delta_{S}(x;x').
    \end{split}
\end{equation}
For the vacuum sector, the time ordering will be relevant. We write $i\Delta_{++}^{V}(x;x')$ in two different representations which turn out to be useful in different contexts,
\begin{equation} \label{propagatorexpression}
    \begin{split}
     & i\Delta_{++}^{V}(x;x') \, = \, \frac{1}{\bar{a}} \int_{k^{\mu}} \frac{-i \, \, e^{i\, k_{\mu}\Delta x^{\mu}}}{-(k^0)^2 + \omega^2(k;\bar{t}) - i \bar{\varepsilon}} \, = \, \frac{(\bar{a}m)^{D-2}}{(2\pi)^{\frac{D}{2}}} \frac{K_{\frac{D-2}{2}}(m\sqrt{\Delta x^2})}{(m \sqrt{\Delta x^2})^\frac{D-2}{2}}.
    \end{split}
\end{equation}
where $K_{n}(z)$ is the modified Bessel function of the second kind and 
\begin{equation} \label{stdistance}
\Delta x^2 = -(|t-t'|-i\varepsilon)^2 \, +\, \bar{a}^2\,  \|\vec{x}- \vec{x}\,'\|^2,
\end{equation}
is the approximated spacetime distance. Here we have introduced two small positive parameters: $\bar{\varepsilon}$, which ensures the right pole structure of the propagator and $\varepsilon$, which regulates light-cone singularities. In what follows, we denote both by $\varepsilon$; their different dimensionality and the context will make it clear which one is being referred to.
This concludes the discussion of the one- and two-point functions of the matter field. Substituting these expressions into the energy-momentum tensor and self-energy terms appearing in \eqref{endpoint} allows us to analyze the backreaction of a Gaussian ULDM initial state on linear cosmological perturbations.

\subsection{Retarded self-energy}
The 3-point graviton self-energy comprises two distinct components: a condensate-dependent term proportional to $i\Delta_{cd}(x;x')\bar{\chi}(x)\bar{\chi}(x')$, and a term involving two matter two-point functions $i\Delta_{cd}(x;x')i\Delta_{dc}(x';x)$. Substituting the decomposition of the two-point function from eq.~(\ref{Wightman function}) yields three contributions from the former and six from the latter. Owing to the property of the state-dependent components introduced in eq.~\eqref{blabla},
\begin{equation}
   i\Delta^{S/N}_{++}(x;x') - i\Delta^{S/N}_{+-}(x;x') = 0,
\end{equation}
the diagrams containing only squeezing or occupation number lines cancel out of the retarded three-point self-energy. Consequently, only four contributions remain relevant for the dynamics of the graviton one-point function:
\begin{equation}
    i\,[{}^{\alpha\beta}\Sigma_{\text{3pt},R}^{\,\rho\sigma}] \sim 
    i\Delta^{V} \times i\Delta^{V} \;+\; 
    i\Delta^{V} \times i\Delta^{N} \;+\; 
    i\Delta^{V} \times i\Delta^{S} \;+\; 
    i\Delta^{V} \times \bar{\chi}\bar{\chi}.
\end{equation}
These terms, together with the 4-point self-energy contribution, are computed in Appendix \ref{Appendix B}. In presenting our results, we focus not on the diagrammatic separation but rather on distinguishing local from non-local terms:
\begin{equation}
    i[^{\alpha \beta}\Sigma_{R}^{\, \rho \sigma}](x;x')= i[^{\alpha \beta}\Sigma_{\text{L}}^{\, \rho \sigma}](x;x') +   i[^{\alpha \beta}\Sigma_{\text{NL}}^{\, \rho \sigma}](x;x'),
\end{equation}
where the dependence on \(x'\) in the local part \(i\Sigma^{\, \text{L}}\) enters only through a delta function \(\delta^{D}(x-x')\). The local terms affect graviton dynamics pointwise and permit analytical treatment, while the non-local terms involve integrations along the past lightcone and generally require numerical methods. Although both contributions are equally important in principle, their separation enables analytical progress in studying the dynamics of $h_{\mu\nu}$. In Chapter 5, we will argue that, under certain assumptions, non-local terms can be interpreted as relativistic corrections to the local terms in the gravitational wave propagation equation. 
\subsubsection{Vacuum contribution}
The one-loop vacuum contribution to the graviton self-energy requires renormalization. To this end, we must add to it the self-energy arising from the counterterm action~\eqref{countertermact}, which is given in eq.~\eqref{countself}. Applying dimensional regularization, we obtain the following choice of coefficients \(c_i\) that renders the one-loop self-energy finite:
\begin{equation} \label{countertermscoeff}
    \begin{split}
        c_1 &= -\frac{D-2}{8(D-3)}\frac{1}{32\pi^2(D^2-1)} \frac{\mu^{D-4}}{D-4} + c_1^f,\\[4pt]
        c_2 &= -\frac{(D-2)^2(D+1)}{16(D-1)}\frac{1}{32\pi^2(D^2-1)} \frac{\mu^{D-4}}{D-4} + c_2^f,\\[4pt]
        c_3 &= 5\,\frac{1}{32\pi^2(D^2-1)} \frac{\mu^{D-4}}{D-4} + c_3^f,\\[4pt]
        c_4 &= -\frac{1}{8\pi^2 D} \frac{\mu^{D-4}}{D-4} + c_4^f,
    \end{split}
\end{equation}
where \(\mu\) is a mass scale introduced to maintain dimensional consistency throughout the dimensional regularization procedure and $c_i^f$ are finite counterterm coefficients, which ought to be  fixed by measurements ($c_3$ and $c_4$ are fixed by measuring the Newton constant and cosmological constant).
With this choice we obtain the following expression for the local contribution to the retarded vacuum self-energy (see Appendix \ref{Appendix B} for details):
\begin{equation}
    \begin{split} \label{vacuumlocalse}
         i[^{\alpha \beta}\Sigma_{V,\text{L}}^{\, \rho \sigma}](x;x') &=-\kappa^2 \bigg[\frac{P^{\alpha (\rho}P^{\sigma) \beta}}{3840 \pi^2}    \, \left( 10 m^2a^2\partial^2 - \partial^4 \right) \mathcal{F}_m   \,-\, \frac{P^{\alpha \beta}P^{\rho \sigma}}{3840 \pi^2}  \, \left( 10 m^2a^2 \partial^2 + 3 \partial^4 \right) \mathcal{F}_m    \\[4pt]
         &
+c_1^f \left( P^{\alpha (\rho}P^{\sigma) \beta} - \frac{1}{3}P^{\alpha \beta}P^{\rho \sigma} \right) \partial^4 \,+\, 2 c_2^f P^{\alpha \beta}P^{\rho \sigma} \partial^4 \,\\[4pt]
          & +\, \frac{1}{2}m^2a^2 c_3^f \left( P^{\alpha (\rho}P^{\sigma) \beta} - P^{\alpha \beta}P^{\rho \sigma} \right) \partial^2 
+ \frac{P^{\alpha (\rho}P^{\sigma)\beta}}{480\pi^2 } \bigg(3m^4a^4 -m^2a^2 \partial^2 +\frac{\partial^4}{8}\bigg)\\[4pt]
          & + \frac{P^{\alpha \beta}P^{\rho \sigma}}{480\pi^2 } \bigg(3m^4a^4 +\frac{5}{2}m^2a^2 \partial^2 + \frac{3}{8}\partial^4\bigg)\bigg] \, i \delta^D(x-x').
    \end{split}
\end{equation}
Here we introduced the transverse projector $P^{\alpha\beta} = \eta^{\alpha \beta} - \frac{\partial^{ \alpha}\partial^{ \beta}}{\partial^2}$ and the constant $\mathcal{F}_m$ obtained from the finite part of the coincident propagator,
\begin{equation} \label{Fm}
    \mathcal{F}_m = \log\left( \frac{m^2}{4\pi \mu^2}\right) + \gamma_E -1, 
\end{equation}
with $\gamma_E$ being the Euler–Mascheroni constant. Furthermore, we have set $c_4^f=-4\mathcal{F}_m/(16\pi)^2$ to cancel the non-transverse terms. As will be presented later in this chapter,  the same choice imposes a null vacuum contribution to the energy-momentum tensor. 
 The dependence on the renormalization scale $\mu$ of the self-energy reflects breaking of the scaling symmetry by the 
quantum fluctuations of the matter field, and generically plagues perturbation theory. This dependence is unphysical and can be removed by RG 
improving the effective action. For simplicity, in this work we shall absorb this dependence in the finite coefficients \(c_i^f\) by their suitable 
 $\mu$-dependent choice.
We now present the result for the retarded non-local contribution. The corresponding Keldysh components, which are useful for the analysis of graviton correlators but do not directly enter the equation for the graviton field~\eqref{endpoint}, can be found in Appendix~\ref{Appendix B}. Let us first define the non-local quantity \( \mathcal{M}^{R}_m(x;x')\) as proportional to the regular part of the difference between the squared propagator and the negative Wightman function:
\begin{equation}
\begin{split}
     \mathcal{M}_m^R (x;x') \!\!=\!\! \frac{i}{128\pi^3} 
\Bigg\{  & \frac{8\bar{m}^4}{m^2|\Delta x^2|}\left[1\!+\!\pi J_1\big(m\sqrt{|\Delta x^2|}\big)Y_1\big(m\sqrt{|\Delta x^2|}\big)\right]  \\
  &\!+\!2\bar{m}^2\partial^2\left[\log\Big(\frac{m^2|\Delta x^2|}{4}\Big)\!+\!2\big(\gamma_E\!-\!1\big)\right] 
\\             
&  
\! \!-\!\partial^4\left[\log\Big(\frac{m^2|\Delta x^2|}{4}\Big)\!-\!1\right]\Bigg\}
 \Big[\Theta\big(\!-\!\Delta x^2\big)\Theta\big(\Delta t\big)\Big]
\,,\qquad
\label{deltam}
\end{split}
\end{equation}
where the derivative operators act on everything to the right.
Here, we called $\bar{m}\equiv m a(\bar{t}) \equiv m \bar{a}$, $\Theta(t)$ is the Heaviside step function, $\Delta x^2 \equiv -(|t-t'|)^2 \, +\,\bar{a}^2\,  \|\vec{x}- \vec{x}\,'\|^2$ is the approximated spacetime distance and
$J_{\nu}(z)$  and $Y_{\nu}(z)$ are the Bessel functions of the first and second kind, respectively.
  The non-local retarded vacuum self-energy can then be expressed as:
\begin{equation} \label{nonlocalvacuumse}
    \begin{split}
         i[^{\alpha \beta}\Sigma_{V,\text{NL}}^{\, \rho \sigma}](x;x')= \kappa^2 &\bigg\{\frac{P^{\alpha (\rho}P^{\sigma) \beta}}{3840 \pi^2}  \left( \partial^2 -4\bar{m}^2\right)^2  + \frac{P^{\alpha \beta}P^{\rho \sigma}}{3840 \pi^2}\left(8 \bar{m}^4 +16\bar{m}^2 \partial^2 +3\partial^4 \right) \\[4pt]
         &+\frac{1}{384 \pi^2} \bigg(\Delta m_1^2 \frac{\partial^{ \rho} \partial^{\sigma}}{\partial^2} \eta^{\alpha \beta} \,+\,\Delta m_2^2 \frac{\partial^{ \alpha} \partial^{ \beta}}{\partial^2} \eta^{\rho \sigma}\bigg) (2\bar{m}^2+ \partial^2)\\[4pt]
         &- \frac{\eta^{\alpha \beta} \eta^{\rho \sigma}}{384 \pi^2}\big[ (\Delta m_1^2+\Delta m_2^2) (2\bar{m}^2+ \partial^2)  - 3 \Delta m_1^2\Delta m_2^2 \big]\bigg\} \, \mathcal{M}_m^R (x;x').
    \end{split}
\end{equation}
The mass difference terms, $\Delta m_1^2(t,t') \equiv \bar{m}^2-m^2a^2 $ and $\Delta m_2^2(t,t') \equiv \bar{m}^2-m^2a'^{\,2}$, originate from the discrepancy between the scale factor at the vertex times $t$ and $t'$ and the mean time $\bar{t}$ used in our propagator approximation. The presence of $\Theta(-\Delta x^2) \Theta(\Delta t)$ in (\ref{deltam}) restricts contributions to the retarded self-energy to inside the past lightcone of $x$, thereby enforcing the required causality in the equation of motion.
\subsubsection{Contributions from occupation number and squeezing}
The retarded self-energy receives contributions from both a non-zero occupation number \(N_{k}\) and a non-vanishing squeezing amplitude \(S_{k}\) of the ULDM field. Each of these contributions contains a local and a non-local component. The details of their computation are presented in Appendix~\ref{Appendix B}. The occupation number contributes to the local self-energy through the following term:
\begin{equation}\label{localsemixing}
\begin{split}
      i[^{\alpha \beta}\Sigma_{N,\text{L}}^{\, \rho \sigma}](x) \,=& \, \kappa^2 \bigg[ \eta^{\alpha \beta} \eta^{\rho\sigma} \bigg(-\frac{I_2^N}{24} +\frac{a^2m^2}{8} I_1^N \bigg)\,+\, \eta^{\alpha (\rho} \eta^{\sigma) \beta} \, \frac{I_2^N }{3}\,\\[4pt]
      &+\, \delta_0^{(\alpha} \eta^{\beta)(\rho}\delta_0^{\sigma)} \bigg( \frac{3}{2}I_2^N + a^2m^2 I_1^N \bigg) +\frac{1}{2}\big( \eta^{\alpha \beta} \delta_0^{\rho} \delta_0^{\sigma} + \eta^{\rho \sigma} \delta_0^{\alpha} \delta_0^{\beta} \big) \bigg( - \frac{I_2^N}{6} \bigg) \\[4pt]
      &+ \delta_0^{\alpha} \delta_0^{\beta} \delta_0^{\rho} \delta_0^{\sigma} \bigg( \frac{2}{3}I_2^N+\frac{a^2m^2}{2} I_1^N \bigg)\bigg]\, i\delta^D(x-x'). \raisetag{2mm}
    \end{split}
      \end{equation} 
The value of this contribution is governed by the integrals $I_i^N$, which depend functionally on the chosen form of the initial occupation number spectrum:
\begin{equation} \label{Iintegralsoccupationnumber}
    \begin{split}
        I^{N}_j  :=\frac{1}{2\pi^2 a(t)}\int_{0}^{\infty} dk\,  \frac{N_k\,\big(k^2\big)^j }{\sqrt{\big(k^2/a^2(t)\big)+m^2}}, \qquad (j=1,2).
    \end{split}
\end{equation}
This contribution exhibits only a weak time dependence, which is governed by the variation of the scale factor $a(t)$.  The squeezing contribution to the local self-energy is: 
\begin{equation}\label{localsesqueezing}
\begin{split}
     i [^{\alpha \beta}\Sigma_{S,\text{L}}^{\, \rho \sigma}](x;x') \,=& \,\kappa^2 \bigg[\eta^{\alpha \beta} \eta^{\rho\sigma} \bigg(\!-\frac{I_2^S}{24} +\frac{a^2m^2}{8} I_1^S \bigg)\,+\, \eta^{\alpha (\rho} \eta^{\sigma) \beta} \bigg(\! -\frac{I_2^S}{6}  - \frac{a^2m^2}{2}I_1^S \bigg)\,\\[4pt]
      &+\, \delta_0^{(\alpha} \eta^{\beta)(\rho}\delta_0^{\sigma)} \bigg(\! -\frac{I_2^S}{2} - a^2m^2 I_1^S \bigg)+\frac{1}{2}\big( \eta^{\alpha \beta} \delta_0^{\rho} \delta_0^{\sigma} + \eta^{\rho \sigma} \delta_0^{\alpha} \delta_0^{\beta} \big) \bigg(\! - \frac{I_2^S}{6} \bigg) \\[4pt]
      & + \delta_0^{\alpha} \delta_0^{\beta} \delta_0^{\rho} \delta_0^{\sigma} \bigg(\! -\frac{I_2^S}{3}-\frac{a^2m^2}{2} I_1^S \bigg)\bigg] \, i \delta^D(x-x'). \raisetag{2mm}
      \end{split}
      \end{equation}
      The squeezing of the different modes contributes to the local graviton self-energy through the integrals $I_i^S$,
       \begin{equation}\label{Iintegralssqueezing}
    \begin{split}
        I^{S}_j (t) :=\, \frac{1}{2\pi^2 a(t)}\int_{0}^{\infty} dk\,  \frac{S_k\,\big(k^2\big)^j }{\sqrt{\big(k^2/a^2\big)+m^2}}\times \cos\big[ 2 \, \Phi(k;t) \, - \, \phi_k \big] , \qquad (j=1,2),
    \end{split}
\end{equation}
which impart an oscillatory time dependence. The frequency of these oscillations for each mode is determined by the hypergeometric function defined in eq.~(\ref{phasesqueezing}).

The non-local self-energy arising from the occupation number and squeezing is conveniently formulated in Wigner space. Consistently with our adiabatic prescription, we apply a partial Wigner transform and disregard the time dependence of the scale factor. This transform is defined as: 
  \begin{equation}
 \,  i [ ^{\alpha \beta}\overline{\Sigma}_{N/S}^{\, \rho \sigma}](k^0,k^2,\bar{t}\,):=  \int  d^4 \!\Delta x \,  i[^{\alpha \beta}\Sigma_{N/S,\text{NL}}^{\, \rho \sigma}](x;x')  \,e^{-i k_{\mu}\Delta x^{\mu}}
.
\end{equation}
The resulting expression is characterized by a set of standard integrals that are functionals of the initial occupation number and squeezing parameters. To express these integrals compactly, we first introduce a sign variable $s_{\pm}^t \in \{-1,1\}$, with $t \in \{1,2,3,4\}$: 
\begin{equation} 
  s_-^t:=(1,1,-1,-1), \qquad s_+^t:=(1,-1,1,-1).
\end{equation}
The set of integrals governing the occupation number contribution is given by:
\begin{equation} \label{nonlocalNintegrals}
    \begin{split}
      &    \bar{J}_{rs}^t(k^0, k^2) = \frac{1}{4\bar{a}^2} \int\frac{d^3q}{(2\pi)^3} \frac{N_+}{\omega_+ \omega_-} \, q^r \left( \frac{\vec{k} \cdot \vec{q}}{k} \right)^{\!\!s} \frac{-i\, s_-^t}{\big(k^0 + i\varepsilon\big) + s_+^t \omega_+ + s_-^t \omega_-};\\
        & \bar{L}_{rs}^t(k^0, k^2) = \frac{1}{4} \int\frac{d^3q}{(2\pi)^3} N_+ \, q^r \left( \frac{\vec{k} \cdot \vec{q}}{k} \right)^{\!\!s} \frac{-i\, s_-^t}{\big(k^0 + i\varepsilon\big) + s_+^t \omega_+ + s_-^t \omega_-};\\
        & \bar{N}_{rs}^t(k^0, k^2) =\frac{1}{4\bar{a}} \int\frac{d^3q}{(2\pi)^3} \frac{N_+}{\omega_+ } \, q^r \left( \frac{\vec{k} \cdot \vec{q}}{k} \right)^{\!\!s} \frac{-i\, s_-^t}{\big(k^0 + i\varepsilon\big) + s_+^t \omega_+ + s_-^t \omega_-};\\
        & \bar{M}_{rs}^t(k^0, k^2) = \frac{1}{4\bar{a}} \int\frac{d^3q}{(2\pi)^3} \frac{N_+}{\omega_-} \, q^r \left( \frac{\vec{k} \cdot \vec{q}}{k} \right)^{\!\!s} \frac{-i\, s_-^t}{\big(k^0 + i\varepsilon\big) + s_+^t \omega_+ + s_-^t \omega_-},
    \end{split}
\end{equation}
while the corresponding set for the squeezing contribution is:
\begin{equation}  \label{nonlocalSintegrals}
    \begin{split}
       & \widetilde{J}_{rs}^t(k^0, k^2,\bar{t}\,) = \frac{1}{4\bar{a}^2} \int\frac{d^3q}{(2\pi)^3} \frac{S_+}{\omega_+ \omega_-} \, q^r \left( \frac{\vec{k} \cdot \vec{q}}{k} \right)^{\!\!s} \frac{-i\, s_-^t}{\big(k^0 + i\varepsilon\big)  + s_-^t \omega_-} \times \exp \big[is_+^t \big( 2 \omega_+\bar{t} - \phi_+\big) \big];\\
        &  \widetilde{L}_{rs}^t(k^0, k^2,\bar{t}\,) = \frac{1}{4} \int\frac{d^3q}{(2\pi)^3} S_+\, q^r \left( \frac{\vec{k} \cdot \vec{q}}{k} \right)^{\!\!s} \frac{-i\, s_-^t}{\big(k^0 + i\varepsilon\big)  + s_-^t \omega_-} \times \exp \big[is_+^t \big( 2 \omega_+\bar{t} - \phi_+\big) \big];\\
        &  \widetilde{N}_{rs}^t(k^0, k^2,\bar{t}\,) = \frac{1}{4\bar{a}} \int\frac{d^3q}{(2\pi)^3} \frac{S_+}{\omega_+ } \, q^r \left( \frac{\vec{k} \cdot \vec{q}}{k} \right)^{\!\!s} \frac{-i\, s_-^t}{\big(k^0 + i\varepsilon\big)  + s_-^t \omega_-} \times \exp \big[is_+^t \big( 2 \omega_+\bar{t} - \phi_+\big) \big];\\
        &  \widetilde{M}_{rs}^t(k^0, k^2,\bar{t}\,) = \frac{1}{4\bar{a}} \int\frac{d^3q}{(2\pi)^3} \frac{S_+}{ \omega_-} \, q^r \left( \frac{\vec{k} \cdot \vec{q}}{k} \right)^{\!\!s} \frac{-i\, s_-^t}{\big(k^0 + i\varepsilon\big)  + s_-^t \omega_-} \times \exp \big[is_+^t \big( 2 \omega_+\bar{t} - \phi_+\big) \big].
    \end{split}  
\end{equation}
In the above expressions, we adopt the following shorthand notation for momentum-dependent functions:
\begin{equation}
    f\!\left(\vec{q} - \vec{k}/{2}\right) \equiv f_- , \qquad 
    f\!\left(\vec{q} + \vec{k}/{2}\right)  \equiv f_+,
\end{equation}
where \(f\) stands for any of \(\omega\), \(N\), \(S\), or \(\phi\).
Using these definitions, both $i\overline{\Sigma}_N$ and $i \overline{\Sigma}_S$ separate into three contributions, classified by their governing integrals\vspace{-1mm}
\begin{equation} \vspace{-1mm}
     i[ ^{\alpha \beta}\overline{\Sigma}_{N/S}^{\, \rho \sigma}] =   i[^{\alpha \beta}\overline{\Sigma}_{(\text{\rom{1}}),N/S}^{\, \rho \sigma}]+i[^{\alpha \beta}\overline{\Sigma}_{(\text{\rom{2}}),N/S}^{\, \rho \sigma}]+i[^{\alpha \beta}\overline{\Sigma}_{(\text{\rom{3}}),N/S}^{\, \rho \sigma}].
\end{equation}
We now present the explicit form of these components, as derived in Appendix~\ref{Appendix B}, beginning with the contribution from the occupation number. The expressions below features tensor structures that separate spatial and temporal components, such as \(\bar{k}^{\mu} \equiv \delta_i^{\mu} k^i\) and \(\overline{\eta}^{\alpha\beta} = \eta^{\alpha\beta} + \delta_0^{\alpha} \delta_0^{\beta}\). This decomposition reflects the preferred foliation of FLRW spacetimes, which is necessary for describing non‑vacuum states. As in the vacuum contribution, these expressions contain mass difference terms proportional to
\(
\Delta m_1^2(t,t') \equiv \bar{m}^2 - m^2 a^2, \) and \(
\Delta m_2^2(t,t') \equiv \bar{m}^2 - m^2 a'^{\,2},
\)
which originate from the discrepancy between the scale factor evaluated at the vertex times \(t\) and \(t'\) and the mean time \(\bar{t}\) used in the propagator approximation. The first contribution is expressed in terms of the integrals \(\bar{J}_{rs}^t\):
\begin{equation} \label{n11}
    \begin{split}
i&[^{\alpha \beta}\overline{\Sigma}_{(\text{\rom{1}}),N}^{\, \rho \sigma}](k^0,k^2) \\[3pt]
&= \frac{\kappa^2}{4}\sum_{t=1}^4\bigg\{ \overline{\eta}^{\alpha \beta} \overline{\eta}^{\rho \sigma} \bigg[ \frac{\bar{J}^t_{40} + 2 \bar{J}^t_{22} +\bar{J}^t_{04}}{4} + \bigg(\frac{k^2}{4} + \bar{m}^2\bigg) \bar{J}^t_{20} - \bigg(\frac{3k^2}{4} + \frac{\Delta m_1^2+\Delta m_2^2}{2}-\bar{m}^2\bigg)\bar{J}^t_{02} \\[3pt]
& \qquad \qquad \qquad + \bigg(\frac{k^4}{16}+ \bar{m}^4 \bigg) \bar{J}^t_{00} + \frac{\Delta m_1^2+\Delta m_2^2}{2}\bigg(\frac{\kappa^2}{4} - \bar{m}^2\bigg) \bar{J}^t_{00} + \frac{\Delta m_1^2\Delta m_2^2 }{2}\bar{J}^t_{00}\bigg] \\[3pt]
&+ \overline{\eta}^{\alpha(\rho} \overline{\eta}^{\sigma) \beta} \bigg[ \frac{\bar{J}^t_{40}-2\bar{J}^t_{22} +\bar{J}^t_{04}}{2} \bigg] \\[3pt]
& + \delta_0^{(\alpha} \overline{\eta}^{\beta)(\rho}\delta_0^{\sigma)} \bigg[ 2 \big( \bar{J}^t_{40} - \bar{J}^t_{22} \big) + 2 \bigg( \frac{k^2}{4}+\bar{m}^2\bigg) \big(\bar{J}^t_{20}-\bar{J}^t_{02} \big)\bigg]\\[3pt]
& + (\overline{\eta}^{\alpha \beta} \delta_0^{\rho} \delta_0^{\sigma} + \overline{\eta}^{\rho \sigma} \delta_0^{\alpha} \delta_0^{\sigma}) \bigg[\frac{\bar{J}^t_{40}-\bar{J}^t_{22}}{2}+\frac{1}{2}\bigg(\frac{3k^2}{4}+ \bar{m}^2\bigg) (\bar{J}^t_{20}-\bar{J}^t_{02}) +  \frac{\Delta m_1^2+\Delta m_2^2}{2} \bar{J}^t_{20}\\[3pt]
&\qquad \qquad \qquad  \qquad \qquad+ \bar{m}^2 \frac{k^2}{2} \bar{J}^t_{00} - \frac{\Delta m_1^2+\Delta m_2^2}{2}\bigg(\frac{\kappa^2}{4} - \bar{m}^2\bigg) \bar{J}^t_{00} - \frac{\Delta m_1^2\Delta m_2^2 }{2}\bar{J}^t_{00} \bigg] \\[3pt]
& + \delta_0^{\alpha} \delta_0^{\beta} \delta_0^{\rho} \delta_0^{\sigma} \bigg[\bar{J}^t_{40} + \bigg(2 \bar{m}^2 -\frac{\Delta m_1^2+\Delta m_2^2}{2}\bigg)\bar{J}^t_{20}-\frac{k^2}{2} \bar{J}^t_{02} + \bigg( \frac{k^4}{16} + \bar{m}^4 \bigg) \bar{J}^t_{00}\\[3pt]
& \quad \qquad \qquad + \frac{\Delta m_1^2+\Delta m_2^2}{2}\bigg(\frac{\kappa^2}{4} - \bar{m}^2\bigg) \bar{J}^t_{00} + \frac{\Delta m_1^2\Delta m_2^2 }{2} \bar{J}^t_{00} \bigg]\\[3pt]
&+ \frac{\bar{k}^{\alpha}\bar{k}^{\beta}\bar{k}^{\rho}\bar{k}^{\sigma}}{k^4}\bigg[\frac{3\bar{J}^t_{40}-30\bar{J}^t_{22}+35\bar{J}^t_{04}}{4}+ \frac{k^2}{2}(\bar{J}^t_{20}-3\bar{J}^t_{02}) + \frac{k^4}{8} \bar{J}^t_{00} \bigg]\\[3pt]
&+\frac{\bar{k}^{\alpha}\bar{k}^{\beta} \overline{\eta}^{\rho \sigma}+ \bar{k}^{\rho}\bar{k}^{\sigma} \overline{\eta}^{\alpha \beta}}{k^2} \bigg[\frac{\bar{J}^t_{40}-5\bar{J}^t_{04}}{4} -\frac{1}{2}\bigg(\frac{k^2}{4}-\bar{m}^2\bigg) \bar{J}^t_{20} + \frac{1}{2}\bigg(\frac{5k^2}{4} - 3\bar{m}^2\bigg) \bar{J}^t_{02} \\[3pt]
& \qquad \qquad \qquad \qquad \qquad - \frac{k^2}{4}\bigg(\frac{k^2}{4}- \bar{m}^2\bigg) \bar{J}^t_{00} \bigg] \\[3pt]
&+ \frac{\bar{k}^{(\alpha}\overline{\eta}^{\beta)(\rho}\bar{k}^{\sigma)}}{k^2} \big[ -\bar{J}^t_{40} + 6 \bar{J}^t_{22}-5\bar{J}^t_{04}\big]\\[3pt]
& + \frac{\bar{k}^{\alpha}\bar{k}^{\beta} \delta_0^{\rho} \delta_0^{\sigma}+ \bar{k}^{\rho}\bar{k}^{\sigma} \delta_0^{\alpha} \delta_0^{\beta}}{k^2} \bigg[\frac{-\bar{J}^t_{40}+3\bar{J}^t_{22}}{2} - \frac{1}{2} \bigg(\frac{k^2}{4}+\bar{m}^2 \bigg) \bar{J}^t_{20} - \frac{1}{2} \bigg(\frac{3k^2}{4}- 3\bar{m}^2\bigg) \bar{J}^t_{02}\\[3pt]
& \qquad \qquad \qquad \qquad \qquad + \frac{k^2}{4} \bigg( \frac{k^2}{4}-\bar{m}^2 \bigg) \bar{J}^t_{00} \bigg]\\[3pt]
&+ \frac{\delta_0^{(\alpha}\bar{k}^{\beta)}\bar{k}^{(\rho}\delta_0^{\sigma)}}{k^2} \bigg[2(-\bar{J}^t_{40} +3\bar{J}^t_{22}) +2\bigg(\frac{k^2}{4}-\bar{m}^2\bigg) \bar{J}^t_{20}- \bigg( \frac{5k^2}{2}-6\bar{m}^2\bigg) \bar{J}^t_{02}+ k^2 \bigg( \frac{k^2}{4}+\bar{m}^2\bigg)\bar{J}^t_{00}\bigg]\\[3pt]
&+ (\overline{\eta}^{\alpha \beta} \delta_0^{\rho} \delta_0^{\sigma} \Delta m_2^2  + \overline{\eta}^{\rho \sigma} \delta_0^{\alpha} \delta_0^{\sigma} \Delta m_1^2) \bigg(\frac{\bar{J}^t_{02}- \bar{J}^t_{20}}{2}\bigg) \\[3pt]
&+\frac{\bar{k}^{\alpha}\bar{k}^{\beta} \overline{\eta}^{\rho \sigma}\Delta m_2^2  + \bar{k}^{\rho}\bar{k}^{\sigma} \overline{\eta}^{\alpha \beta} \Delta m_1^2 }{k^2}\bigg(\frac{ -\bar{J}^t_{20} + 3 \bar{J}^t_{02}}{2}-\frac{k^2}{4} \bar{J}^t_{00} \bigg)\\[3pt]
&+\frac{\bar{k}^{\alpha}\bar{k}^{\beta} \delta_0^{\rho} \delta_0^{\sigma} \Delta m_2^2  + \bar{k}^{\rho}\bar{k}^{\sigma} \delta_0^{\alpha} \delta_0^{\beta}  \Delta m_1^2 }{k^2}\bigg(\frac{ \bar{J}^t_{20} - 3 \bar{J}^t_{02}}{2}+ \frac{k^2}{4} \bar{J}^t_{00} \bigg)\bigg\}. \raisetag{8mm}
    \end{split}
\end{equation}
The second contribution involves the terms with $\bar{L}_{rs}^t$, defined in \eqref{nonlocalNintegrals}:
\begin{equation} \label{n12}
    \begin{split}
       i[^{\alpha \beta}\overline{\Sigma}&_{(\text{\rom{2}}),N}^{\, \rho \sigma}](k^0,k^2)=\frac{\kappa^2}{4}\sum_{t=1}^4(s_+^t s_-^t)\bigg\{ \overline{\eta}^{\alpha \beta} \overline{\eta}^{\rho \sigma} \bigg[ \bar{L}^t_{02} - \bigg(\frac{k^2}{4} +\frac{\Delta m_1^2 + \Delta m_2^2}{2}-\bar{m}^2\bigg) \bar{L}^t_{00} \bigg] \\[1pt]
& + (\overline{\eta}^{\alpha \beta} \delta_0^{\rho} \delta_0^{\sigma} + \overline{\eta}^{\rho \sigma} \delta_0^{\alpha} \delta_0^{\sigma})\bigg[ \frac{-\bar{L}^t_{20}+\bar{L}^t_{02}}{2} +\frac{  \Delta m_1^2+\Delta m_2^2}{2} \bar{L}^t_{00}\bigg]\\[1pt]
& + \delta_0^{(\alpha} \overline{\eta}^{\beta)(\rho}\delta_0^{\sigma)} \big[2(-\bar{L}^t_{20}+\bar{L}^t_{02})\big] + \delta_0^{\alpha} \delta_0^{\beta} \delta_0^{\rho} \delta_0^{\sigma} \bigg[ - \bar{L}^t_{20} + \bigg(\frac{k^2}{4}+\frac{\Delta m_1^2 + \Delta m_2^2}{2}-\bar{m}^2\bigg) \bar{L}^t_{00}\bigg] \\[1pt]
&+\frac{\bar{k}^{\alpha}\bar{k}^{\beta} \overline{\eta}^{\rho \sigma}+ \bar{k}^{\rho}\bar{k}^{\sigma} \overline{\eta}^{\alpha \beta}}{k^2}\bigg[ \frac{\bar{L}^t_{20}-3 \bar{L}^t_{02}}{2}+ \frac{k^2}{4} \bar{L}^t_{00} \bigg] + \frac{\bar{k}^{\alpha}\bar{k}^{\beta} \delta_0^{\rho} \delta_0^{\sigma}+ \bar{k}^{\rho}\bar{k}^{\sigma} \delta_0^{\alpha} \delta_0^{\beta}}{k^2}\bigg[ \frac{\bar{L}^t_{20}-3 \bar{L}^t_{02}}{2}+ \frac{k^2}{4} \bar{L}^t_{00} \bigg]\\[1pt]
& +  \frac{\delta_0^{(\alpha}\bar{k}^{\beta)}\bar{k}^{(\rho}\delta_0^{\sigma)}}{k^2} \big[ 2(\bar{L}^t_{20}-3\bar{L}^t_{02}) + k^2 \bar{L}^t_{00} \big] + (\overline{\eta}^{\alpha \beta} \delta_0^{\rho} \delta_0^{\sigma} \Delta m_2^2  + \overline{\eta}^{\rho \sigma} \delta_0^{\alpha} \delta_0^{\sigma} \Delta m_1^2) \big[ - \bar{L}^t_{00} \big] \bigg\}.
    \end{split}
\end{equation}
The third and final contribution involves the terms with $\bar{N}_{rs}^t$ and $\bar{M}_{rs}^t$:
\begin{equation} \label{n13}
\begin{split}
   i[^{\alpha \beta}\overline{\Sigma}&_{(\text{\rom{3}}),N}^{\, \rho \sigma}](k^0,k^2)=\frac{\kappa^2}{4}\sum_{t=1}^4(- s_-^t) \bigg\{ \frac{\bar{k}^{(\alpha}\overline{\eta}^{\beta)(\rho}\delta_0^{\sigma)} + \bar{k}^{(\rho}\overline{\eta}^{\sigma)(\alpha} \delta_0^{\beta)}}{k} \big[-2 (\bar{N}^t_{21}-\bar{N}^t_{03})\big] \\
& + \frac{\overline{\eta}^{\alpha \beta} \bar{k}^{(\rho}\delta_0^{\sigma)} + \overline{\eta}^{\rho \sigma} \bar{k}^{(\alpha}\delta_0^{\beta)} }{k} \big[ -(\bar{N}^t_{21}-\bar{N}^t_{03}) + \frac{k}{2} (\bar{N}^t_{20}-\bar{N}^t_{02})+\bar{m}^2 k \bar{N}^t_{00}\big]\\
&+ \frac{\delta_0^{\alpha} \delta_0^{\beta}  \bar{k}^{(\rho}\delta_0^{\sigma)} + \delta_0^{\rho} \delta_0^{\sigma} \bar{k}^{(\alpha}\delta_0^{\beta)} }{k}\bigg[ -2\bar{N}^t_{21} + k\bar{N}^t_{02} + 2 \bigg(\frac{k^2}{4}-\bar{m}^2\bigg) \bar{N}^t_{01}+ \frac{k^3}{4}\bar{N}^t_{00} \bigg] \\
&+ \frac{\bar{k}^{\alpha}\bar{k}^{\beta}\bar{k}^{(\rho} \delta_0^{\sigma)} + \bar{k}^{\rho}\bar{k}^{\sigma} \bar{k}^{(\alpha} \delta_0^{\beta)}}{k^3} \bigg[ (3\bar{N}^t_{21}-5\bar{N}^t_{03}) + \frac{k}{2} (\bar{N}^t_{20}-3\bar{N}^t_{02})+ \frac{k^2}{2} \bar{N}^t_{01} + \frac{k^3}{4} \bar{N}^t_{00} \bigg] \bigg\}\\
& + \big( \Delta m_2^2 \,  \overline{\eta}^{\rho \sigma} \frac{\bar{k}^{(\alpha}\delta_0^{\beta)}}{k} + \Delta m_1^2 \,  \overline{\eta}^{\alpha \beta} \frac{\bar{k}^{(\rho}\delta_0^{\sigma)}}{k}\big) \bigg[-\bar{N}^t_{01} - \frac{k}{2} \bar{N}^t_{00} \bigg]\\
& + \big( \Delta m_2^2 \,  \delta_0^{\rho} \delta_0^{\sigma} \frac{\bar{k}^{(\alpha}\delta_0^{\beta)}}{k} + \Delta m_1^2 \, \delta_0^{\alpha} \delta_0^{\beta} \frac{\bar{k}^{(\rho}\delta_0^{\sigma)}}{k}\big)\bigg[\bar{N}^t_{01} + \frac{k}{2} \bar{N}^t_{00} \bigg]\\[1pt]
&\qquad \qquad \qquad   + \frac{\kappa^2}{4}\sum_{t=1}^4( s_+^t) \bigg\{ \frac{\bar{k}^{(\alpha}\overline{\eta}^{\beta)(\rho}\delta_0^{\sigma)} + \bar{k}^{(\rho}\overline{\eta}^{\sigma)(\alpha} \delta_0^{\beta)}}{k} \big[-2 (\bar{M}^t_{21}-\bar{M}^t_{03})\big] \\
& + \frac{\overline{\eta}^{\alpha \beta} \bar{k}^{(\rho}\delta_0^{\sigma)} + \overline{\eta}^{\rho \sigma} \bar{k}^{(\alpha}\delta_0^{\beta)} }{k} \big[ -(\bar{M}^t_{21}-\bar{M}^t_{03}) - \frac{k}{2} (\bar{M}^t_{20}-\bar{M}^t_{02})-\bar{m}^2 k \bar{M}^t_{00}\big]\\
&+ \frac{\delta_0^{\alpha} \delta_0^{\beta}  \bar{k}^{(\rho}\delta_0^{\sigma)} + \delta_0^{\rho} \delta_0^{\sigma} \bar{k}^{(\alpha}\delta_0^{\beta)} }{k}\bigg[ -2\bar{M}^t_{21} + k\bar{M}^t_{02} + 2 \bigg(\frac{k^2}{4}-\bar{m}^2\bigg)\bar{M}^t_{01} - \frac{k^3}{4}\bar{M}^t_{00} \bigg] \\
&+ \frac{\bar{k}^{\alpha}\bar{k}^{\beta}\bar{k}^{(\rho} \delta_0^{\sigma)} + \bar{k}^{\rho}\bar{k}^{\sigma} \bar{k}^{(\alpha} \delta_0^{\beta)}}{k^3} \bigg[ (3\bar{M}^t_{21}-5\bar{M}^t_{03}) -\frac{k}{2} (\bar{M}^t_{20}-3\bar{M}^t_{02})+ \frac{k^2}{2} \bar{M}^t_{01} - \frac{k^3}{4} \bar{M}^t_{00} \bigg]\\
& + \big( \Delta m_2^2 \,  \overline{\eta}^{\rho \sigma} \frac{\bar{k}^{(\alpha}\delta_0^{\beta)}}{k} + \Delta m_1^2 \, \overline{\eta}^{\alpha \beta} \frac{\bar{k}^{(\rho}\delta_0^{\sigma)}}{k}\big) \bigg[-\bar{M}^t_{01} + \frac{k}{2} \bar{M}^t_{00} \bigg]\\
& + \big( \Delta m_2^2 \, \delta_0^{\rho} \delta_0^{\sigma}  \frac{\bar{k}^{(\alpha}\delta_0^{\beta)}}{k} + \Delta m_1^2 \,  \delta_0^{\alpha} \delta_0^{\beta} \frac{\bar{k}^{(\rho}\delta_0^{\sigma)}}{k}\big) \bigg[\bar{M}^t_{01} - \frac{k}{2} \bar{M}^t_{00} \bigg]\bigg\}. \raisetag{8mm}
\end{split}
\end{equation}
The contribution from squeezing admits an analogous decomposition. The component involving the standard integrals $\widetilde{J}_{rs}^t$ defined in \eqref{nonlocalSintegrals} is:
\begin{equation} \label{s11}
    \begin{split}
i&[^{\alpha \beta}\overline{\Sigma}_{(\text{\rom{1}}),S}^{\, \rho \sigma}](k^0,k^2,\bar{t}) \\[0pt]
&= \frac{\kappa^2}{4}\sum_{t=1}^4\bigg\{ \overline{\eta}^{\alpha \beta} \overline{\eta}^{\rho \sigma} \bigg[ \frac{-3 \widetilde{J}^t_{40} + 2 \widetilde{J}^t_{22} +\widetilde{J}^t_{04}}{4} - \bigg(\frac{\kappa^2}{4} + \bar{m}^2\bigg) \widetilde{J}^t_{20} + \bigg(\frac{k^2}{4} -\frac{\Delta m_1^2 + \Delta m_2^2}{2} +\bar{m}^2\bigg)\widetilde{J}^t_{02} \\[3pt]
&-\frac{\bar{m}^2k^2}{2} \widetilde{J}^t_{00} +  \frac{\Delta m_1^2 + \Delta m_2^2}{2} \bigg(\frac{k^2}{4}- \bar{m}^2 \bigg) \widetilde{J}^t_{00} +  \frac{\Delta m_1^2  \Delta m_2^2}{2} \widetilde{J}^t_{00} \bigg] \\[3pt]
&+ \overline{\eta}^{\alpha(\rho} \overline{\eta}^{\sigma) \beta} \bigg[ \frac{\widetilde{J}^t_{40}-2\widetilde{J}^t_{22} +\widetilde{J}^t_{04}}{2} \bigg] \, + \,  \delta_0^{(\alpha} \overline{\eta}^{\beta)(\rho}\delta_0^{\sigma)} \big[2k(\widetilde{J}^t_{03}-\widetilde{J}^t_{21}\big) \big]\\[3pt]
& + (\overline{\eta}^{\alpha \beta} \delta_0^{\rho} \delta_0^{\sigma} + \overline{\eta}^{\rho \sigma} \delta_0^{\alpha} \delta_0^{\sigma}) \bigg[\frac{-\widetilde{J}^t_{40}-\widetilde{J}^t_{22}}{2}-\frac{1}{2}\bigg(\frac{k^2}{4}- \Delta m_1^2 - \Delta m_2^2+ 3\bar{m}^2\bigg) \widetilde{J}^t_{20}+\frac{1}{2}\bigg(\frac{5k^2}{4}-\bar{m}^2\bigg)\widetilde{J}^t_{02} \\[3pt]
& \qquad \qquad \qquad \qquad \qquad - \bigg(\frac{k^4}{16}+\bar{m}^4\bigg) \widetilde{J}^t_{00}  -  \frac{\Delta m_1^2 + \Delta m_2^2}{2} \bigg(\frac{k^2}{4}- \bar{m}^2 \bigg) \widetilde{J}^t_{00} -  \frac{\Delta m_1^2  \Delta m_2^2}{2} \widetilde{J}^t_{00} \bigg] \bigg] \\[3pt]
& + \delta_0^{\alpha} \delta_0^{\beta} \delta_0^{\rho} \delta_0^{\sigma} \bigg[- \frac{\kappa^2}{2} (\widetilde{J}^t_{20}-\widetilde{J}^t_{02})- \frac{\Delta m_1^2 + \Delta m_2^2}{2} \widetilde{J}^t_{20} -\frac{\bar{m}^2k^2}{2} \widetilde{J}^t_{00} \\[3pt]
& \qquad \qquad \qquad +  \frac{\Delta m_1^2 + \Delta m_2^2}{2} \bigg(\frac{k^2}{4}- \bar{m}^2 \bigg) \widetilde{J}^t_{00} +  \frac{\Delta m_1^2  \Delta m_2^2}{2} \widetilde{J}^t_{00} \bigg]\\[3pt]
&+ \frac{\bar{k}^{\alpha}\bar{k}^{\beta}\bar{k}^{\rho}\bar{k}^{\sigma}}{k^4}\bigg[\frac{3\widetilde{J}^t_{40}-30\widetilde{J}^t_{22}+35\widetilde{J}^t_{04}}{4}+ \frac{k^2}{2}(\widetilde{J}^t_{20}-3\widetilde{J}^t_{02}) + \frac{k^4}{8} \widetilde{J}^t_{00} \bigg]\\[3pt]
&+\frac{\bar{k}^{\alpha}\bar{k}^{\beta} \overline{\eta}^{\rho \sigma}+ \bar{k}^{\rho}\bar{k}^{\sigma}\overline{\eta}^{\alpha \beta}}{k^2} \bigg[\frac{\widetilde{J}^t_{40}-5\widetilde{J}^t_{04}}{4} -\frac{1}{2}\bigg(\frac{k^2}{4}-\bar{m}^2\bigg) \widetilde{J}^t_{20} + \frac{1}{2}\bigg(\frac{5k^2}{4} - 3\bar{m}^2\bigg) \widetilde{J}^t_{02} \\[3pt]
& \qquad \qquad \qquad \qquad \qquad - \frac{k^2}{4}\bigg(\frac{k^2}{4}- \bar{m}^2\bigg) \widetilde{J}^t_{00} \bigg] \\[3pt]
&+ \frac{\bar{k}^{(\alpha}\overline{\eta}^{\beta)(\rho}\bar{k}^{\sigma)}}{k^2} \big[ -\widetilde{J}^t_{40} + 6 \widetilde{J}^t_{22}-5\widetilde{J}^t_{04}\big]\\[3pt]
& + \frac{\bar{k}^{\alpha}\bar{k}^{\beta} \delta_0^{\rho} \delta_0^{\sigma}+ \bar{k}^{\rho}\bar{k}^{\sigma} \delta_0^{\alpha} \delta_0^{\beta}}{k^2} \bigg[\frac{-\widetilde{J}^t_{40}+3\widetilde{J}^t_{22}}{2} - \frac{1}{2} \bigg(\frac{k^2}{4}+\bar{m}^2 \bigg) \widetilde{J}^t_{20} - \frac{3}{2} \bigg(\frac{k^2}{4}- \bar{m}^2\bigg) \widetilde{J}^t_{02}\\[3pt]
& \qquad \qquad \qquad \qquad \qquad + \frac{k^2}{4} \bigg( \frac{k^2}{4}-\bar{m}^2 \bigg) \widetilde{J}^t_{00} \bigg]\\[3pt]
&+ \frac{\delta_0^{(\alpha}\bar{k}^{\beta)}\bar{k}^{(\rho}\delta_0^{\sigma)}}{k^2} \big[ 2k(3\widetilde{J}^t_{21}-3\widetilde{J}^t_{03} +2\bar{m}^2 \widetilde{J}^t_{01}\big]\\[3pt]
&+ (\overline{\eta}^{\alpha \beta} \delta_0^{\rho} \delta_0^{\sigma} \Delta m_2^2  + \overline{\eta}^{\rho \sigma} \delta_0^{\alpha} \delta_0^{\sigma} \Delta m_1^2) \bigg(\frac{\bar{J}^t_{02}- \bar{J}^t_{20}}{2}\bigg) \\[3pt]
&+\frac{\bar{k}^{\alpha}\bar{k}^{\beta} \overline{\eta}^{\rho \sigma}\Delta m_2^2  + \bar{k}^{\rho}\bar{k}^{\sigma} \overline{\eta}^{\alpha \beta} \Delta m_1^2 }{k^2}\bigg(\frac{ -\bar{J}^t_{20} + 3 \bar{J}^t_{02}}{2}-\frac{k^2}{4} \bar{J}^t_{00} \bigg)\\[3pt]
&+\frac{\bar{k}^{\alpha}\bar{k}^{\beta} \delta_0^{\rho} \delta_0^{\sigma} \Delta m_2^2  + \bar{k}^{\rho}\bar{k}^{\sigma} \delta_0^{\alpha} \delta_0^{\beta}  \Delta m_1^2 }{k^2}\bigg(\frac{ \bar{J}^t_{20} - 3 \bar{J}^t_{02}}{2}+ \frac{k^2}{4} \bar{J}^t_{00} \bigg)\bigg\}. \raisetag{8mm}
\end{split}
\end{equation}
The second involves the terms with $\widetilde{L}_{rs}^t$:
\begin{equation}
    \begin{split}
      i[^{\alpha \beta}\overline{\Sigma}&_{(\text{\rom{2}}),S}^{\, \rho \sigma}](k^0,k^2,\bar{t}) =\frac{\kappa^2}{4}\sum_{t=i}^4 (-s_-^ts_+^t)\bigg\{ \overline{\eta}^{\alpha \beta} \overline{\eta}^{\rho \sigma} \bigg[ \frac{\Delta m_2^2 - \Delta m_1^2}{2} \widetilde{L}^t_{00}  \bigg] +   \delta_0^{\alpha} \delta_0^{\beta} \delta_0^{\rho} \delta_0^{\sigma} \bigg[ - \frac{\Delta m_2^2 - \Delta m_1^2}{2} \widetilde{L}^t_{00}  \bigg]\\[1pt]
       & + (\overline{\eta}^{\alpha \beta} \delta_0^{\rho} \delta_0^{\sigma} + \overline{\eta}^{\rho \sigma} \delta_0^{\alpha} \delta_0^{\sigma}) \bigg[  \frac{\Delta m_1^2 - \Delta m_2^2}{2} \widetilde{L}^t_{00}  \bigg]+(\overline{\eta}^{\alpha \beta} \delta_0^{\rho} \delta_0^{\sigma} - \overline{\eta}^{\rho \sigma} \delta_0^{\alpha} \delta_0^{\beta})\bigg[ \frac{\widetilde{L}^t_{20}+\widetilde{L}^t_{02}}{2} - \bigg(\frac{k^2}{4}-\bar{m}^2 \bigg)\widetilde{L}^t_{00} \bigg]\\[1pt]
&+\frac{\bar{k}^{\alpha}\bar{k}^{\beta} \overline{\eta}^{\rho \sigma}- \bar{k}^{\rho}\bar{k}^{\sigma} \overline{\eta}^{\alpha \beta}}{k^2}\bigg[ \frac{\widetilde{L}^t_{20}-3 \widetilde{L}^t_{02}}{2}+ \frac{k^2}{4} \widetilde{L}^t_{00} \bigg] + \frac{\bar{k}^{\alpha}\bar{k}^{\beta} \delta_0^{\rho} \delta_0^{\sigma}-\bar{k}^{\rho}\bar{k}^{\sigma} \delta_0^{\alpha} \delta_0^{\beta}}{k^2}\bigg[ \frac{\widetilde{L}^t_{20}-3 \widetilde{L}^t_{02}}{2}+ \frac{k^2}{4} \widetilde{L}^t_{00} \bigg]\\[1pt]
&+( \Delta m_1^2 \, \overline{\eta}^{\alpha \beta} \delta_0^{\rho} \delta_0^{\sigma} - \Delta m_2^2 \, \overline{\eta}^{\rho \sigma} \delta_0^{\alpha} \delta_0^{\beta})\big[ -\widetilde{L}^t_{00}\big] \bigg\}.
    \end{split}
\end{equation}
The third and final involves the terms with $\widetilde{N}_{rs}^t$ and $\widetilde{M}_{rs}^t$:
\begin{equation}\label{s13}
\begin{split}
    i[^{\alpha \beta}\overline{\Sigma}&_{(\text{\rom{3}}),S}^{\, \rho \sigma}](k^0,k^2,\bar{t}) =\frac{\kappa^2}{4}\sum_{t=1}^4( s_-^t) \bigg\{ \frac{\bar{k}^{(\alpha}\overline{\eta}^{\beta)(\rho}\delta_0^{\sigma)} + \bar{k}^{(\rho}\overline{\eta}^{\sigma)(\alpha} \delta_0^{\beta)}}{k} \big[2 (\widetilde{N}^t_{21}-\widetilde{N}^t_{03})\big] \\
& + \frac{\overline{\eta}^{\alpha \beta} \bar{k}^{(\rho}\delta_0^{\sigma)} + \overline{\eta}^{\rho \sigma} \bar{k}^{(\alpha}\delta_0^{\beta)} }{k} \bigg[ -\widetilde{N}^t_{21}+\widetilde{N}^t_{03} + \frac{k}{2} (\widetilde{N}^t_{20}-3\widetilde{N}^t_{02})+ 2\bigg(\frac{k^2}{4}-\bar{m}^2\bigg)\widetilde{N}^t_{01}+ \frac{k^3}{4}\widetilde{N}^t_{00} \bigg]\\
&+ \frac{\delta_0^{\alpha} \delta_0^{\beta}  \bar{k}^{(\rho}\delta_0^{\sigma)} + \delta_0^{\rho} \delta_0^{\sigma} \bar{k}^{(\alpha}\delta_0^{\beta)} }{k}\bigg[ k (\widetilde{N}^t_{20}-\widetilde{N}^t_{02})+\bar{m}^2k\widetilde{N}^t_{00} \bigg] \\
&+ \frac{\bar{k}^{\alpha}\bar{k}^{\beta}\bar{k}^{(\rho} \delta_0^{\sigma)} + \bar{k}^{\rho}\bar{k}^{\sigma} \bar{k}^{(\alpha} \delta_0^{\beta)}}{k^3} \bigg[ -3\widetilde{N}^t_{21}+5\widetilde{N}^t_{03} - \frac{k}{2} (\widetilde{N}^t_{20}-3\widetilde{N}^t_{02})- \frac{k^2}{2} \widetilde{N}^t_{01} - \frac{k^3}{4} \widetilde{N}^t_{00} \bigg] \\
& + \big(  \Delta m_1^2 \, \overline{\eta}^{\alpha \beta} \frac{\bar{k}^{(\rho}\delta_0^{\sigma)}}{k}+ \Delta m_2^2 \,  \overline{\eta}^{\rho \sigma} \frac{\bar{k}^{(\alpha}\delta_0^{\beta)}}{k}  \big) \bigg[\bar{N}^t_{01} + \frac{k}{2} \bar{N}^t_{00} \bigg]\\
& + \big(\Delta m_1^2 \,  \delta_0^{\alpha} \delta_0^{\beta} \frac{\bar{k}^{(\rho}\delta_0^{\sigma)}}{k}+  \Delta m_2^2 \, \delta_0^{\rho} \delta_0^{\sigma}  \frac{\bar{k}^{(\alpha}\delta_0^{\beta)}}{k} \big) \bigg[- \bar{N}^t_{01} - \frac{k}{2} \bar{N}^t_{00} \bigg]\bigg\} \\[3pt]
&\qquad \qquad \qquad \quad + \frac{\kappa^2}{4}\sum_{t=1}^4(- s_+^t) \bigg\{ \frac{\bar{k}^{(\alpha}\overline{\eta}^{\beta)(\rho}\delta_0^{\sigma)} - \bar{k}^{(\rho}\overline{\eta}^{\sigma)(\alpha} \delta_0^{\beta)}}{k} \big[-2 (\widetilde{M}^t_{21}-\widetilde{M}^t_{03})\big] \\
& + \frac{\overline{\eta}^{\alpha \beta} \bar{k}^{(\rho}\delta_0^{\sigma)} - \overline{\eta}^{\rho \sigma} \bar{k}^{(\alpha}\delta_0^{\beta)} }{k} \bigg[ \widetilde{M}^t_{21}+\widetilde{M}^t_{03} + \frac{k}{2} (\widetilde{M}^t_{20}-3\widetilde{M}^t_{02}) -2 \bigg(\frac{k^2}{4}-\bar{m}^2\bigg)\widetilde{M}^t_{01}+ \frac{k^3}{4}\widetilde{M}^t_{00} \bigg]\\
&+ \frac{\delta_0^{\alpha} \delta_0^{\beta}  \bar{k}^{(\rho}\delta_0^{\sigma)} -\delta_0^{\rho} \delta_0^{\sigma} \bar{k}^{(\alpha}\delta_0^{\beta)} }{k}\big[ k(\widetilde{M}^t_{20}-\widetilde{M}^t_{02}) +\bar{m}^2k\widetilde{M}^t_{00}   \big] \\
&+ \frac{\bar{k}^{\alpha}\bar{k}^{\beta}\bar{k}^{(\rho} \delta_0^{\sigma)} - \bar{k}^{\rho}\bar{k}^{\sigma} \bar{k}^{(\alpha} \delta_0^{\beta)}}{k^3} \bigg[ 3\widetilde{M}^t_{21}-5\widetilde{M}^t_{03} -\frac{k}{2} (\widetilde{M}^t_{20}-3\widetilde{M}^t_{02})+ \frac{k^2}{2} \widetilde{M}^t_{01} - \frac{k^3}{4} \widetilde{M}^t_{00} \bigg] \\
& + \big(  \Delta m_1^2 \, \overline{\eta}^{\alpha \beta} \frac{\bar{k}^{(\rho}\delta_0^{\sigma)}}{k}- \Delta m_2^2 \,  \overline{\eta}^{\rho \sigma} \frac{\bar{k}^{(\alpha}\delta_0^{\beta)}}{k}  \big) \bigg[-\bar{M}^t_{01} + \frac{k}{2} \bar{M}^t_{00} \bigg]\\
& + \big(\Delta m_1^2 \,  \delta_0^{\alpha} \delta_0^{\beta} \frac{\bar{k}^{(\rho}\delta_0^{\sigma)}}{k}-  \Delta m_2^2 \, \delta_0^{\rho} \delta_0^{\sigma}  \frac{\bar{k}^{(\alpha}\delta_0^{\beta)}}{k} \big) \bigg[\bar{M}^t_{01} - \frac{k}{2} \bar{M}^t_{00} \bigg]\bigg\}.
\end{split}
\end{equation}
Summing these contributions yields the Wigner transform of the full non-local self-energy generated by the occupation number and squeezing of the initial state.
\subsubsection{Condensate contribution}
The self-energy contribution from the condensate exhibits greater complexity due to the absence of the mid-point mass approximation used for the propagator. However, this expression simplifies considerably because the condensate is independent of spatial coordinates. Notably, the condensate
contribution to the 4-point self-energy does not contribute to the equation of motion for the graviton field \eqref{endpoint}. 
Thus, the local self-energy consists entirely of local terms derived from the 3-point self-energy:
\begin{equation} \label{secondensatelocal}
    \begin{split}
        i[^{\alpha \beta}\Sigma_{\chi,\text{L}}^{\, \rho \sigma}](x,x') =&i\kappa^2\delta^D(x-x')\frac{\phi_0^2}{4a}  a^2m^2\sin^2(mt- \theta_0 )\\[2pt]
        &\times\big[  \eta^{\alpha \beta} \eta^{\rho \sigma} + 2\, \big( \eta^{\alpha \beta}  \delta_0^{\rho} \delta_0^{\sigma} + \eta^{\rho \sigma}\delta_0^{\alpha} \delta_0^{\beta}\big) +4\,\delta_0^{\alpha} \delta_0^{\beta} \delta_0^{\rho} \delta_0^{\sigma}  \big].
    \end{split}
\end{equation}
The non-local component separates into two distinct contributions dependent on the time difference and the average time, denominated, respectively, $i\Sigma_{\chi}^{\Delta t}$ and $i\Sigma_{\chi}^{\bar{t}}$. This structure originates from the decomposition of the condensate product:
\begin{equation}
\begin{split}
 \bar{\chi}(t) \times \bar{\chi}(t') &= \frac{\phi_0}{\sqrt{a}} \cos(mt-\theta_0) \times \frac{\phi_0}{\sqrt{a'}} \cos(mt'-\theta_0)\\
&= \frac{\phi_0^2}{2\sqrt{ aa'}}\big[ \cos(m \Delta t) +\cos(2m\bar{t} - 2 \theta_0) \big].
\end{split}
\end{equation}
We employ the same strategy used for the vacuum contribution, extracting spatial derivatives such that they operate outside the self-energy integral when inserted into the equation of motion (\ref{endpoint}). This allows the integral to be computed first, significantly simplifying the calculation. The expression for the term dependent on the time difference is: \vspace{1mm}
\begin{equation} \label{timediffterm}
    \begin{split}
        i[^{\alpha \beta}&\Sigma_{\chi,\Delta t }^{ \, \rho \sigma}](x;x') \, \\[3pt]
        &=\kappa^2\, \frac{m^2 \phi_0^2}{8} \bigg\{ \, \delta_0^{\alpha} \delta_0^{\beta} \delta_0^{\rho} \delta_0^{\sigma}  \big(\bar{m}^2 -\nabla^2 + m^2a^2 +2 m^2a'a\big) \, +\, \overline{\eta}^{\alpha \beta}\overline{\eta}^{\rho \sigma} \, \big(\bar{m}^2 -\nabla^2 - m^2a^2\big)\\[2pt]
        &+ \big(\overline{\eta}^{\alpha \beta} \delta_0^{\rho} \delta_0^{\sigma} + \overline{\eta}^{\rho \sigma} \delta_0^{\alpha} \delta_0^{\beta}\big) \big(\bar{m}^2 -\nabla^2 + m^2a^2\big) \,+ \, \big( \overline{\eta}^{\alpha \beta} \bar{\partial}^{(\rho}\delta_0^{\sigma)} + \overline{\eta}^{\rho \sigma} \bar{\partial}^{(\alpha}\delta_0^{\beta)}\big) \, \big(2  \partial_0 \big)  \\[2pt]
        & + \big( \delta_0^{\alpha} \delta_0^{\beta} \bar{\partial}^{(\rho}\delta_0^{\sigma)} + \delta_0^{\rho} \delta_0^{\sigma} \bar{\partial}^{(\alpha}\delta_0^{\beta)}\big) \, \big(2 \partial_0\big) \, - \, 2\,   \big(m^2a^2 \, \overline{\eta}^{\alpha \beta} \delta_0^{\rho} \delta_0^{\sigma} + m^2aa' \, \overline{\eta}^{\rho \sigma} \delta_0^{\alpha} \delta_0^{\beta}\big) \\[2pt]
        &  - \, 4 \, \delta_0^{(\alpha} \bar{\partial}^{\beta)}\bar{\partial}^{(\rho}\delta_0^{\sigma)} \, \bigg\} \, \times \, \big[  \sqrt{aa'} \cos(m \Delta t) \times i\Delta^{\! R}(x;x') \big]\\[2pt]
        & \quad+\kappa^2 \, \frac{m^2 \phi_0^2}{8}  \bigg\{ \, \delta_0^{\alpha} \delta_0^{\beta} \delta_0^{\rho} \delta_0^{\sigma}  \big(-ma - m a'\big)\partial_0  \, +\, \overline{\eta}^{\alpha \beta}\overline{\eta}^{\rho \sigma} \, \big(ma + ma'\big)\partial_0 \\[2pt]
        &+ \big(\overline{\eta}^{\alpha \beta} \delta_0^{\rho} \delta_0^{\sigma} + \overline{\eta}^{\rho \sigma} \delta_0^{\alpha} \delta_0^{\beta}\big) \big(-ma - m a'\big)\partial_0 \,+ \, \big( \overline{\eta}^{\alpha \beta} \bar{\partial}^{(\rho}\delta_0^{\sigma)} + \overline{\eta}^{\rho \sigma} \bar{\partial}^{(\alpha}\delta_0^{\beta)}\big) \,\big(2  ma\big) \\[2pt]
        & + \big( \delta_0^{\alpha} \delta_0^{\beta} \bar{\partial}^{(\rho}\delta_0^{\sigma)} + \delta_0^{\rho} \delta_0^{\sigma} \bar{\partial}^{(\alpha}\delta_0^{\beta)}\big) \big( 2 ma \big)\, + \,    \big(ma \, \overline{\eta}^{\alpha \beta} \delta_0^{\rho} \delta_0^{\sigma} + ma' \, \overline{\eta}^{\rho \sigma} \delta_0^{\alpha} \delta_0^{\beta}\big) \big(2 \, \partial_0\big) \\[2pt]
        &-2 \, \big( ma \, \overline{\eta}^{\alpha \beta} \bar{\partial}^{(\rho}\delta_0^{\sigma)} + ma' \, \overline{\eta}^{\rho \sigma} \bar{\partial}^{(\alpha}\delta_0^{\beta)}\big) \, + \, 2 \, \big(ma\,  \delta_0^{\alpha} \delta_0^{\beta} \bar{\partial}^{(\rho}\delta_0^{\sigma)} + \, ,ma'\, \delta_0^{\rho} \delta_0^{\sigma} \bar{\partial}^{(\alpha}\delta_0^{\beta)}\big)\bigg\}\\[2pt]
        & \qquad \quad \qquad \qquad \qquad  \times \, \big[ \sqrt{aa'} \sin(m \Delta t) \times i\Delta^{\! R}(x;x') \big], \raisetag{4mm}
    \end{split}
\end{equation}
where $\bar{\partial}^{\mu} \equiv \delta_i^{\mu} \partial^{i}$ denotes the spatial derivative operator. 

The contribution dependent on the average time $\bar{t}$ is given by: \vspace{2mm}
\begin{equation} \label{avrgtimeterm} 
    \begin{split}
        i[^{\alpha \beta}&\Sigma_{\chi,\bar{t} }^{\, \rho \sigma}](x;x') \,\\[3pt]
        &=\,-\kappa^2\, \frac{m^2 \phi_0^2}{8} \bigg\{ \, \delta_0^{\alpha} \delta_0^{\beta} \delta_0^{\rho} \delta_0^{\sigma}  \big(\bar{m}^2 -\nabla^2 + m^2a^2 -2 m^2a'a\big) \, +\, \overline{\eta}^{\alpha \beta}\overline{\eta}^{\rho \sigma} \, \big(\bar{m}^2 -\nabla^2 - m^2a^2\big)\\[2pt]
        &+ \big(\overline{\eta}^{\alpha \beta} \delta_0^{\rho} \delta_0^{\sigma} + \overline{\eta}^{\rho \sigma} \delta_0^{\alpha} \delta_0^{\beta}\big) \big(\bar{m}^2 -\nabla^2 + m^2a^2\big) \,+ \, \big( \overline{\eta}^{\alpha \beta} \bar{\partial}^{(\rho}\delta_0^{\sigma)} + \overline{\eta}^{\rho \sigma} \bar{\partial}^{(\alpha}\delta_0^{\beta)}\big) \big( 2  \partial_0 \big) \\[2pt]
        & + \big( \delta_0^{\alpha} \delta_0^{\beta} \bar{\partial}^{(\rho}\delta_0^{\sigma)} + \delta_0^{\rho} \delta_0^{\sigma} \bar{\partial}^{(\alpha}\delta_0^{\beta)}\big) \big( 2  \partial_0 \big)\, - \, 2\,   \big(m^2a^2 \, \overline{\eta}^{\alpha \beta} \delta_0^{\rho} \delta_0^{\sigma} - m^2aa' \, \overline{\eta}^{\rho \sigma} \delta_0^{\alpha} \delta_0^{\beta}\big) \\[2pt]
        & \, - \, 4 \, \delta_0^{(\alpha} \bar{\partial}^{\beta)}\bar{\partial}^{(\rho}\delta_0^{\sigma)} \, \bigg\} \, \times \, \big\{ \sqrt{aa'} \cos[2(m \bar{t} - \theta_0)] \times i\Delta^{\!R}(x;x') \big\}\\[2pt]
          &\quad  +\kappa^2 \, \frac{m^2 \phi_0^2}{8}  \bigg\{ \, \delta_0^{\alpha} \delta_0^{\beta} \delta_0^{\rho} \delta_0^{\sigma}  \big(ma - m a'\big)\partial_0  \, +\, \overline{\eta}^{\alpha \beta}\overline{\eta}^{\rho \sigma} \, \big(ma' -ma \big)\partial_0 \\[2pt]
        &+ \big(\overline{\eta}^{\alpha \beta} \delta_0^{\rho} \delta_0^{\sigma} + \overline{\eta}^{\rho \sigma} \delta_0^{\alpha} \delta_0^{\beta}\big) \big(ma - m a'\big)\partial_0 \,+ \, \big( \overline{\eta}^{\alpha \beta} \bar{\partial}^{(\rho}\delta_0^{\sigma)} + \overline{\eta}^{\rho \sigma} \bar{\partial}^{(\alpha}\delta_0^{\beta)}\big) \big(- 2  ma\big) \\[2pt]
        & + \big( \delta_0^{\alpha} \delta_0^{\beta} \bar{\partial}^{(\rho}\delta_0^{\sigma)} + \delta_0^{\rho} \delta_0^{\sigma} \bar{\partial}^{(\alpha}\delta_0^{\beta)}\big) \big(- 2 ma \big)\, + \,    \big(ma \, \overline{\eta}^{\alpha \beta} \delta_0^{\rho} \delta_0^{\sigma} - ma' \, \overline{\eta}^{\rho \sigma} \delta_0^{\alpha} \delta_0^{\beta}\big) \big(-2 \, \partial_0\big) \\[2pt]
        &+ \, 2 \, \big( ma \, \overline{\eta}^{\alpha \beta} \bar{\partial}^{(\rho}\delta_0^{\sigma)} - \, ma'\,  \overline{\eta}^{\rho \sigma} \bar{\partial}^{(\alpha}\delta_0^{\beta)}\big)   - \, 2 \, \big(ma\, \delta_0^{\alpha} \delta_0^{\beta} \bar{\partial}^{(\rho}\delta_0^{\sigma)} -\, ma' \delta_0^{\rho} \delta_0^{\sigma} \bar{\partial}^{(\alpha}\delta_0^{\beta)}\big) \bigg\}\\[2pt]
        &  \qquad \quad \qquad \qquad \qquad   \times \, \big\{ \sqrt{aa'} \sin[2(m \bar{t} - \theta_0)] \times i\Delta^{\!R}(x;x') \big\}.
\end{split}
\end{equation}
The retarded causality structure is encoded in the retarded propagator, which we express in a form that makes this structure explicit:
\begin{equation}
    \begin{split}
    & i \Delta^R (x;x')=\frac{1}{\bar{a}} \int \frac{d^4k}{(2\pi)^4} \frac{i \, e^{ik_{\mu}\Delta x^{\mu}}}{\big(k^0 + i\varepsilon \big)^2 - \omega^2(k;\bar{t})} \\[4pt]
    & =\frac{i\bar{m}^2}{4 \pi} \bigg( \frac{J_1(m\sqrt{|\Delta x^2|})}{m\sqrt{|\Delta x^2|}} +\frac{\partial^2}{ 2\bar{m}^2} \bigg) \big[ \Theta(- \Delta x^2) \Theta (\Delta t) \big]  . 
    \end{split}
\end{equation}
This concludes our analysis of the self-energy induced by a generic Gaussian initial state of the dark matter field in the adiabatic regime, under the mid-point mass approximation. 
\subsection{Energy-momentum tensor}
The energy-momentum tensor, being linear in the matter two-point function, decomposes naturally into four contributions arising from the initial state. The state-dependent terms are parameterized by the occupation number, squeezing parameters and the condensate. In contrast, the vacuum contribution depends on the coincident vacuum propagator and thus requires renormalization:
\begin{equation}
    \begin{split}
       a^{D-2}\big< \mathcal{T}^{\,\alpha \beta}_{V}\big>_{\mathcal{C}}(x) &= \bigg[ \partial^{(\alpha} \partial'^{\beta)}   -\frac{1}{2}\eta^{\alpha \beta} \bigg( \partial \cdot \! \partial' + m^2 a^2 \bigg) \bigg] i\Delta_{++}^{V}(x;x') \bigg|_{x'\to x} \\[4pt]
        &=  -\eta^{\alpha \beta} \frac{a^2m^2}{D} i\Delta_{++}^{V}(x;x) = -m^4\eta^{\alpha \beta}  \left( \frac{a^D}{8D \,  \pi^2}\frac{\mu^{D-4}}{D-4} + \frac{a^4}{64\pi^2} \mathcal{F}_m \right).
    \end{split}
\end{equation}
 To preserve the predictive power of our theory, we must include the contribution to the energy-momentum tensor derived from the counterterm action defined in eq. (\ref{countertermact}). At leading adiabatic order and neglecting boundary terms, one can show that the contributions from counterterms proportional to $R$, $R^2$, and $C_{\mu\nu\rho\sigma}C^{\mu\nu\rho\sigma}$ vanish. The only remaining contribution comes from the cosmological constant counterterm. Substituting the expression for the coefficient $c_4$ from eq.~(\ref{vacuumlocalse}) yields:
\begin{equation}
\begin{split}
      a^{D-2}\big<\mathcal{T}^{\, \alpha \beta}_{\, \text{ct}}\big>(x)& = 2 \, \frac{\delta}{\delta h_{\alpha \beta}(x)} \bigg[ \int d^Dx' \, a^{D} \, \frac{h}{2}  m^4 \left( \frac{1}{8\pi^2D} \frac{\mu^{D-4}}{D-4}+ 4\frac{\mathcal{F}_m} {(16\pi)^2} \right) \bigg] \\[4pt]
     &=  \, m^4 \eta^{\alpha \beta} \left( \frac{a^{D}}{8\pi^2D} \frac{\mu^{D-4}}{D-4}+ 4a^4\frac{\mathcal{F}_m} {(16\pi)^2} \right).
\end{split}
\end{equation}
Therefore, this choice for the counterterm action, which renormalizes the vacuum self-energy and  enforces the transversality of its local component, also ends up nullifying this vacuum term entirely. The contributions dependent on the occupation number and squeezing are readily expressed as functionals of the $I_i^{N/S}$ integrals introduced in eqs.~(\ref{Iintegralsoccupationnumber}) and (\ref{Iintegralssqueezing}). Specifically, the occupation number contribution is given by
\begin{equation}
    \begin{split}
          a^{2}\big<\mathcal{T}^{\,\alpha \beta}_N \big>_{\mathcal{C}} (x) &=  \bigg[ \partial^{(\alpha} \partial'^{\beta)}   -\frac{1}{2}\eta^{\alpha \beta} \bigg( \partial \cdot \! \partial' + m^2 a^2 \bigg) \bigg] i\Delta_N(x;x') \bigg|_{x'\to x}\\[4pt]
          &=\bigg[ -\delta_0^{\alpha} \delta_0^{\beta} \big(\nabla^2-a^2m^2\big) -  \big( \eta^{\alpha \beta} + \delta_0^{\alpha} \delta_0^{\beta} \big) \frac{\nabla^2}{3}\bigg] i\Delta_{N}(x;x') \bigg|_{x'\to x} \\[4pt]
         & =  \delta_0^{\alpha} \delta_0^{\beta} \bigg( \frac{4}{3} I^{N}_2 + a^2m^2 I^{N}_1 \bigg) + \eta^{\alpha \beta} \, \frac{I^{N}_2}{3}  ,
    \end{split}
\end{equation}
while the squeezing contribution takes the form
\begin{equation}
    \begin{split}
          a^{2}\big<\mathcal{T}^{\,\alpha \beta}_S \big>_{\mathcal{C}}(x) &=  \bigg[ \partial^{(\alpha} \partial'^{\beta)}   -\frac{1}{2}\eta^{\alpha \beta} \bigg( \partial \cdot \! \partial' + m^2 a^2 \bigg) \bigg]  i\Delta_S(x;x') \bigg|_{x'\to x}\\[4pt]
          &=\bigg[ \delta_0^{\alpha} \delta_0^{\beta} \big(\nabla^2-a^2m^2\big) -  \big( \eta^{\alpha \beta} + \delta_0^{\alpha} \delta_0^{\beta} \big) \frac{\nabla^2}{3} +\eta^{\alpha \beta} \big(\nabla^2 -a^2m^2 \big)\bigg] i\Delta_S(x;x') \bigg|_{x'\to x} \\[4pt]
         & =  - \delta_0^{\alpha} \delta_0^{\beta} \bigg( \frac{2}{3} I^{S}_2 (t) + a^2m^2 I^{S}_1 (t)\bigg) - \eta^{\alpha \beta} \bigg( \frac{2}{3} I^{S}_2 (t)+ a^2m^2 I^{S}_1 (t)\bigg).
    \end{split}
\end{equation}
A notable feature of the above expression is that the energy density of the ULDM field remains unaffected by the squeezing amplitude $S_k$, as evidenced by the vanishing expectation value $\big< \mathcal{T}^{\,00}_S \big>_{\mathcal{C}} = 0$. The squeezing amplitude does, however, affect the pressure of the field through the periodic functions \(I_j^S\), which introduce an oscillatory contribution. We conclude this chapter by presenting the condensate contribution to the energy-momentum tensor:
\begin{equation}
\begin{split}
   a^{2}\mathcal{T}_{\,\text{cl}}^{\,\alpha \beta}(x) &=
      \delta_0^{\alpha} \delta_0^{\beta} \big(\partial_0 \bar{\chi}\big)^2 - \frac{1}{2}\eta^{\alpha \beta} \bigg[ m^2 a^2\bar{\chi}^2   - \big(\partial_0\bar{\chi}\big)^2  \bigg] \\[4pt]
&  = \delta_0^{\alpha} \delta_0^{\beta} \frac{am^2\phi_0^2}{2} \big\{1 - \cos\big[ 2(mt-\theta_0)\big] \big\} - \eta^{\alpha \beta} \frac{am^2\phi_0^2}{2}\cos\big[ 2(mt-\theta_0)\big].
    \end{split}
\end{equation}
With these results established, we now possess all necessary tools to analyze the equation of motion for the graviton field~(\ref{endpoint}) during the matter-dominated epoch, which will be the focus of the following chapter.

\section{Equations for tensorial perturbations and scalar potentials}
Building on the previous chapter's derivation of the self-energy and energy-momentum tensor as functionals of the initial dark matter state parameters, $N_k$, $S_k$, $\phi_k$, $\phi_0$, and $\theta_0$, we now investigate how these parameters affect the evolution of the gravitational perturbations. Substituting our results into the linearized graviton equation of motion~(\ref{endpoint}) yields an integro-differential equation containing both local and non-local terms. While non-local contributions present interesting numerical challenges, our analysis will focus primarily on the local terms, whose direct interpretation provides clearer physical insight into perturbation dynamics. A proper analysis of the equation requires gauge fixing to be imposed to $h_{\mu\nu}$. For reference, we first state the complete non-gauge-fixed perturbation equation with all local terms explicit: 
\begin{equation}\label{nongaugefixedfinalequation} \vspace{1mm}
    \begin{split}
         &a^2 \bigg\{[{}^{\alpha \beta }\mathcal{L}_{(\eta)}^{\, \rho \sigma}] +  \mathcal{H} \big[ \eta^{\alpha \beta} \eta^{\rho \sigma} \partial_0 + \eta^{\rho \sigma} \delta_0^{(\alpha} \partial^{\beta)} + \eta^{\alpha \beta} \delta_0^{(\rho}\partial^{\sigma)} \\
         & \qquad  \qquad  \qquad \qquad - \eta^{\alpha ( \rho} \eta^{\sigma) \beta} \partial_0  - \delta_0^{(\alpha} \eta^{\beta)(\rho} \partial^{\sigma)} -  \delta_0^{(\rho} \eta^{\sigma)(\alpha} \partial^{\beta)} \big]  \bigg\} h_{\rho \sigma} (x)\\[6pt]
&= \kappa^2  \bigg\{ \, \eta^{\alpha \beta} \eta^{\rho \sigma} \bigg( \frac{I_2^N}{24} + \frac{a^2m^2}{8} I_1^N - \frac{5}{24} I^S_2  - \frac{a^2m^2}{8}I^S_1 + \frac{\phi_0^2}{4a} a^2m^2 \sin^2\big(mt- \theta_0 \big) \bigg) \\[4pt]
& \qquad + \big( \eta^{\alpha \beta}\delta_0^{\rho}\delta_0^{\sigma}+  \eta^{\rho \sigma}\delta_0^{\alpha}\delta_0^{\beta} \big) \bigg[ \frac{I_2^N}{6} + \frac{a^2m^2}{8} I_1^N - \frac{I_2^S}{3} - \frac{3a^2m^2}{8}I_1^S\\
& \qquad \qquad \qquad \qquad  \qquad \qquad \! \quad  + \frac{\phi_0^2}{2a}a^2m^2 \bigg( 1-\frac{5}{4} \cos^2\big(mt- \theta_0 \big)\bigg)\bigg]\\[4pt]
&\qquad + \delta_0^{\alpha}\delta_0^{\beta} \delta_0^{\rho}\delta_0^{\sigma} \bigg( \frac{2}{3}I_2^N + \frac{a^2m^2}{2} I_1^N - \frac{I_2^S}{3} - \frac{a^2m^2}{2} I_1^S + \frac{\phi_0^2}{a} a^2m^2 \sin^2\big(mt- \theta_0 \big)\bigg) \\[4pt]
&\qquad +  \eta^{\alpha (\rho} \eta^{\sigma) \beta} \bigg( \frac{I_2^N}{6} +  \frac{I_2^S}{6} \bigg) \,+ \,  \delta_0^{(\alpha} \eta^{\beta)(\rho}\delta_0^{\sigma)}  \bigg( \frac{I_2^N}{6} +  \frac{I_2^S}{6} \bigg)\\[4pt]
&\qquad -  \frac{P^{\alpha (\rho}P^{\sigma) \beta}}{3840 \pi^2} \big[24 m^4a^4 + \big(10 \mathcal{F}_m-8+1920\pi^2 c_3^f \big) m^2a^2\, \partial^2 \\
& \qquad \qquad \qquad \qquad \quad \qquad \, \, + \big( -\mathcal{F}_m +1 +3840\pi^2 c_1^f \big) \partial^4 \big] \\[4pt]
&\qquad +  \frac{P^{\alpha \beta}P^{\rho \sigma}}{3840 \pi^2}\big[ 24 m^4a^4 + \big( 10 \mathcal{F}_m-20+1920\pi^2 c_3^f \big) m^2a^2\, \partial^2 \\
& \qquad \qquad \qquad \qquad \quad \qquad \! \! + \big( 3\mathcal{F}_m - 3 +1280\pi^2 (c_1^f -6 c_2^f) \big) \partial^4 \big] \bigg\} h_{\rho \sigma}(x)\\
         & + \int_{x'} [^{\alpha \beta} \Sigma_{\text{NL}}^{\, \rho \sigma}] (x;x') h_{\rho \sigma}(x'),
    \end{split}
     \raisetag{17pt}
\end{equation}
where $\int_{x'} = \int d^4x'$ and $\int_{k} = \int \frac{d^4k}{(2\pi)^4}$;
the local integrals \(I_j^N\) and \(I_j^S\), which encode the dependence on occupation number and squeezing respectively, are defined in equations \eqref{Iintegralsoccupationnumber} and \eqref{Iintegralssqueezing}; the renormalization-scale-dependent constant \(\mathcal{F}_m\) is given in \eqref{Fm}; the coefficients \(c_i^f\) denote the finite counterterms from the action \eqref{countertermact}; and finally \(P^{\alpha\beta} = \eta^{\alpha\beta} - \partial^{\alpha}\partial^{\beta}/\partial^2\) is the transverse projector. 

The dynamics of the graviton are more conveniently studied in conformal time \(\eta\), whereas the evolution of the ULDM quantities is more naturally expressed in cosmic time \(t\). The relation between the two variables is given by:
\begin{equation}
    \eta - \eta_{\mathrm{eq}} = \int_{t_{\mathrm{eq}}}^{t} \frac{dt'}{a(t')} = \frac{3t}{a} - \frac{3t_{\mathrm{eq}}}{a_{\mathrm{eq}}}.
\end{equation}
To convert the \(t\)-dependence to a \(\eta\)-dependence, we define a shifted variable
\begin{equation} \label{tildeeta}
    \tilde{\eta} := \eta - \eta_{\mathrm{eq}} + \frac{3t_{\mathrm{eq}}}{a_{\mathrm{eq}}},
\end{equation}
so that the relation between the two quantities becomes straightforward:
\begin{equation}
    t = \frac{a}{3} \left( \eta - \eta_{\mathrm{eq}} + \frac{3t_{\mathrm{eq}}}{a_{\mathrm{eq}}} \right) \equiv \frac{a}{3} \, \tilde{\eta}.
\end{equation}
Throughout this chapter, we keep this substitution implicit to avoid overburdening the notation. Its explicit use will be employed in the quantitative computations of the next chapter. In the foliation of spacetime defined by the FLRW coordinates, the cosmological perturbation field decomposes into scalar, vector, and tensor components, classified by their transformation properties under spatial rotations~\cite{bertschinger2000cosmologicalperturbationtheorystructure, Mukhanov:1990me},
\begin{equation}
\begin{split}
    ds^2 = a^2 (\eta)  &\left[ -\left(1+2\phi\right) d\eta^2 + 2\left( \partial_iB +S_i\right) d\eta dx^i \right. \\[4pt]
    &+ \left. \left( (1-2\psi) \delta_{ij} + 2 \partial_i \partial_j E + 2\partial_{(i}F_{j)} + h_{ij}^{\text{TT}}\right) dx^idx^j \right].
\end{split}
\end{equation}
Here, \(B\), \(E\), \(\phi\), and \(\psi\) are scalar perturbations; \(S_i\) and \(F_i\) are transverse vectors satisfying \(\partial^i S_i = \partial^i F_i = 0\); and \(h_{ij}^{\text{TT}}\) is a traceless, transverse tensor:
\begin{equation} \label{TTprop}
    \partial^i h_{ij}^{\text{TT}} = 0 , \qquad \delta^{ij} h_{ij}^{\text{TT}} = 0.
\end{equation}
General relativity is invariant under diffeomorphisms, which introduce four gauge degrees of freedom that can be eliminated via an infinitesimal coordinate transformation. We use this freedom to set \(B\), \(E\), and \(S_i\) to zero. At linear order, the evolution equations for the remaining vector field \(F_i\) decouple, and such perturbations are generally suppressed in an expanding universe (\(F_i \propto a^{-1}\)). For these reasons, we neglect the dynamics of this field and adopt the following metric decomposition:
\begin{equation} \vspace{1mm}
    ds^2 = a^2(\eta) \left[ -(1+2\phi)\, d\eta^2 + \bigl( (1-2\psi)\delta_{ij} + h_{ij}^{\text{TT}} \bigr) dx^i dx^j \right].
\end{equation}
The four remaining degrees of freedom (\(\phi\), \(\psi\), and the two polarisations of \(h_{ij}^{\mathrm{TT}}\)) are physical and define the longitudinal gauge. In this gauge, the scalar perturbations \(\phi\) and \(\psi\) coincide with the gauge‑invariant Bardeen potentials \cite{PhysRevD.22.1882}, which are coupled to the energy density and pressure inhomogeneities of the matter content and are responsible for the formation of large‑scale structures \cite{1980lssu.book.....P}. The tensor field \(h_{ij}^{\mathrm{TT}}\) describes primordial gravitational waves, which are generated by anisotropic stress in the early universe \cite{buonanno2004tasilecturesgravitationalwaves} and constitute a key prediction of inflation. 

The study of these fields is of particular interest because their predictions can be tested against a wide range of cosmological observables, including those associated with the cosmic microwave background (CMB). Scalar perturbations imprint themselves on CMB temperature anisotropies and the E-mode polarization spectrum. Tensor perturbations, by contrast, leave a distinctive signature in the B-mode polarization spectrum \cite{Kamionkowski_1997, Zaldarriaga_1997, Jiang_2025, steier2025unbiasedprimordialgravitationalwave}. In the following sections, we write down the equations of motion for both scalar and tensor components, with a particular focus on identifying resonance phenomena induced by the squeezing parameters of the ULDM initial state.

\subsection{Gravitational wave equation}
The properties of the gravitational wave field \eqref{TTprop} greatly simplify its equation of motion:
\begin{equation}\label{TTfullequation}
    \begin{split}
        &\frac{a^2}{2} \left(\,  \partial^2 -\, 2\mathcal{H}\,  \partial^0    \right) \, h_{ij}^{\text{TT}} (x)\,= \, \kappa^2 \, \bigg( \, \frac{I_2^N }{6} \, + \, \frac{ I_2^S}{6}\bigg) \, h_{ij}^{TT} (x)\\[3pt]
         &+ \frac{\kappa^2}{8i}\sum_{t=1}^4\int_{x',k} \, e^{i k_{\mu} \Delta x^{\mu}} \left[  \left( \bar{J}^t_{40} -2 \bar{J}^t_{22} + \bar{J}^t_{04} \right) +  \left( \widetilde{J}^t_{40} -2 \widetilde{J}^t_{22} + \widetilde{J}^t_{04} \right) \right] h_{ij}^{\text{TT}} (x')\\[2pt]
        &- \frac{\kappa^2}{3840 \pi^2} \big[24 m^4a^4 + \big(10 \mathcal{F}_m-8+1920\pi^2 c_3^f \big) m^2a^2\, \partial^2 + \big( -\mathcal{F}_m +1 +3840\pi^2 c_1^f \big) \partial^4 \big] h_{ij}^{\text{TT}}(x)\\[2pt]
       &+\frac{\kappa^2}{ \, i\,240}  \int_{x'} \, \big(\partial^2  - 4 m^2 \bar{a}^2 \big)^2\mathcal{M}^R_m(x;x') \, h_{ij}^{\text{TT}} (x'). 
    \end{split}
\end{equation}
Here $\bar{J}^t_{rs}$ and $\widetilde{J}^t_{rs}$ encode the non-local dependence on occupation number and squeezing respectively, and are defined in equations \eqref{nonlocalNintegrals} and \eqref{nonlocalSintegrals}. The non-local function \(\mathcal{M}_m^R\) is proportional to the regularized part of the difference between the squared vacuum Feynman propagator and the negative Wightman function; it is given explicitly in eq.~\eqref{deltam}.

 Equation (\ref{TTfullequation}) reveals several key features of the gravitational wave dynamics. First, the spatially homogeneous matter condensate does not enter the equation; corrections arise exclusively from the matter two-point function, necessitating a full quantum field-theoretic treatment. We emphasize that this cancellation is possible because the condensate four-point self-energy, which would otherwise appear in eq.~\eqref{endpoint}, cancels against contributions from the gravitational sector. This follows from the symmetries of the background spacetime and can therefore be understood as a manifestation of the Noether-Ward identity~\cite{Prokopec:2025jrd}.

In the high-occupation regime relevant for cosmological dark matter (\( N_k \gg 1 \)), terms proportional to \( N_k \) dominate over \( \mathcal{O}(\hbar) \) vacuum fluctuations by many orders of magnitude. For pure squeezed states, the same hierarchy applies to \( \mathcal{O}(S_k) \) terms due to the identity \( S_{k}^2 = N_{k}(N_{k}+1) \). Consequently, vacuum contributions can therefore be safely neglected when studying the dominant backreaction and mode enhancement effects (for their impact on gravitational potentials, see \cite{Jimu:2024xqm}). 

The state-dependent local contributions, appearing in the first line, consist of two distinct terms: a slowly-varying graviton mass term proportional to the occupation number $N_k$ (whose time dependence enters solely through the adiabatically evolving scale factor), and an oscillatory mass term driven by the squeezing amplitude $S_k$. Transforming to Fourier space and removing the Hubble friction term through the substitution
\begin{equation} 
\widetilde{h}_{ij}^{\text{TT}}(\vec{k};\eta) = a^{-1}(\eta) \xi_{ij}(\vec{k};\eta),
\end{equation}
the equation of motion for the rescaled graviton field simplifies to:
\begin{equation} \label{xiresonanceequation}
\left( \partial_0^2 + \omega_{\xi}^2(k; \eta)  + \frac{\kappa^2}{3 a^2} I_2^S(\eta) \right) \xi_{ij}(\vec{k};\eta) = \frac{1}{a^2(t)}\int_{t_{\text{eq}}}^{\infty} \, dt' \,\frac{\xi_{ij}(\vec{k};t')}{a(t')} \int dk^0 \, \frac{e^{-i k^{0}\Delta t} }{2\pi} \,
\overline{\Sigma}^{\text{TT}}(k^0;\vec{k};\bar{t}) \, ,
\end{equation}
where we keep the time variable \(t\) on the right-hand side to facilitate estimation of these terms; consequently, the scale factor appears inside the integrand.  Here, we define the typical angular frequency of the field modes:
\begin{equation}
\omega_{\xi}^2(k; \eta) := k^2 + \, \frac{\kappa^2}{ 3 \, a^2 } \, I_2^N  \, - \frac{a''}{a}.
\end{equation}
where primed quantities denote derivatives with respect to conformal time. The non-local self-energy contributions relevant to the TT gauge are expressed compactly in equation \eqref{xiresonanceequation} as:
\begin{equation}
   \overline{\Sigma}^{\text{TT}}(k^0;\vec{k};t) = \frac{i\kappa^2}{4}\sum_{t=1}^4  \left[  \left( \bar{J}^t_{40} -2 \bar{J}^t_{22} + \bar{J}^t_{04} \right) +  \left( \widetilde{J}^t_{40}  -2 \widetilde{J}^t_{22} + \widetilde{J}^t_{04} \right) \right].
\end{equation}
We emphasize that an exact treatment of these terms can only be achieved numerically. In the following chapter, under specific assumptions on the squeezing and occupation number spectra, we will show that these terms are suppressed relative to the local contributions at higher order in the relativistic expansion, justifying a perturbative treatment of the non-local terms.

Focusing on the dominant local contributions, the source term \(I_2^S(t)\) exhibits periodic behavior inherited from the periodicity of its integrand. Consequently, the homogeneous part of eq.~\eqref{xiresonanceequation} reduces to a Mathieu equation \cite{Hill1886, Mathieu1868}. The dynamics of this system are therefore properly analyzed within the framework of Floquet theory, suitably adapted to account for a slowly evolving background. Its solutions may develop instabilities due to squeezing-induced parametric resonance, leading to resonant growth for specific modes. Such enhancements could leave observable imprints, for instance in the B-mode polarization spectrum of the CMB. For a review of Mathieu's equations and their band structures, we refer to \cite{magnus2004hill}. Observational constraints on such features could therefore test the hypothesis that ULDM in a squeezed state constituted the dominant dark matter component throughout the matter-dominated era. 

Making quantitative predictions about these phenomena requires assumptions about the initial state of the ULDM field. In the following chapter, we carry out this analysis explicitly for power-law occupation number and squeezing spectra.
\subsection{Equation for scalar potentials}
Scalar perturbations about an FLRW background encode information about the local curvature and the Newtonian gravitational potential on comoving slices \cite{Mukhanov:1990me}. In the longitudinal gauge, \(\phi\) and \(\psi\) are generally determined by constraint equations.\footnote{If \(\psi\) is generated during inflation and couples to dynamical matter perturbations, its equation of motion becomes dynamical in this gauge.} Following the same approach as for tensor perturbations, we neglect vacuum terms due to the high-occupation regime relevant for cosmological dark matter. We extract three independent equations from the perturbed field equations: the $(0,0)$ component of eq.~(\ref{nongaugefixedfinalequation}), its combination with the spatial trace, and the $(0,i)$ component. These form a closed set of equations for the gravitational potentials $\phi$ and $\psi$. Expressing only the local terms explicitly, the $(0,0)$ component gives:
\begin{equation} \label{potentials1}
 \begin{split}
    & a^2\, \nabla^2 \, \psi \, = \,\kappa^2 \, \phi\, \bigg[ \frac{3}{8} I^N_2  + \frac{3a^2m^2}{8} I_1^N \, +\frac{1}{8}I_2^S + \frac{a^2m^2}{8} I_1^S\bigg] \\[4pt]
    &+ \, \kappa^2 \, 3 \psi\, \bigg[  \frac{1}{8} I^N_2  -\frac{1}{8}I_2^S - \frac{a^2m^2}{4} I_1^S + \frac{\phi_0}{16a} a^2m^2 \big(1 - 3\cos\big[2(mt- \theta_0)\big] \big)\bigg] \\[4pt]
    & + \, \int_{x'} \, [{}^{00}\Sigma^{\, 00}_{\text{NL}}] \, (x;x') \, \big( \phi + \psi \big) (x')\,+ \, \int_{x'} \, [{}^{00}\Sigma_{\text{NL}, \mu}^{ \, \mu}] \, (x;x') \, \psi(x') \,.
 \end{split}
\end{equation}
The combination with the spatial trace yields:
\begin{equation}
 \begin{split}
    & a^2\,\big( \, \nabla^2 \, \phi \, + \, 3 \,(\partial_0 +\, 2 \, \mathcal{H}) \partial_0\, \psi \big)\\[4pt]
    &= \,\kappa^2 \, \phi\, \bigg[ \frac{3}{4} I^N_2  + \frac{3a^2m^2}{8} I_1^N \, -\frac{1}{4}I_2^S -\frac{a^2m^2}{8} I_1^S  + \frac{\phi_0}{16a} a^2m^2 \big(7-9\cos\big[2(mt- \theta_0)\big] \big)\bigg] \\[4pt]
    &+ \, \kappa^2 \, 3 \psi\,\bigg[ \frac{5}{12} I^N_2  + \frac{3a^2m^2}{8} I_1^N \, -\frac{7}{12}I_2^S -\frac{5a^2m^2}{8} I_1^S  + \frac{\phi_0}{16a} a^2m^2 \big(7-9\cos\big[2(mt- \theta_0)\big] \big) \bigg]\\[4pt]
    & + \,   \int_{x'} \, \big( \, [{}^{00}\Sigma^{\, 00}_{\text{NL}}] + \sum_i [{}^{ii}\Sigma^{\, 00}_{\text{NL}}]\, \big)\, (x;x')  \, \big( \phi + \psi \big) (x')\,\\
    &+ \, \int_{x'} \, \big(  \, [{}^{00}\Sigma^{\, \mu}_{\text{NL}, \mu}] + \sum_i [{}^{ii}\Sigma^{\, \mu}_{\text{NL}, \mu}] \, \big)  \, (x;x') \, \psi(x') \,, \raisetag{4mm}
 \end{split}
\end{equation}
while the $(0,i)$ component becomes:
\begin{equation} \label{potential2}
 \begin{split}
    & a^2\, \big[ \, \partial_0  \partial_i \, \big(\phi+\psi \big) \, +\,2\, \mathcal{H}\, \partial_i \, \psi \big]  = \\[4pt]
    & - \, \int_{x'} \, [{}^{0i}\Sigma^{\, 00}_{\text{NL}}] \, (x;x') \, \big( \phi + \psi \big) (x')\,- \, \int_{x'} \, [{}^{0i}\Sigma^{\,\mu}_{\text{NL}, \mu}] \, (x;x') \, \psi(x') \,.
 \end{split}
\end{equation}
As expected, unlike primordial gravitational waves, scalar potentials are affected by the presence of a homogeneous ULDM condensate. The oscillatory terms arising from both the condensate and the squeezing of the matter field open the possibility of resonant solutions. 

These equations emerge from our perturbative approach and are proposed as an alternative to the framework used in \cite{Khmelnitsky_2014} for the analysis of pulsar timing array observables, where setting \(a = 1\) recovers the perturbed Minkowski framework.

Given its complexity, the system of equations presented in this section requires a detailed analysis, which can be carried out under specific assumptions on the occupation number spectrum \(N_k\) and the squeezing spectrum \(S_k\) of the ULDM field. In the present work, we focus on the study of gravitational wave propagation, leaving the analysis of scalar potential evolution for future investigation.

\section{Quantitative analysis of the gravitational waves equation}
To construct solutions of equation \eqref{xiresonanceequation}, we must introduce an ansatz for the occupation number spectrum \(N_k\) and the squeezing amplitude spectrum \(S_k\). In this work, we model these spectral functions as power laws with a Gaussian ultraviolet cutoff. Explicitly, we adopt the ansatz
\begin{equation} \label{ansatzspectra}
\begin{split}
N_k &= N_0(k_*) \left(\frac{k}{k_*}\right)^{\! n_{N}} \! \! \exp\left(-\frac{k^2}{k_{\text{UV}} ^2}\right), \\[4pt]
S_k &= S_0(k_*)\left(\frac{k}{k_*}\right)^{\! n_{S}} \! \! \exp\left(-\frac{k^2}{k_{\text{UV}} ^2}\right),  
\end{split}
\end{equation}
where \( k_* \) is a fiducial momentum scale, \( k_{\text{UV}} \) is the physical UV‑cutoff scale, and the exponential factor suppresses modes with \( k \gg k_{\text{UV}} \). The fiducial scale \( k_* \) is arbitrary and can be absorbed into the definitions of the amplitudes \( N_0(k_*) \) and \( S_0(k_*) \). The cutoff \( k_{\text{UV}} \) is set by the production mechanism of the ultralight dark matter; for the present analysis we assume that the field is predominantly non‑relativistic at matter–radiation equality \cite{Kopp_2018}, which implies the condition
\begin{equation} \label{nonrelativisticcondition}
   \frac{ k_{\text{UV}}}{a_{\text{eq}} }\, \ll \,  m.
\end{equation}
The value of \(\phi_k\) depends sensitively on the ULDM production mechanism and its subsequent evolution up to the onset of matter domination. For isocurvature inflationary production, the initial phase is zero and remains frozen until the mode \(k\) becomes sub‑horizon and \(m \sim H\), after which it begins to oscillate under the assumption of negligible interactions with the thermal bath. For production via gravitational reheating, by contrast, the initial squeezing phase is determined by the detailed evolution of the mode functions during the event. In this work, we remain agnostic about the specific production and evolution history of the ULDM field during inflation and radiation domination. Since a non‑zero phase shift does not significantly affect the parametric resonance phenomena discussed in the following sections, we set \(\phi_k = 0\) for simplicity. Generalization to the specific \(\phi_k\) spectra associated with different scenarios can be readily implemented from our baseline case.

This parametrization introduces four dimensionless spectral parameters: the amplitudes \( N_0,\, S_0 \) and the indices \( n_N,\, n_S \). These are partially constrained by the present-day ULDM energy density and pressure, as discussed in the next section. 

With these spectra, the local integrals \eqref{Iintegralsoccupationnumber} and \eqref{Iintegralssqueezing} are solved explicitly; details are given in Appendix \ref{Appendix C}. To leading order in the non-relativistic expansion, the occupation-number integral contributing to the graviton mass reads
\begin{equation}\label{main:solutionlocalintegralN}
   \frac{1}{3a^2} I_2^N \;\approx \frac{1}{12\pi^2 a^3 m} \,
      \frac{N_0}{k_*^{n_N}} \,
      k_{\text{UV}}^{5+n_N} \,
      \Gamma\!\left( \frac{n_N + 5}{2}\right).
\end{equation}
For the squeezing integral, we approximate the amplitude to leading order in the non-relativistic expansion while retaining the first-order correction in the phase. To ensure that the cumulative error introduced by this phase approximation remains negligible, we must impose a stricter condition on the cutoff scale than that of eq.~\eqref{nonrelativisticcondition}:
\begin{equation} \label{phasecondition}
    \frac{k_{\text{UV}}}{a_{\text{eq}}} \ll \sqrt{m H_{\text{eq}}}.
\end{equation}
This constraint can be relaxed by including higher-order phase corrections, asymptotically recovering the original non-relativistic condition~\eqref{nonrelativisticcondition}, as detailed in Appendix~\ref{Appendix C}.
For squeezing spectra whose cutoff scale satisfies condition \eqref{phasecondition}, the local integral contributing to the graviton mass takes the following form:
\begin{equation}
    \begin{split}
      \frac{1}{3a^2}   I_2^S &\approx  \frac{1}{12\pi^2 a^3m} \, \frac{S_0}{k_*^{n_S}}\, k_{\text{UV}}^{5+n_S} \, \Gamma \, \bigg(\frac{n_S +5}{2}\bigg)   \\[2pt]
 &  \quad \times
  \, \cos\bigg\{\bigg(2m +\frac{3(n_S +5)}{2} \frac{k_{\text{UV}}^2}{ma^2} \bigg)\, t \, - \varphi_2\bigg] \bigg\},
    \end{split}
\end{equation}
where we have defined:
\begin{equation}
    \varphi_{j} := 2mt_{\text{eq}}+ \frac{3(n_S +2j+1)}{2} \frac{k_{\text{UV}}^2t_{\text{eq}}}{ma^2_{\text{eq}}}.
\end{equation}
This completes the evaluation of the left-hand side of eq. \eqref{xiresonanceequation}. A proper treatment of the right-hand side can only be achieved by numerical means. However, to estimate its magnitude, we have evaluated these integrals analytically under a set of approximations, detailed in Appendix \ref{Appendix D}. For this purpose, we model the rescaled tensor perturbation \(\xi_{ij}(\vec{k})\) using an adiabatic ansatz consisting of a slowly varying amplitude \(f_{ij}(\vec{k};t)\) and a slowly evolving angular frequency \(\mathcal{W}(\vec{k};t)\):
\begin{equation} \label{main:ansatz}
\begin{split}
\xi_{ij}(\vec{k};t) &= f_{ij}(\vec{k};t) \exp\!\left(-i \, \Omega_0(\vec{k}) \int_{t_{\text{eq}}}^{t} dt'\, a^{\alpha}(t') \right) \\[4pt]
&= f_{ij}(\vec{k};t) \exp\!\Big\{-i \Big[ \mathcal{W}(\vec{k};t)t - \mathcal{W}(\vec{k};t_{\text{eq}}) t_{\text{eq}} \Big] \Big\},
\end{split}
\end{equation}
where we have defined
\begin{equation} \label{main:adiabaticphase}
\mathcal{W}(\vec{k};t) := \Omega_0(\vec{k}) \, \frac{3a^{\alpha}(t)}{2\alpha + 3}.
\end{equation} 
Expanding to first order in the amplitude around \(t\), we obtain an estimate of the various contributions to the right-hand side of eq.~\eqref{xiresonanceequation}:
\begin{equation}
\begin{split}
\frac{1}{a^2(t)}\int_{t_{\text{eq}}}^{\infty} dt' \,\frac{\xi_{ij}(\vec{k};t') }{a(t')}\int \frac{dk^0}{2\pi} \, e^{-i k^{0}\Delta t} \,
\overline{\Sigma}_{TT}(k^0;\vec{k};\bar{t}) \, &\approx - \kappa^2 \big[ M^2_N(\vec{k},t) + M^2_S(\vec{k},t) \big]\xi_{ij}(\vec{k};t) \\
&\quad + \kappa^2 \big[ \gamma_N(\vec{k},t)+\gamma_S(\vec{k},t)\big] \partial_t f_{ij}(\vec{k};t) .
\end{split}
\end{equation}
Here, the occupation-number and squeezing contributions to the non-local mass $ M^2_N$ and $ M^2_S$ are defined in eqs.~\eqref{mass-nonlocal-N} and \eqref{mass-nonlocal-S}, respectively, while the corresponding friction terms, $\gamma_N$ and $\gamma_S$, are given in eqs.~\eqref{friction-nonlocal-N} and \eqref{friction-nonlocal-S}. Within this approximation, the non-local mass contributions are found to be suppressed relative to the local ones. Defining the squeezing mass $m_S$ as the angular frequency that governs squeezing-induced oscillations in cosmic time $t$,
\begin{equation}
    m_S(t) := m + \frac{3 (n_S+7)}{4} \frac{k_{\text{UV}}^2}{m a^2(t)},
\end{equation}
we find that the non-local mass contributions exhibit two distinct behaviors, depending on whether the graviton frequency \(\mathcal{W}\) lies near one of four characteristic frequencies. In our approximation, the latter are given by
\begin{equation}
    \mathcal{W}^{N}_{s_1 s_2}(k;t) = s_1 m + s_2 \, \omega(k;t),
\end{equation}
for the occupation-number-dependent mass term, and by
\begin{equation}
    \mathcal{W}^{S}_{s_1 s_2}(k;t) = s_1 m_S(t) + s_2 \,\omega(k;t),
\end{equation}
for the squeezing-dependent one, where $s_1, s_2 \in \{-1;1\}$ are sign variables. The suppression of the non-local terms relative to the local graviton mass scales as:
\begin{equation} \label{scalingnonlocalterms}
   a^2 \frac{M^2_{N/S}}{I_2^{N/S}} \sim 
    \begin{cases}
        \dfrac{k_{\text{UV}}^2}{a^2 m^2}, & \text{for } \mathcal{W} \neq \mathcal{W}^{N/S}_{s_1 s_2}, \quad s_1, s_2 \in \{-1, 1\}; \\[10pt]
        \dfrac{k_{\text{UV}}^2 (t - t_{\text{eq}})}{a^2 m}, & \text{for } \mathcal{W} \approx \mathcal{W}^{N/S}_{s_1 s_2}, \quad s_1, s_2 \in \{-1, 1\}.
    \end{cases}
\end{equation}
Assuming an adiabatic evolution characterized by \(\partial_t f_{ij} \sim H f_{ij}\), the friction term receives the same suppression. Consequently, provided that condition~\eqref{phasecondition} holds and that our approximations capture the amplitude with reasonable accuracy, we may treat the right-hand side of eq.~\eqref{xiresonanceequation} as a relativistic correction and adopt the perturbative ansatz:
\begin{equation}
\xi_{ij}(\vec{k};t) = \xi_{ij}^{\mathrm{L}}(\vec{k};t) + \xi_{ij}^{\mathrm{NL}}(\vec{k};t),
\end{equation}
where \(\xi^{\mathrm{L}}\) denotes the solution in the absence of non-local contributions and is assumed to dominate over the correction \(\xi^{\mathrm{NL}}\). In the present work, we focus on finding a solution for the dominant component  \(\xi^{ \mathrm{L}}\) and on the amplification it experiences through parametric resonance induced by the squeezing of ULDM field modes for specific initial states. 

\subsection{Energy considerations}
In this section, we present explicit expressions for the background-level pressure and energy density of the ULDM field. These quantities are essential for relating the parameters of the occupation number spectrum to present-day cosmological observables. At the present time \(t = t_0\) the energy density of the ULDM field takes the following form:
\begin{equation} \label{1stequation}
    \begin{split}
        \rho_{\mathrm{DM},0} &= \big\langle \rho \big\rangle_{\mathcal{C}}(t_0) + \rho_{\mathrm{cl}}(t_0) \\[4pt]
        &= I_2^N(t_0) + m^2 I_1^N(t_0) + \frac{m^2\phi_0^2}{2} \\[4pt]
        &= \frac{m}{4\pi^2}\,k_{\mathrm{UV}}^{3+n_N} \, \frac{N_0}{k_*^{n_N}} \, \Gamma \!\left(\frac{n_N+3}{2} \right) \left(1+ \frac{n_N+3}{2} \frac{k_{\mathrm{UV}}^2}{m^2} \right) + \frac{m^2\phi_0^2}{2} \\[3pt]
        &\approx \frac{m}{4\pi^2}\,k_{\mathrm{UV}}^{3+n_N} \, \frac{N_0}{k_*^{n_N}} \, \Gamma \!\left(\frac{n_N+3}{2} \right) + \frac{m^2\phi_0^2}{2},
    \end{split}
\end{equation}
where in the second line we have used the result from \eqref{main:solutionlocalintegralN}, and in the third line we have dropped higher-order relativistic corrections. As a consistency check of our non-relativistic expansion, we note that the ULDM energy density evolves in time as that of a non-relativistic fluid:
\begin{equation}
    \rho_{\mathrm{DM}}(t) = \frac{\rho_{\mathrm{DM},0}}{a^3(t)}.
\end{equation}
The present-day pressure of the field requires a more complicated expression:
\begin{equation}
    \begin{split}
        P_{\mathrm{DM},0} &= \big\langle P \big\rangle_{\mathcal{C}}(t_0) + P_{\mathrm{cl}}(t_0) \\[4pt]
        &= \frac{I_2^N(t_0)}{3} - \frac{2}{3} I_2^S(t_0) - m^2 I_1^S(t_0) - \frac{m^2 \phi_0^2}{2} \cos(2mt_0 -  \theta_0) \\[5pt]
        &= \frac{1}{12\pi^2 m}\,k_{\mathrm{UV}}^{5+n_N} \, \frac{N_0}{k_*^{n_N}} \, \Gamma \!\left(\frac{n_N+5}{2} \right) - \frac{m^2 \phi_0^2}{2} \cos(2mt_0 - \theta_0) \\[4pt]
        & - \frac{1}{6\pi^2 m}\,k_{\mathrm{UV}}^{5+n_S} \, \frac{S_0}{k_*^{n_S}} \, \Gamma \!\left(\frac{n_S+5}{2} \right) \cos\!\left[ \left(2m+\frac{3 (n_S+5)}{2} \frac{k_{\mathrm{UV}}^2}{m} \right) t_0 \right] \\[4pt]
        & - \frac{m}{4\pi^2 }\,k_{\mathrm{UV}}^{3+n_S} \, \frac{S_0}{k_*^{n_S}} \, \Gamma \!\left(\frac{n_S+3}{2} \right) \cos\!\left[ \left(2m+\frac{3 (n_S+3)}{2} \frac{k_{\mathrm{UV}}^2}{m} \right) t_0 \right].
    \end{split}
\end{equation}
Since cosmological observables depend on the time-averaged energy-momentum tensor, we retain only the slowly varying part of \(P_{\mathrm{DM},0}\) when relating field parameters to present-day observables:
\begin{equation} \label{2ndequation}
    P_{\mathrm{DM},0} = \frac{I_2^N(t_0)}{3} = \frac{1}{12\pi^2 m}\,k_{\mathrm{UV}}^{5+n_N} \, \frac{N_0}{k_*^{n_N}} \, \Gamma \!\left(\frac{n_N+5}{2} \right).
\end{equation}
Like the energy density, the time-averaged pressure scales as that of a non-relativistic fluid:\footnote{Oscillatory contributions to the pressure are not included in this scaling.}
\begin{equation}
    P_{\mathrm{DM}}(t) = \frac{P_{\mathrm{DM},0}}{a^5(t)}.
\end{equation}
The equation of state parameter for the ULDM field $w_{\mathrm{DM}}$ is then given by
\begin{equation}
    w_{\mathrm{DM}}(t) := \frac{P_{\mathrm{DM}}}{\rho_{\mathrm{DM}}}(t) = \frac{1}{a^2(t)} \frac{k_{\mathrm{UV}}^2}{3m^2}\frac{n_N+3}{2} \left(1 - \frac{\Omega_{\mathrm{cl}}}{\Omega_{\mathrm{DM}}}\right) \equiv \frac{w_{\mathrm{DM},0}}{a^2(t)}.
\end{equation}
A crucial observation is that for dark matter to be consistent with the formation of the inhomogeneities observed in the CMB, $w_{\mathrm{DM}}$ must already be small by the time of matter-radiation equality \cite{Kopp_2018}. This imposes the condition:
\begin{equation} \label{wcondition}
    \frac{w_{\mathrm{DM},0}}{a^2_{\mathrm{eq}}} \ll 1.
\end{equation}
To relate the parameters governing the occupation spectrum of ULDM to cosmological parameters we introduce the critical density at present time, \(\rho_{\mathrm{crit}}\), and write:
\begin{equation}
    \begin{split}
        \rho_{\mathrm{DM},0} &= \Omega_{\mathrm{DM}} \, \rho_{\mathrm{crit}}, \\
        P_{\mathrm{DM},0} &= w_{\mathrm{DM},0} \, \Omega_{\mathrm{DM}} \, \rho_{\mathrm{crit}},
    \end{split}
\end{equation}
where \(\Omega_{\mathrm{DM}} \equiv \rho_{\mathrm{DM},0}/\rho_{\mathrm{crit}}\) is the present-day density parameter of ULDM, and \(\Omega_{\mathrm{cl}} \equiv \rho_{\mathrm{cl},0}/\rho_{\mathrm{crit}}\) denotes the fraction contributed by the condensate. 
Inverting the system formed by eqs. \eqref{1stequation} and \eqref{2ndequation} then yields expressions for the amplitude \(N_0\) and the spectral index \(n_N\) in terms of \(\Omega_{\mathrm{DM}}\), \(w_{\mathrm{DM},0}\), \(\Omega_{\mathrm{cl}}\), and the cutoff scale $k_{\text{UV}}$:
\begin{equation} \label{parameterspressure}
 \begin{split}
     N_0 &= \Omega_{\mathrm{DM}} \, \rho_{\mathrm{crit}} \left( 1 - \frac{\Omega_{\mathrm{cl}}}{\Omega_{\mathrm{DM}}} \right) \frac{4 \pi^2}{m k_{\mathrm{UV}}^3} \, \left\{ \Gamma\!\left[ \frac{3 w_{\mathrm{DM},0} \, m^2}{k_{\mathrm{UV}}^2} \left(1 - \frac{\Omega_{\mathrm{cl}}}{\Omega_{\mathrm{DM}}}\right)^{\!-1} \right]\right\}^{\!-1}, \\[6pt]
     n_N &= \frac{6 w_{\mathrm{DM},0} \, m^2}{k_{\mathrm{UV}}^2} \left(1 - \frac{\Omega_{\mathrm{cl}}}{\Omega_{\mathrm{DM}}}\right)^{\! -1} - 3.
 \end{split}
\end{equation} 
Since the squeezing terms do not contribute to the time-averaged pressure or energy density, the parameters \(S_0\) and \(n_S\) remain unconstrained by these relations. They may instead be fixed by assuming a specific initial state, such as a pure squeezed state (\(S_0 \approx N_0\), \(n_S \approx n_N\)) or a state with no squeezing (\(S_0 = 0\)).

With these relations in hand, we now proceed to a quantitative assessment of the ULDM field's impact on gravitational wave propagation.

\subsection{Equation of motion for specific states}
In this section, we present the equation of motion for the dominant contribution to the rescaled gravitational wave field $\xi_{ij}^{\text{L}}(k;t)$, sourced by different initial quantum states of the ULDM field. To maintain generality, we include a potentially non-negligible condensate energy density, $\Omega_{\chi}$; the condensate-free limit is recovered by taking $\Omega_{\chi} \to 0$. We examine four representative scenarios, each corresponding to a distinct class of initial conditions for the Gaussian state.
\subsubsection{Classical condensate}
This scenario corresponds to the standard misalignment production mechanism, where the ULDM field is initially in a highly coherent state, well-approximated by a classical condensate.The condensate amplitude drives the background dynamics but does not source gravitational waves, leaving \(\xi_{ij}^{\mathrm{L}}(k;\eta)\) to evolve as a free wave on an unperturbed FLRW background:
\begin{equation} \label{condensateGW}
    \left( \frac{d^2}{d\eta^2} + k^2 - \frac{a''}{a} \right) \xi_{ij}^{\mathrm{L}}(k;\eta) = 0 .
\end{equation}
During matter domination, we have \(a''/a = 2/\eta^2\), and the solution can be expressed in terms of the spherical Bessel functions of the first and second kind, \(j_1(x)\) and \(y_1(x)\), respectively. To distinguish it from the solution during radiation domination, which is used to match initial conditions, we denote the matter-era gravitational wave solution as \(h_{ij}^{\mathrm{MD}}(k;\eta)\):
\begin{equation} \label{MDsol}
   a^{-1}(\eta) \xi_{ij}^{\mathrm{L}}(k;\eta) \equiv  h_{ij}^{\mathrm{MD}}(k;\eta) = \frac{1}{k\eta} \big[ A(k) j_1(k\eta) + B(k) y_1(k \eta) \big].
\end{equation}
The coefficients \(A(k)\) and \(B(k)\) in the matter-era solution are determined by matching to the radiation-era solution at the time of matter–radiation equality. During radiation domination we have \(a''/a = 0\), yielding the solution
\begin{equation}
    h_{ij}^{\mathrm{RD}}(k;\eta) = \frac{1}{k\eta} \big[ C(k) \sin(k\eta) + D(k) \cos(k\eta) \big].
\end{equation}
At the end of inflation, \(\eta = \eta_I\), all modes relevant for current observations are super‑horizon, i.e. \(k\eta_I \ll 1\). Since \(\cos(k\eta)/(k\eta)\) diverges in this limit, the coefficient \(D(k)\) must vanish to keep the solution finite. Denoting the frozen super‑horizon amplitude inherited from inflation as \(h_I(k)\), we obtain \(C(k) = h_I(k)\), therefore yielding:
\begin{equation} \label{RDsol}
    h_{ij}^{\mathrm{RD}}(k;\eta) = h_I(k)\,\frac{\sin(k\eta)}{k\eta}.
\end{equation}
The amplitude \(h_I(k)\) can be constrained via measurements of the inflationary tensor power spectrum, parameterized by the scalar amplitude \(A_S\) at the pivot scale \(k_* = 0.05\,\mathrm{Mpc}^{-1}\), the tensor‑to‑scalar ratio \(r\), and the tensor spectral index \(n_t\):
\begin{equation}
    h_I^2(k) = \frac{2\pi^2}{k^3} \, r A_S \left(\frac{k}{k_*}\right)^{\! n_t}.
\end{equation}
As reported in \cite{ppplank}, however, only the upper bound
\begin{equation}
    r < 0.036 \qquad (95\%\ \mathrm{C.L.})
\end{equation}
is currently available, precluding a precise determination of \(h_I(k)\). With this observational link established, we match the solutions \eqref{MDsol} and \eqref{RDsol}, together with their derivatives, at \(\eta_{\mathrm{eq}}\) to determine the coefficients \(A(k)\) and \(B(k)\):
\begin{equation}
    \begin{split}
        A(k) &= h_I(k)\left(1 + \sin^2(k \eta_{\mathrm{eq}}) + \frac{\sin(2k \eta_{\mathrm{eq}})}{k \eta_{\mathrm{eq}}}\right), \\[4pt]
        B(k) &= -h_I(k)\left(k \eta_{\mathrm{eq}} + \frac{\sin(2k \eta_{\mathrm{eq}})}{2} - \frac{2 \sin^2(k \eta_{\mathrm{eq}})}{k \eta_{\mathrm{eq}}}\right).
    \end{split}
\end{equation}
Modes that are super‑horizon at matter–radiation equality are of particular interest, as they avoid the suppression experienced by sub‑horizon modes during radiation domination. In the limit \(k \eta_{\mathrm{eq}} \ll 1\), we obtain \(A(k) = 3 h_I(k)\) and \(B(k) = 0\). These modes therefore depend only on the spherical Bessel function \(j_1(x)\):
\begin{equation}
    h_{ij}^{\mathrm{MD}}(k;\eta)\big|_{k\eta_{\mathrm{eq}}\ll1} = 3 h_I(k) \, \frac{j_1(k\eta)}{k\eta}.
\end{equation}
Note that such modes may still become sub‑horizon during matter domination and can therefore exhibit the resonance phenomena discussed in Section~5.2.3. \\
Once a mode with wavenumber \(k\) enters the horizon, its evolution can be approximated by the dominant asymptotic contribution:
\begin{equation}
    \begin{split}\label{freesubhorsolution}
        h_{ij}^{\mathrm{MD}}(k;\eta)\big|_{k\eta \gg 1} & \approx -\frac{1}{k^2 \eta^2} \big[A(k) \cos(k\eta) + B(k)   \sin(k\eta)\big]\\[7pt]
        &= \frac{1}{k^2 \eta^2} \big[ \alpha^*(k) e^{ik\eta} + \alpha(k) e^{-ik\eta} \big],
    \end{split}
    \end{equation}
where we introduced:
\begin{equation} \label{alphak}
    \alpha(k):= -\frac{1}{2} \big(A(k) +i B(k)\big).
\end{equation}

\subsubsection{Mixed state}
This scenario describes a state with zero initial squeezing (\(S_0 = 0\)) but a non-zero occupation number spectrum. Such a state may arise, for instance, if particle production occurs during radiation domination, where the absence of an extended period of accelerated expansion would prevent significant squeezing from developing. The state is consequently mixed (non-pure) due to decoherence or interactions during the production process. The equation of motion reduces to that of a free wave propagating on an FLRW background, supplemented by an additional correction term:
\begin{equation}
    \begin{aligned}[t]
        &\left[ \frac{d^2}{d\eta^2} + k^2 -\frac{2}{\eta^2} + \kappa^2 \frac{N_0}{k_*^{n_N}} \frac{k_{\text{UV}}^{5+n_N}}{12\pi^2 a^3 m} \Gamma\!\left(\frac{n_N +5}{2}\right) \right] \xi_{ij}^{\mathrm{L}}(k;\eta)=0.
    \end{aligned}
\end{equation}
Using the the definition of \(P_{\text{DM},0}\) from eq. \eqref{2ndequation}, we can rewrite this equation as
\begin{equation}
    \left[ \frac{d^2}{d\eta^2} + k^2 -\frac{2}{\eta^2}  + \kappa^2\frac{P_{\text{DM},0}}{a^3} \right] \xi_{ij}^{\mathrm{L}}(k;\eta) = 0.
\end{equation}
Our description of how the ULDM field influences gravitational wave propagation relies on approximations that presuppose the sub‑horizon condition. Under this assumption, the equation of motion can be treated adiabatically, leading to the following solution for sub-horizon modes:
\begin{equation} \label{notfreesubhozsol}
\begin{split}
   h_{ij}^{\mathrm{MD}}(k;\eta)\big|_{k\eta \gg 1} = \frac{1}{k^2 \eta^2}\bigg(1+\frac{\kappa^2 P_{\mathrm{DM},0}}{a^3k^2}\bigg)^{\! -\frac{1}{2}}\bigg[ &  \alpha(k) e^{-i  k \eta_{\text{eq}}} \exp\bigg(\!\! - \! ik\int^\eta_{\eta_{\text{eq}}} \sqrt{1+\frac{\kappa^2 P_{\text{DM},0}}{a^3k^2}}d\eta' \, \bigg)  \\[2pt]
   & \!+ \alpha^*(k)  e^{i k \eta_{\text{eq}}} \exp\bigg(ik\int^\eta_{\eta_{\text{eq}}} \sqrt{1+\frac{\kappa^2 P_{\text{DM},0}}{a^3k^2}}d\eta' \bigg) \bigg],
   \end{split}
\end{equation}
where \(\alpha(k)\) has been defined in eq.~\eqref{alphak} and we have included a phase shift of \(k \eta_{\mathrm{eq}}\) to recover the solution \eqref{freesubhorsolution} in the limit \(P_{\mathrm{DM},0} \to 0\). At leading order, the relative change in the amplitude \(\mathcal{A}(k;\eta)\) and the shift in the phase \(\Phi(k;\eta)\) with respect to the unperturbed solution~\eqref{freesubhorsolution} at a given time \(\eta\) are given by
\begin{equation}
    \frac{\Delta \mathcal{A}(k;\eta)}{\mathcal{A}(k;\eta)} = -\frac{\kappa^2 P_{\mathrm{DM},0}}{2k^2 a^3(\eta)}, \qquad 
    \Delta \Phi(k;\eta) = - \frac{\kappa^2 P_{\mathrm{DM},0}}{10k}\left( \frac{\eta}{a^3(\eta)} - \frac{\eta_{\mathrm{eq}}}{a^3_{\mathrm{eq}}}\right).
\end{equation}
Owing to the smallness of \(\kappa^2 \simeq 3.4 \times 10^{-55}\,\mathrm{eV}^{-2}\)
and the non-relativistic condition~\eqref{wcondition}, these corrections are negligible for sub‑horizon modes. To estimate their upper bound, we adopt the values \(\rho_{\mathrm{DM},0} \sim 10^{-11}\,\mathrm{eV}^4\) and \(H_{\mathrm{eq}} \sim 2.3 \times 10^{-28}\,\mathrm{eV}\). Using the relation \(\eta = 2/(aH)+ \mathrm{const}\) (valid during matter domination) together with condition~\eqref{wcondition}, we obtain
\begin{equation}
    \frac{\Delta \mathcal{A}(k;\eta)}{\mathcal{A}(k;\eta)} \ll (k\eta)^{-2}, \qquad 
    \Delta \Phi(k;\eta) \ll (10 k\eta)^{-1},
\end{equation}
which remain small for modes with \(k\eta \gg 1\).

\subsubsection{Pure squeezed state}

This case describes a state generated during an early period of accelerated expansion (e.g. inflation), which imparts significant squeezing to the field modes. Assuming minimal subsequent decoherence from interactions with other sectors (e.g., radiation, thermal bath), the state remains approximately pure. For a pure Gaussian state, the squeezing and occupation number spectra are related by $S^2_k = N_k(N_k + 1)$. In the high-occupation regime ($N_k \gg 1$), this implies $S^2_k \approx N_k^2$, and consequently the spectral parameters are related as $S_0 \approx N_0$ and $n_S \approx n_N$. The equation of motion for this scenario takes the following form,
\begin{eqnarray} \label{geneqq}
        &&\hskip -0.5cm\bigg\{  \frac{d^2}{d\tilde{\eta}^2} + k^2 -\frac{2}{\eta^2} 
+ \kappa^2 \frac{N_0}{k_*^{n_N}} \frac{k_{\text{UV}}^{5+n_N}}{12\pi^2 a^3 m} \Gamma\!\left(\frac{n_N +5}{2}\right) 
\\
&&\hskip 5cm
\times\bigg(1 +   \cos\bigg[\bigg(2m \, +\,\,\frac{3(n_N+5)}{2}\frac{k_{\text{UV}}^2}{ma^2}\bigg) \, \frac{a\tilde{\eta}}{3}\bigg] \bigg)\bigg\}  \xi_{ij}^{\mathrm{L}}(k;\tilde{\eta})=0,
\nonumber
\end{eqnarray}
where the shifted conformal coordinate \(\tilde{\eta}\) is defined in eq.~\eqref{tildeeta}, and the term \(-2/\eta^2\) is understood as a function of \(\tilde{\eta}\). 
Substituting the values of \(N_0\) and \(n_N\) from eq.~\eqref{parameterspressure}, and using the definition of the present‑day ULDM pressure from eq.~\eqref{2ndequation}, we rewrite the equation as
\begin{equation}
\begin{split}
    \left\{ \frac{d^2}{d\tilde{\eta}^2} + k^2 -\right.&\frac{2}{\eta^2} + \,\frac{\kappa^2P_{\text{DM},0}}{a^3(\eta)} \\
    & \left. \times \left\{1 \!+\!   \cos\left[\left(1 \, -\frac{3}{2}\frac{k_{\text{UV}}^2}{m^2a^2} \!-\! \frac{9w_{\mathrm{DM}}}{2a^2}\bigg(1-\frac{\Omega_{cl}}{\Omega_{\mathrm{DM}}} \bigg)^{\, -1} \right) \,\frac{2am\tilde{\eta}}{3}\right]\right\}\right\}  \xi_{ij}^{\mathrm{L}}(k;\tilde{\eta}) \, =\,0,
    \end{split}
\end{equation}
To proceed, we define the effective squeezing mass \(m_S\), the time-dependent frequency \(\omega_{\xi}(k;\tilde{\eta})\) of the perturbation mode, and the amplitude \(g_1\) of the oscillatory term:
\begin{equation} \label{mathieuparameters}
\begin{split}
m_S(\tilde{\eta}) &:= \frac{a(\tilde{\eta})m}{3} \left[1 - \frac{3}{2} \frac{k_{\text{UV}}^2}{m^2 a^2(\tilde{\eta})} -\frac{9 w_{\mathrm{DM}}}{2 a^2(\tilde{\eta})} \left(1 - \frac{\Omega_{\chi}}{\Omega_{\mathrm{DM}}}\right)^{-1} \right], \\[4pt]
\omega^2_{\xi}(k;\tilde{\eta}) &:= k^2 - \frac{2}{\eta^2} + \frac{\kappa^2P_{\text{DM},0}}{a^3(\tilde{\eta})} , \\[4pt]
g_1(\tilde{\eta}) &:= \frac{1}{2}\frac{\kappa^2P_{\text{DM},0}}{a^3(\tilde{\eta})} ,
\end{split}
\end{equation}
where we denoted $a(\tilde{\eta}) \equiv a(\eta(\tilde{\eta}))$.
Crucially, the conditions \eqref{nonrelativisticcondition} and \eqref{wcondition} guarantee that the deviation \(m_S(t)-m\) stays small over the whole relevant time domain. Rewriting the equation of motion in terms of these quantities yields a Mathieu equation with time-dependent coefficients:
\begin{equation}\label{mathieu}
\left[ \frac{d^2}{d\tilde{\eta}^2} + \omega^2_{\xi}(k;\tilde{\eta}) + 2g_1(\tilde{\eta}) \cos\!\big(2m_S(\tilde{\eta})\, \tilde{\eta} \big) \right] \xi_{ij}^{\mathrm{L}}(k;\tilde{\eta}) = 0.
\end{equation}
When the coefficients \(\omega_{\xi}\), \(m_S\) and \(g_1\) are constant, this equation is known to exhibit parametric resonance, leading to exponentially growing solutions \cite{bogoliubov1961asymptotic}. As established in the previous section, our framework assumes the sub‑horizon condition \(k\eta \gg 1\), which justifies an adiabatic treatment of the time dependence of the parameters. In particular, this ensures \(|\partial_{\tilde{\eta}}{\omega}_{\xi}| \ll \omega^2_{\xi}\), while \(|\partial_{\tilde{\eta}}{m}_S| \ll m_S^2\) follows directly from the definition of the adiabatic parameter in eq.~\eqref{adiabatic}. These conditions allow us to apply the generalization of Floquet theory to slowly varying coefficients developed in \cite{Shtanov_1995}, whose validity precisely requires adiabatic evolution of the characteristic frequencies. 

Crucially, the amplitude \(g_1(t)\) of the oscillating mass term is suppressed by \(\kappa^2\) and the smallness of the ULDM pressure, which is bounded by condition~\eqref{wcondition}. Consequently, the system lies in the narrow resonance regime, where the oscillatory term in eq.~\eqref{mathieu} can be treated perturbatively. This requires:
\begin{equation}\label{narrow resonance condition}
|g_1| \ll \omega^2_{\xi}(k),
\end{equation}
which is satisfied for sub-Hubble modes. 
The solutions to this equation are given by Floquet theory adapted to slowly varying time-dependent backgrounds. For the purposes of this paper, we consider only the lowest resonance band, centered at \(\omega_{\text{res}} = m_S\) with width \(\sim |g_1|\); higher frequency bands contribute at higher order in \(\kappa^2 P_{\mathrm{DM},0}\) and are neglected. The detuning parameter, which quantifies the deviation from exact resonance, is defined for sub‑horizon modes as:
\begin{equation}
\Delta(\eta) := \omega^2_{\xi}(k;\eta) - \omega_{\text{res}}^2(\eta) = k^2 + \frac{\kappa^2 P_{\mathrm{DM},0}}{a^3(\eta)} - \frac{2}{\eta^2} - m_S^2(\eta),
\end{equation}
where we have restored the explicit dependence on \(\eta\), as the shift will drop out of the expressions in what follows.
The two frequencies governing $\Delta$ exhibit distinct dominant time dependencies for sub-horizon modes:
\begin{equation}
\omega_{\xi}(k;\eta) \approx \text{const}, \qquad \omega_{\text{res}} \propto a.
\end{equation}
A given mode \(k\) therefore enters the resonance band at time \(\eta_1(k)\) and exits at \(\eta_2(k)\), where these times satisfy
\begin{equation}
g_1^2(\eta_i) - \Delta^2(\eta_i) = 0, \qquad (i = 1,2).
\end{equation}
The net amplification of the mode is given by the exponential of the integrated positive Floquet exponent \(\mu_+\) over the resonance band:
\begin{equation}
\xi_{\text{enh}}(k) := \exp\!\left( \int_{\eta_1}^{\eta_2} \mu_+(\eta) \, d\eta \right),
\end{equation}
where
\begin{equation}
\mu_+(\eta) := \frac{1}{2\omega_{\text{res}}(\eta)} \sqrt{g_1^2(\eta) - \Delta^2(\eta)}
= \frac{1}{2m_S(\eta)} \sqrt{ \frac{\kappa^4 P^2_{\text{DM},0}}{4a^6(\eta)}- \left( k^2 + \frac{\kappa^2P_{\text{DM},0}}{a^3(\eta)}-\frac{2}{\eta^2} - m_S^2(\eta) \right)^2 }.
\end{equation}
In the narrow resonance regime, the interval \(|\eta_2 - \eta_1|\) is short compared to the Hubble time, and the frequency range \(\Delta \omega := \omega_{\xi}(k) - \omega_{\text{res}}\) is small. We may therefore treat slowly varying quantities as constant during the resonance crossing and change integration variable to \(x = 2\Delta \omega \, \omega_{\text{res}} / |g_1|\). Expanding around \(\Delta \omega = 0\) yields
\begin{equation} \label{enhanc}
\begin{split}
\xi_{\text{enh}}(k) &\approx \exp\!\left( \frac{g_1^2(\eta_{\text{res}})}{4\omega_{\text{res}}^3 \mathcal{H}(\eta_{\text{res}})} \int_{-1}^{1} \sqrt{1-x^2} \, dx \right) \\[2pt]
&= \exp\!\left( \frac{\pi g_1^2(\eta_{\text{res}})}{8\omega_{\text{res}}^3\mathcal{H}(\eta_{\text{res}})} \right) \\[2pt]
&= \exp\!\left( \frac{ \pi}{32} \frac{\kappa^4 P_{\text{DM},0}^2}{m_S^3(\eta_{\text{res}}) \mathcal{H}(\eta_{\text{res}}) a^{6}(\eta_{\text{res}})} \right),
\end{split}
\end{equation}
where the resonant time \(\eta_{\text{res}}(k)\) is defined as the time where the typical frequency of a certain mode corresponds to the resonant frequency of the system,
\begin{equation}
k^2 = m_S^2(\eta_{\text{res}}) + \frac{2}{\eta_{\text{res}}^2} -\frac{\kappa^2P_{\text{DM},0}}{a^3(\eta_{\text{res}})}.
\end{equation}
The resulting enhanced solution will therefore take the following form:
\begin{equation}
    h_{ij}^{\text{res}}(k;\eta) \approx h_{ij}^{\mathrm{MD}}(k;\eta) \Big[  \Theta\big(\eta_{\mathrm{res}}(k)-\eta\big) + \xi_{\mathrm{enh}}(k) \Theta\big(\eta-\eta_{\mathrm{res}}(k)\big) \Big],
\end{equation}
where $h_{ij}^{\mathrm{MD}}(k;\eta)$ is given in \eqref{notfreesubhozsol}. To estimate the magnitude of this effect across modes, we adopt the leading-order approximations \(m_S \approx am/3\) and \(\omega_{\xi}(k) \approx k\), valid in the non-relativistic and narrow resonance regimes. This gives
\begin{equation}
\xi_{\text{enh}}(k) \approx \exp\!\left( \frac{27\pi}{32} \frac{\kappa^4 P^2_{\text{DM},0} \, m^{\frac{11}{2}} z_{\text{eq}}^{\frac{3}{2}}}{k^{\frac{17}{2}} H_{\text{eq}}} \right).
\end{equation}
Resonant amplification only occurs for modes that enter the band after matter-radiation equality, i.e., those satisfying 
\begin{equation}\label{boundsenhancing}\vspace{3mm}
k \gtrsim m a_{\text{eq}} \sim 10^{-28} \text{--} 10^{-25}~\mathrm{eV},
\end{equation}
corresponding to comoving wavelengths \(\lambda \lesssim 10^3 \text{--} 10^6~\mathrm{pc}\). These modes are well inside the horizon throughout the period of interest. For modes with \(k \sim m a_{\text{eq}}\), the enhancement is maximized. To estimate the upper bound for the enhancement due to the squeezing-induced resonance effect, we adopt the values \(\kappa^2 \sim 1.7 \times 10^{-55}\,\mathrm{eV}^{-2}\) and \(H_{\mathrm{eq}} \sim 2.3 \times 10^{-28}\,\mathrm{eV}\), yielding:
\begin{equation}
\xi_{\text{enh}}(k) \lesssim \exp\!\left( 10^{15} \text{--} 10^{24}~\mathrm{eV}^{-8} P_{\mathrm{DM},0}^2 \right).
\end{equation}
Taking as an upper bound the scenario in which dark matter pressure and energy density were equal at matter-radiation equality, the present-day pressure would be of order \(P_{\mathrm{DM},0} \sim 10^{-18}\,\mathrm{eV}^4\). This yields a completely negligible enhancement for sub-horizon modes:
\begin{equation} \label{finalresult}
\xi_{\text{enh}}(k) -1 \lesssim 10^{-12}.
\end{equation}
\subsubsection{Mixed squeezed state}
This scenario incorporates both non-zero squeezing and finite state mixing, arising from interactions, decoherence, or non-adiabatic effects during or after production. For any physical Gaussian state, the squeezing spectrum satisfies \(S_k^2 \leq N_k(N_k + 1)\). In the high‑occupation limit and for power‑law spectra, this inequality becomes
\begin{equation} \label{constraint}
    S_0 \,\bigg( \frac{k}{k_*}\bigg)^{\! n_S} \leq N_0 \, \bigg( \frac{k}{k_*}\bigg)^{\! n_n} , \qquad k \leq k_{\text{UV}},
\end{equation}
where the inequality allows for adjustable purity. The resulting Mathieu equation generalizes eq.~\eqref{geneqq}:
\begin{equation}
    \begin{split}
        &\left\{ \frac{d^2}{d\tilde{\eta}^2} + k^2 - \frac{2}{\eta^2} + \kappa^2 \frac{N_0}{k_*^{n_N}} \frac{k_{\text{UV}}^{5+n_N}}{12\pi^2 a^3 m} \Gamma\!\left(\frac{n_N+5}{2}\right) \right. \\[3pt]
        & \, \, \, \left. + \kappa^2 \frac{S_0}{k_*^{n_S}} \frac{k_{\text{UV}}^{5+n_S}}{12\pi^2 a^3 m} \Gamma\!\left(\frac{n_S+5}{2}\right) 
        \cos\!\left[ \left(2m + \frac{3(n_S+5)}{2}\frac{k_{\text{UV}}^2}{m a^2}\right) \frac{a\tilde{\eta}}{3} \right] \right\} \xi_{ij}^{\mathrm{L}}(k;\tilde{\eta}) = 0.
    \end{split}
\end{equation}
The frequency \(\omega_{\xi}\) remains the same as in the previous section, while the amplitude \(g_1\) of the oscillatory term is bounded by eq.~\eqref{constraint} to be smaller than that of the pure squeezed state. Consequently, the resonance band width is reduced, and so is the resulting enhancement \(\xi_{\text{enh}}\) experienced by the growing solutions, as seen from eq.~\eqref{enhanc}. The pure squeezed state thus provides an upper bound on the enhancement attainable in the mixed squeezed state, and the bound presented in eq.~\eqref{finalresult} therefore also applies to this case.

\section{Conclusions}
We have presented a first‑principles quantum field theoretic framework for describing how a general Gaussian initial state of ultralight scalar dark matter (ULDM) affects the evolution of linear cosmological perturbations during the matter‑dominated era. Starting from the two‑particle‑irreducible (2PI) effective action on the Schwinger‑Keldysh contour and adopting the adiabatic (WKB) approximation for the matter sector, we obtained renormalized evolution equations for the tensorial and scalar perturbations in the longitudinal gauge. The most important novel contribution to 
the effective action is the one-loop graviton self-energy induced by the massive minimally coupled scalar field, which mimics ultralight dark matter 
(ULDM). For the purposes of this work, we compute the self-energy at the leading adiabatic order, meaning that our approximations apply when the scalar mass $m$ is much larger than the expansion rate of the Universe $H(t)$, which is amply satisfied throughout matter era. We renormalize 
the self-energy by using dimensional regularization and we assume that dark matter is in a general Gaussian state, 
but make no further simplifications.

The analysis proceeded in three main stages. First, we obtained a closed equation of motion for the graviton field \eqref{endpoint} in which the 4‑point self‑energy of the condensate precisely cancels against the corresponding local terms of the Einstein tensor. This cancellation can be interpreted as a consequence of the Noether-Ward identity \cite{Prokopec:2025jrd} and implies that any effect of the condensate on the propagation of primordial gravitational wave modes disappears already at the linear level,  leaving only the two‑point correlators of the ULDM field to imprint themselves on the graviton dynamics. Second, we computed the one‑loop energy‑momentum tensor and the retarded graviton self‑energy for a generic Gaussian ULDM initial state within the adiabatic approximation, treating non‑local contributions through the mid‑point prescription. The vacuum sector was renormalized through the action \eqref{countertermact} and the choice of counterterms in \eqref{countertermscoeff}, while the state‑dependent contributions were expressed in terms of momentum integrals over the occupation number and squeezing spectra, both for the local \eqref{localsemixing}, \eqref{localsesqueezing} and non‑local \eqref{n11}–\eqref{s13} parts. Third, we fixed the gauge to the longitudinal form, where the transverse‑traceless sector simplifies to a Mathieu equation \eqref{xiresonanceequation}, featuring a slowly varying graviton mass driven by the occupation number and an oscillatory mass induced by the squeezing. The non‑local contribution to this equation, which we estimated analytically in Appendix \ref{Appendix D}, is suppressed relative to the local terms through the scaling presented in \eqref{scalingnonlocalterms} under the assumption that the condition \eqref{phasecondition} holds and can be treated as relativistic corrections. A fully rigorous inclusion of these non‑local terms can only be achieved through a dedicated numerical treatment, which would also serve as an independent test of the analytic suppression estimates obtained here.

Our analysis of the GW equation yields two central results for the evolution of the primordial tensor spectrum during the matter‑dominated era.
\begin{itemize}
   \item The homogeneous ULDM condensate decouples completely from the graviton equation of motion; tensor modes are therefore insensitive to the classical misalignment amplitude and are shaped exclusively by the two‑point statistics of the dark matter field.  Consequently, no observable gravitational wave imprint can arise from a purely coherent treatment of the scalar field.
   \item The time‑periodic effective mass induced by mode squeezing can induce a resonant growth of specific \(k\)-modes.  The maximum enhancement is set by the dark matter quantum pressure, and imposing the condition of non‑relativistic ULDM at matter–radiation equality yields an upper bound on the relative resonant growth of \(\sim 10^{-12}\) for masses \(m \sim 10^{-24}\text{--}10^{-21}\,\mathrm{eV}\).  This is far too small to be detectable by current or foreseeable CMB or gravitational‑wave experiments.  Mixed squeezed states produce even weaker resonances, with the pure squeezed state providing the absolute upper bound.
\end{itemize}
For scalar perturbations we have derived a closed set of equations for the Bardeen potentials \eqref{potentials1}-\eqref{potential2}. Unlike the tensor sector, the evolution of scalar potentials is influenced by the ULDM condensate as well as by occupation number and squeezing. The resulting system exhibits oscillatory source terms that could, in principle, drive resonance phenomena analogous to those studied for gravitational waves. A detailed analysis of these scalar dynamics is left for a separate investigation, as it constitutes an important prediction of the model and may reveal distinctive signatures in large‑scale structure and CMB anisotropies. Importantly, in the Minkowski limit \(a = 1\), these equations differ qualitatively from those employed in \cite{Khmelnitsky_2014}, since they contain only linear terms in the potentials. This suggests that the bounds derived from that study, including those reported in \cite{Porayko_2014, Porayko_2018}, may warrant re‑examination.

The present work is intended as a first step toward a rigorous quantum field theoretic description of gravitational perturbations coupled to dark matter; it does not aspire to be exhaustive but focuses on the simplest consistent model, a free scalar minimally coupled to Hilbert-Einstein gravity at linear order, and it adopts a set of physically motivated approximations to obtain explicit quantitative results. Consequently, there is ample room for future improvements and extensions: including the treatment of super‑horizon modes, relaxing the adiabatic assumption, incorporating the transition from radiation to matter domination, or adding self‑interactions, non‑minimal couplings, and non‑Gaussianities. A numerical treatment of the full non‑local renormalized self‑energy would also refine the predictions and test the robustness of the analytic estimates provided here.
When viewed in a broader context, this work contributes towards understanding the evolution of gravitational perturbations 
from primordial inflation, where they were created, via radiation~\cite{Ota:2022xni,Ota:2023iyh,Frob:2025sfq,Sasaki:2025zao,Glavan:2026wwl,Liu:2026, BaghouzianFennemaProkopec:2026}
and matter era towards the late dark energy epoch that goes beyond the tree-level treatment of matter. 

\acknowledgments
 T.P. acknowledges support by the Delta ITP
consortium, a program of the Netherlands’ Organisation for Scientific Research (NWO), which is funded by the Dutch Ministry of Education, Culture
and Science (OCW) – NWO project number 24.001.027; by the NWA ORC
2023 consortium grant: Cosmic emergence: from abstract simplicity to complex diversity, and by The Magnetic Universe grant – NWO project number
OCENW.XL.23.147.

\appendix

\section{Differential operators and interaction vertices} \label{appendixA}
In this appendix we write the full expression for the vertices and differential operators appearing in the main text, which are obtained by deriving the classical Schwinger-Keldysh action $S_{SK}$. 

The perturbative matter action gives rise to interaction vertices that encode the coupling between the matter field and gravitational perturbations. In the Schwinger–Keldysh (in-in) formalism, these vertices carry a Keldysh index \(l = \pm\), labeling the forward (\(+\)) and backward (\(-\)) branches of the closed time contour, which correspond to forward and backward time evolution, respectively. Consequently, the matter vertices also acquires this index. Practically, the only effect is a relative sign: vertices on the forward branch coincide with the usual in-out ones, while those on the backward branch come with the opposite sign. 

The interaction vertices originate from the first- and second-order expansions of the matter action in the graviton perturbation $h_{\mu\nu}$. Specifically, the term $\epsilon S_m^{(1)}[a,\bar{\chi}_c, h_{\mu \nu,c}]$ generates three-point vertices coupling two matter field lines to a single graviton line. Similarly, the $\epsilon^2 S_m^{(2)}[a,\bar{\chi}_c, h_{\mu \nu,c}]$ term gives rise to four-point vertices involving two matter fields and two gravitons. In addition, the presence of a non-vanishing matter condensate, $\bar{\chi}_c \neq 0$, yields an extra contribution to the three-point vertex arising from the second-order term $\epsilon^2 S_m^{(2)}[\bar{\chi}_c, a, h_{\mu \nu,c}]$. This contribution is proportional to the background field and describes an effective three-point interaction mediated by the expectation value of the matter field. Although this vertex will not be used in our analysis since diagrams involving it contribute only at order $\mathcal{O}(\epsilon^4)$, beyond the perturbative order considered here, we include it in the present section for completeness.

We now turn to the explicit expressions for the amputated interaction vertices derived from the matter sector. Because the interaction terms in the action contain derivatives acting on the matter field, special care is required when constructing the corresponding Feynman diagrams to ensure the correct implementation of the derivative structure. To handle this systematically, we assign numerical labels ($1$, $2$) to each matter-field leg of a vertex. These labels track which derivative acts on which external line, thereby preserving the proper momentum dependence of each interaction in the amputated vertices. With this convention established, we present both the diagrammatic and analytic forms of the amputated interaction vertices relevant to our theory:
\begin{center}
\begin{minipage}[t]{0.48\textwidth}
\vspace{-0.5em}
\centering
$\vcenter{\hbox{%
\begin{fmfgraph*}(80,40)
    \fmfleft{ul,dl}
    \fmfright{r}
    \fmf{plain}{ul,d1}
    \fmf{plain}{dl,d1}
    \fmf{gluon}{d1,r}
    \fmfdot{d1}
    \fmfv{label=$x,,c$,label.dist=2mm, label.angle=65}{d1}
    \fmfv{label=$1$}{ul}
    \fmfv{label=$2$}{dl}
    \fmfv{label=\hspace{-0.8cm}\vspace{0.8cm}$\alpha\beta$}{r}
  \end{fmfgraph*}}}\; \propto  \, \epsilon [iV_{(3),c}^{\alpha\beta}](x|1;2); $
\end{minipage}
\hfill
\begin{minipage}[t]{0.48\textwidth}
\vspace{-0.5em}
\centering
$\vcenter{\hbox{%
\begin{fmfgraph*}(80,40)
    \fmfleft{ur,dl}
    \fmfright{ul,dr}
    \fmf{plain}{ul,d1}
    \fmf{plain}{dl,d1}
    \fmf{gluon}{d1,ur}
    \fmf{gluon}{d1,dr}
    \fmfdot{d1}
    \fmfv{label=$x,,c$, label.dist=3mm, label.angle=90}{d1}
    \fmfv{label=$1$}{ul}
    \fmfv{label=$2$}{dl}
    \fmfv{label=$\alpha\beta$,label.side=left}{ur}
    \fmfv{label=$\rho\sigma$,label.side=right}{dr}
  \end{fmfgraph*}}} \;\propto  \epsilon^2 \,[{}^{\alpha\beta}iV_{(4),c}^{\rho\sigma}](x;x'|1;2)$
\end{minipage}
\end{center}
\vspace{1.2em}
\begin{flalign*}
& \vcenter{\hbox{%
  \begin{fmfgraph*}(80,40)
    \fmfleft{ul,dl}
    \fmfright{r}
    \fmf{gluon}{ul,d1}
    \fmf{gluon}{dl,d1}
    \fmf{plain}{d1,r}
    \fmfdot{d1}
    \fmfv{label=$x,,c$,label.dist=2mm, label.angle=65}{d1}
    \fmfv{label=\hspace{-0cm}\vspace{0cm}$\alpha\beta$}{dl}
    \fmfv{label=$\rho \sigma$}{ul}
  \end{fmfgraph*}%
}} \; \propto  \, \epsilon^2 [{}^{\alpha\beta} iV_{(3b),c}^{\rho\sigma}][\bar{\chi}_c](x;x')
. \end{flalign*}
where:
\begin{equation} \label{verticesreal}
    \begin{split}
      [iV_{(3),c}^{\alpha\beta}](x|1;2) &= -i\, c  \left[ -\partial_{1}^{(\alpha} \partial_{2}^{\beta)} - \delta_0^{\alpha}\delta_0^{\beta}\mathcal{H}^2 \left( \frac{D-2}{2} \right)^2 + \frac{D-2}{2} \mathcal{H} \delta_0^{(\alpha} (\partial_{2}^{\beta)} + \partial_{1}^{\beta)})  \right.\\[3pt]
      &  + \left. \frac{1}{2}\eta^{\alpha \beta} \left( m^2a^2 + \partial_1 \cdot \! \partial_2 - \mathcal{H}^2 \left( \frac{D-2}{2} \right)^2 +  \frac{D-2}{2} \mathcal{H} (\partial_0^1 + \partial_0^2 )\right) \right];
      \end{split}
      \end{equation}
      \begin{equation}
    \begin{split}
     [{}^{\alpha\beta} iV_{(3b),c}^{\rho\sigma}][&\bar{\chi}_c](x;x') = - 2i \, c\,  \delta^D(x-x')\\
     &\times\left\{\partial^{(\alpha}\bar{\chi}_c(x)\eta^{\beta)(\rho} \partial^{\sigma)} -  \frac{1}{4} \eta^{\alpha \beta} \partial^{(\rho}\bar{\chi}_c(x) \partial^{\sigma)}- \frac{1}{4} \eta^{\rho \sigma} \partial^{(\alpha}\bar{\chi}_c(x) \partial^{\beta)} \right.  \\[3pt]
      &   \left.+ \mathcal{H}^2 \left( \frac{D-2}{2} \right)^{\!2} \left[ \delta_0^{(\alpha} \eta^{\beta)(\rho}  \delta_0^{\sigma)}   - \frac{1}{4} \eta^{\alpha \beta} \delta_0^{\rho}\delta_0^{\sigma}  - \frac{1}{4} \eta^{\rho \sigma} \delta_0^{\alpha} \delta_0^{\beta} \right] \bar{\chi}_c(x) 
       \right.  \\
       &  
       -  \frac{D-2}{4} \mathcal{H} \bigg[ \delta_0^{(\alpha} \eta^{\beta)(\rho}(\partial^{\sigma)}\bar{\chi}_c(x) 
        + \bar{\chi}_c(x) \partial^{\sigma)})  + \delta_0^{(\rho} \eta^{\sigma)(\alpha}(\partial^{\beta)}\bar{\chi}_c(x)
        + \bar{\chi}_c(x)\partial^{\beta)}) 
       \\[3pt]
      &  - \frac{1}{2} \eta^{\alpha \beta} \delta_0^{(\rho} (\partial^{\sigma)}\bar{\chi}_c(x)+ \bar{\chi}_c(x)\partial^{\sigma)}) 
      - \frac{1}{2} \eta^{\rho \sigma} \delta_0^{(\alpha} (\partial^{\beta)}\bar{\chi}_c(x)+  \bar{\chi}_c(x) \partial^{\beta)}) \bigg]
      \\[3pt]
      &  + \frac{1}{8}\left(\eta^{\alpha \beta} \eta^{\rho \sigma} - 2\eta^{\alpha (\rho} \eta^{\sigma) \beta}\right) \bigg[\bar{\chi}_c(x) m^2a^2 
       + \partial\bar{\chi}_c(x) \cdot \! \partial   \\[3pt]
      & \left. -\mathcal{H}^2 \left( \frac{D-2}{2} \right)^{\!2} \! \! \bar{\chi}_c(x)
    +  \frac{D-2}{2} \mathcal{H} (\partial_0 \bar{\chi}_c(x) + \bar{\chi}_c(x) \partial_0 )\bigg] \right\}; 
      \end{split}
      \end{equation}
      \begin{equation}
    \begin{split}
      [{}^{\alpha\beta}i&V_{(4),c}^{\rho\sigma}](x;x'|1;2) = - i c \,\delta^D(x-x')\\[3pt]
      &\times  \left\{ \frac12 \partial_{1}^{(\alpha}\eta^{\beta)(\rho} \partial_{2}^{\sigma)} + \frac12 \partial_{2}^{(\alpha}\eta^{\beta)(\rho} \partial_{1}^{\sigma)} -  \frac{1}{4} \eta^{\alpha \beta} \partial_{1}^{(\rho} \partial_{2}^{\sigma)}- \frac{1}{4} \eta^{\rho \sigma} \partial_{1}^{(\alpha} \partial_{2}^{\beta)} \right. \\[3pt]
      &
      + \mathcal{H}^2 \left( \frac{D-2}{2} \right)^{\!2}\left[ \delta_0^{(\alpha} \eta^{\beta)(\rho}  \delta_0^{\sigma)}    - \frac{1}{4} \eta^{\alpha \beta} \delta_0^{\rho}\delta_0^{\sigma}  - \frac{1}{4} \eta^{\rho \sigma} \delta_0^{\alpha} \delta_0^{\beta} \right] -  \frac{D-2}{4} \mathcal{H}  \\[3pt]
      & \times \bigg[ \delta_0^{(\alpha} \eta^{ \beta)(\rho}(\partial_{1}^{\sigma)}+ \partial_{2}^{\sigma)}) 
      + \delta_0^{(\rho} \eta^{\sigma)(\alpha}(\partial_{1}^{\beta)}+ \partial_{2}^{\beta)})   - \frac{1}{2} \eta^{\alpha \beta} \delta_0^{(\rho} (\partial_{1}^{\sigma)}+ \partial_{2}^{\sigma)}) - \frac{1}{2} \eta^{\rho \sigma} \delta_0^{(\alpha} (\partial_{1}^{\beta)}+ \partial_{2}^{\beta)}) \bigg] \\[3pt]
      &
      \left.  +  \left(\frac{1}{8}\eta^{\alpha \beta} \eta^{\rho \sigma} - \frac{1}{4}\eta^{\alpha (\rho} \eta^{\sigma) \beta}\right)  \left[m^2a^2 + \partial_1 \cdot \!\partial_2  - \mathcal{H}^2 \left( \frac{D-2}{2} \right)^2 +  \frac{D-2}{2} \mathcal{H} (\partial_1^0 + \partial_2^0 )\right] \right\}.
    \end{split}
\end{equation}
We continue by writing the expressions for the differential operators of the theory. The in-in Lichnerowitz operator for a FLRW background is obtained by deriving the $\mathcal{O}(\epsilon^2)$ gravitational action:
\begin{equation} \label{appendixA:gravitonoperator}
    \begin{split}
        [^{\mu \nu}\mathcal{L}^{\rho \sigma}_{cd}](x;x')  &=  \frac{\delta^2 S_{SK,g}^{(2)}}{\epsilon^2 \delta h_{\mu \nu,c}(x) \delta h_{\rho \sigma,d}(x')}\\
        &= -\,  c \,\delta_{cd} \,  \frac{a^{D-2}}{\kappa^2} \bigg\{ 2(D-2) \big[(D-3)\mathcal{H}^2 + 2 \mathcal{H}' \big] \bigg( \frac{1}{8} \eta^{\mu \nu} \eta^{\rho \sigma} - \frac{1}{4} \eta^{\mu (\rho} \eta^{\sigma) \nu} \bigg)  \\[3pt]
        &  +  2 (D-2) \big( \mathcal{H}^2 - \mathcal{H}'\big)  \bigg( \delta_0^{(\mu} \eta^{\nu) ( \rho} \delta_0^{\sigma)} - \frac{1}{4} \eta^{\mu \nu} \delta_0^{\rho} \delta_0^{\sigma} - \frac{1}{4}  \eta^{\rho \sigma} \delta_0^{\mu} \delta_0^{\nu} \bigg) + \big( \eta^{\mu \nu } \delta_0^{\rho} \delta_0^{\sigma} \\[3pt]
       &   + \eta^{\rho \sigma} \delta_0^{\mu} \delta_0^{\nu} \big)\frac{D-2}{4} \big[ \mathcal{H}' + (D-2)\mathcal{H}^2\big] + \mathcal{H}\frac{D-2}{2} \big[ \eta^{\mu \nu} \eta^{\rho \sigma} \partial_0 + \eta^{\rho \sigma} \delta_0^{(\mu} \partial^{\nu)} \\[3pt]
       & + \eta^{\mu \nu} \delta_0^{(\rho}\partial^{\sigma)}  - \eta^{\mu ( \rho} \eta^{\sigma) \nu} \partial_0  - \delta_0^{(\mu} \eta^{\nu)(\rho} \partial^{\sigma)} -  \delta_0^{(\rho} \eta^{\sigma)(\mu} \partial^{\nu)} \big] + [{}^{\mu \nu}\mathcal{L}_{(\eta)}^{\, \rho \sigma}] \bigg\} \delta^D (x-x'),
    \end{split}
\end{equation}
where $[{}^{\mu \nu}\mathcal{L}_{(\eta)}^{\, \rho \sigma}] $ is the amputated flat-space Lichnerowicz operator
\begin{equation}\label{appendixA:flatoperator}
    [{}^{\mu \nu}\mathcal{L}_{(\eta)}^{\, \rho \sigma}] (x)= \frac{1}{2} \big(- \eta^{\mu \nu}\eta^{\rho \sigma} \partial^2 + \eta^{\rho \sigma} \partial^{\mu}\partial^{\nu} + \eta^{\mu \nu} \partial^{\rho} \partial^{\sigma} + \eta^{\mu ( \rho} \eta^{\sigma) \nu} \partial^2 - 2 \partial^{(\mu}\eta^{\nu) (\rho} \partial^{\sigma)} \big). 
\end{equation}
We conclude this appendix by presenting the corresponding operator for the matter field. Its expression follows directly from the vertices introduced above:
\begin{equation} \label{appendixA:matteroperator}
    \begin{split}
         \mathcal{D}_{cd}(x;x')&:= \sum_{i=0}^2  \frac{\delta^2 S_{SK,m}^{(i)}}{\delta \bar{\chi}_c(x) \delta \bar{\chi}_d(x')} \\
         &= \, \delta_{cd}\,c  \, \delta^D(x-y) \, \bigg[ \partial^2 - a^2m^2 + \mathcal{H}^2 \left(\frac{D-2}{2}\right)^2 + \mathcal{H}' \frac{D-2}{2} \bigg] \, \\
       & -\, i \epsilon \, \delta_{cd} \, \int_{w} \,  h_{\alpha \beta,c} (w) \,  [iV_{(3),c}^{\alpha\beta}](w|1;2) \, \delta^D(w-y|1) \, \delta^D(w-x|2) \, \\
      & - \, i \epsilon^2\, \delta_{cd} \, \int_{w_1,w_2} \! \! \! \! \! \! \! \! \! \! \! h_{\alpha \beta,c} (w_1)\, h_{\rho \sigma,c} (w_2) \, [{}^{\alpha\beta} iV_{(4),c}^{\rho\sigma}](w_1;w_2|1;2) \, \delta^D(w_1-y|1) \, \delta^D(w_1-x|2) \,.
    \end{split}
\end{equation}

\vskip 1cm

\section{Computation of the retarded self-energy} \label{Appendix B}

In this appendix, we perform most of the calculations that lead to the expressions
for the self-energy and energy-momentum tensor presented in Chapter 3. The following derivation
must be understood within the adiabatic regime established in the main text, which justifies neglecting
derivatives of the scale factor.

\subsection{Vacuum contribution}
This section will treat the vacuum contribution to the retarded self-energy.

\subsubsection{Computation of the ++ component}
We are going to use the expression for the vacuum Feynman propagator derived in (\ref{propagatorexpression}):
\begin{equation}
    i \Delta_{ V}(x;x')  = \, \frac{(\bar{a}m)^{D-2}}{(2\pi)^{\frac{D}{2}}} \frac{K_{\frac{D-2}{2}}(y_{++})}{(y_{++})^\frac{D-2}{2}},
\end{equation}
where, in this section, we denote \(i\Delta_{++}^{V} \equiv i \Delta_V\), as this is the only quantity
we will need. Here $K_\nu(z)$ is the modified Bessel function of the second kind (Macdonald function) and $y$ is the
$i\varepsilon$-shifted space-time distance defined by
\begin{equation} \label{y++}
    y^2_{++}(x;x'):= m^2 \Delta x_{++}^2,\quad \Delta x_{++}^2= -(|t - t'|- i \varepsilon)^2 + \bar{a}^2\| \vec{x}- \vec{x}\,' \|^2 \qquad (\varepsilon>0)
\,.\quad
\end{equation}
The Macdonald function $K_\nu(z)$ can also be expanded as a series~\cite{Mancha_2021},
\begin{equation}
\frac{K_{\frac{D-2}{2}}(y)}{y^{\frac{D-2}{2}}} = \Gamma\left(1 - \frac{D}{2}\right) \left(\frac{y}{2}\right)^{\frac{D}{2}} \sum_{n=0}^{\infty} \frac{(y/2)^{2n}}{\left(\frac{D}{2}\right)_n n!}
+ \Gamma\left(\frac{D}{2} - 1\right) \left(\frac{y}{2}\right)^{\frac{D}{2}-1} \sum_{n=0}^{\infty} \frac{(y/2)^{2n+2-D}}{(2 - D)_n n!},
\end{equation}
where $(\alpha)_n:= \frac{\Gamma \left(\alpha + n\right)}{\Gamma(\alpha)}$ denotes the Pochhammer symbol. 
We start our analysis by discussing the 4-point contribution. Using the definition
of such contribution \eqref{definition4ptse} and the expression for the adiabatic vertices
\eqref{adiabatic vertices} we obtain:
\begin{equation}
    \begin{split}
        i[^{\alpha \beta}\Sigma_{\text{4pt},++}^{\, \rho \sigma}](x;x')&= \, i \kappa^2 \delta^D(x-x') \left[  \left(\frac{1}{8}\eta^{\alpha \beta} \eta^{\rho \sigma} - \frac{1}{4}\eta^{\alpha (\rho} \eta^{\sigma) \beta}\right)   \left(m^2a^2 + \partial \cdot \partial'\right)\right. \\
         & \left.+ \frac12 \partial'^{(\alpha}\eta^{\beta)(\rho} \partial^{\sigma)}+ \frac12 \partial^{(\alpha}\eta^{\beta)(\rho} \partial'^{\sigma)} -   \frac{1}{4} \eta^{\alpha \beta} \partial^{(\rho} \partial'^{\sigma)} -  \frac{1}{4} \eta^{\rho \sigma} \partial^{(\alpha} \partial'^{\beta)} \right] i\Delta_V(x;x').
    \end{split}
\end{equation}
The vacuum propagator depends on the space-time points $x,x'$ solely by their difference
$\Delta x^2_{++}$. Consequently, the derivative of this object with respect to $x'$ satisfies a reflection identity,
$\partial'^{\mu} i\Delta_V = -\partial^{\mu} i\Delta_V $. Using this property we obtain the following identities:
\begin{equation}
    \begin{split}
       & \lim_{x'\to x} \left(m^2a^2 + \partial \cdot \partial'\right) i\Delta_V(x;x') = 0; \\
       &\lim_{x'\to x} \, \partial^{\mu}\partial'^{\nu} i\Delta_V(x;x') = - \eta^{\mu \nu} \frac{(am)^D}{D(4 \pi)^{D/2}} \Gamma\left(1-\frac{D}{2}\right) = -  \eta^{\mu \nu} \frac{(am)^2}{D} i\Delta_V(x;x).
    \end{split}
\end{equation}
Knowing this, we can write:
\begin{equation} \label{4pointcontr}
          i[^{\alpha \beta}\Sigma_{\text{4pt},++}^{\, \rho \sigma}](x;x')= - i\delta^D(x-x') \frac{(am)^2}{2D}  \left[ 2 \eta^{\alpha ( \rho} \eta^{\sigma)\beta} - \eta^{\alpha \beta} \eta^{\rho \sigma} \right] i\Delta_V(x;x).
\end{equation}
The 3-point contribution will have a much more complicated expression. We have:
\begin{equation}
    \begin{split}
       i[^{\alpha \beta}\Sigma&_{\text{3pt},++}^{\, \rho \sigma}](x;x')\\
       &:=\frac{\kappa^2}{2}\left[ \partial^{\alpha)}\partial'^{(\rho}  i\Delta_V(x;x') \partial'^{\sigma)}\partial^{(\beta}   i\Delta_V(x;x')-\frac{1}{2}\eta^{\alpha \beta}  \partial_{\lambda}  \partial'^{(\rho}i\Delta_V(x;x') \partial'^{\sigma)} \partial^{\lambda}  i\Delta_V(x;x')   \right. \\
        &-  \frac{1}{2}\eta^{\rho \sigma} \partial^{\alpha)}\partial_{\lambda}'  i\Delta_V(x;x') \partial'^{\lambda}\partial^{(\beta}  i\Delta_V(x;x') + \frac{1}{4} \eta^{\rho \sigma} \eta^{\alpha \beta}  \partial_{\gamma}\partial_{\lambda}'  i\Delta_V(x;x') \partial'^{\lambda} \partial^{\gamma} i\Delta_V(x;x')  \\
        & + \frac{1}{2}\eta^{\alpha \beta}m^2a^2\left(-\partial'^{(\rho} i\Delta_V(x;x') \partial'^{\sigma)} i\Delta_V(x;x')  + \frac{1}{2} \eta^{\rho \sigma}
\partial_{\lambda}'   i\Delta_V(x;x')  \partial'^{\lambda}   i\Delta_V(x;x') \right)  \\
&+ \frac{1}{2}\eta^{\rho \sigma}m^2a'^{\,2} \left( - \partial^{(\alpha} i\Delta_V(x;x') \partial^{\beta)}  i\Delta_V(x;x') + \frac{1}{2} \eta^{\alpha \beta}
\partial_{\lambda}  i\Delta_V(x;x')   \partial^{\lambda}   i\Delta_V(x;x') \right) \\
& + \left. \frac{1}{4} \eta^{\alpha \beta}\eta^{\rho \sigma} m^4 a^2a'^{\,2}  i\Delta_V^2(x;x') \right],
    \end{split}
\end{equation}
where we introduced the notation $a\equiv a(t)$, $a'\equiv a(t')$.

To find an explicit expression for this term, we have to find a proper expression for:
\begin{equation}
    \begin{split}
    &  \mathcal{A}^{\alpha \beta} (x;x') := \partial^{\alpha} i\Delta_V(x;x') \partial^{\beta}  i\Delta_V(x;x');\\
& \mathcal{B}^{\alpha \beta \rho \sigma} (x;x') :=\partial^{\alpha)}\partial^{(\rho}  i\Delta_V(x;x') \partial^{\sigma)}\partial^{(\beta}  i\Delta_V(x;x') .
    \end{split}
\end{equation}
Due to the symmetries of these expressions, we know that the most general form they will have is:
\begin{equation}
    \begin{split} 
    \mathcal{A}^{\alpha \beta} (x;x')&= F_1 \eta^{\mu \nu} + \frac{\partial^{\alpha}\partial^{\beta}}{\partial^2} F_2;\\
    \mathcal{B}^{\alpha \beta \rho \sigma} (x;x') &= G_1 \eta^{\alpha \beta} \eta^{\rho \sigma} + G_2 \eta^{\alpha (\rho}\eta^{\sigma) \beta} + \left(\eta^{\alpha \beta} \frac{\partial^{\rho}\partial^{\sigma}}{\partial^2} + \eta^{\rho \sigma} \frac{\partial^{\alpha}\partial^{\beta}}{\partial^2}\right) G_3 +\\
    &+ 2 \eta^{\alpha)(\rho}\frac{\partial^{\sigma)}\partial^{(\beta}}{\partial^2} G_4+ \frac{\partial^{\alpha}\partial^{\beta}\partial^{\rho}\partial^{\sigma} }{\partial^4}G_5.
    \end{split}
\end{equation}
To obtain the coefficients $F_i, G_j$ we can contract such expressions with the inverse of
the tensor structures that appear in such expressions. We obtain the following terms:
\begin{equation} \label{coeff-rel}
\begin{split}
   &\partial_{\mu}i\Delta_V\partial^{\mu}i\Delta_V:=R_1=DF_1 + F_2; \\
   &\frac{\partial^{\alpha}\partial^{\beta}}{\partial^2} \partial_{\alpha} i\Delta_V \partial_{\beta}  i\Delta_V:=R_2 = F_1+F_2;\\
    &\partial_{\mu}\partial_{\nu} i \Delta_V \partial^{\mu}\partial^{\nu} i \Delta_V := S_1 = D^2 G_1+ D G_2+2 DG_3 +2G_4 + G_5; \\
   & \eta^{\alpha (\rho}\eta^{\sigma) \beta} \partial_{\alpha)}\partial_{(\rho}  i\Delta_V \partial_{\sigma)}\partial_{(\beta}  i\Delta_V := S_2 = DG_1 + \frac{D}{2} (D+1) G_2 + 2 G_3 + (D+1)G_4 + G_5; \\
   &\! \! \left(\eta^{\alpha \beta} \frac{\partial^{\rho}\partial^{\sigma}}{\partial^2} + \eta^{\rho \sigma} \frac{\partial^{\alpha}\partial^{\beta}}{\partial^2}\right)  \partial_{\alpha)}\partial_{(\rho}  i\Delta_V \partial_{\sigma)}\partial_{(\beta}  i\Delta_V := S_3 = 2DG_1 + 2G_2 \\
   & \hskip9cm + 2 (D+1) G_3 + 4G_1+ 2 G_5;\\
   &2 \eta^{\alpha)(\rho}\frac{\partial^{\sigma)}\partial^{(\beta}}{\partial^2}\partial_{\alpha)}\partial_{(\rho}  i\Delta_V\partial_{\sigma)}\partial_{(\beta}  i\Delta_V:= S_4 = 2G_1 + (D+1)G_2+ 4 G_3 + (D+3)G_4 + 2 G_5; \\
   &\frac{\partial^{\alpha}\partial^{\beta}\partial^{\rho}\partial^{\sigma} }{\partial^4} \partial_{\alpha)}\partial_{(\rho}  i\Delta_V \partial_{\sigma)}\partial_{(\beta}  i\Delta_V := S_5 = G_1+ G_2+ 2G_3 + 2G_4 + G_5.
\end{split}
\end{equation}
We can invert these relationship and find $G_i, F_j$ as functions of $S_i,R_j$, obtaining:
\begin{equation}
    \begin{split}
        &F_1=\frac{1}{D-1}(R_1-R_2);\\
    &F_2= \frac{1}{D-1}(DR_2-R_1);\\
    & G_1=\frac{1}{(D^2-1)(D-2)}[DS_1-2S_2 - DS_3 +2 S_4+(D-2)S_5];\\
    & G_2=\frac{1}{(D^2-1)(D-2)}[-2S_1+2(D-1)S_2 + 2S_3 -2(D-1) S_4+2(D-2)S_5];\\
    & G_3=\frac{1}{(D^2-1)(D-2)}[-DS_1+2S_2 +\frac{1}{2}(D+2)(D-1)S_3 -2 S_4-(D^2-4)S_5];\\
    & G_4=\frac{1}{(D^2-1)(D-2)}[2S_1-2(D-1)S_2 - 2S_3 +(D^2+D-4) S_4-2(D^2-4)S_5];\\
    & G_5=\frac{1}{(D^2-1)(D-2)}[(D-2)S_1+2(D-2)S_2 - (D^2-4)S_3\\
    & \hskip4cm -2(D^2-4) S_4+(D+4)(D^2-4)S_5].
    \end{split}
\end{equation}
In this way, we have reduced the problem to determining the coefficients \(S_i\) and \(R_j\).
Using properties of the vacuum propagator, namely
\begin{equation}
    \begin{split}
        &\partial^2 i\Delta_V(x;x')= \bar{m}^2 i\Delta_V(x;x') + i \delta^D(x-x');\\
        & \partial^{\mu}(i\Delta_V(x;x')) i \delta^D(x-x')= \frac12 \partial^{\mu}(i\Delta_V(x;x)) i \delta^D(x-x') =0 ;\\
        &\partial^{\mu}\partial^{\nu}(i\Delta_V(x;x')) i \delta^D(x-x') = - \partial^{\mu}\partial'{}^{\nu}(i\Delta_V(x;x')) i \delta^D(x-x')
     \\
     & \hskip4.8cm= \frac{(am)^2}{D} \eta^{\mu \nu} i\Delta_V(x;x) i \delta^D(x-x'),
    \end{split}
\end{equation}
where we have defined \(\bar{m} \equiv m a(\bar{t})\), together with the general identity
\begin{equation}
    \partial_{\mu}f(x)\partial^{\mu}g(x) = \frac{1}{2} \partial^2(f(x)g(x))-\frac{1}{2} f(x) \partial^2 g(x)- \frac{1}{2} g(x) \partial^2 f(x),
\end{equation}
valid for any differentiable functions \(f\) and \(g\), we can compute these coefficients.
The result we obtain is the following:
\begin{equation}
    \begin{split}
      &R_1= \left(\frac{1}{2}\partial^2- \bar{m}^2\right)(i\Delta_V(x;x'))^2 -  i\Delta_V(x;x) i\delta^D(x-x');\\
        &R_2= \frac{1}{4}\partial^2(i\Delta_V(x;x'))^2 - \frac{1}{2} i\Delta_V(x;x) i\delta^D(x-x');\\
        &S_1= \left(\frac{1}{4}\partial^4- \bar{m}^2 \partial^2 + \bar{m}^4\right)   (i\Delta_V(x;x'))^2 +\left( -\frac{1}{2}\partial^2+2m^2a^2\right)(i\Delta_V(x;x) i\delta^D(x-x'));  \\
        &S_2= \left(\frac{1}{8}\partial^4- \frac{1}{2}\bar{m}^2\partial^2 + \bar{m}^4\right)   (i\Delta_V(x;x'))^2 +\left( -\frac{1}{4}\partial^2+2m^2a^2\right)(i\Delta_V(x;x) i\delta^D(x-x')); \\
        &S_3= \left(\frac{1}{4}\partial^4- \frac{1}{2}\bar{m}^2\partial^2 \right)   (i\Delta_V(x;x))^2 +\left( -\frac{1}{2}\partial^2+\frac{D+2}{D}m^2a^2\right)(i\Delta_V(x;x) i\delta^D(x-x')); \\
        &S_4= \frac{1}{8}\partial^4  (i\Delta_V(x;x'))^2 +\left( -\frac{1}{4}\partial^2+\frac{D+2}{D}m^2a^2\right)(i\Delta_V(x;x) i\delta^D(x-x')); \\
        &S_5= \frac{1}{16}\partial^4  (i\Delta_V(x;x'))^2 +\left( -\frac{1}{8}\partial^2+\frac{3}{2D}m^2a^2\right)(i\Delta_V(x;x) i\delta^D(x-x')).
    \end{split}
\end{equation}
Taking into account that, owing to the symmetries of FLRW spacetime,
\(\partial^{\mu} i\Delta_V(x;x') = -\partial'^{\mu} i\Delta_V(x;x')\), we now have all the ingredients
needed to compute the \(++\) component of the self-energy.
The three-point contribution reads as follows:
\begin{equation} \label{nonrenself}
    \begin{split}
         i[^{\alpha \beta}\Sigma&_{\text{3pt},++}^{\, \rho \sigma}](x;x')=  \kappa^2 \bigg\{ \frac{P^{\alpha(\rho}P^{\sigma)\beta}}{16(D^2-1)}\bigl\{\left(\partial^2 - 4\bar{m}^2\right)^2[i\Delta_V(x;x')]^2 \\[4pt]
        & -\left[8(D-2)a^2m^2+2\partial^2\right]( i\Delta_V(x;x) i\delta^D(x-x')) \bigr\}  \\[4pt]
        & + \frac{P^{\alpha\beta}P^{\rho\sigma}}{32(D^2-1)} \left\{ \left[  16\bar{m}^4 + 8 D \bar{m}^2\partial^2 + (D^2-2D-2)\partial^4 \right] [i\Delta_V(x;x')]^2    \right.\\[4pt]
        & - \left. \left[8(D-2)m^2a^2+ 2(D^2-2D-2)\partial^2\right] ( i\Delta_V(x;x) i\delta^D(x-x')) \right\}  \\[4pt]
        & + \left(2 \eta^{\alpha(\rho}\eta^{\sigma)\beta} - \eta^{\alpha \beta} \eta^{\rho \sigma}\right) \frac{m^2a^2}{4D}  i\Delta_V(x;x) i\delta^D(x-x')\\[4pt]
         &+ \frac{1}{16(D^2-1)} \bigg( \eta^{\alpha \beta} \frac{\partial^{\rho} \partial^{\sigma}}{\partial^2} \, \Delta m_1^2 \, + \, \eta^{\rho \sigma}\frac{\partial^{\alpha}\partial^{\beta}}{\partial^2} \, \Delta m_2^2 \bigg)\\[4pt]
        & \qquad  \quad \qquad \qquad \times (D+1)(4\bar{m}^2 + (D-2) \partial^2)\,  \big[i\Delta_V(x;x')\big]^2 \\[4pt]
        & + \frac{1}{16(D^2-1)} \eta^{\alpha \beta} \eta^{\rho \sigma}  \big[ -(D+1)(\Delta m_1^2 + \Delta m_2^2) (4\bar{m}^2 + (D-2) \partial^2)\\[2pt]
        & \qquad \qquad  \qquad \qquad \qquad + 2(D^2-1) \Delta m_1^2\, \Delta m_2^2\big]  \,\big[i\Delta_V(x;x')\big]^2\bigg\},
        \end{split}
        \end{equation}
where $P^{\alpha\beta} = \eta^{\alpha \beta} - \frac{\partial^{\alpha}\partial^{\beta}}{\partial^2}$ is
the transverse projector. Here we introduced a short-hand notation for the difference of the squared masses evaluated
at the mid-point $\bar{x}$ and at the vertices $x,x'$: $\Delta m_1^2 \equiv  \bar{m}^2 -m^2a^2$ and
$\Delta m_2^2 \equiv  \bar{m}^2 -m^2a'^{\,2}$. Summing this expression to the 4-point contribution
in \eqref{4pointcontr} we get the full $++$ component of the self-energy:
\begin{equation} \label{nonregse}
        \begin{split}
   i[^{\alpha \beta}\Sigma_{V,++}^{\, \rho \sigma}](x;x')&=\kappa^2 \bigg\{ \, \frac{P^{\alpha(\rho}P^{\sigma)\beta}}{16(D^2-1)}\bigl\{\left(\partial^2 - 4m^2a^2\right)^2[i\Delta_V(x;x')]^2 \\[4pt]
        & -\left[8(D-2)m^2a^2+2\partial^2\right]( i\Delta_V(x;x) i\delta^D(x-x')) \bigr\}  \\[4pt]
        & + \frac{P^{\alpha\beta}P^{\rho\sigma}}{32(D^2-1)} \left\{ \left[  16m^4a^4 + 8 D m^2a^2\partial^2 + (D^2-2D-2)\partial^4 \right] [i\Delta_V(x;x')]^2    \right.\\[4pt]
        & -  \left[8(D-2)m^2a^2+ 2(D^2-2D-2)\partial^2\right] ( i\Delta_V(x;x) i\delta^D(x-x')) \big\}  \\[4pt]
        & - \left(2 \eta^{\alpha(\rho}\eta^{\sigma)\beta} - \eta^{\alpha \beta} \eta^{\rho \sigma}\right) \frac{m^2a^2}{4D}  i\Delta_V(x;x) i\delta^D(x-x')\\[4pt]
        & + \frac{1}{16(D^2-1)} \bigg( \eta^{\alpha \beta} \frac{\partial^{\rho} \partial^{\sigma}}{\partial^2} \, \Delta m_1^2 \, + \, \eta^{\rho \sigma}\frac{\partial^{\alpha}\partial^{\beta}}{\partial^2} \, \Delta m_2^2 \bigg)\\[4pt]
        & \qquad  \quad \qquad \qquad \times (D+1)(4\bar{m}^2 + (D-2) \partial^2)\,  \big[i\Delta_V(x;x')\big]^2 \\[4pt]
        & + \frac{1}{16(D^2-1)} \eta^{\alpha \beta} \eta^{\rho \sigma}  \big[ -(D+1)(\Delta m_1^2 + \Delta m_2^2) (4\bar{m}^2 + (D-2) \partial^2)\\[4pt]
        & \qquad \qquad  \qquad \qquad \qquad + 2(D^2-1) \Delta m_1^2\, \Delta m_2^2\big]  \,\big[i\Delta_V(x;x')\big]^2\bigg\} .
    \end{split}
\end{equation}

\subsubsection{Self-energy renormalization}
The expression in (\ref{nonrenself}) contains divergent terms and thus requires renormalization.
Fortunately, this is the only component of the self-energy that necessitates such treatment.
As a result, this section completes the full renormalization of the graviton self-energy.\footnote{
The \( -- \) component also contains divergences and must therefore be renormalized.
However, this component is not independent of the one under consideration, as the identity
\( i\Delta_{++} = [i\Delta_{--}]^{\ast} \) always holds. Consequently, renormalizing the \( ++ \)
component automatically ensures the renormalization of its complex conjugate. }
Our expression (\ref{nonrenself}) contains the propagator evaluated at coincidence,
$i\Delta_V(x;x)$, which is divergent in $D=4$. It also contain the integrated squared propagator
$[i\Delta_V(x;x')]^2$, which likewise exhibits divergences. To extract the  divergences in eq.~\eqref{nonregse}, we study the dependence of $[i\Delta_V]^2$
on the variable $y_{++}$, which we will call $y$ for brevity in this chapter (the $++$ subscript
will then be reintroduced to differentiate it from the similar variable $y_{+-}$).

Since the range of integration will cross the points where $y=0$, every power of $y$ strictly smaller than $-1$
(in $D=4$) will produce a divergent term that will have to be normalized.
To obtain a finite expression, we have to regularize these two types of divergencies through
the introduction of counter-terms in the Einstein-Hilbert action. To do so, we have to first
isolate and localise such divergent terms from our expression (\ref{nonregse}).

We begin with the propagator at coincidence, which is easily divided in finite and divergent term
in the limit $D \to 4$:
\begin{equation}
    \begin{split}
        i\Delta_V(x;x)&= \frac{(am)^{D-2}}{(4\pi)^{D-2}}\Gamma\left( 1-\frac{D}{2}\right)= \frac{a^{D-2}m^2}{16\pi^2}\left[ \frac{2 \mu^{D-4}}{D-4} + \mathcal{F}_m\right],\\[4pt]
        \text{where} \quad \mathcal{F}_m&= \log\left( \frac{m^2}{4\pi \mu^2}\right) + \gamma_E -1.
    \end{split}
\end{equation}
The mass scale $\mu$ is introduced to maintain dimensional consistency, such that $ i\Delta_V(x;x)$ maintains $\mu$-independence.
We can then divide $i\Delta_V(x;x)$ such to explicit the divergent part:
\begin{equation}\label{coincidence}
    \begin{split}
        &i\Delta_V(x;x) =i\Delta_V^{div}(x;x) + i\Delta_V^{f}(x;x);\\[5pt]
        &i\Delta_V^{div}(x;x):= \frac{a^{D-2}m^2}{16\pi^2} \frac{2 \mu^{D-4}}{D-4};\\[5pt]
        & i\Delta_V^{f}(x;x):= \frac{a^2m^2}{16\pi^2} \mathcal{F}_m. 
    \end{split}
\end{equation}
Now it is the turn of $[i\Delta_V(x;x')]^2$. We know that:
\begin{equation}
    \begin{split} \label{seriesprop}
       i\Delta_V(x;x')&= \frac{(\bar{a}m)^{D-2}}{(4\pi)^{D/2}}  \Gamma\left( 1-\frac{D}{2}\right) \sum_{n=0}^{\infty} \frac{(\frac{y}{4})^n}{(\frac{D}{2})_n n!} + \frac{(\bar{a}m)^{D-2}}{(4\pi)^{D/2}} \Gamma\left( \frac{D}{2}-1\right) \sum_{n=0}^{\infty} \frac{(\frac{y}{4})^{n+1-\frac{D}{2}}}{(2-\frac{D}{2})_n n!}  \\[4pt]
       &= \frac{(\bar{a}m)^{D-2}}{(4\pi)^{D/2}}  \Gamma\left( \frac{D}{2}-1\right) \Big(\frac{y}{4}\Big)^{1-\frac{D}{2}} 
   + \frac{(\bar{a}m)^{D-2}}{(4\pi)^{D/2}}  f(y), \quad \, \, \\[4pt]
       \end{split}
       \end{equation}
where we defined:
       \begin{equation}
        f(y):=\frac{\Gamma\left( \frac{D}{2}-1\right)}{(2-\frac{D}{2})}  \sum_{n=0}^{\infty} \frac{(\frac{y}{4})^{n+2-\frac{D}{2}}}{(3-\frac{D}{2})_n (n+1)!} + \Gamma\left( 1-\frac{D}{2}\right)\sum_{n=0}^{\infty} \frac{(\frac{y}{4})^n}{(\frac{D}{2})_n n!},
\end{equation}
such that $f(y)$ has a finite limit in $D=4$,
\begin{equation}
    f(y)\bigg|_{D=4}= \sum_{n=0}^{\infty} \frac{1}{n!(n+1)!} \left(\frac{y}{4} \right)^n\left[  \log \left( \frac{y}{4}\right) - \psi(n+1)- \psi(n+2) \right],
\end{equation}
where we introduced the digamma function $\psi(z):= (d/dz) \log \Gamma(z)$.
This expression has been isolated from the total contribution in order to make the divergent part manifest. The part that requires renormalization is proportional to \((\Delta x^2)^{D-2}\). To extract this divergence in a form independent of \(y\), we will employ a set of identities. We begin by considering the propagator for a massless field, \(i\Delta_0\), which can be obtained by setting $m=0$ in the expansion~\eqref{seriesprop}:
\begin{equation}
        i\Delta_0(x;x')= \frac{\Gamma \left(\frac{D}{2}-1 \right)}{4\pi^{D/2}} \big(\Delta x^2_{++} \big)^{1-\frac{D}{2}}.
\end{equation}
Since such objects obey the following equation of motion:
\begin{equation}
\begin{split} \label{eqret1}
    \partial^2 i \Delta_0(x;x')&= i \delta^D(x-x'),
\end{split}
\end{equation}
we have the following identity:
\begin{equation} \label{eqret2}
\begin{split}
    \frac{\partial^2}{a^2m^2} \left( \frac{y}{4} \right)^{1-\frac{D}{2}} &= \frac{(4\pi)^{\frac{D}{2}}}{\Gamma\left(\frac{D}{2}-1\right) (am)^D} i \delta^D(x-x').
\end{split}
\end{equation}
We also notice that we can write:
\begin{equation}
\left(\frac{y}{4} \right)^{-\alpha} = \frac{-2}{(\alpha-1)(D-2\alpha)} \frac{\partial^2}{\bar{m}^2} \left( \frac{y}{4} \right)^{1-\alpha}.
\end{equation}
Combining these two results we have:
\begin{equation}
    \begin{split}
       & \left(\frac{y}{4} \right)^{2-D} = \frac{2}{(D-4)(D-3)} \frac{\partial^2}{\bar{m}^2} \left( \frac{y}{4} \right)^{3-D} \\[4pt]
       &= \frac{2}{(D-4)(D-3)} \frac{\partial^2}{\bar{m}^2} \left[  \left( \frac{y}{4} \right)^{3-D} -  \left( \frac{y}{4} \right)^{\frac{2-D}{2}} + \left( \frac{y}{4} \right)^{\frac{2-D}{2}} \right] \\[4pt]
        &= \frac{2}{(D-4)(D-3)} \frac{\partial^2}{\bar{m}^2} \left\{ \frac{4}{y} \left[  \left( \frac{y}{4} \right)^{4-D} -  \left( \frac{y}{4} \right)^{\frac{4-D}{2}} \right] \right\} \\
        & +  \frac{2}{(D-4)(D-3)}\frac{(4\pi)^{\frac{D}{2}}}{\Gamma\left(\frac{D}{2}-1\right) (am)^D} i \delta^D(x-x').
    \end{split}
\end{equation}
Now, we have that:
\begin{equation}
    \left[  \left( \frac{y}{4} \right)^{4-D} -  \left( \frac{y}{4} \right)^{\frac{4-D}{2}} \right] \xrightarrow[]{D \to 4} - \frac{1}{2} (D-4) \log \left( \frac{y}{4} \right),
\end{equation}
so calling:
\begin{equation}
    g(y):=- \frac{\partial^2}{\bar{m}^2} \frac{\log \left( y/4 \right)}{y/4},
\end{equation}
in $D=4$, we see how $g(y)$ will be part of our finite contribution. We have then localised
the divergent part of the  squared propagator. Indeed, we can write:
\begin{equation}
    \left( \frac{y}{4} \right)^{2-D}  \! \! =  g(y) +  \frac{2}{(D-4)(D-3)}\frac{(4\pi)^{\frac{D}{2}}}{\Gamma\left(\frac{D}{2}-1\right) (am)^D} i \delta^D(x-x'),
\end{equation}
so that we have:
\begin{equation} \label{check}
    \begin{split}
       &  i\Delta_V(x;x')=\frac{\bar{m}^{D-2}}{(4\pi)^{D/2}} \Gamma\left(\frac{D}{2}-1\right)  \left( \frac{y}{4} \right)^{1-\frac{D}{2}} + \frac{\bar{m}^{D-2}}{(4\pi)^{D/2}} f(y) \\[5pt]
        \Rightarrow \big[ &  i\Delta_V(x;x')\big]^2 = \frac{\bar{m}^{4}}{(4\pi)^{4}} \left( g(y) + f^2(y) + \frac{8}{y}f(y) \right) + \frac{2(am)^{D-4}}{(4\pi)^{\frac{D}{2}}} \frac{\Gamma\left(\frac{D}{2}-1\right)}{(D-3)(D-4)} i \delta^D(x-x').
    \end{split}
\end{equation}
To express such a quantity in the proper way to be renormalized we expand the following terms
in the limit $D\to 4$:
\begin{equation}
    \begin{split}
        &\frac{m^{D-4}}{(4\pi)^{\frac{D}{2}}} =  \frac{\mu^{D-4}}{(4\pi)^{2}}\left(   \frac{m^2}{4\pi\mu^2}\right)^{\frac{D-4}{2}} = \frac{\mu^{D-4}}{(4\pi)^{2}}\left[ 1 + \frac{D-4}{2} \log \left( \frac{m^2}{4\pi \mu^2}\right) \right] ;\\[4pt]
        & \frac{\Gamma\left(\frac{D}{2}-1\right)}{\frac{D}{2}-2} = \Gamma\left(\frac{D}{2}-2\right) = \frac{2}{D-4} + \gamma_E;\\[4pt]
        &\frac{1}{D-3} = 1 - \frac{D-4}{D-3}.
    \end{split}
\end{equation}
Plugging this terms in (\ref{check}), we arrive to the following expression:
\begin{equation}
    \begin{split} \label{bigM}
       & [i\Delta_V(x;x')]^2= \frac{(a \mu)^{D-4}}{16 \pi^2}  \bigg( \frac{2}{D-4} + \mathcal{F}_m -1 \bigg)  i \delta^D(x-x') + \mathcal{M}_m(y);\\[6pt]
       \text{where} \quad &\mathcal{M}_m(y):= \frac{\bar{m}^4}{(4\pi)^4}\left( g(y) + f^2(y) + \frac{8}{y}f(y) \right).
    \end{split}
\end{equation}
In the same way as we did for (\ref{coincidence}), we can split this expression in the divergent
and finite part:\vspace{2mm}
\begin{equation} \label{noncoinc}
    \begin{split}
        &[i\Delta_V(x;x')]^2 = [i\Delta_V(x;x')]^2_{div}+ [i\Delta_V(x;x')]^2_f;\\[5pt]
        &[i\Delta_V(x;x')]^2_{div}:= \frac{a^{D-4}}{16\pi^2} \frac{2 \mu^{D-4}}{D-4}\, i\delta^D(x-x')= \frac{1}{a^2m^2}i\Delta_V^{div}(x;x) i\delta^D(x-x') ;\\[5pt]
        & [i\Delta_V(x;x')]^2_f:= \mathcal{M}_m(y) + \frac{1}{16\pi^2} \big( \mathcal{F}_m-1 \big) i \delta^D(x-x'). \\[3pt]
    \end{split}
\end{equation}
Plugging (\ref{coincidence}) and (\ref{noncoinc}) into (\ref{nonregse}), we identify the
divergent part of $i\Sigma_{++}$, which will read \vspace{2mm}
\begin{equation}
    \begin{split} 
       i[^{\alpha \beta}\Sigma_{div, ++}^{\, \rho \sigma}] (x;x')&=\frac{\kappa^2 a^{D-4}}{32\pi^2 (D^2-1)} \bigg[ -(D-4)m^4a^4\left( P^{\alpha \beta}P^{\rho \sigma}+ 2 P^{\alpha (\rho}P^{\sigma) \beta}\right)\\[3pt]
        &+ \left( \frac{1}{4}(-D^2 + 6D+2)  P^{\alpha \beta}P^{\rho \sigma}- \frac{5}{2}P^{\alpha (\rho}P^{\sigma) \beta}\right) m^2a^2 \partial^2  \\[3pt]
        &\left. + \left( \frac{1}{8}(D^2-2D-2) P^{\alpha \beta}P^{\rho \sigma} + \frac{1}{4} P^{\alpha (\rho}P^{\sigma) \beta}\right) \partial^4\right]\, \frac{\mu^{D-4}}{D-4}i\delta^D(x-x')  \\[3pt]
        &-\kappa^2(2\eta^{\alpha (\rho}\eta^{\sigma) \beta}- \eta^{\alpha \beta}\eta^{\rho \sigma}) \frac{a^Dm^4}{32D \pi^2} \frac{\mu^{D-4}}{D-4}i\delta^D(x-x'),\\[3pt]
    \end{split}
\end{equation}
where terms proportional to \(\Delta m_1^2\) and \(\Delta m_2^2\) vanish because they multiply a local expression, i.e., \((\bar{a}^2 - a^2) \delta^D(x-x') = 0\).
To maintain the predictive power of our theory, such divergencies have to be eliminated by
the counter term action introduced in equation (\ref{countertermact}).
By applying functional derivatives to such action, we obtain its contribution to the self-energy\footnote{
Since $i\Sigma_{\text{ct}}$ arises from a functional derivative of an action involving a single Keldysh index,
\( \delta S_{ct}[h_d]/\delta h_c \propto \delta_{cd} \), only components with matching indices
can be non-zero. Consequently, the \( ++ \) component is the only independent non-vanishing term,
aside from the \( -- \) component, which is related to it by complex conjugation. 
}
\begin{equation} \label{countself}
    \begin{split}
        i[^{\alpha \beta}\Sigma_{\text{ct}, ++}^{\, \rho \sigma}](x;x') &= \kappa^2 a^{D-4} \left[ c_1 \frac{2(D-3)}{D-2}\left( P^{\alpha (\rho}P^{\sigma) \beta} - \frac{1}{D-1}P^{\alpha \beta}P^{\rho \sigma} \right)\partial^4  \right. \\[3pt]
        & + 2c_2 P^{\alpha \beta}P^{\rho \sigma}\partial^4 + \frac{1}{2} c_3 m^2 a^2 \left( P^{\alpha (\rho}P^{\sigma) \beta} - P^{\alpha \beta}P^{\rho \sigma} \right) \partial^2 \\[3pt]
        &  - (2\eta^{\alpha (\rho}\eta^{\sigma) \beta}- \eta^{\alpha \beta}\eta^{\rho \sigma}) \frac{c_4}{4} m^4 a^4 \bigg] i \delta^D(x-x').
    \end{split}
\end{equation}
To match the two expression we will put the divergent part of the self-energy in a similar form:
\begin{equation}
    \begin{split}
        i[^{\alpha \beta}\Sigma_{div, ++}^{\, \rho \sigma}](x;x')&=\frac{\kappa^2 a^{D-4}}{32\pi^2 (D^2-1)} \left[ -(D-4)m^4a^4\left( P^{\alpha \beta}P^{\rho \sigma}+ 2 P^{\alpha (\rho}P^{\sigma) \beta}\right) \right. \\[3pt]
        & -\frac{10}{4}   \left( P^{\alpha (\rho}P^{\sigma) \beta} \right.  -\left. P^{\alpha \beta}P^{\rho \sigma} \right) m^2a^2 \partial^2 - \frac{(D-4)(D-2)}{4} P^{\alpha \beta}P^{\rho \sigma} \\[3pt]
        &+  \frac{1}{4} \bigg( P^{\alpha (\rho}P^{\sigma) \beta}  -  \frac{1}{D-1}P^{\alpha \beta}P^{\rho \sigma} \bigg)\partial^4 \\
        & \left. + \frac{(D-2)^2(D+1)}{8(D-1)} P^{\alpha \beta}P^{\rho \sigma} \partial^4 \right]\, \frac{\mu^{D-4}}{D-4}i\delta^D(x-x')  \\[3pt]
        &-\kappa^2 (2\eta^{\alpha (\rho}\eta^{\sigma) \beta}- \eta^{\alpha \beta}\eta^{\rho \sigma}) \frac{a^4m^4}{32D \pi^2} a^{D-4} \frac{\mu^{D-4}}{D-4}i\delta^D(x-x').
    \end{split}
\end{equation}
The following terms do not diverge in the limit $D\to 4$ and therefore do not need to be regularized:
\begin{equation}
    \frac{\kappa^2a^{D-4}(D-4)}{32\pi^2 (D^2-1)} \left[-m^4a^4 \left( P^{\alpha (\rho}P^{\sigma) \beta}+2 P^{\alpha \beta}P^{\rho \sigma} \right) - \frac{1}{2} P^{\alpha \beta}P^{\rho \sigma} m^2a^2 \partial^2\right] \frac{\mu^{D-4}}{D-4} i \delta^D(x-x').
\end{equation}
Without such terms, we can completely cancel the divergent component of the self-energy by setting:
\begin{equation}
    \begin{split}
        &c_1=- \frac{D-2}{8(D-3)}\frac{1}{32\pi^2(D^2-1)} \frac{\mu^{D-4}}{D-4}+c_1^f;\\[4pt]
        &c_2= -\frac{(D-2)^2}{16(D-1)}\frac{1}{32\pi^2(D-1)} \frac{\mu^{D-4}}{D-4}+c_2^f;\\[4pt]
        &c_3= \frac{5}{32\pi^2(D^2-1)} \frac{\mu^{D-4}}{D-4}+c_3^f;\\[4pt]
        &c_4= -\frac{1}{8\pi^2D} \frac{\mu^{D-4}}{D-4}+c_4^f,
    \end{split}
\end{equation}
where $c_i^f$ encode the finite part of the coefficients.
We end up with a renormalized contribution to the self-energy that can be split in the sum
of the local and non-local contributions. The latter yields
\begin{equation} \label{++comp}
    \begin{split}
         i[^{\alpha \beta}\Sigma_{V,\text{NL}}^{\, \, \rho \sigma}](x;x')&= \kappa^2 \bigg\{\frac{P^{\alpha (\rho}P^{\sigma) \beta}}{240}  \left( \partial^2 -4\bar{m}^2\right)^2  + \frac{P^{\alpha \beta}P^{\rho \sigma}}{240}\left(8 \bar{m}^4 +16\bar{m}^2 \partial^2 +3\partial^4 \right) \\[4pt]
         &+\frac{1}{24} \bigg(\Delta m_1^2 \frac{\partial^{\rho} \partial^{\sigma}}{\partial^2} \eta^{\alpha \beta} \,+\,\Delta m_2^2 \frac{\partial^{\alpha } \partial^{\beta}}{\partial^2} \eta^{\rho \sigma}\bigg) (2\bar{m}^2+ \partial^2)\\[4pt]
         &- \frac{\eta^{\alpha \beta} \eta^{\rho \sigma}}{24}\big[ (\Delta m_1^2+\Delta m_2^2) (2\bar{m}^2+ \partial^2)  - 3 \Delta m_1^2\Delta m_2^2 \big]\bigg\} \, \mathcal{M}_m (y),
    \end{split}
\end{equation}
while the former reads as follow:
\begin{equation}
    \begin{split} \label{++comp2}
            i[^{\alpha \beta}\Sigma_{V,\text{L}}^{\, \, \rho \sigma}](x;x') &=-\kappa^2 \bigg\{\frac{P^{\alpha (\rho}P^{\sigma) \beta}}{3840 \pi^2}    \, \left( 10 m^2a^2\partial^2 - \partial^4 \right) \mathcal{F}_m   \,-\, \frac{P^{\alpha \beta}P^{\rho \sigma}}{3840 \pi^2}  \, \left( 10 m^2a^2 \partial^2 + 3 \partial^4 \right) \mathcal{F}_m    \\[4pt]
         & 
+c_1^f \left( P^{\alpha (\rho}P^{\sigma) \beta} - \frac{1}{3}P^{\alpha \beta}P^{\rho \sigma} \right) \partial^4 \,+\, 2 c_2^f P^{\alpha \beta}P^{\rho \sigma} \partial^4 \, \\[4pt]
& +\, \frac{1}{2}m^2a^2 c_3^f \left( P^{\alpha (\rho}P^{\sigma) \beta} - P^{\alpha \beta}P^{\rho \sigma} \right) \partial^2 
 + \frac{P^{\alpha (\rho}P^{\sigma)\beta}}{480\pi^2 } \bigg(3m^4a^4 -m^2a^2 \partial^2 +\frac{\partial^4}{8}\bigg) \\[4pt]
 &+ \frac{P^{\alpha \beta}P^{\rho \sigma}}{480\pi^2 } \bigg(3m^4a^4 +\frac{5}{2}m^2a^2 \partial^2 + \frac{3}{8}\partial^4\bigg)\bigg\} \, i \delta^D(x-x').
    \end{split}
\end{equation}
where we have set $c_4^f=-4\mathcal{F}_m/(16\pi)^2$ to ensure transversality of the local terms.
As we are going to see later, the same choice will cause the vacuum sector of the one-loop energy-momentum tensor
to vanish. As expected, a suitable choice of the finite coefficients \(c_i^f\) renders the self-energy
\eqref{++comp2} independent of the mass scale \(\mu\) introduced in dimensional regularization.

\subsubsection{Wightman contribution to the vacuum self-energy}

We now compute the $+-$ component of the vacuum self-energy. As discussed in the main text, only the three-point self-energy contributes to this component; the $+-$ component of the four-point self-energy vanishes.
Employing the definition \eqref{definition3ptse} and the expression for the adiabatic vertices
\eqref{adiabatic vertices} we obtain:
\begin{equation}
   \begin{split}
   i[^{\alpha \beta}\Sigma&_{V,+-}^{\, \rho \sigma}](x;x')\\
   &:=\frac{\kappa^2}{2}\left[ \partial^{\alpha)}\partial'^{(\rho}  i \Delta_V^{\!+-}(x;x') \partial'^{\sigma)}\partial^{(\beta}   i \Delta_V^{\!+-}(x;x')-\frac{1}{2}\eta^{\alpha \beta}  \partial_{\lambda}  \partial'^{(\rho} i \Delta_V^{\!+-}(x;x') \partial'^{\sigma)} \partial^{\lambda}  i \Delta_V^{\!+-}(x;x')   \right. \\
    &-  \frac{1}{2}\eta^{\rho \sigma} \partial^{\alpha)}\partial_{\lambda}'  i \Delta_V^{\!+-}(x;x') \partial'^{\lambda}\partial^{(\beta}  i \Delta_V^{\!+-}(x;x') + \frac{1}{4} \eta^{\rho \sigma} \eta^{\alpha \beta}  \partial_{\gamma}\partial_{\lambda}'  i \Delta_V^{\!+-}(x;x') \partial'^{\lambda} \partial^{\gamma} i \Delta_V^{\!+-}(x;x')  \\
    & + \frac{1}{2}\eta^{\alpha \beta}m^2a^2\left(-\partial'^{(\rho} i \Delta_V^{\!+-}(x;x') \partial'^{\sigma)} i \Delta_V^{\!+-}(x;x')  + \frac{1}{2} \eta^{\rho \sigma}
\partial_{\lambda}'   i \Delta_V^{\!+-}(x;x')  \partial'^{\lambda}   i \Delta_V^{\!+-}(x;x') \right)  \\
&+ \frac{1}{2}\eta^{\rho \sigma}m^2a'^{\,2} \left( - \partial^{(\alpha} i \Delta_V^{\!+-}(x;x') \partial^{\beta)}  i \Delta_V^{\!+-}(x;x') + \frac{1}{2} \eta^{\alpha \beta}
\partial_{\lambda}  i \Delta_V^{\!+-}(x;x')   \partial^{\lambda}   i \Delta_V^{\!+-}(x;x') \right) \\
& + \left. \frac{1}{4} \eta^{\alpha \beta}\eta^{\rho \sigma} m^4 a^2a'^{\,2} \big[ i {\Delta_V^{\!+-}}(x;x')\big]^2 \right],
\end{split}
\end{equation}
We evaluate this term with the same procedure used for the $++$ component, so by extracting derivatives.
The only difference lies in the equation of motion satisfied by the different components of the Keldysh propagator: the Wightman function \(i\Delta_V^{+-}(x;x')\) satisfies the homogeneous propagator equation, \((\partial^2 - m^2 a^2)i\Delta_V^{+-}(x;x') = 0\).
This means that the $+-$ components of the self-energy will not have any coincident contribution
proportional to the delta function:
\begin{equation} \label{+-selfenergy1}
    \begin{split}
  i[^{\alpha \beta}\Sigma_{V,+-}^{\, \rho \sigma}](x;x')&=\kappa^2 \bigg\{ - \frac{P^{\alpha(\rho}P^{\sigma)\beta}}{16(D^2-1)} \left. \left(\partial^2 - 4m^2a^2\right)^2[i\Delta_V^{+-}(x;x')]^2   \right. \\[4pt]
        & -  \frac{P^{\alpha\beta}P^{\rho\sigma}}{32(D^2-1)}  \left[  16m^4a^4 + 8 Dm^2a^2\partial^2 + (D^2-2D-2)\partial^4 \right] [i\Delta_V^{+-}(x;x')]^2 \\[4pt]
         & - \frac{1}{16(D^2-1)} \bigg( \eta^{\alpha \beta} \frac{\partial^{\rho} \partial^{\sigma}}{\partial^2} \, \Delta m_1^2 \, + \, \eta^{\rho \sigma}\frac{\partial^{\alpha}\partial^{\beta}}{\partial^2} \, \Delta m_2^2 \bigg)\\[4pt]
        & \qquad  \quad \qquad \qquad \times (D+1)(4\bar{m}^2 + (D-2) \partial^2)\,  \big[i\Delta_V^{+-}(x;x')\big]^2 \\[4pt]
        & - \frac{1}{16(D^2-1)} \eta^{\alpha \beta} \eta^{\rho \sigma}  \big[ -(D+1)(\Delta m_1^2 + \Delta m_2^2) (4\bar{m}^2 + (D-2) \partial^2)\\[4pt]
        & \qquad \qquad  \qquad \qquad \qquad + 2(D^2-1) \Delta m_1^2\, \Delta m_2^2\big]  \,\big[i\Delta_V^{+-}(x;x')\big]^2\bigg\} .
    \end{split}
\end{equation}
This implies that this contribution is free from divergences associated with the coincident propagator, and, as we will briefly see, is in fact devoid of divergences entirely. To proceed with the analysis, we require an expression for the Wightman function, which differs from the Feynman propagator only by a slight modification of the spacetime distance variable on which it depends. The negative Wightman function is expressed in terms of:
\begin{equation} \label{y+-}
     y^2_{+-}(x;x'):=m^2 \Delta x_{+-}^2,\quad \Delta x_{+-}^2= -(t - t'+i \varepsilon)^2 + \bar{a}^2\| \vec{x}- \vec{x}\,' \|^2 = \Delta x^2 +\text{sgn}(-\Delta t) \,i\varepsilon,
\end{equation}
therefore, their evaluation will proceed in mostly the same way. We repeat the same steps used
to compute $[i\Delta^{++}_{V}]^2$, with the only difference that equations (\ref{eqret1})
and (\ref{eqret2}) need to be modified. Indeed, to get the right result, we need to consider
in this case the massless Wightman function:
\begin{equation}
        i\Delta^{+ -}_0(x;x')= \frac{\Gamma \left(\frac{D}{2}-1 \right)}{(4\pi)^{D/2}} \big(\Delta x_{+ -}^2 \big)^{1-\frac{D}{2}}.
\end{equation}
Since such object obeys the following equation of motion:
\begin{equation}
\begin{split}
    \partial^2 i \Delta_0^{+-}(x;x')&= 0,
\end{split}
\end{equation}
we will have the following identity:
\begin{equation}
\begin{split}
     \frac{\partial^2}{\bar{m}^2} \left( \frac{y_{+-}}{4} \right)^{1-\frac{D}{2}} &= 0.
\end{split}
\end{equation}
With such a result, we obtain that $(i\Delta^{+-}_V(x;x'))^2$ has no divergent contribution,
as it reads:
\begin{equation}
\begin{split}
    [i  \Delta_V^{+-}(x;x')]^2 &= \frac{\bar{m}^{4}}{(4\pi)^{4}} \left( g(y_{+-}) + f^2(y_{+-}) + \frac{8}{y_{+-}}f(y_{+-}) \right) \,=  \, \mathcal{M}_m(y_{+-}),
\end{split}
\end{equation}
where we have introduced the function \(\mathcal{M}_m\) defined in eq.~\eqref{bigM}.
The $+-$ component of the vacuum contribution to the self-energy is then obtained by substituting
this expression in equation (\ref{+-selfenergy1}) and by setting in it $D=4$,
\begin{equation} \label{+-comp}
    \begin{split}
        i[^{\alpha \beta}\Sigma_{V,+-}^{\, \rho \sigma}](x;x')&=  - \kappa^2 \bigg\{\frac{P^{\alpha (\rho}P^{\sigma) \beta}}{240}  \left( \partial^2 -4\bar{m}^2\right)^2  + \frac{P^{\alpha \beta}P^{\rho \sigma}}{240}\left(8 \bar{m}^4 +16\bar{m}^2 \partial^2 +3\partial^4 \right) \\[4pt]
         &+\frac{1}{24} \bigg(\Delta m_1^2 \frac{\partial^{\rho} \partial^{\sigma}}{\partial^2} \eta^{\alpha \beta} \,+\,\Delta m_2^2 \frac{\partial^{\alpha } \partial^{\beta}}{\partial^2} \eta^{\rho \sigma}\bigg) (2\bar{m}^2+ \partial^2)\\[4pt]
         &- \frac{\eta^{\alpha \beta} \eta^{\rho \sigma}}{24}\big[ (\Delta m_1^2+\Delta m_2^2) (2\bar{m}^2+ \partial^2)  - 3 \Delta m_1^2\Delta m_2^2 \big]\bigg\} \, \mathcal{M}_m(y_{+-}).
    \end{split}
\end{equation}

\subsubsection{Vacuum retarded self-energy}
\label{Vacuum retarded self-energy}

We can proceed to compute the vacuum contribution to the retarded self-energy:
\begin{equation}
    \begin{split}
         i[^{\alpha \beta}\Sigma_{V,R}^{\, \rho \sigma}] =   i[^{\alpha \beta}\Sigma_{V,++}^{\, \rho \sigma}] +    i[^{\alpha \beta}\Sigma_{V,+-}^{\, \rho \sigma}].
    \end{split}
\end{equation}
The local contribution \eqref{++comp2} remains unchanged. Summing the expressions (\ref{++comp})
and (\ref{+-comp}) we obtain the final expression for the non-local term expressed in
\eqref{nonlocalvacuumse}, where we have defined,
\begin{equation}
\begin{split}
     &\mathcal{M}_m^R (x;x')  := \mathcal{M}_m (y_{++}) -\mathcal{M}_m(y_{+-})\\[5pt]
    &=\frac{\bar{m}^4}{(4 \pi)^4}\bigg[ \big(g(y_{++})-g(y_{+-})\big)
     + \left(\frac{8}{y_{++}}f(y_{++})  - \frac{8}{y_{+-}}f(y_{+-})\right)
    +\big( f^2(y_{++})-f^2(y_{+-}) \big)
    \bigg],
    \label{retarded Mm}
\end{split}
\end{equation}
where we reintroduced the $++$ subscript on $y$ to better differentiate it from $y_{+-}$.
The latter variables have been defined in equations (\ref{y++}) and (\ref{y+-}).
The focus is now on the evaluation of this term. In particular we need to find a way
to express the difference between these functions evaluated at these slightly different variables.
Since the latter differ just by a infinitesimal imaginary part, the difference will be non-null
only when the $i \varepsilon$ procedure will be relevant for the evaluation.
Thus, we take some time to analyze the dependence on the $i \epsilon$ prescription
of the terms that appear in the relevant functions.
We start with evaluating $\log(\mu^2 \Delta x^2 \pm i\varepsilon)$, in which $\mu^2$ is a generic
multiplicative factor with the dimension of a mass squared, and $\mu^2 \Delta x^2 \, \in \, \mathbb{R}$.
The logarithm of a complex variable is a palindrome function. To properly evaluate it,
we need to cut a branch line in the complex plane and choose a Riemann surface.
The branch line will correspond to the negative real axis, and we will work on the principal branch.
This is the same as saying that we consider the argument of the complex numbers to be contained
in the range $-\pi \leq \arg(z) \leq \pi$. In this constructions, we will have:
\begin{equation}
   \log(\mu^2 \Delta x^2 \pm i\varepsilon) =  \begin{cases}
     \log(\mu^2 \Delta x^2)  & \text{if } \Delta x^2 > 0; \\
    \log(-\mu^2 \Delta x^2) \pm i \pi             & \text{if } \Delta x^2 < 0,
\end{cases}
\end{equation}
which can be written in a compact way as:
\begin{equation}
    \log(\mu^2 \Delta x^2 \pm i\varepsilon) = \log(|\mu^2 \Delta x^2|)\pm \Theta(-\Delta x^2) \, i\pi
    \,.
\end{equation}
Let us first consider the first contribution in eq.~(\ref{retarded Mm}),
\begin{eqnarray}
\!
 g(y_{++})\!-\!g(y_{+-}) \!&\!=\!&\! -\frac{\partial^4}{\bar{m}^4}
 \left\{\left[\frac12\log^2\Big(\frac{y_{++}}{4}\Big)\!-\!\log\Big(\frac{y_{++}}{4}\Big)\right]
   \!-\! \left[\frac12\log^2\Big(\frac{y_{+-}}{4}\Big)\!-\!\log\Big(\frac{y_{+-}}{4}\Big)\right]\right\}
\!\!
\nonumber\\
&\!\!=\!\!& -\frac{1}{\bar{m}^2}\partial^2\frac{1}{\bar{m}^2}\partial^2
 \left\{\left[\log\Big(\frac{|y|}{4}\Big)\!-\!1\right]2\pi i\Theta\big(\!-\!\Delta x^2\big)\Theta\big(\Delta t\big)
             \right\}
             \,.\quad
\label{retarded Mm 1}
\end{eqnarray}
The second contribution in eq.~(\ref{retarded Mm}) can be evaluated as follows,
\begin{eqnarray}
\frac{8}{y_{++}}f(y_{++})  - \frac{8}{y_{+-}}f(y_{+-})
&\!\!=\!\!& 
\frac{2}{\bar{m}^2}\partial^2\bigg\{\left[\frac12\log^2\Big(\frac{y_{++}}{4}\Big)
                       \!+\!2\big(\gamma_E\!-\!1\big)\log\Big(\frac{y_{++}}{4}\Big)\right]
 \nonumber\\
&& \hskip 1cm
   -\, \left[\frac12\log^2\Big(\frac{y_{+-}}{4}\Big)\!+\!2\big(\gamma_E\!-\!1\big)\log\Big(\frac{y_{+-}}{4}\Big)\right]\bigg\}
 \nonumber\\
&& \hskip -1.5cm
+\,2\sum_{n=0}^\infty\frac{1}{(n+1)!(n+2)!} \left[\left(\frac{y_{++}}{4}\right)^{\!n}
                \log\left(\frac{y_{++}}{4}\right)
                -\left(\frac{y_{+-}}{4}\right)^{\!n}\log\left(\frac{y_{+-}}{4}\right)\right] 
\nonumber\\             
&&  \hskip -1.5cm
=\,\frac{2}{\bar{m}^2}\partial^2\bigg\{\left[\log\Big(\frac{|y|}{4}\Big)\!+\!2\big(\gamma_E\!-\!1\big)\right]
                   2\pi i\Theta\big(\!-\!\Delta x^2\big)\Theta\big(\Delta t\big)
                   \bigg\}
 \nonumber\\
&& \hskip -1.cm
+\,2\sum_{n=0}^\infty\frac{(-1)^n}{(n+1)!(n+2)!}\left(\frac{|y|}{4}\right)^{\!n}
                \left[2\pi i\Theta\big(\!-\!\Delta x^2\big)\Theta\big(\Delta t\big)\right] 
\nonumber\\             
&&  \hskip -1.5cm
=\,\frac{2}{\bar{m}^2}\partial^2\bigg\{\left[\log\Big(\frac{|y|}{4}\Big)\!+\!2\big(\gamma_E\!-\!1\big)\right]
                   2\pi i\Theta\big(\!-\!\Delta x^2\big)\Theta\big(\Delta t\big)
                   \bigg\}
 \nonumber\\
&& \hskip -1.cm
+\,\frac{8}{|y|}\left[1\!-\!\frac{2J_1\big(\sqrt{|y|}\big)}{\sqrt{|y|}}\right]
                \left[2\pi i\Theta\big(\!-\!\Delta x^2\big)\Theta\big(\Delta t\big)\right]
\,.\qquad\;
\label{retarded Mm 2}
\end{eqnarray}
Finally, the last term in eq.~(\ref{retarded Mm}) can be evaluates to,
\begin{eqnarray}
 f^2(y_{++})-f^2(y_{+-}) &\!\!=\!\!&  \sum_{n=0}^\infty\sum_{n'=0}^\infty\frac{1}{n!(n+1)!}\frac{1}{n'!(n'+1)!}
             \left(\frac{y_{++}}{4}\right)^{\!n}
                   \left(\frac{y_{++}}{4}\right)^{\!n'}
\bigg[\log^2\left(\frac{y_{++}}{4}\right)
\nonumber\\
&& \hskip 0.cm
  -\,\big[\psi(n+1)\!+\!\psi(n+2)\!+\!\psi(n'+1)\!+\!\psi(n'+2)\big]\log\left(\frac{y_{++}}{4}\right)
             \bigg] 
\nonumber\\
&& \hskip -1.cm
 -\,\sum_{n=0}^\infty\sum_{n'=0}^\infty\frac{1}{n!(n+1)!}\frac{1}{n'!(n'+1)!}\left(\frac{y_{+-}}{4}\right)^{\!n}
                   \left(\frac{y_{+-}}{4}\right)^{\!n'}
\bigg[\log^2\left(\frac{y_{+-}}{4}\right)
\nonumber\\
&& \hskip 0.cm
  -\,\big[\psi(n+1)\!+\!\psi(n+2)\!+\!\psi(n'+1)\!+\!\psi(n'+2)\big]\log\left(\frac{y_{+-}}{4}\right)
             \bigg] 
\qquad
\nonumber\\
&& \hskip -1.cm
=\, 2\sum_{n=0}^\infty\frac{(-1)^n}{n!(n+1)!}\left(\frac{|y|}{4}\right)^{\!n}
              \times\sum_{n'=0}^\infty\frac{(-1)^{n'}}{n'!(n'+1)!}
                   \left(\frac{|y|}{4}\right)^{\!n'}
\nonumber\\
&& \hskip 0.cm
\times\bigg[\log\left(\frac{|y|}{4}\right)  \!-\!\psi(n'+1)\!-\!\psi(n'+2)
             \bigg]  \left[2\pi i\Theta\big(\!-\!\Delta x^2\big)\Theta\big(\Delta t\big)\right] 
\nonumber\\
&& \hskip -1.cm
=\, \frac{4J_1\big(\sqrt{|y|}\big)}{\sqrt{|y|}}\left[\frac{4}{|y|}\!+\!\frac{2\pi Y_1\big(\sqrt{|y|}\big)}{\sqrt{|y|}} \right]
        \left[2\pi i\Theta\big(\!-\!\Delta x^2\big)\Theta\big(\Delta t\big)\right]
\,,\qquad
\label{retarded Mm 3}
\end{eqnarray}
where $J_{\nu}(z)$ and $Y_{\nu}(z)$ are the Bessel functions of the first and second kind, respectively. Upon combining all three contributions one obtains,
\begin{eqnarray}
     \mathcal{M}_m^R (x;x') &\!\!=\!\!& \frac{i}{128\pi^3} 
\Bigg\{\! \!-\!\partial^4\left[\log\Big(\frac{|y|}{4}\Big)\!-\!1\right]
  \!+\!2\bar{m}^2\partial^2\left[\log\Big(\frac{|y|}{4}\Big)\!+\!2\big(\gamma_E\!-\!1\big)\right]
 \nonumber\\             
&&  \hskip 1.55cm
+\, \frac{8\bar{m}^4}{|y|}\left[1\!+\!\pi J_1\big(\sqrt{|y|}\big)Y_1\big(\sqrt{|y|}\big)\right]\Bigg\}
 \Big[\Theta\big(\!-\!\Delta x^2\big)\Theta\big(\Delta t\big)\Big]
\,,\qquad
\label{retarded Mm: final}
\end{eqnarray}
where the derivative operators act on everything to the right.

\subsection{Contribution arising from the occupation spectrum}

Since it satisfies a homogeneous equation, the contribution to the Keldysh propagator arising from
the occupation spectrum of the ULDM field takes the same value for the \(++\) and \(+-\) components.
We therefore denote it by \(i \Delta_N\) and its expression is obtained by applying the middle-point
approximation to the expression given in \eqref{Wightman function}:
\begin{equation}
      i\Delta_{N}(x;x') = \frac{1}{\bar{a}}\int_{\vec{q}} \frac{e^{i\vec{q}\cdot\vec{\Delta x}}}{\omega(q;\bar{t})} N_{q}\cos[\omega (q;\bar{t}) \Delta t] ,
\end{equation}
where we employ the notation $\bar{a}\equiv a(\bar{t})$.
We use this expression to evaluate its contribution to the graviton self-energy, restoring the exact
adiabatic phase $\Phi(q;t)$ for the local contribution.

\subsubsection{4-point contribution}

Employing the definition \eqref{definition4ptse} and the expression for the adiabatic vertices
\eqref{adiabatic vertices} we can write the 4-point self-energy as:
\begin{equation}
    \begin{split}
         i[^{\alpha \beta}\Sigma_{N, \text{4pt}}^{ \, \rho \sigma}](x;x')& = i\kappa^2 \delta^D(x-x')  \bigg[  \bigg( \frac{1}{8}\eta^{\alpha \beta} \eta^{\rho \sigma} - \frac{1}{4} \eta^{\alpha(\rho} \eta^{\sigma) \beta}  \bigg)\bigg( \partial\cdot \! \partial' + m^2 a^2 \bigg)\\
&+\frac12 \eta^{\alpha)(\rho} \partial'^{\sigma)}\partial_{(\beta} +  \frac12 \eta^{\alpha)(\rho} \partial^{\sigma)}\partial'^{(\beta} - \frac{1}{4} \eta^{\alpha \beta} \partial^{(\rho} \partial'^{\sigma)} - \frac{1}{4} \eta^{\rho \sigma} \partial^{(\alpha} \partial'^{\beta)}    \bigg]i\Delta_{N}(x;x') .
    \end{split}
\end{equation}
Throughout this chapter we use the notation $q^2 \equiv \lVert \vec{q\,}\rVert^2$. When evaluated at coincidence,
the 2-point function $i \Delta_N$ and its Laplacian can be expressed through the local integrals
$I_j^N$ introduced in \eqref{Iintegralsoccupationnumber}:
\begin{equation}
\begin{split}
       &i\Delta_{N}(x;x) =  \frac{1}{2\pi^2 a(t)}\int_{0}^{\infty} dq\,  \frac{N_q\,q^2 }{\sqrt{\big(q^2/a^2\big)+m^2}}=I_1^N;\\[5pt]
       & \nabla^2 i\Delta_{N}(x;x) \equiv \sum_j \partial^2_j i\Delta_{N}(x;x')\bigg|_{x' \to x} =   -\frac{1}{2\pi^2 a(t)}\int_{0}^{\infty} dq\,  \frac{N_q\,q^4 }{\sqrt{\big(q^2/a^2\big)+m^2}}=-I_2^N.
\end{split}
\end{equation}
To obtain an useful expression for the 4-point self-energy we will use the following properties
of the 2-point function $i\Delta_N$ at leading adiabatic order:
\begin{equation} \label{mixprop}
    \begin{split}
       & \partial'^{\mu} i\Delta_{N}(x;x')= - \partial^{\mu} i\Delta_{N}(x;x');\\[5pt]
       & \partial^2_0 i\Delta_{N}(x;x')\bigg|_{x' \to x}   = -a^2m^2 I_1^N -I_2^N ;\\[5pt]
       &\partial^0 \partial^j i\Delta_{N}(x;x')\bigg|_{x' \to x}  = i \int \frac{d^3 q}{(2\pi )^3} N_q\, q^j \, \sin[\omega(q;\bar{t})\Delta t]\bigg|_{\Delta t \to 0} \, =0;\\[5pt]
       & \partial^j \partial^k i\Delta_{N}(x;x')\bigg|_{x' \to x} = - \frac{1}{a}\int \frac{d^3 q}{(2\pi )^3} \frac{N_q}{\omega(q;t)} q^j q^k  = -\frac{\delta^{jk}}{3} I_2^N,
    \end{split}
\end{equation}
where the equality in the last line comes from the fact that, when $j\ne k$, the integrand is asymmetric
with respect to the reflection of a single component, and from its isotropy in $\vec{q}$.
Employing these identities we compute the 4-point self-energy:
\begin{equation} \label{app:4ptN}
  \begin{split}
      i[^{\alpha \beta}\Sigma_{N, \text{4pt}}^{ \, \rho \sigma}](x;x')&=i \kappa^2 \delta^D(x-x') \bigg[ \frac{1}{3} \bigg(\eta^{\alpha(\rho}\eta^{\sigma) \beta} -\frac{1}{2} \eta^{\alpha \beta} \eta^{\rho \sigma} \bigg)I^{N}_2  \\
      &+\bigg(\delta_0^{(\alpha} \eta^{\beta)(\rho}\delta_0^{\sigma)} - \frac{1}{4}\eta^{\alpha \beta} \delta_0^{\rho} \delta_0^{\sigma} -  \frac{1}{4}\eta^{\rho \sigma} \delta_0^{\alpha} \delta_0^{\beta} \bigg) \bigg(\frac{4}{3}I^N_2 + a^2 m^2 I^N_1\bigg)  \bigg] .
  \end{split}
\end{equation}
This contribution, summed with the local terms coming from the 3-point self-energy, will constitute
the graviton local self-energy arising from the occupation spectrum of the ULDM field $i\Sigma^L_N$.

\subsubsection{3-point contribution}

The 3-point contribution to the retarded self-energy is obtained using the definition
\eqref{retardedseproper3pt} and employing the expression for the adiabatic vertices
introduced in \eqref{adiabatic vertices}. The contribution arising from a non-zero occupation spectrum
reads as follow:
\begin{equation}
    \begin{split} \label{++init22}
            i [^{\alpha \beta}\Sigma&_{N,\text{3pt},R}^{\, \rho \sigma}](x;x')\\
            &=\frac{\kappa^2}{2} \bigg[ \, \, \partial^{\alpha)}\partial'^{(\rho}  i \Delta^R(x;x') \partial'^{\sigma)}\partial^{(\beta}   i\Delta_{N}(x;x')-\frac{1}{2}\eta^{\alpha \beta}  \partial'^{\rho)} \partial_{\lambda} i \Delta^R(x;x') \partial^{\lambda}  \partial'^{(\sigma} i\Delta_{N}(x;x')   \\[3pt]
        &-  \frac{1}{2}\eta^{\rho \sigma} \partial_{\lambda}' \partial^{(\alpha} i \Delta^R(x;x') \partial^{\beta)}\partial'^{\lambda}  i\Delta_{N}(x;x') + \frac{1}{4} \eta^{\rho \sigma} \eta^{\alpha \beta}  \partial_{\gamma}\partial_{\lambda}'  i \Delta^R(x;x') \partial'^{\lambda} \partial^{\gamma} i\Delta_{N}(x;x')  \\[3pt]
        & + \frac{1}{2}\eta^{\alpha \beta}m^2a^2  \left(-\partial'^{(\rho} i \Delta^R(x;x') \partial'^{\sigma)} i\Delta_{N}(x;x')  + \frac{1}{2} \eta^{\rho \sigma}
\partial_{\lambda}'   i \Delta^R(x;x')  \partial'^{\lambda}   i\Delta_{N}(x;x') \right) \\[3pt]
&+ \frac{1}{2}\eta^{\rho \sigma}m^2a'^{\,2} \left( - \partial^{(\alpha} i \Delta^R(x;x') \partial^{\beta)}  i\Delta_{N}(x;x') + \frac{1}{2} \eta^{\alpha \beta}
\partial_{\lambda}  i \Delta^R(x;x')   \partial^{\lambda}   i\Delta_{N}(x;x') \right) \\[3pt]
& + \frac{1}{4} \eta^{\alpha \beta}\eta^{\rho \sigma} m^4 a^2a'^{\,2} \, i \Delta^R(x;x') \, i\Delta_{N}(x;x')\bigg] .
    \end{split}
\end{equation}
Here $i \Delta^R$ is the vacuum retarded propagator:
\begin{equation} \label{vacuumretardedpropagator}
    i \Delta^R (x;x')=\frac{1}{\bar{a}} \int \frac{d^4k}{(2\pi)^4} \frac{i \, e^{i k_{\mu}\Delta x^{\mu}}}{\big(k^0 + i\varepsilon \big)^2 - \omega^2(k;\bar{t})}.
\end{equation}
The property $\partial^{\mu} i\Delta_{V/N}(x,x')=- \partial'^{\mu} i\Delta_{V/N}(x,x') $ allows us
to replace all derivatives over $x'$ with derivatives over $x$. This expression is more conveniently
computed in Wigner space:
\begin{equation}
  i\, [^{\alpha \beta}\overline{\Sigma}_{N}^{\, \rho \sigma}](k^0,k^2) := \int d^4 \Delta x \; i\, [^{\alpha \beta}\Sigma_{N,\text{3pt},R}^{\, \rho \sigma}](x;x') \, e^{-i k_{\mu}\Delta x^{\mu}},
\end{equation}
where, within the adiabatic regime, we treat the scale factor as constant in the Fourier transform,
i.e., we do not integrate over its time dependence. The position-space self-energy can be recovered
via an inverse Wigner transform:
\begin{equation}
     i[^{\alpha \beta}\Sigma_{N,\text{3pt},R}^{\, \rho \sigma}](x;x') = \int \frac{d^4 k}{(2\pi)^4} \, \,  e^{i k_{\mu}\Delta x^{\mu}}  \,\,i[ ^{\alpha \beta}\overline{\Sigma}_{N}^{\, \rho \sigma}](k^0,k^2).
\end{equation}
The initial focus of our derivation will be to separate the tensor structures of our expression
and identify the terms that contribute to each of these.
We start by expressing the sum of the terms containing 4 derivatives. To express these terms
with less invasive notation, we will introduce the following convention:
\begin{equation}
    [^{\alpha \beta} \smash{\overline{\Omega}}^{\rho \sigma}](k^0,k^2) =  \int d^4 \Delta x \, \partial^{\alpha} \partial^{\beta}  i \Delta^R(x;x') \partial^{\rho}  \partial^{\sigma}  i\Delta_{N}(x;x')\, e^{-i k_{\mu}\Delta x^{\mu}}.
\end{equation}
By dividing temporal derivatives from spatial ones, we can therefore write
\begin{equation}
    \begin{split}
 [^{\alpha) (\rho} \smash{\overline{\Omega}}^{\sigma)(\beta}] & = \delta_0^{\alpha} \delta_0^{\beta}  \delta_0^{\rho} \delta_0^{\sigma} \, [^{00} \smash{\overline{\Omega}}^{00}] + (\delta_i^{(\alpha} \delta_0^{\beta)}  \delta_0^{\rho}  \delta_0^{\sigma} + \delta_0^{\alpha} \delta_0^{\beta}  \delta_i^{(\rho}  \delta_0^{\sigma)}) \, ( [^{i0} \smash{\overline{\Omega}}^{00}] + [^{00} \smash{\overline{\Omega}}^{i0}]) \\
 & + (\delta_i^{(\alpha} \delta_j^{\beta)}  \delta_0^{\rho}  \delta_0^{\sigma} + \delta_0^{\alpha} \delta_0^{\beta}  \delta_i^{(\rho}  \delta_j^{\sigma)}) \,  [^{0(i} \smash{\overline{\Omega}}^{j)0}]  \, + \, \delta_i^{(\alpha} \delta_0^{\beta)}  \delta_j^{(\rho} \delta_0^{\sigma)} \, ( [^{ij} \smash{\overline{\Omega}}^{00}] + 2 [^{0(i} \smash{\overline{\Omega}}^{j)0}] + [^{00} \smash{\overline{\Omega}}^{ij}])\\
 &+ (\delta_i^{(\alpha} \delta_j^{\beta)}  \delta_l^{(\rho} \delta_0^{\sigma)} + \delta_l^{(\alpha} \delta_0^{\beta)}  \delta_i^{(\rho} \delta_j^{\sigma)})  \, ( [^{l(i} \smash{\overline{\Omega}}^{j)0}] + [^{0(i} \smash{\overline{\Omega}}^{j)l}]) \, + \,  \delta_i^{(\alpha} \delta_j^{\beta)}  \delta_l^{(\rho} \delta_n^{\sigma)} \,  [^{n)(i} \smash{\overline{\Omega}}^{j)(l}]  \,,
    \end{split}
\end{equation}
where repeated indices are summed. Contracting this expression with $\eta^{\alpha \beta}, \, \eta^{\rho \sigma}$
and $\eta^{\alpha \beta} \eta^{\rho \sigma}$ we obtain $[_{\lambda} ^{(\rho} \smash{\overline{\Omega}}^{\sigma)  \lambda}]$,
$[_{ \lambda}^{(\alpha} \smash{\overline{\Omega}}^{\beta)\lambda }]$ and $[_{\gamma \lambda} \smash{\overline{\Omega}}^{\lambda \gamma}]$,
necessary to find the expression for the 4-derivative contribution, which reads:
\begin{equation} \label{4derm}
    \begin{split}
         \int  d^4 \Delta x \, &e^{-i k_{\mu}\Delta x^{\mu}} \! \bigg( \! \! \partial^{\alpha)} \partial'^{(\rho}  i \Delta^{\!R}(x;x') \partial'^{\sigma)}  \partial^{(\beta}  i\Delta_{N}(x;x')-\frac{1}{2}\eta^{\alpha \beta}  \partial'^{\rho)} \partial_{\lambda} i \Delta^{\!R}(x;x') \partial^{\lambda}  \partial'^{(\sigma} i\Delta_{N}(x;x')   \\
        &-  \frac{1}{2}\eta^{\rho \sigma} \partial_{\lambda}' \partial^{(\alpha} i \Delta^{\!R}(x;x') \partial^{\beta)}\partial'^{\lambda}  i\Delta_{N}(x;x') + \frac{1}{4} \eta^{\rho \sigma} \eta^{\alpha \beta}  \partial_{\gamma}\partial_{\lambda}'  i \Delta^{\!R}(x;x')\partial'^{\lambda} \partial^{\gamma} i\Delta_{N}(x;x')  \bigg)\\[6pt]
        &= \frac{1}{4}\,
\delta_0^{\alpha}\,\delta_0^{\beta}\,\delta_0^{\rho}\,\delta_0^{\sigma}
\Bigl(
  2\,\delta_{ij}\,[^{0(i}\smash{\overline{\Omega}}^{\,j)0}]
+ [^{00} \smash{\overline{\Omega}}^{00}]
+ \delta_{ij}\,\delta_{\ell n}\,[^{\ell)(i}\smash{\overline{\Omega}}^{\,j)(n}]
\Bigr)
\\[0.5ex]
&\quad
+\;\frac12\,
\bigl(
  \delta_{\ell}^{(\alpha}\,\delta_{0}^{\beta)}\,\delta_0^{\rho}\,\delta_0^{\sigma}
+ \delta_0^{\alpha}\,\delta_0^{\beta}\,\delta_{\ell}^{(\rho}\,\delta_{0}^{\sigma)}
\bigr)
\Bigl[
  \delta_{ij}\bigl([^{\ell(i}\smash{\overline{\Omega}}^{\,j)0}] + [^{0(i}\smash{\overline{\Omega}}^{\,j)\ell}]\bigr)
  +  [^{\ell0}\smash{\overline{\Omega}}^{00}] + [^{00}\smash{\overline{\Omega}}^{\ell0}]
\Bigr]
\\[0.5ex]
&\quad
+\;\frac12\,
\bigl(
  \delta_i^{\alpha}\,\delta_j^{\beta}\,\delta_0^{\rho}\,\delta_0^{\sigma}
+ \delta_0^{\alpha}\,\delta_0^{\beta}\,\delta_i^{\rho}\,\delta_j^{\sigma}
\bigr)
\,
\bigl( \delta_{\ell n}
   [^{\ell)(i}\smash{\overline{\Omega}}^{\,j)(n}]
 + [^{0(i}\smash{\overline{\Omega}}^{j)0}]
\bigr)
\\[0.5ex]
&\quad
+\;\delta_{i}^{(\alpha}\,\delta_{0}^{\beta)}\,\delta_{j}^{(\rho}\,\delta_{0}^{\sigma)}
\bigl(
   [^{ij}\smash{\overline{\Omega}}^{00}]
 + 2\,[^{0(i}\smash{\overline{\Omega}}^{\,j)0}]
 + [^{00}\smash{\overline{\Omega}}^{ij}]
\bigr)
\\[0.5ex]
&\quad
+\;
\bigl(
  \delta_i^{\alpha}\,\delta_j^{\beta}\,\delta_{\ell}^{(\rho}\,\delta_{0}^{\sigma)}
+ \delta_{\ell}^{(\alpha}\,\delta_{0}^{\beta)}\,\delta_i^{\rho}\,\delta_j^{\sigma}
\bigr)
\bigl(
  [^{\ell(i}\smash{\overline{\Omega}}^{\,j)0}]
+ [^{0(i)}\smash{\overline{\Omega}}^{\,j)\ell}]
\bigr)
\\[0.5ex]
&\quad
+\;
\delta_i^{\alpha}\,\delta_n^{\beta}\,\delta_{\ell}^{\rho}\,\delta_j^{\sigma}\,
[^{\ell)(i}\smash{\overline{\Omega}}^{\,j)(n}]
\\[1ex]
&\quad
+\;\frac14\,
\bigl(\overline{\eta}^{\alpha\beta}\,\delta_0^{\rho}\,\delta_0^{\sigma}
      + \overline{\eta}^{\rho\sigma}\,\delta_0^{\alpha}\,\delta_0^{\beta}\bigr)
\Bigl(
  [^{00}\smash{\overline{\Omega}}^{00}]
 - \delta_{ij}\,\delta_{\ell n}\,[^{\ell)(i}\smash{\overline{\Omega}}^{\,j)(n}]
\Bigr)
\\[0.5ex]
&\quad
+\;\frac14\,
\overline{\eta}^{\alpha\beta}\,\overline{\eta}^{\rho\sigma}
\Bigl(
  -2\,\delta_{ij}\,[^{0(i}\smash{\overline{\Omega}}^{\,j)0}]
  + [^{00}\smash{\overline{\Omega}}^{00}]
  + \delta_{ij}\,\delta_{\ell n}\,[^{\ell)(i}\smash{\overline{\Omega}}^{\,j)(n}]
\Bigr)
\\[0.5ex]
&\quad
+\;\frac12\,
\bigl(\overline{\eta}^{\alpha\beta}\,\delta_{\ell}^{(\rho}\,\delta_{0}^{\sigma)}
      + \overline{\eta}^{\rho\sigma}\,\delta_{\ell}^{(\alpha}\,\delta_{0}^{\beta)}\bigr)
\Bigl[
  [^{\ell0}\smash{\overline{\Omega}}^{00}]
 + [^{00}\smash{\overline{\Omega}}^{\ell0}]
 - \delta_{ij}\bigl([^{\ell(i}\smash{\overline{\Omega}}^{\,j)0}] + [^{0(i}\smash{\overline{\Omega}}^{\,j)\ell}]\bigr)
\Bigr]
\\[0.5ex]
&\quad
+\;\frac12\,
\bigl(\overline{\eta}^{\alpha\beta}\,\delta_{i}^{(\rho}\,\delta_{j}^{\sigma)}
      + \overline{\eta}^{\rho\sigma}\,\delta_{i}^{(\alpha}\,\delta_{j}^{\beta)}\bigr)
\bigl(
  [^{0(i}\smash{\overline{\Omega}}^{\,j)0}]
  - \delta_{\ell n}\,[^{\ell)(i}\smash{\overline{\Omega}}^{\,j)(n}]
\bigr).\end{split}
\end{equation}
where we have defined a new tensor structure $\overline{\eta}^{\alpha \beta}:= \eta^{\alpha \beta} + \delta_0^{\alpha} \delta_0^{\beta}$
which acts as the spatial part of the Minkowski metric. As we did for the 4-derivative terms,
we will call
\begin{equation}
    \begin{split}
        & [^{\alpha} \smash{\overline{\Omega}}^{\beta}](k^0,k^2) =  \int  d^4 \Delta x \, \partial^{\alpha}   i \Delta^R(x;x') \partial^{\beta} i\Delta_{N}(x;x')\, e^{-i k_{\mu}\Delta x^{\mu}};\\
        &\overline{\Omega}(k^0,k^2) =  \int  d^4 \Delta x \,    i \Delta^R(x;x')  i\Delta_{N}(x;x')\, e^{-i k_{\mu}\Delta x^{\mu}}.
    \end{split}
\end{equation}
Separating spatial derivatives from temporal ones, we write
\begin{equation}
    [^{(\alpha} \smash{\overline{\Omega}}^{\beta)}] = \delta_0^{\alpha} \delta_0^{\beta} \, [^{0} \smash{\overline{\Omega}}^{0}] + \delta_{i}^{(\alpha} \delta_{0}^{\beta)}\, ( [^{0} \smash{\overline{\Omega}}^{i}] + [^{i} \smash{\overline{\Omega}}^{0}]) + \delta_{i}^{(\alpha} \delta_{j}^{\beta)}\, [^{(i} \smash{\overline{\Omega}}^{j)}].
\end{equation}
Contracting this expression with $\eta_{\alpha \beta}$ we obtain $[^{\lambda} \smash{\overline{\Omega}}_{\lambda}]$.
We can now write the tensor structure of the remaining terms of the momentum-space self-energy contribution:
\begin{equation}\label{2derm}
    \begin{split}
        \int  d^4 \Delta x \, & e^{-i k_{\mu}\Delta x^{\mu}}\,  \\
        \times&  \left[ \frac{1}{2}\eta^{\alpha \beta}m^2a^2  \left(-\partial'^{(\rho} i \Delta^R(x;x') \partial'^{\sigma)} i\Delta_{N}(x;x')  + \frac{1}{2} \eta^{\rho \sigma}
\partial_{\lambda}'   i \Delta^R(x;x')  \partial'^{\lambda}   i\Delta_{N}(x;x') \right) \right.\\
&+ \frac{1}{2}\eta^{\rho \sigma}m^2a'^{\,2} \left( - \partial^{(\alpha} i \Delta^R(x;x') \partial^{\beta)}  i\Delta_{N}(x;x') + \frac{1}{2} \eta^{\alpha \beta}
\partial_{\lambda}  i \Delta^R(x;x')   \partial^{\lambda}   i\Delta_{N}(x;x') \right) \\
&+ \left. \frac{1}{4} \eta^{\alpha \beta}\eta^{\rho \sigma} m^4 a^2a'^{\,2} \, i \Delta^R(x;x') \, i\Delta_{N}(x;x')\right] \\[6pt]
&=-\frac14\bigl(\delta_0^{\alpha}\,\delta_0^{\beta}\,\overline{\eta}^{\rho\sigma}
            +\delta_0^{\rho}\,\delta_0^{\sigma}\,\overline{\eta}^{\alpha\beta}\bigr)
\,\big[ \bar{m}^2
\Bigl(
  2\,\delta_{ij}\,[^{(i}\smash{\overline{\Omega}}^{\,j)}]
  +\bar{m}^2\,\overline\Omega
\Bigr) + \Delta m_1^2\Delta m_2^2 \, \overline\Omega
\\[0.5ex]
 & \qquad \qquad \qquad \qquad \qquad \qquad - (\Delta m_1^2+ \Delta m_2^2) \big( -[^{0}\smash{\overline{\Omega}}^{0}]  +\,\delta_{ij}\,[^{(i}\smash{\overline{\Omega}}^{\,j)}]
 +\bar{m}^2\, \overline\Omega \Bigr)  \big] \\[0.5ex]
&\quad
-\frac12
\bigl(
  \overline{\eta}^{\rho\sigma}\,\delta_{i}^{(\alpha}\,\delta_{0}^{\beta)}
 +\overline{\eta}^{\alpha\beta}\,\delta_{i}^{(\rho}\,\delta_{0}^{\sigma)}
\bigr)\,
\bar{m}^2\,
\bigl(
  [^{i}\smash{\overline{\Omega}}^{0}]
 +[^{0}\smash{\overline{\Omega}}^{i}]
\bigr)
\\[0.5ex]
&\quad
-\frac12
\bigl(
  \overline{\eta}^{\rho\sigma}\,\delta_i^{\alpha}\,\delta_j^{\beta}
 +\overline{\eta}^{\alpha\beta}\,\delta_i^{\rho}\,\delta_j^{\sigma}
\bigr)\,
\bar{m}^2\,
[^{(i}\smash{\overline{\Omega}}^{j)}]
\\[0.5ex]
&\quad
+\frac14\,
\delta_0^{\alpha}\,\delta_0^{\beta}\,\delta_0^{\rho}\,\delta_0^{\sigma}
\,  \big[\bar{m}^2
\Bigl(
  2\,[^{0}\smash{\overline{\Omega}}^{0}]
 +2\,\delta_{ij}\,[^{(i}\smash{\overline{\Omega}}^{\,j)}]
 +\bar{m}^2\,[^{0}\smash{\overline{\Omega}}^{0}]
\Bigr)
\\[0.5ex]
 & \qquad \qquad \qquad \quad  - (\Delta m_1^2+ \Delta m_2^2) \big( [^{0}\smash{\overline{\Omega}}^{0}]  +\,\delta_{ij}\,[^{(i}\smash{\overline{\Omega}}^{\,j)}]
 +\bar{m}^2\, \overline\Omega \Bigr) + \Delta m_1^2\Delta m_2^2 \overline\Omega \, \big] \\[0.5ex]
&\quad
+\frac12
\bigl(
  \delta_0^{\alpha}\,\delta_0^{\beta}\,\delta_{i}^{(\rho}\,\delta_{0}^{\sigma)}
 +\delta_{i}^{(\alpha}\,\delta_{0}^{\beta)}\,\delta_0^{\rho}\,\delta_0^{\sigma}
\bigr)\,
\bar{m}^2\,
\bigl(
  [^{i}\smash{\overline{\Omega}}^{0}]
 +[^{0}\smash{\overline{\Omega}}^{i}]
\bigr)
\\[0.5ex]
&\quad
+\frac12
\bigl(
  \delta_0^{\alpha}\,\delta_0^{\beta}\,\delta_i^{\rho}\,\delta_j^{\sigma}
 +\delta_i^{\alpha}\,\delta_j^{\beta}\,\delta_0^{\rho}\,\delta_0^{\sigma}
\bigr)\,
\bar{m}^2\,
[^{(i}\smash{\overline{\Omega}}^{j)}]
\\[0.5ex]
&\quad
+\frac14\,
\overline{\eta}^{\alpha\beta}\,\overline{\eta}^{\rho\sigma}
\,\big[ \bar{m}^2
\Bigl(
   -2\,[^{0}\smash{\overline{\Omega}}^{0}]
 +2\,\delta_{ij}\,[^{(i}\smash{\overline{\Omega}}^{\,j)}]
 +\bar{m}^2\, \overline\Omega \Bigr)\\
 & \qquad \qquad \qquad - (\Delta m_1^2+ \Delta m_2^2) \big( -[^{0}\smash{\overline{\Omega}}^{0}]  +\,\delta_{ij}\,[^{(i}\smash{\overline{\Omega}}^{\,j)}]
 +\bar{m}^2\, \overline\Omega \Bigr) + \Delta m_1^2\Delta m_2^2 \overline\Omega \, \big] \\[0.5ex]
 & \quad + \frac12\bigl(\Delta m_2^2 \,  \delta_0^{\alpha}\,\delta_0^{\beta}\,\overline{\eta}^{\rho\sigma} + \Delta m_1^2 \, \delta_0^{\rho}\,\delta_0^{\sigma}\,\overline{\eta}^{\alpha\beta}\bigr) [^{0}\smash{\overline{\Omega}}^{0}]\\[0.5ex]
 &\quad +\frac{1}{2} \bigl(
   \Delta m_2^2 \, \overline{\eta}^{\rho\sigma}\,\delta_{i}^{(\alpha}\,\delta_{0}^{\beta)}
 + \Delta m_1^2 \, \overline{\eta}^{\alpha\beta}\,\delta_{i}^{(\rho}\,\delta_{0}^{\sigma)}
\bigr)\, \bigl(
  [^{i}\smash{\overline{\Omega}}^{0}]
 +[^{0}\smash{\overline{\Omega}}^{i}]
\bigr)\\[0.5ex]
&\quad -\frac{1}{2} \bigl(
   \Delta m_2^2 \, \delta_0^{\rho} \delta_0^{\sigma} \,\delta_{i}^{(\alpha}\,\delta_{0}^{\beta)}
 + \Delta m_1^2 \, \delta_0^{\alpha} \delta_0^{\beta} \,\delta_{i}^{(\rho}\,\delta_{0}^{\sigma)}
\bigr)\, \bigl(
  [^{i}\smash{\overline{\Omega}}^{0}]
 +[^{0}\smash{\overline{\Omega}}^{i}]
\bigr)\\[0.5ex]
 &\quad +\frac{1}{2} \bigl(
   \Delta m_2^2 \, \overline{\eta}^{\rho\sigma}\,\delta_i^{\alpha}\,\delta_j^{\beta}
 + \Delta m_1^2 \, \overline{\eta}^{\alpha\beta}\,\delta_i^{\rho}\,\delta_j^{\sigma}
\bigr)\,  [^{(i}\smash{\overline{\Omega}}^{j)}]\\[0.5ex]
&\quad -\frac{1}{2} \bigl(
   \Delta m_2^2 \, \delta_0^{\rho} \delta_0^{\sigma} \,\delta_i^{\alpha}\,\delta_j^{\beta}
 + \Delta m_1^2 \, \delta_0^{\alpha} \delta_0^{\beta} \,\delta_i^{\rho}\,\delta_j^{\sigma}
\bigr)\, 
  [^{(i}\smash{\overline{\Omega}}^{j)}].
    \end{split}
\end{equation}
Summing (\ref{4derm}) and (\ref{2derm}) and dividing by 2 gives us the expression of the Wigner-space self-energy.

To get a better grasp of the effect of the mixing on the dynamic of the graviton one point function,
we aim at separating local term when possible. In this expression, local contributions are to be found
in terms containing $[^{00} \smash{\overline{\Omega}}^{\alpha \beta}]$. The latter is proportional to
$\partial_0^2 \, i \Delta^R$, which contains both a local and a non-local contribution due to
the time ordering of the Feynman propagator:
\begin{equation}
 \partial_0^2 \,  i \Delta^R(x;x') \partial^{\alpha}  \partial^{\beta}  i\Delta_{N}(x;x') =\left[-i\delta^D(x-x') + \left( \nabla^2 - a^2m^2 \right)i \Delta^R(x;x')\right] \partial^{\alpha}  \partial^{\beta}  i\Delta_{N}(x;x').
\end{equation}
To understand how this separation work, let us compute $[^{00}\, \smash{\overline{\Omega}}^{\,00}]$:
\begin{equation}
\begin{split}
&i \Delta^R(x; x') \, i\Delta_{N}(x'; x) 
= \frac{1}{\bar{a}^2}\int \frac{d^4 q\, d^3 q'}{(2\pi)^7} \,
\frac{-i}{-(q^0+i\varepsilon)^2 + \omega^2(q)}
\, \frac{N_{q'}}{2\omega(q')} \\[4pt]
&\quad \qquad \qquad \qquad \qquad \times e^{i(\vec{q} + \vec{q}') \cdot \Delta \vec{x}} 
\left[ e^{i(\omega(q') - q^0)\Delta t} + e^{-i(\omega(q') + q^0)\Delta t} \right]\\[9pt]
\Rightarrow \, &\partial_0^2 i \Delta^R \, \partial_0^2 i\Delta_{N} 
=\bar{a}^2\int \frac{d^4 q\, d^3 q'}{(2\pi)^7} \,
\frac{-i}{-(q^0+i\varepsilon)^2 + \omega^2(q) } \,
\frac{N_{q'}}{2\omega(q')} \\[4pt]
&\quad \qquad \qquad \qquad \qquad  \times e^{i(\vec{q} + \vec{q}') \cdot \Delta \vec{x}} 
\, \omega^2(q') (q^0)^2 
\left[ e^{i(\omega(q') - q^0)\Delta t} + e^{-i(\omega(q') + q^0)\Delta t} \right]\\[9pt]
\Rightarrow \, &\, [^{00}\, \smash{\overline{\Omega}}^{\,00}](k^0, \vec{k}) 
= \int d^4 \Delta x \, e^{-i k \cdot \Delta x} 
\left( \partial_0^2 i \Delta^R \, \partial_0^2 i\Delta_{N} \right) \\[4pt]
&\quad \qquad \qquad =\bar{a}^2 \int \frac{d^4 q\, d^3 q'}{(2\pi)^3} \,
\frac{-i}{-(q^0+i\varepsilon)^2 + \omega^2(q)} \,
\frac{N_{q'}}{2\omega(q')}
\, \omega^2(q') (q^0)^2 \\[4pt]
& \qquad \qquad \qquad \qquad  \times \, \delta^3(\vec{q} + \vec{q}' - \vec{k})  \,  
\left[ \delta(k^0 + q^0 - \omega(q')) + \delta(k^0 - q^0 - \omega(q')) \right].
\end{split}
\end{equation}
We apply the delta functions and change the integration variable to $\vec{p} = \vec{q}- \vec{k}/2 \equiv \vec{q}$.
We introduce a new notation to distinguish functions that depend on $\vec{q}- \vec{k}/2$ from those that depend on $\vec{q}+ \vec{k}/2$:
\begin{equation} \label{notationmomentum}
    f\bigg(\vec{q}- \frac{\vec{k}}{2}\bigg) \equiv f_- , \qquad  f\bigg(\vec{q}+ \frac{\vec{k}}{2}\bigg) \equiv f_+,
\end{equation}
where $f$ denotes either $\omega$ or $N$.
Applying these changes we obtain:
\begin{equation} \label{0000}
\begin{split}
[^{00}\, \smash{\overline{\Omega}}^{\,00}](k^0, \vec{k}) 
&= \bar{a}^2 \int \frac{d^3 q}{(2\pi)^3} \frac{N_{+}}{2\omega_{+}}\, \omega_{+}^2 \, (-i) 
\left[ \frac{(k^0 + \omega_{+} )^2}{-(k^0 + \omega_{+}+i\varepsilon)^2 + \omega_{-}^2 }
     + \frac{(k^0 - \omega_{+})^2}{-(k^0 - \omega_{+}+i\varepsilon)^2 + \omega_{-}^2} \right] \\[3pt]
&=i \bar{a}^2\int \frac{d^3 q}{(2\pi)^3} \frac{N_{+}}{\omega_{+}} \, \omega_{+}^2 +\bar{a}^2 \int \frac{d^3 q}{(2\pi)^3} \frac{N_{+} }{2\omega_{+}} \omega_{+}^2 \omega_{-}^2
     \\[3pt]
     &  \qquad \qquad \qquad \qquad \qquad   \times \left[ \frac{-i}{-(k^0 + \omega_{+}+i\varepsilon)^2 + \omega_{-}^2 } \right. + \left. \frac{-i}{-(k^0 - \omega_{+}+i\varepsilon)^2 + \omega_{-}^2 } \right] \\[3pt]
&= i \int \frac{d^3 q}{(2\pi)^3} \frac{N_q}{\omega_q} 
\, (q^2 + a^2m^2) \\[3pt]
&\quad +\bar{a}^2\int \frac{d^3 q}{(2\pi)^3} \frac{N_{+}}{4 \omega_{+} \omega_{-}} 
\, \omega_{+}^2 \omega_{-}^2 
\left[ \frac{-i}{k^0 + \omega_{+} + \omega_{-} + i\varepsilon} 
       + \frac{-i}{k^0 - \omega_{+} + \omega_{-} + i\varepsilon} \right. \\[3pt]
       & \left. \qquad \qquad \qquad \qquad \qquad \qquad \qquad  +\frac{i}{k^0 + \omega_{+} - \omega_{-} + i\varepsilon} 
       + \frac{i}{k^0 - \omega_{+} - \omega_{-} + i\varepsilon} \right],
\end{split}
\end{equation}
where the first term acts locally on the graviton field:
\begin{equation}
\begin{split}
   & i \, \int_{k} \frac{e^{i k \cdot \Delta x}}{(2\pi)^4} \,  \int \frac{d^3 q}{(2\pi)^3} \frac{N_q}{\omega(q)} \, (q^2 +a^2 m^2) = i  \delta^4(x-x')  \int \frac{d^3 q}{(2\pi)^3} \frac{N_q}{\omega(q)} \, (q^2 +a^2 m^2) \\[7pt]
   &=  - i  \delta^4(x-x') \,\frac{1}{a} \partial_0^2 \, i\Delta_{N} (x-x') =   i  \delta^4(x-x') \left( a^2 m^2 \, I^{N}_1 +  I^{N}_2\right).
\end{split}
\end{equation}
This is the furthest we can go in the evaluation of these integrals without assuming a specific
spectrum $N_q$. To continue our investigation we express the self-energy as a function of a limited
set of integrals shown in \eqref{nonlocalNintegrals} dependent on the exact form of the spectrum $N_q$
that is considered. 
Using the latter definition, we can rewrite equation (\ref{0000}):
\begin{equation}\label{confront}
    \begin{split}
        [^{00}\, \smash{\overline{\Omega}}^{\,00}](k^0, \vec{k}) &= \, i \, \left( a^2 m^2 \, I^{N}_1 +  I^{N}_2\right) \\[5pt]
        &+ \sum_{t=1}^4 \left[\bar{J}^t_{40}-k^2 \, \bar{J}^t_{02}+ 2 \left( \frac{k^2}{4}+ \bar{m}^2\right) \bar{J}^t_{20} + \left( \frac{k^2}{4}+ \bar{m}^2\right)^2 \bar{J}^t_{00} \right],
    \end{split}
\end{equation}
We can apply the same steps used to calculate $[^{00}\, \smash{\overline{\Omega}}^{\,00}]$ to evaluate all the
other components of $[^{\alpha \beta}\, \smash{\overline{\Omega}}^{\,\rho \sigma}]$ in terms of standard integrals.
To do so, we will need the following integral identities:
\begin{equation} \label{intest}
    \begin{split}
        & \int_{\vec{q}} f(\vec{q}, \vec{k} ) \, q^i  = \, \frac{k^i}{k} \int_{\vec{q}} f(\vec{q}, \vec{k} ) \,\frac{\vec{k} \cdot \vec{q}}{k} ;\\
        & \int_{\vec{q}} f(\vec{q}, \vec{k} ) \, q^i q^j = \frac{\delta^{ij}}{2} \int_{\vec{q}} f(\vec{q}, \vec{k} ) \left[ q^2- \left( \frac{\vec{k} \cdot \vec{q}}{k} \right)^2 \right] + \frac{k^i k^j}{2 k^2} \int_{\vec{q}} f(\vec{q}, \vec{k} ) \left[ -q^2+3 \left( \frac{\vec{k} \cdot \vec{q}}{k} \right)^2 \right];\\
        & \int_{\vec{q}} f(\vec{q}, \vec{k} ) \, q^i q^j q^l = \frac{k^i \delta^{jl} + k^j \delta^{li} + k^l \delta^{ij} }{2k} \int_{\vec{q}} f(\vec{q}, \vec{k} ) \left[ q^2 \, \frac{\vec{k} \cdot \vec{q}}{k} - \left( \frac{\vec{k} \cdot \vec{q}}{k} \right)^3 \right]\\
        & \qquad \qquad \qquad \qquad + \frac{k^ik^jk^l}{2k^3}  \int_{\vec{q}} f(\vec{q}, \vec{k} ) \left[ -3 \, q^2 \, \frac{\vec{k} \cdot \vec{q}}{k} +5 \left( \frac{\vec{k} \cdot \vec{q}}{k} \right)^3 \right];\\
        &         \int_{\vec{q}} f(\vec{q}, \vec{k} ) \, q^i q^j q^l q^n = A_1(k^0, k^2)   \ \big( \delta^{ij}\delta^{ln} + 2 \, \delta^{i(l}\delta^{n)j} \big) \\
        & \qquad \qquad \qquad \qquad+ A_2 (k^0, k^2) \frac{k^i k^j \delta^{nl} + k^l k^n \delta^{ij}+4 k^{(i}\delta^{j)(l}k^{n)}}{k^2} + A_3(k^0,k^2) \frac{k^ik^jk^lk^n}{k^4}; 
    \end{split}
\end{equation}
where in the last line the coefficients $A_i(k^0, k^2)$ read
\begin{equation}
    \begin{split}
        A_1(k^0, k^2)&= \frac{1}{8}  \int_{\vec{q}} f(\vec{q}, \vec{k} ) \left[ q^4  -2 \, q^2 \, \left( \frac{\vec{k} \cdot \vec{q}}{k} \right)^2 + \left( \frac{\vec{k} \cdot \vec{q}}{k} \right)^4 \right];\\
                A_2(k^0, k^2)&= \frac{1}{8}  \int_{\vec{q}} f(\vec{q}, \vec{k} ) \left[- q^4  +6 \, q^2 \, \left( \frac{\vec{k} \cdot \vec{q}}{k} \right)^2 -5 \left( \frac{\vec{k} \cdot \vec{q}}{k} \right)^4 \right];\\
                        A_3(k^0, k^2)&= \frac{1}{8}  \int_{\vec{q}} f(\vec{q}, \vec{k} ) \left[ 3q^4  -30 \, q^2 \, \left( \frac{\vec{k} \cdot \vec{q}}{k} \right)^2 + 35\left( \frac{\vec{k} \cdot \vec{q}}{k} \right)^4 \right].
    \end{split}
\end{equation}
We now have all the tools necessary to evaluate the remaining components, which will read:
\begin{equation} \label{omega1}
\begin{split}
[^{i)(l \,} \smash{\overline{\Omega}}^{n)(j}] &= \sum_t \Bigg\{ 
\left( \delta^{ij} \delta^{ln} + 2\delta^{i(l} \delta^{n)j} \right) 
\left( \bar{J}_{40}^t - 2 \bar{J}_{22}^t + \bar{J}_{04}^t \right) + \frac{k^{(i}\delta^{j)(l} k^{n)}}{k^2} \frac{ -\bar{J}_{40}^t + 6\bar{J}_{22}^t - 5 \bar{J}_{04}^t }{2} \\
&\quad 
-\left( \frac{k^i k^j \delta^{ln}}{k^2} +\frac{ k^l k^n \delta^{ij}}{k^2} \right) \frac{1}{8} \left[ 
\bar{J}_{40}^t - 6 \bar{J}_{22}^t + 5 \bar{J}_{04}^t 
+ k^2 (\bar{J}_{20}^t -  \bar{J}_{02}^t)  \right] \\
& \quad + \frac{k^i k^j k^l k^n}{k^4}\frac{1}{8} \left[ 
3 \bar{J}_{40}^t - 30 \bar{J}_{22}^t + 35 \bar{J}_{04}^t 
+ k^2 (2\bar{J}_{20}^t - 6 \bar{J}_{02}^t) 
+ \frac{ k^4}{2} \bar{J}_{00}^t  \right] \bigg\};\\[3pt]
[^{0(i \,} \smash{\overline{\Omega}}^{\, j) l}] &= \sum_t  \left( s_-^t \right) \Bigg\{
\frac{\delta^{l(i} k^{j)}}{k} \left( \bar{N}_{21}^t - \bar{N}_{03}^t \right)
+ \frac{k^l \delta^{ij}}{k} \frac{1}{4} \left[ 
2 \bar{N}_{21}^t - 2 \bar{N}_{03}^t +k (\bar{N}_{20}^t - \bar{N}_{02}^t) \right] \\
&\quad + \frac{k^i k^j k^l}{k^3} \frac{1}{4} \left[ 
-6 \bar{N}_{21}^t + 10 \bar{N}_{03}^t + k (3 \bar{N}_{02}^t - \bar{N}_{20}^t) 
- k^2\bar{N}_{01}^t - \frac{k^3}{2} \bar{N}_{00}^t \right] 
\Bigg\} ; \\[3pt]
[^{i0\,} \smash{\overline{\Omega}}^{\,00}] &= \sum_t (s_-^t )\, \frac{k^i}{k} \left[  \bar{N}_{21}^t + k \frac{2 \bar{N}_{02}^t - \bar{N}_{20}^t}{2} 
- \left( \frac{k^2}{4} - \bar{m}^2 \right) \bar{N}_{01}^t - \frac{k}{2} \left( \frac{k^2}{4} + \bar{m}^2 \right) \bar{N}_{00}^t 
\right] ; \\[3pt]
[^{l(i \,} \smash{\overline{\Omega}}^{\, j) 0}]&= \sum_t (-s_+^t) \Bigg\{ 
\frac{\delta^{l(i} k^{j)}}{k} \left( \bar{M}_{21}^t - \bar{M}_{03}^t \right)
+ \frac{k^l \delta^{ij}}{k} \frac{1}{4} \left[ 
2 \bar{M}_{21}^t - 2 \bar{M}_{03}^t - k (\bar{M}_{20}^t - \bar{M}_{02}^t) \right] \\
&\quad + \frac{k^i k^j k^l}{k^3} \frac{1}{4} \left[ 
-6 \bar{M}_{21}^t + 10 \bar{M}_{03}^t - k (3 \bar{M}_{02}^t - \bar{M}_{20}^t) 
- k^2 \bar{M}_{01}^t + \frac{k^3}{2} \bar{M}_{00}^t \right]
\Bigg\} ;\\[3pt]
[^{00\,} \smash{\overline{\Omega}}^{\,i0}] &= \sum_t (-s_+^t) \, \frac{k^i}{k} \Bigg[ 
\bar{M}_{21}^t + \frac{k}{2} \bar{M}_{20}^t - k \bar{M}_{02}^t 
- \left( \frac{k^2}{4} - \bar{m}^2 \right) \bar{M}_{01}^t 
+ \frac{k}{2} \left( \frac{k^2}{4} + \bar{m}^2 \right) \bar{M}_{00}^t 
\Bigg] ; \\[3pt]
[^{0(i \,} \smash{\overline{\Omega}}^{\, j)0 }]&= \sum_t \bigl(-\,s_-^t\,s_+^t\bigr)\left[
\delta^{ij}\,\frac{\bar J_{20}^t - \bar J_{02}^t}{2}
\;+\;\frac{k^i\,k^j}{k^2}\;\frac{1}{2}\Bigl(3\,\bar J_{02}^t - \bar J_{20}^t - \frac{k^2}{2}\,\bar J_{00}^t\Bigr)
\right] ; \\[3pt]
[^{ij \,} \smash{\overline{\Omega}}^{\,00 }] + [^{00 \,} \smash{\overline{\Omega}}^{\,ij }] &= \, i \, \frac{I_2^{N}} {3} \, + \sum_t \Biggl\{
\delta^{ij}\Bigl[\bar J_{40}^t - \bar J_{22}^t + \Bigl(\frac{k^2}{4} + \bar{m}^2\Bigr)\bigl(\bar J_{20}^t - \bar J_{02}^t\bigr)\Bigr] \\
&\quad
+ \frac{k^i\,k^j}{k^2}\Bigl[
3\,\bar J_{22}^t - \bar J_{40}^t 
+ \frac{1}{2}\Bigl(\frac{k^2}{4} + \bar{m}^2\Bigr)\bar J_{00}^t
+ \bar{m}^2\bigl(3\,\bar J_{02}^t - \bar J_{20}^t\bigr)
+ \frac{k^2}{4}\bigl(\bar J_{20}^t - 5\,\bar J_{02}^t\bigr)
\Bigr]
\Biggr\}\;;
\end{split}
\end{equation}
\begin{equation}\label{omega2}
    \begin{split} 
\overline{\Omega} \, &= \sum_t \bar J_{00}^t; \qquad [^{0\,}\smash{\overline{\Omega}}^{\,0 }] = \sum_t \bigl(-\,s_+^t\,s_-^t\bigr)\,\bar L_{00}^t\;;  \\[3pt]
[^{0\,} \smash{\overline{\Omega}}^{\,i }]&= \sum_t \bigl(\,s_-^t)\;\frac{k^i}{k}\,\Bigl(\bar N_{01}^t + \frac{k}{2}\,\bar{N}_{00}^t\Bigr); \qquad [^{i\,} \smash{\overline{\Omega}}^{\,0 }] = \sum_t \,(-s_+^t)\,\frac{k^i}{k}\,\Bigl(\bar M_{01}^t - \frac{k}{2}\,\bar M_{00}^t\Bigr)\;;  \\[3pt]
[^{(i\,} \smash{\overline{\Omega}}^{\,j) }]&= \sum_t \left[
\delta^{ij}\,\frac{\bar J_{20}^t - \bar J_{02}^t}{2}
\;+\;\frac{k^i\,k^j}{k^2}\;\frac{1}{4}\Bigl(\bar J_{02}^t - 2\,\bar J_{20}^t - k^2\,\bar J_{00}^t\Bigr)
\right]\;. \\[4pt]
\end{split}
\end{equation}
Plugging these expressions in (\ref{2derm}) and (\ref{4derm}), and summing the result with the contribution
coming from the 4-point self-energy in \eqref{app:4ptN}, we obtain the final expression of the local
self-energy arising from the occupation spectrum, shown in \eqref{localsemixing}, and of its non-local contribution.
In the latter, we separate the terms that depend on different kinds of standard integrals
\begin{equation}
   i[^{\alpha \beta}\overline{\Sigma}_{N}^{\, \rho \sigma}](k^0,k^2) =i[^{\alpha \beta}\overline{\Sigma}_{(\text{\rom{1}}),N}^{\, \rho \sigma}](k^0,k^2) + i[^{\alpha \beta}\overline{\Sigma}_{(\text{\rom{2}}),N}^{\, \rho \sigma}](k^0,k^2) +i[^{\alpha \beta}\overline{\Sigma}_{(\text{\rom{3}}),N}^{\, \rho \sigma}](k^0,k^2) .
\end{equation}
The results obtained for these components are shown in equations (\ref{n11})-(\ref{n13}).
\subsection{Contribution arising from the squeezing spectrum}

We now evaluate the change in the graviton self-energy induced by squeezing of the matter field's
quantum state. The derivations in this section are conceptually analogous to those of the previous section,
but the results differ because the squeezing contribution to the propagator, and hence to the self-energy, exhibits a distinct time dependence. A squeezed state is characterized by an oscillatory time evolution
in phase space, which is encoded in the dependence of the two-point function \(i\Delta_S\) on the average
time \(\bar{t} = (t+t')/2\). Using the mid-point approximation, we obtain
\begin{equation}
     i\Delta_{S}(x;x') = \frac{1}{\bar{a}} \int \frac{d^3 q}{(2\pi)^3} \frac{S_q}{\omega(q;\bar{t})} e^{i \vec{q}\cdot \vec{\Delta x}} \cos\bigl[2\omega (q;\bar{t})\bar{t} - \phi_q\bigr], 
\end{equation}
where the squeezing amplitude \(S_q\) and phase \(\phi_q\) were introduced in Chapter 3, and we employ the notation $\bar{a} \equiv a(\bar{t})$.

\subsubsection{4-point contribution}

The coincident squeezing 2-point function and its Laplacian can be written in terms of the local
squeezing integrals $I_j^S$ defined in \eqref{Iintegralssqueezing}:
\begin{equation}
    \begin{split}
        &i\Delta_{S}(x;x)= \frac{1}{2\pi^2 a(t)}\int_{0}^{\infty} dq\,  \frac{S_q\,q^2 }{\sqrt{\big(q^2/a^2\big)+m^2}}\times \cos\big[ 2 \, \Phi(q;t) \, - \, \phi_q \big]=I_1^S;\\
        & \nabla^2 i\Delta_{S}(x;x)\equiv \sum_j \partial^2_j i\Delta_{S}(x;x')\bigg|_{x' \to x} \\
        & \qquad \, \, \,   \quad \qquad = - \frac{1}{2\pi^2 a(t)}\int_{0}^{\infty} dq\,  \frac{S_q\,q^2 }{\sqrt{\big(q^2/a^2\big)+m^2}}\times \cos\big[ 2 \, \Phi(q;t) \, - \, \phi_q \big] =-I_2^S,
    \end{split}
    \raisetag{15pt}
\end{equation}
where $\Phi$ is the exact adiabatic phase shown in \eqref{adiabatic phase og}.
Due to its different time dependence, the squeezing two-point function satisfies slightly different
identities than the analogous contribution due to the occupation spectrum shown in \eqref{mixprop}:
\begin{equation} \label{sqprop}
\begin{split}
       & \partial'^{ j} i\Delta_{S}(x;x')= - \partial^{j} i\Delta_{S}(x;x');\\[4pt]
       & \partial'^{0} i\Delta_{S}(x;x')= \partial^{0} i\Delta_{S}(x;x');\\[4pt]
       & \partial^2_0 i\Delta_{S}(x;x')\bigg|_{x' \to x}= -a^2m^2 I_1^S - I_2^S;\\[4pt]
       &\partial^0 \partial^j i\Delta_{S}(x;x')\bigg|_{x' \to x} =0;\\[4pt]
       & \partial^j \partial^k i\Delta_{S}(x;x')\bigg|_{x' \to x} = -\frac{\delta^{jk}}{3} I_2^S.
\end{split}
\end{equation}
With these identities, we can now proceed to the evaluation of the four-point self-energy defined in
\eqref{definition4ptse}:
\begin{equation} \label{app:4ptS}
    \begin{split}
         i[^{\alpha \beta}\Sigma_{S,\text{4pt}}^{\, \rho \sigma}](x;x')& = i\kappa^2\delta^D(x-x')  \left[  \left(\frac{1}{8}\eta^{\alpha \beta} \eta^{\rho \sigma} - \frac{1}{4}\eta^{\alpha (\rho} \eta^{\sigma) \beta}\right)   \left(m^2a^2 + \partial \cdot \partial'\right)\right. \\
         & \left.+ \frac12 \partial'^{(\alpha}\eta^{\beta)(\rho} \partial^{\, \sigma)}+ \frac12 \partial^{\, (\alpha}\eta^{\beta)(\rho} \partial'^{\sigma)} -   \frac{1}{4} \eta^{\alpha \beta} \partial^{\, (\rho} \partial'^{\sigma)} -  \frac{1}{4} \eta^{\rho \sigma} \partial^{\, (\alpha} \partial'^{\beta)} \right]i\Delta_{S}(x;x') \\[6pt]
&=i\kappa^2 \delta^D(x-x') \bigg[\frac{1}{2}\bigg(\eta^{\alpha(\rho}\eta^{\sigma) \beta} -\frac{1}{2} \eta^{\alpha \beta} \eta^{\rho \sigma} \bigg) \bigg(-\frac{1}{3}I^{S}_2 (t) - a^2 m^2 I^{S}_1 (t) \bigg) \\[3pt]
& \bigg(\delta_0^{(\alpha} \eta^{\beta)(\rho}\delta_0^{\sigma)} - \frac{1}{4}\eta^{\alpha \beta} \delta_0^{\rho} \delta_0^{\sigma} -  \frac{1}{4}\eta^{\rho \sigma} \delta_0^{\alpha} \delta_0^{\beta} \bigg)  \bigg(-\frac{2}{3}I^{S}_2 (t)  - a^2 m^2 I^{S}_1 (t) \bigg)\bigg].
    \end{split}
\end{equation}
This contribution will be summed to the local terms of the 3-point self-energy to constitute the local
squeezing self-energy $i\Sigma_S^{\text{L}}$.

\subsubsection{3-point contribution}

The 3-point contribution to the retarded self-energy is obtained using the definition
\eqref{retardedseproper3pt} and employing the expression for the adiabatic vertices introduced in
\eqref{adiabatic vertices}. The contribution arising from a non-zero squeezing spectrum reads as follow:
\begin{equation} \label{++init}
\begin{split}
            i [^{\alpha \beta}\Sigma&_{S,\text{3pt},R}^{\, \rho \sigma}](x;x')
          \\[3pt]
          &=\, \frac{\kappa^2}{2} \bigg[\,\partial^{\, \alpha)}\partial'^{(\rho}  i \Delta^R(x;x') \partial'^{\sigma)} \partial^{\, \beta)}   i\Delta_{S}(x;x')-\frac{1}{2}\eta^{\alpha \beta}  \partial'^{\rho)} \partial_{\, \lambda} i \Delta^R(x;x') \partial^{\, \lambda}  \partial'^{(\sigma} i\Delta_{S}(x;x')  \\[3pt]
       & -  \frac{1}{2}\eta^{\rho \sigma} \partial'_{\lambda} \partial^{\, (\alpha} i \Delta^R(x;x') \partial^{\, \beta)} \partial'^{\lambda}  i\Delta_{S}(x;x') + \frac{1}{4} \eta^{\rho \sigma} \eta^{\alpha \beta}  \partial_{ \gamma}\partial'_{\lambda}  i \Delta^R(x;x') \partial'^{\lambda} \partial^{\, \gamma} i\Delta_{S}(x;x')  \\[3pt]
        & + \frac{1}{2}\eta^{\alpha \beta}m^2a^2 \left(-\partial'^{(\rho} i \Delta^R(x;x') \partial'^{\sigma)} i\Delta_{S}(x;x')  + \frac{1}{2} \eta^{\rho \sigma}
\partial'_{\lambda}   i \Delta^R(x;x')  \partial'^{\lambda}   i\Delta_{S}(x;x') \right) \\[3pt]
&+ \frac{1}{2}\eta^{\rho \sigma}m^2a'^{\,2} \left( - \partial^{\, (\alpha} i \Delta^R(x;x') \partial^{\, \beta)}  i\Delta_{S}(x;x') + \frac{1}{2} \eta^{\alpha \beta}
\partial_{\, \lambda}  i \Delta^R(x;x')   \partial^{\, \lambda}   i\Delta_{S}(x;x') \right) \\[3pt]
& +  \frac{1}{4} \eta^{\alpha \beta}\eta^{\rho \sigma} m^4 a^2a'^{\,2} \, i \Delta^R(x;x') \, i\Delta_{S}(x;x')\bigg] ,
    \end{split}
\end{equation}
where $i\Delta^R$ is the vacuum retarded propagator, given explicitly in \eqref{vacuumretardedpropagator}.

This expression is more conveniently analyzed in Wigner space. Accordingly, the quantity we will compute is:
\begin{equation}
  i[^{\alpha \beta}\overline{\Sigma}_{S}^{\, \rho \sigma}](k^0,k^2):=  \int  d^4 \Delta x \,  i[^{\alpha \beta}\Sigma_{S,\text{3pt},R}^{\, \rho \sigma}](x;x') \,e^{-i k_{\mu}\Delta x^{\mu}}, 
\end{equation}
from which we can obtain the position-space self-energy by performing a reverse Fourier transformation:
\begin{equation}
     i[^{\alpha \beta}\Sigma_{S,\text{3pt},R}^{\, \rho \sigma}](x;x') = \int \frac{d^4 k}{(2\pi)^4} \, \,  e^{i k_{\mu}\Delta x^{\mu}}  \,\,i[^{\alpha \beta}\overline{\Sigma}_{S}^{\, \rho \sigma}](k^0,k^2).
\end{equation}
The goal now is to put this expression into a form that is useful for studying the dynamics of
cosmological perturbations. First, we separate the contributions proportional to different tensor structures.
This separation will be performed differently than in the previous section, due to the different time
dependence of the squeezing term of the propagator $i\Delta_{S}$. The latter depends on the average
time $\bar{t}$. As a consequence, deriving this propagator with respect to $t$ or $t'$ produces the same
result, as is evident from the properties shown in (\ref{sqprop}). This result traces a difference with
the 2-point functions encountered so far, for which we had $\partial^0 i\Delta = - \partial'^0 i\Delta$.
This means that we have to be more careful when treating prime derivatives, as spatial ones will have
properties different from temporal ones. Similarly to what we did for the occupation number term,
we will introduce the following notations:
\begin{equation}
\begin{split}
    &[^{\alpha \beta} \smash{\widetilde{\Omega}}^{\rho \sigma}](k^0,k^2) :=  \int  d^4 \Delta x \, \partial^{\, \alpha} \partial^{\, \beta}  i \Delta^R(x;x') \partial^{\, \rho}  \partial^{\, \sigma}  i\Delta_{S}(x;x')\, e^{-i k_{\mu}\Delta x^{\mu}};\\[3pt]
    &  [^{\alpha} \smash{\widetilde{\Omega}}^{\beta}](k^0,k^2) :=  \int  d^4 \Delta x \, \partial^{\, \alpha}  i \Delta^R(x;x')   \partial^{\, \beta}  i\Delta_{S}(x;x')\, e^{-i k_{\mu}\Delta x^{\mu}}; \\[3pt]
    &\widetilde{\Omega}(k^0,k^2) :=  \int  d^4 \Delta x \,   i \Delta^R(x;x')  \, i\Delta_{S}(x;x')\, e^{-i k_{\mu}\Delta x^{\mu}}.
\end{split}
\end{equation}
By dividing temporal derivatives from spatial ones, we can write:
\begin{equation}
    \begin{split}
       &\int  d^4 \Delta x \, \partial^{\, \alpha)} \partial'^{(\rho}  i \Delta^R(x;x') \partial'^{\sigma)}  \partial^{\, (\beta}  i\Delta_{S}(x;x')\, e^{-i k_{\mu}\Delta x^{\mu}}\\[6pt]
       &=-\delta_0^{\alpha} \delta_0^{\beta}  \delta_0^{\rho} \delta_0^{\sigma} \, [^{00} \smash{\widetilde{\Omega}}^{00}] - (\delta_{i}^{(\alpha} \delta_0^{\beta)}  \delta_0^{\rho}  \delta_0^{\sigma} + \delta_0^{\alpha} \delta_0^{\beta}  \delta_{i}^{(\rho}  \delta_0^{\sigma)})  \,  [^{i0} \smash{\widetilde{\Omega}}^{00}] - (\delta_{i}^{(\alpha} \delta_0^{\beta)}  \delta_0^{\rho}  \delta_0^{\sigma} - \delta_0^{\alpha} \delta_0^{\beta}  \delta_{i}^{(\rho}  \delta_0^{\sigma)})  \, [^{00} \smash{\widetilde{\Omega}}^{i0}] \\[3pt]
 & - (\delta_{i}^{(\alpha} \delta_{j}^{\beta)}  \delta_0^{\rho}  \delta_0^{\sigma} - \delta_0^{\alpha} \delta_0^{\beta}  \delta_{i}^{(\rho}  \delta_{j}^{\sigma)})  \, [^{0(i} \smash{\overline{\Omega}}^{j)0} ]  \, + \, \delta_{i}^{(\alpha} \delta_0^{\beta)}  \delta_{j}^{(\rho}  \delta_0^{\sigma)} \, ( - [^{ij} \smash{\widetilde{\Omega}}^{00}] +  [^{00} \smash{\widetilde{\Omega}}^{ij}])\\[3pt]
 &+ (\delta_{i}^{(\alpha} \delta_{j}^{\beta)}  \delta_{l}^{(\rho}  \delta_0^{\sigma)} + \delta_{l}^{(\alpha} \delta_0^{\beta)}  \delta_{i}^{(\rho}  \delta_{j}^{\sigma)})  \,  [^{0(i} \smash{\widetilde{\Omega}}^{j)l}]  - (\delta_{i}^{(\alpha} \delta_{j}^{\beta)}  \delta_{l}^{(\rho}  \delta_0^{\sigma)} - \delta_{l}^{(\alpha} \delta_0^{\beta)}  \delta_{i}^{(\rho}  \delta_{j}^{\sigma)})   [^{l(i} \smash{\widetilde{\Omega}}^{j)0} ]\\[3pt]
 &+ \,  \delta_{i}^{(\alpha} \delta_{j}^{\beta)}  \delta_{l}^{(\rho}  \delta_{n}^{\sigma)} \,  [^{n)(i} \smash{\widetilde{\Omega}}^{j)(l}]  \,. 
    \end{split}
\end{equation}
For the two-derivatives term, we will have:
\begin{equation}
    \begin{split}
      &  \int  d^4 \Delta x \, \partial^{\, (\alpha}  i \Delta^R(x;x')   \partial^{\, \beta)}  i\Delta_{S}(x;x')\, e^{-i k_{\mu}\Delta x^{\mu}} \\
      &\qquad \qquad = \delta_0^{\alpha} \delta_0^{\beta} \, [^{0} \smash{\widetilde{\Omega}}^{0}] + \delta_{i}^{(\alpha} \delta_0^{\beta)}\, ( [^{0} \smash{\widetilde{\Omega}}^{i}] + [^{i} \smash{\widetilde{\Omega}}^{0}]) + \delta_{i}^{(\alpha} \delta_{j}^{\beta)}\,  [^{(i} \smash{\widetilde{\Omega}}^{j)}],
      \end{split}
      \end{equation}
      and similarly for derivation with respect to $x'$:
      \begin{equation}
          \begin{split}
       & \int  d^4 \Delta x\, \partial'^{(\rho}  i \Delta^R(x;x')   \partial'^{\sigma)}  i\Delta_{S}(x;x')\, e^{- i k_{\mu}\Delta x^{\mu}} \\
       &\qquad \qquad = -\delta_0^{\rho} \delta_0^{\sigma} \, [^{0} \smash{\widetilde{\Omega}}^{0}] + \delta_{i}^{(\rho} \delta_0^{\sigma)}\, ( [^{0} \smash{\widetilde{\Omega}}^{i}] - [^{i} \smash{\widetilde{\Omega}}^{0}]) + \delta_{i}^{(\rho} \delta_{j}^{\sigma)}\, [^{(i} \smash{\widetilde{\Omega}}^{j)}].
    \end{split}
\end{equation}
Using this result, we can write the tensor structure of the total contribution to the Wigner space self-energy coming from the 4-derivatives terms,
\begin{equation} \label{4ders}
    \begin{split}
         \int  d^4 \Delta x \, &e^{- i k_{\mu}\Delta x^{\mu}}\,\\
        \times &\bigg(  \partial^{\, \alpha)} \partial'^{(\rho}  i \Delta^R(x;x') \partial'^{\sigma)}  \partial^{\, (\beta}  i\Delta_{S}(x;x')-\frac{1}{2}\eta^{\alpha \beta}  \partial'^{(\rho} \partial_{\, \lambda} i \Delta^R(x;x') \partial^{ \lambda}  \partial'^{\sigma)} i\Delta_{S}(x;x') -  \\
        &  -  \frac{1}{2}\eta^{\rho \sigma} \partial'_{\lambda} \partial^{\, (\alpha} i \Delta^R(x;x') \partial^{\, \beta)} \partial'^{\lambda}  i\Delta_{S}(x;x') + \frac{1}{4} \eta^{\rho \sigma} \eta^{\alpha \beta}  \partial_{\, \gamma}\partial'_{\lambda}  i \Delta^R(x;x') \partial'^{\lambda} \partial^{\, \gamma} i\Delta_{S}(x;x')  \bigg)\\[4pt]
&\quad = \frac14\,
\bigl(\delta_0^{\alpha}\,\delta_0^{\beta}\,\delta_0^{\rho}\,\delta_0^{\sigma}
      +\;\overline{\eta}^{\alpha\beta}\,\overline{\eta}^{\rho\sigma}\bigr)
\Bigl(
  -\,[^{00}\smash{\widetilde{\Omega}}^{00}]
  \;+\;\delta_{ij}\,\delta_{\ell n}\;[^{\ell)( i}\smash{\widetilde{\Omega}}^{j)(n}]
\Bigr)
\\[0.5ex]
&\quad
-\;\frac14\,
\bigl(\overline{\eta}^{\rho\sigma}\,\delta_0^{\alpha}\,\delta_0^{\beta} +\;\overline{\eta}^{\alpha\beta}\,\delta_0^{\rho}\,\delta_0^{\sigma}\bigr)
\Bigl(
   [^{00}\smash{\widetilde{\Omega}}^{00}]
  +\;\delta_{ij}\,\delta_{\ell n}\;[^{\ell)( i}\smash{\widetilde{\Omega}}^{j)(n}]
\Bigr)
\\[0.5ex]
&\quad
-\;\frac12\,
\bigl(\delta_0^{\alpha}\,\delta_0^{\beta}\,\delta_{\ell}^{(\rho}\,\delta_0^{\sigma)}
+\;\delta_{\ell}^{(\alpha}\,\delta_0^{\beta)}\,\delta_0^{\rho}\,\delta_0^{\sigma}\bigr)
\Bigl(
   [^{\ell 0}\smash{\widetilde{\Omega}}^{00}]
  -\;\delta_{ij}\;[^{0(i}\smash{\widetilde{\Omega}}^{j)\ell}]
\Bigr)
\\[0.5ex]
&\quad
-\;\frac12\,
\bigl(\overline{\eta}^{\alpha\beta}\,\delta_{\ell}^{(\rho}\,\delta_0^{\sigma)}
+\;\overline{\eta}^{\rho\sigma}\,\delta_{\ell}^{(\alpha}\,\delta_0^{\beta)}\bigr)
\Bigl(
   [^{\ell 0}\smash{\widetilde{\Omega}}^{00}]
  +\;\delta_{ij}\;[^{0(i}\smash{\widetilde{\Omega}}^{j)\ell}]
\Bigr)
\\[0.5ex]
&\quad
-\;\frac12\,
\bigl(\delta_{\ell}^{(\alpha}\,\delta_0^{\beta)}\,\overline{\eta}^{\rho\sigma}
      -\;\delta_{\ell}^{(\rho}\,\delta_0^{\sigma)}\,\overline{\eta}^{\alpha\beta}\bigr)
\Bigl(
   [^{00}\smash{\widetilde{\Omega}}^{\ell 0}]
  +\;\delta_{ij}\;[^{\ell(i}\smash{\widetilde{\Omega}}^{j)0}]
\Bigr)
\\[0.5ex]
&\quad
-\;\frac12\,
\bigl(\delta_{\ell}^{(\alpha}\,\delta_0^{\beta)}\,\delta_0^{\rho}\,\delta_0^{\sigma}
      -\;\delta_0^{\alpha}\,\delta_0^{\beta}\,\delta_{\ell}^{(\rho}\,\delta_0^{\sigma)}\bigr)
\Bigl(
   [^{00}\smash{\widetilde{\Omega}}^{\ell 0}]
  +\;\delta_{ij}\;[^{\ell(i}\smash{\widetilde{\Omega}}^{j)0}]
\Bigr)
\\[0.5ex]
&\quad
+\;\delta_{i}^{(\alpha}\,\delta_0^{\beta)}\,\delta_{j}^{(\rho}\,\delta_0^{\sigma)}
\Bigl(
  -\,[^{i j}\smash{\widetilde{\Omega}}^{00}]
  +\,[^{00}\smash{\widetilde{\Omega}}^{ij}]
\Bigr)
\\[0.5ex]
&\quad
-\;\frac12\,
\bigl(\delta_{i}^{(\alpha}\,\delta_{j}^{\beta)}\,\delta_0^{\rho}\,\delta_0^{\sigma}
      -\;\delta_0^{\alpha}\,\delta_0^{\beta}\,\delta_{i}^{(\rho}\,\delta_{j}^{\sigma)}\bigr)
[^{0(i}\smash{\widetilde{\Omega}}^{j)0}]
\\[0.5ex]
&\quad
-\;\frac12\,
\bigl(\delta_{i}^{(\alpha}\,\delta_{j}^{\beta)}\,\overline{\eta}^{\rho\sigma}
      -\;\overline{\eta}^{\alpha\beta}\,\delta_{i}^{(\rho}\,\delta_{j}^{\sigma)}\bigr)
[^{0(i}\smash{\widetilde{\Omega}}^{j)0}]
\\[0.5ex]
&\quad
-\;\frac12\,
\bigl(\overline{\eta}^{\rho\sigma}\,\delta_0^{\alpha}\,\delta_0^{\beta}
      -\;\overline{\eta}^{\alpha\beta}\,\delta_0^{\rho}\,\delta_0^{\sigma}\bigr)
\delta_{ij}\;[^{0(i}\smash{\widetilde{\Omega}}^{j)0}]
\\[0.5ex]
&\quad
+\;\bigl(\delta_{i}^{(\rho}\,\delta_{j}^{\sigma)}\,\delta_{\ell}^{(\alpha}\,\delta_0^{\beta)}   +\delta_{i}^{(\alpha}\,\delta_{j}^{\beta)}\,\delta_{\ell}^{(\rho}\,\delta_0^{\sigma)}\bigr)
[^{0(i}\smash{\widetilde{\Omega}}^{j)\ell}]
\\[0.5ex]
&\quad
+\;\bigl(\delta_{i}^{(\rho}\,\delta_{j}^{\sigma)}\,\delta_{\ell}^{(\alpha}\,\delta_0^{\beta)}
      -\;\delta_{i}^{(\alpha}\,\delta_{j}^{\beta)}\,\delta_{\ell}^{(\rho}\,\delta_0^{\sigma)}\bigr)
[^{\ell(i}\smash{\widetilde{\Omega}}^{j)0}]
\\[0.5ex]
&\quad
-\;\frac12\,
\bigl(\overline{\eta}^{\rho\sigma}\,\delta_{i}^{(\alpha}\,\delta_{j}^{\beta)}
+\overline{\eta}^{\alpha\beta}\,\delta_{i}^{(\rho}\,\delta_{j}^{\sigma)}\bigr)
\delta_{\ell n}\;[^{\ell)( i}\smash{\widetilde{\Omega}}^{j)(n}]
\\[0.5ex]
&\quad
+\;\frac12\,
\bigl(\delta_0^{\rho}\,\delta_0^{\sigma}\,\delta_{i}^{(\alpha}\,\delta_{j}^{\beta)}      +\delta_0^{\alpha}\,\delta_0^{\beta}\,\delta_{i}^{(\rho}\,\delta_{j}^{\sigma)}\bigr)
\delta_{\ell n}\;[^{\ell)( i}\smash{\widetilde{\Omega}}^{j)(n}]
\\[0.5ex]
&\quad
+\;\delta_{i}^{(\alpha}\,\delta_{j}^{\beta)}\,\delta_{\ell}^{(\rho}\,\delta_{n}^{\sigma)}\;
[^{\ell)( i}\smash{\widetilde{\Omega}}^{j)(n}]
\;,
\end{split}
\end{equation}
and the remaining terms:
\begin{equation}\label{2ders}
    \begin{split}
        \int  d^4 \Delta &x \,  e^{-i k_{\mu}\Delta x^{\mu}}\, \\
        & \left[ \frac{1}{2}\eta^{\alpha \beta}m^2a^2  \left(-\partial'^{(\rho} i \Delta^R(x;x') \partial'^{\sigma)} i\Delta_{S}(x;x')  + \frac{1}{2} \eta^{\rho \sigma}
\partial'_{\lambda}   i \Delta^R(x;x')  \partial'^{\lambda}   i\Delta_{S}(x;x') \right) \right.\\
&+ \frac{1}{2}\eta^{\rho \sigma}m^2a'^{\,2} \left( - \partial^{\, (\alpha} i \Delta^R(x;x') \partial^{\, \beta)}  i\Delta_{S}(x;x') + \frac{1}{2} \eta^{\alpha \beta}
\partial_{\, \lambda}  i \Delta^R(x;x')   \partial^{\, \lambda}   i\Delta_{S}(x;x') \right) \\
& + \left. \frac{1}{4} \eta^{\alpha \beta}\eta^{\rho \sigma} m^4 a^2a'^{\,2}  \, i \Delta^R(x;x') \, i\Delta_{S}(x;x')\right] \\[7pt]
&=-\frac14\bigl(\delta_0^{\alpha}\,\delta_0^{\beta}\,\overline{\eta}^{\rho\sigma}  +\delta_0^{\rho}\,\delta_0^{\sigma}\,\overline{\eta}^{\alpha\beta}\bigr)
\,\big[\bar{m}^2
\Bigl(
  2\,\delta_{ij}\,[^{(i}\smash{\widetilde{\Omega}}^{j)}]
  +\bar{m}^2\,\widetilde{\Omega}
\Bigr)  + \Delta m_2^2 \Delta m_1^2 \, \widetilde{\Omega}
\\
& \qquad \qquad \qquad + (\Delta m_2^2 - \Delta m_1^2) \,[^{0}\smash{\widetilde{\Omega}}^{0}] - (\Delta m_2^2 + \Delta m_1^2)  \Bigl(2\,\delta_{ij}\,[^{(i}\smash{\widetilde{\Omega}}^{j)}]
 +\bar{m}^2\,[^{0}\smash{\widetilde{\Omega}}^{0}]
\Bigr) \big] \\[0.5ex]
& \quad -\frac12\bigl(\delta_0^{\alpha}\,\delta_0^{\beta}\,\overline{\eta}^{\rho\sigma}
            -\delta_0^{\rho}\,\delta_0^{\sigma}\,\overline{\eta}^{\alpha\beta}\bigr)
\,\bar{m}^2
 \,[^{0}\smash{\widetilde{\Omega}}^{0}]{}
\\[0.5ex]
&\quad
-\frac12
\bigl(  \overline{\eta}^{\rho\sigma}\,\delta_{i}^{(\alpha}\,\delta_0^{\beta)}
+\overline{\eta}^{\alpha\beta}\,\delta_{i}^{(\rho}\,\delta_0^{\sigma)}
\bigr)\,
\bar{m}^2\,
  [^{0}\smash{\widetilde{\Omega}}^{i}]
-\frac12
\bigl(  \overline{\eta}^{\rho\sigma}\,\delta_{i}^{(\alpha}\,\delta_0^{\beta)}
 -\overline{\eta}^{\alpha\beta}\,\delta_{i}^{(\rho}\,\delta_0^{\sigma)}
\bigr)\,\bar{m}^2\, [^{i}\smash{\widetilde{\Omega}}^{0}]
\\[0.5ex]
&\quad
+\frac12
\bigl(  \delta_0^{\alpha}\,\delta_0^{\beta}\,\delta_{i}^{(\rho}\,\delta_0^{\sigma)} +\delta_{i}^{(\alpha}\,\delta_0^{\beta)}\,\delta_0^{\rho}\,\delta_0^{\sigma}
\bigr)\,
\bar{m}^2\,
[^{0}\smash{\widetilde{\Omega}}^{i}]
+\frac12
\bigl( \delta_0^{\alpha}\,\delta_0^{\beta}\,\delta_{i}^{(\rho}\,\delta_0^{\sigma)}
 -\delta_{i}^{(\alpha}\,\delta_0^{\beta)}\,\delta_0^{\rho}\,\delta_0^{\sigma}
\bigr)\,
\bar{m}^2\,
  [^{i}\smash{\widetilde{\Omega}}^{0}]
\\[0.5ex]
&\quad
-\frac12
\bigl(  \overline{\eta}^{\rho\sigma}\,\delta_{i}^{\alpha}\,\delta_{j}^{\beta} +\overline{\eta}^{\alpha\beta}\,\delta_{i}^{\rho}\,\delta_{j}^{\sigma}
\bigr)\,
\bar{m}^2\,
[^{(i}\smash{\widetilde{\Omega}}^{j)}]
\\[0.5ex]
&\quad
+\frac14\,
\delta_0^{\alpha}\,\delta_0^{\beta}\,\delta_0^{\rho}\,\delta_0^{\sigma}
\, \big[\bar{m}^2
\Bigl(
2\,\delta_{ij}\,[^{(i}\smash{\widetilde{\Omega}}^{j)}]
 +\bar{m}^2\,[^{0}\smash{\widetilde{\Omega}}^{0}]
\Bigr)  + \Delta m_2^2 \Delta m_1^2 \, \widetilde{\Omega}
\\
& \qquad  \qquad \qquad \quad - (\Delta m_2^2 - \Delta m_1^2) \,[^{0}\smash{\widetilde{\Omega}}^{0}] - (\Delta m_2^2 + \Delta m_1^2) \Bigl(2\,\delta_{ij}\,[^{(i}\smash{\widetilde{\Omega}}^{j)}]
 +\bar{m}^2\,[^{0}\smash{\widetilde{\Omega}}^{0}]
\Bigr) \big] \\[0.5ex]
&\quad
+\frac12
\bigl(  \delta_0^{\alpha}\,\delta_0^{\beta}\,\delta_{i}^{\rho}\,\delta_{j}^{\sigma} +\delta_{i}^{\alpha}\,\delta_{j}^{\beta}\,\delta_0^{\rho}\,\delta_0^{\sigma}
\bigr)\,
\bar{m}^2\,
[^{(i}\smash{\widetilde{\Omega}}^{j)}]
\\[0.5ex]
&\quad
+\frac14\,
\overline{\eta}^{\alpha\beta}\,\overline{\eta}^{\rho\sigma}
\,\big[ \bar{m}^2
\Bigl(2\,\delta_{ij}\,[^{(i}\smash{\widetilde{\Omega}}^{j)}]
 +\bar{m}^2\,[^{0}\smash{\widetilde{\Omega}}^{0}]
\Bigr)  + \Delta m_2^2 \Delta m_1^2 \, \widetilde{\Omega}\\
& \qquad \qquad \qquad + (\Delta m_2^2 - \Delta m_1^2) \,[^{0}\smash{\widetilde{\Omega}}^{0}] - (\Delta m_2^2 + \Delta m_1^2) \Bigl(2\,\delta_{ij}\,[^{(i}\smash{\widetilde{\Omega}}^{j)}]
 +\bar{m}^2\,[^{0}\smash{\widetilde{\Omega}}^{0}]
\Bigr) \big] \\[0.5ex]
& \quad + \frac12\bigl(\Delta m_2^2 \,  \delta_0^{\alpha}\,\delta_0^{\beta}\,\overline{\eta}^{\rho\sigma} - \Delta m_1^2 \, \delta_0^{\rho}\,\delta_0^{\sigma}\,\overline{\eta}^{\alpha\beta}\bigr)\,  [^{0}\smash{\widetilde{\Omega}}^{0}]\\[0.5ex]
 &\quad +\frac{1}{2} \bigl(
   \Delta m_2^2 \, \overline{\eta}^{\rho\sigma}\,\delta_{i}^{(\alpha}\,\delta_0^{\beta)}
 + \Delta m_1^2 \, \overline{\eta}^{\alpha\beta}\,\delta_{i}^{(\rho}\,\delta_0^{\sigma)}
\bigr)\, 
 [^{0}\smash{\widetilde{\Omega}}^{i}]
\\[0.5ex]
&\quad -\frac{1}{2} \bigl(
   \Delta m_2^2 \, \delta_0^{\rho} \delta_0^{\sigma} \,\delta_{i}^{(\alpha}\,\delta_0^{\beta)}
 + \Delta m_1^2 \, \delta_0^{\alpha} \delta_0^{\beta} \,\delta_{i}^{(\rho}\,\delta_0^{\sigma)}
\bigr)\, 
 [^{0}\smash{\widetilde{\Omega}}^{i}]
\\[0.5ex]
 &\quad +\frac{1}{2} \bigl(
   \Delta m_2^2 \, \overline{\eta}^{\rho\sigma}\,\delta_{i}^{(\alpha}\,\delta_0^{\beta)}
 - \Delta m_1^2 \, \overline{\eta}^{\alpha\beta}\,\delta_{i}^{(\rho}\,\delta_0^{\sigma)}
\bigr)\, 
 [^{i}\smash{\widetilde{\Omega}}^{0}]
\\[0.5ex]
&\quad -\frac{1}{2} \bigl(
   \Delta m_2^2 \, \delta_0^{\rho} \delta_0^{\sigma} \,\delta_{i}^{(\alpha}\,\delta_0^{\beta)}
 - \Delta m_1^2 \, \delta_0^{\alpha} \delta_0^{\beta} \,\delta_{i}^{(\rho}\,\delta_0^{\sigma)}
\bigr)\, 
 [^{i}\smash{\widetilde{\Omega}}^{0}]
\\[0.5ex]
 &\quad +\frac{1}{2} \bigl(
   \Delta m_2^2 \, \overline{\eta}^{\rho\sigma}\,\delta_{i}^{\alpha}\,\delta_{j}^{\beta}
 + \Delta m_1^2 \, \overline{\eta}^{\alpha\beta}\,\delta_{i}^{\rho}\,\delta_{j}^{\sigma}
\bigr)\,  [^{(i}\smash{\widetilde{\Omega}}^{j)}]\\[0.5ex]
&\quad -\frac{1}{2} \bigl(
   \Delta m_2^2 \, \delta_0^{\rho} \delta_0^{\sigma} \,\delta_{i}^{\alpha}\,\delta_{j}^{\beta}
 + \Delta m_1^2 \, \delta_0^{\alpha} \delta_0^{\beta} \,\delta_{i}^{\rho}\,\delta_{j}^{\sigma}
\bigr)\, 
  [^{(i}\smash{\widetilde{\Omega}}^{j)}].
    \end{split}
\end{equation}
Summing (\ref{4ders}) and (\ref{2ders}) gives us the expression of the self-energy evaluated in Wigner space. To complete the evaluation of this term, we are left with computing the components of $[^{\alpha \beta} \smash{\widetilde{\Omega}}^{\rho \sigma}]$. We identify the local contributions in terms that contain $[^{00} \smash{\widetilde{\Omega}}^{\alpha \beta}]$. The latter are proportional to $\partial_0^2 \, i \Delta^R$, which contains both a local and a non-local contribution due to the time ordering of the Feynman propagator. We show this by computing $[^{00} \smash{\widetilde{\Omega}}^{00}]$. Using the integral expression (\ref{vacuumretardedpropagator}) for the retarded propagator, we obtain:
\begin{equation}
\begin{split}
&i \Delta^R(x; x') \, i\Delta_{S}(x; x') 
= \frac{1}{\bar{a}^2}\int \frac{d^4 q\, d^3 q'}{(2\pi)^7} \,
\frac{-i}{-(q^0+ i\varepsilon)^2 + \omega^2(q) }
\, \frac{S_{q'}}{2\omega(q')} \\[3pt]
&\quad \qquad \qquad \qquad \qquad \times  e^{-iq^0 \Delta t} e^{i(\vec{q} + \vec{q}') \cdot \Delta \vec{x}} 
\left[ e^{i\, (2\omega(q') \bar{t}- \phi_{q'})} +  e^{-i\,( 2\omega_(q') \bar{t}- \phi_{q'})} \right]\\[9pt]
\Rightarrow &\partial_0^2 i \Delta^R \, \partial_0^2 i\Delta_{S} 
= \bar{a}^2\int \frac{d^4 q\, d^3 q'}{(2\pi)^7} \,
\frac{-i}{-(q^0+i\varepsilon)^2 + \omega_q^2 } \,
\frac{S_{q'}}{2\omega(q')} \, (q^0)^2 \omega^2(q') \\[3pt]
&\quad \qquad \qquad \qquad \qquad \times  e^{-iq^0 \Delta t} e^{i(\vec{q} + \vec{q}') \cdot \Delta \vec{x}} 
\left[ e^{i\, (2\omega(q') \bar{t}- \phi_{q'})} +  e^{-i\,( 2\omega(q') \bar{t}- \phi_{q'})} \right]\\[9pt]
\Rightarrow \,& \, [^{00} \smash{\widetilde{\Omega}}^{00}](k^0, \vec{k}) 
= \int d^4 \Delta x \, e^{-i k \cdot \Delta x} 
\left( \partial_0^2 i \Delta^R \, \partial_0^2 i\Delta_{S} \right) \\[3pt]
&\quad \qquad \qquad= \bar{a}^2\int \frac{d^4 q\, d^3 q'}{(2\pi)^7} \,
\frac{-i}{-(q^0+ i\varepsilon)^2 + \omega^2(q) } \,
\frac{S_{q'}}{2\omega(q')} \, (q^0)^2 \omega^2(q') \\[3pt]
&\quad \qquad \qquad \qquad \qquad \times \, \delta^3(\vec{q} + \vec{q}' - \vec{k}) \delta(k^0-q^0) \,  
\left[ e^{i\, (2\omega(q') \bar{t}- \phi_{q'})} +  e^{-i\,( 2\omega(q') \bar{t}- \phi_{q'})} \right].
\end{split}
\end{equation}
We apply the delta functions and change the integration variable to $\vec{p} = \vec{q}- \vec{k}/2 \equiv \vec{q}$:
\begin{equation} \label{0000s}
\begin{split}
[^{00} \smash{\widetilde{\Omega}}^{00}](k^0, \vec{k})  &= \bar{a}^2\int \frac{d^3 q}{(2\pi)^3} \,  \frac{S_+}{2\omega_+}\omega_+^2 \, (-i) 
\left[ \frac{(k^0 )^2}{-(k^0+i\varepsilon )^2 + \omega_-^2 }\right] \left[ e^{i\, (2\omega_+\bar{t}- \phi_+)} +  e^{-i\, (2\omega_+\bar{t}- \phi_+)} \right] \\[3pt]
&= i\bar{a}^2 \int \frac{d^3 q}{(2\pi)^3} \frac{S_+}{\omega_+}  \, \omega_+^2 \, \cos[2\omega(q;\bar{t}) \bar{t}- \phi(q)] \\[3pt]
&+\bar{a}^2 \int \frac{d^3 q}{(2\pi)^3} \frac{S_+ }{2\omega_+} \omega_+^2 \omega_-^2
   \left[ \frac{-i}{-(k^0+i\varepsilon)^2 + \omega_-^2 } \right] \left[ e^{i\, (2\omega_+\bar{t}- \phi_+)} +  e^{-i\, (2\omega_+\bar{t}- \phi_+)} \right] \\[3pt]
&= i \int \frac{d^3 q}{(2\pi)^3} \frac{S_q}{\omega(q;\bar{t})} 
\, (q^2 + a^2m^2) \, \cos[2\omega(q;\bar{t}) \bar{t}- \phi_q] \\[3pt]
&\quad + \bar{a}^2\int \frac{d^3 q}{(2\pi)^3} \frac{S_+}{4 \omega_+ \omega_-} 
\, \omega_+^2 \omega_-^2 
\left[ \frac{-i}{k^0 +  \omega_- + i\varepsilon} 
       + \frac{i}{k^0  -\omega_- + i\varepsilon} \right] \\[2pt]
       &  \qquad \qquad \qquad \qquad \qquad \qquad  \times \left[ e^{i\, (2\omega_+\bar{t}- \phi_+)} +  e^{-i\, (2\omega_+\bar{t}- \phi_+)} \right],
\end{split}
\end{equation}
where we used the notation for momentum-dependent function introduced in \eqref{notationmomentum}.
Here the first term will be the local contribution, since
\begin{equation}
\begin{split}
   &  \, \int_{k} \frac{e^{i k \cdot \Delta x}}{(2\pi)^4} \,  \int \frac{d^3 q}{(2\pi)^3} \frac{S_q}{\omega(q;\bar{t})} 
\,i (q^2 + a^2m^2) \, \cos[2(\omega_q \bar{t}- \phi(q))]\\[5pt]
   &=  - i  \delta^4(x-x') \, \partial_0^2 \, i\Delta_{S} (x,x') =   i  \delta^4(x-x') \left( a^2 m^2 \, I^{S}_1  +  I^{S}_2\right).
\end{split}
\end{equation}
Employing the standard integrals introduced in \eqref{nonlocalSintegrals} we therefore obtain:
 \begin{equation} 
    \begin{split}
        [^{00} \smash{\widetilde{\Omega}}^{00}](k^0, \vec{k}) &= \, i \, \left( \bar{m}^2\, I^{S}_1+  I^{S}_2\right) \\
        &+ \sum_{t=1}^4 \left[\widetilde{J}^t_{40}-k^2 \, \widetilde{J}^t_{02}+ 2 \left( \frac{k^2}{4}+ \bar{m}^2\right) \widetilde{J}^t_{20} + \left( \frac{k^2}{4}+ \bar{m}^2 \right)^2 \widetilde{J}^t_{00} \right].
    \end{split}
\end{equation}
We notice how this expression is equal to the one found for the mixing contribution in equation (\ref{confront}) once we substitute $I^{N}_i\to I^{S}_i$ and $\bar{J_{rs}^t}(k^0, k^2) \to \widetilde{J}_{rs}^t (k^0,k^2, \bar{t})$. This will continue to be true for all components of $[^{\alpha \beta} \smash{\widetilde{\Omega}}^{\rho \sigma}]$, since their derivation will be completely analogous to the previous case, just with a different integral structure. This means that all the results obtained for $[^{\alpha \beta} \smash{\overline{\Omega}}^{\rho \sigma}]$ in equations (\ref{omega1})-(\ref{omega2}) can be applied to $[^{\alpha \beta} \smash{\widetilde{\Omega}}^{\rho \sigma}]$ once we substitute the relative standard integrals. 
Now we have everything we need to compute the 3-point self-energy. The local contribution reads:
\begin{equation}
\begin{split}
      i[^{\alpha \beta}\Sigma_{S,\text{3pt},\text{L}}^{\, \rho \sigma}](x;x')& =i \kappa^2\delta^D(x-x') \bigg\{\frac{1}{3} \delta_0^{(\alpha} \eta^{\beta)(\rho}\delta_0^{\sigma)} I^{S}_2   \\
      &-  \frac{1}{4}\left[ \eta^{\alpha \beta}\eta^{\rho \sigma} + 2 (\eta^{\alpha \beta} \delta_0^{\rho} \delta_0^{\sigma} +  \eta^{\rho \sigma} \delta_0^{\alpha} \delta_0^{\beta}) + 4 \delta_0^{\alpha} \delta_0^{\beta}\delta_0^{\rho} \delta_0^{\sigma}\right] \bigg(I^{S}_2   + a^2 m^2 I^{S}_1\bigg) \bigg\},
      \end{split}
      \end{equation}
    which will be summed to the 4-point contribution in \eqref{app:4ptS} to give the full local contribution, presented in \eqref{localsesqueezing}.
      The non-local part is better expressed in Wigner space. 
Using the same formalism as the previous section, we are going to distinguish between the contributions coming from different classes of standard integrals:
\begin{equation}
   i[^{\alpha \beta}\overline{\Sigma}_{S}^{\, \rho \sigma}](k^0,k^2,\bar{t}) =i[^{\alpha \beta}\overline{\Sigma}_{S,(\text{\rom{1}})}^{\, \rho \sigma}](k^0,k^2,\bar{t}) +  i[^{\alpha \beta}\overline{\Sigma}_{S,(\text{\rom{2}})}^{\, \rho \sigma}](k^0,k^2,\bar{t}) + i[^{\alpha \beta}\overline{\Sigma}_{S,(\text{\rom{3}})}^{\, \rho \sigma}](k^0,k^2,\bar{t}).
\end{equation}
The results obtained for these components are shown in equations (\ref{s11})-(\ref{s13}).

\subsection{Contribution arising from the condensate}
We now proceed to evaluate the change in the retarded self-energy of the graviton field due to the presence of a non-zero condensate, introduced in eq.~\eqref{retardedseproper3pt}. Since the occupation number and squeezing do not contribute to the retarded propagator, such a contribution to the self-energy will read
\begin{equation} \label{initialstep}
\begin{split}
  i[^{\alpha \beta}\Sigma&_{\chi,\text{3pt},R}^{\, \rho \sigma}](x;x')\\
  &=\kappa^2 \left[ \partial^{\, \alpha)}\partial'^{(\rho}  i\Delta^{\, R}(x;x') \partial'^{\sigma)}\partial^{(\beta}  \bar{\chi}(t)\bar{\chi}(t')   -\frac{1}{2}\eta^{\alpha \beta}  \partial'^{\rho)} \partial_{\, \lambda} i\Delta^{\, R}(x;x')  \partial^{\, \lambda} \partial'^{(\sigma} \bar{\chi}(t) \bar{\chi}(t')    \right. \\[3pt]
        & -  \frac{1}{2}\eta^{\rho \sigma} \partial^{\, \alpha)}\partial'_{\lambda}  i\Delta^{\, R}(x;x') \partial'^{\lambda} \partial^{( \beta} \bar{\chi}(t)\bar{\chi}(t')  + \frac{1}{4} \eta^{\rho \sigma} \eta^{\alpha \beta}  \partial_{\, \gamma}\partial'_{\lambda}  i\Delta^{\, R}(x;x')  \partial^{\, \gamma} \partial'^{\lambda} \bar{\chi}(t) \bar{\chi}(t')   \\[3pt]
        &+ \frac{1}{2}\eta^{\alpha \beta}m^2a^2   \left(-\partial'^{(\rho} i\Delta^{\, R}(x;x') \partial'^{\sigma)}\bar{\chi}(t) \bar{\chi}(t')  + \frac{1}{2} \eta^{\rho \sigma}
\partial'_{\lambda}  i\Delta^{\, R}(x;x')   \partial'^{\lambda}\bar{\chi}(t) \bar{\chi}(t')  \right) \\[3pt]
&+ \frac{1}{2}\eta^{\rho \sigma}m^2a'^{\,2} \left( - \partial^{( \alpha} i\Delta^{\, R}(x;x')  \partial^{\, \beta)}  \bar{\chi}(t) \bar{\chi}(t') + \frac{1}{2} \eta^{\alpha \beta}
\partial_{\, \lambda}  i\Delta^{\, R}(x;x')   \partial^{\, \lambda}   \bar{\chi}(t)  \bar{\chi}(t')\right) \\[3pt]
& + \left. \frac{1}{4} \eta^{\alpha \beta}\eta^{\rho \sigma} m^4 a^2 a'^{\, 2} i\Delta^{\, R}(x;x') \bar{\chi}(t)\bar{\chi}(t')\right] ,
\end{split}    
\end{equation}
where \(i\Delta^R\) is the vacuum retarded propagator, given explicitly in \eqref{vacuumretardedpropagator}.
We notice how the product of the condensate is a sum of a $\Delta t= t-t'$ dependent term and a $\bar{t}=(t+t')/2$ dependent one. Since these two terms need to be treated differently (as they have a different sign when derived with respect to $t'$) we will distinguish between them
\begin{equation} \label{distinction}
\begin{split}
    &\bar{\chi}(t)\bar{\chi}(t')= \bar{\chi}\bar{\chi}(\Delta t) + \bar{\chi}\bar{\chi}(\bar{t});\\[3pt]
    &\bar{\chi}\bar{\chi}(\Delta t):=\frac{\phi_0^2}{2\sqrt{aa'}} \cos (m \Delta t);\\[3pt]
   &\bar{\chi}\bar{\chi}(\bar{t}):=\frac{\phi_0^2}{2\sqrt{aa'}} \cos [2(m \bar{t}- \theta_0 )].
\end{split}
\end{equation}
  We use the separation of the product of the condensates defined in (\ref{distinction}) to obtain two different contributions with different dependence on time. The latter are obtained by substituting individually the two different contributions in (\ref{initialstep}). The sum of these terms will give the total self-energy.
Similarly to the previous cases we have studied, such expression will possess a local and a non-local part, due to the presence of $\partial_0^2 i \Delta^R$. We separate such terms from the rest of the self-energy, and write 
\begin{equation}
     i [^{\alpha \beta}\Sigma_{\chi,\text{3pt},R}^{\, \rho \sigma}](x;x')=  i[^{\alpha \beta}\Sigma_{\chi,\text{L}}^{\, \rho \sigma}](x;x') +  i [^{\alpha \beta}\Sigma_{\chi,\Delta t}^{\, \rho \sigma}](x;x')+  i [^{\alpha \beta}\Sigma_{\chi,\bar{t}}^{\, \rho \sigma}](x;x').
\end{equation}
A quick computation then brings us to the following results for these terms:
\begin{equation}
    \begin{split}
      i[^{\alpha \beta}\Sigma_{\chi,\text{L}}^{\, \rho \sigma}](x,x') =&i\kappa^2 \delta^D(x-x')\frac{a^2m^2}{4a} \phi_0^2 \sin^2(mt- \theta_0)\\
     & \times \big[ \delta_0^{\alpha} \delta_0^{\beta} \delta_0^{\rho} \delta_0^{\sigma}  + \overline{\eta}^{\alpha \beta} \overline{\eta}^{\rho \sigma} + \big( \overline{\eta}^{\alpha \beta}  \delta_0^{\rho} \delta_0^{\sigma} + \overline{\eta}^{\rho \sigma}\delta_0^{\alpha} \delta_0^{\beta}\big)\big],
    \end{split}
\end{equation}
for the local term, while the component dependent on the time separation reads as follows:
\begin{equation}
   \begin{split}
        i &[^{\alpha \beta}\Sigma_{\chi,\Delta t}^{\, \rho \sigma}](x;x') = -\kappa^2 \frac{\sqrt{aa'} m^2}{4} \phi_0^2 \cos(m \Delta t) \\[3pt]
        &\times \bigg[ \overline{\eta}^{\alpha \beta} \overline{\eta}^{\rho \sigma}\frac{-\bar{m}^2-m^2aa'-\nabla^2}{2} + \delta_0^{\alpha} \delta_0^{\beta} \delta_0^{\rho} \delta_0^{\sigma}  \frac{-\bar{m}^2-m^2aa'- \nabla^2}{2} \\[3pt]
       & +  \big( \overline{\eta}^{\alpha \beta}  \delta_0^{\rho} \delta_0^{\sigma} + \overline{\eta}^{\rho \sigma}\delta_0^{\alpha} \delta_0^{\beta}\big)\,  \frac{\nabla^2 +m^2aa' }{2}\,+  \big( \overline{\eta}^{\alpha \beta}  \delta_0^{(\rho} \delta_{j}^{\sigma)} + \overline{\eta}^{\rho \sigma}\delta_0^{(\alpha} \delta_{j}^{\beta)}\big) \,  (-\partial_{0} \partial^{j}) \, \\[3pt]
       &  + \big( \delta_0^{\alpha} \delta_0^{\beta}  \delta_0^{(\rho} \delta_{j}^{\sigma)} + \delta_0^{\rho} \delta_0^{\sigma} \delta_0^{(\alpha} \delta_{j}^{\beta)}\big) (- \partial_{0} \partial^{j}) \,  + \delta_{j}^{(\alpha} \delta_0^{\beta)} \delta_{k}^{(\rho} \delta_0^{\sigma)}  \, 2\partial^{j} \partial^{k}\bigg]i\Delta^{\! R}(x;x')\\[3pt]
      & \qquad \qquad \qquad \qquad - \kappa^2\frac{\sqrt{aa'} m^2}{4} \phi_0^2 \sin(m \Delta t)\\[3pt]
       &\times \bigg[ \overline{\eta}^{\alpha \beta} \overline{\eta}^{\rho \sigma} \frac{-am\, -a'm}{2} \partial_{0}  + \delta_0^{\alpha} \delta_0^{\beta} \delta_0^{\rho} \delta_0^{\sigma}  \frac{am\, + a'm}{2} \partial_{0}  \\[3pt]
       & +  \big( am\,  \overline{\eta}^{\alpha \beta}  \delta_0^{(\rho} \delta_{j}^{\sigma)} + a'\!m \,\overline{\eta}^{\rho \sigma}\delta_0^{(\alpha} \delta_{j}^{\beta)}\big) \,  \partial^{j}  + \big( am \, \delta_0^{\alpha} \delta_0^{\beta}  \delta_0^{(\rho} \delta_{j}^{\sigma)} +a'\!m \,\delta_0^{\rho} \delta_0^{\sigma} \delta_0^{(\alpha} \delta_{j}^{\beta)}\big) (-\partial^{j})\\[3pt]
       &+ \big( am\,  \overline{\eta}^{\alpha \beta}  \delta_0^{\rho} \delta_0^{\sigma} + a'm \,\overline{\eta}^{\rho \sigma}\delta_0^{\alpha} \delta_0^{\beta}\big) (-\partial_{0}) +  \big(\,  \overline{\eta}^{\alpha \beta}  \delta_0^{\rho} \delta_0^{\sigma} + \,\overline{\eta}^{\rho \sigma}\delta_0^{\alpha} \delta_0^{\beta}\big)\frac{am\, + a'm}{2} \partial_{0} \bigg]i\Delta^{\! R}(x;x').
    \end{split}
\end{equation}
The last term evolves with the average time $\bar{t}$:
\begin{equation}
   \begin{split}
       i [^{\alpha \beta}&\Sigma_{\chi,\bar{t}}^{\, \rho \sigma}](x;x') =\kappa^2  \frac{\sqrt{aa'} m^2}{4} \phi_0^2 \cos[2(m \bar{t}-\theta_0)] \\[3pt]
        & \times\bigg[ \overline{\eta}^{\alpha \beta} \overline{\eta}^{\rho \sigma} \frac{-\bar{m}^2+m^2aa'+\nabla^2 }{2}\, + \delta_0^{\alpha} \delta_0^{\beta} \delta_0^{\rho} \delta_0^{\sigma}   \frac{-\bar{m}^2+m^2aa'+\nabla^2 }{2}\,   \\[3pt]
       & + \big( \overline{\eta}^{\alpha \beta}  \delta_0^{\rho} \delta_0^{\sigma} + \overline{\eta}^{\rho \sigma}\delta_0^{\alpha} \delta_0^{\beta}\big) \frac{-m^2aa' -\bar{m}^2+\nabla^2}{2} +  \big( \overline{\eta}^{\alpha \beta}  \delta_0^{(\rho} \delta_{j}^{\sigma)} + \overline{\eta}^{\rho \sigma}\delta_0^{(\alpha} \delta_{j}^{\beta)}\big) \,  (-\partial_{0} \partial^{j})   \\[3pt]
       & + \big( \delta_0^{\alpha} \delta_0^{\beta}  \delta_0^{(\rho} \delta_{j}^{\sigma)} + \delta_0^{\rho} \delta_0^{\sigma} \delta_0^{(\alpha} \delta_{j}^{\beta)}\big)  (-\partial_{0} \partial^{j}) + \delta_{j}^{(\alpha} \delta_0^{\beta)} \delta_{k}^{(\rho} \delta_0^{\sigma)}  \, 2\partial^{j} \partial^{k} \bigg]i\Delta^{\! R}(x;x')\\[3pt]
      & \qquad \qquad  \qquad - \kappa^2 \frac{\sqrt{aa'} m^2}{4} \phi_0^2 \sin[2(m \bar{t}- \theta_0)] \\[3pt]
       &\times \bigg[ \overline{\eta}^{\alpha \beta} \overline{\eta}^{\rho \sigma} \frac{am\, -a'm}{2} \partial_{0}  + \delta_0^{\alpha} \delta_0^{\beta} \delta_0^{\rho} \delta_0^{\sigma}  \frac{-am\, + a'm}{2} \partial_{0}  \\[3pt]
       & +  \big(  am\,  \overline{\eta}^{\alpha \beta}  \delta_0^{(\rho} \delta_{j}^{\sigma)} -a'm \,\overline{\eta}^{\rho \sigma}\delta_0^{(\alpha} \delta_{j}^{\beta)}\big) \, (- \partial^{j})  + \big( am \, \delta_0^{\alpha} \delta_0^{\beta}  \delta_0^{(\rho} \delta_{j}^{\sigma)} -a'm \,\delta_0^{\rho} \delta_0^{\sigma} \delta_0^{(\alpha} \delta_{j}^{\beta)}\big) \,\partial^{j} \,\\[3pt]
       &+ \big( am\,  \overline{\eta}^{\alpha \beta}  \delta_0^{\rho} \delta_0^{\sigma} - a'm \,\overline{\eta}^{\rho \sigma}\delta_0^{\alpha} \delta_0^{\beta}\big) \, \partial_{0} +  \big(\,  \overline{\eta}^{\alpha \beta}  \delta_0^{\rho} \delta_0^{\sigma} + \,\overline{\eta}^{\rho \sigma}\delta_0^{\alpha} \delta_0^{\beta}\big)\frac{-am\, + a'm}{2} \partial_{0} \bigg]i\Delta^{\! R}(x;x').
    \end{split}
\end{equation}
We rewrite these expressions to cast each term as a total derivative. This allows the derivatives to be pulled outside the integral over $x'$, thereby simplifying the integration process.
To achieve this, we make use of the following identities:
\begin{equation}
    \begin{split}
        &\cos(mt) \, \partial_{\, 0} \, i\Delta^R\, = \, \partial_{\, 0} [\cos(mt) \, i\Delta^R] + \, am \, \sin(mt) \, i\Delta^R;\\[5pt]
        &\sin(mt) \, \partial_{\, 0} \, i\Delta^R\, = \, \partial_{\, 0} [\sin(mt) \, i\Delta^R] - \, am \, \cos(mt) \, i\Delta^R,
    \end{split}
\end{equation}
where we have used \(\partial_0 t = a\). Substituting these identities into the expression for the self-energy, we obtain the final contributions shown in equations (\ref{timediffterm}) and (\ref{avrgtimeterm}).

\vskip 1cm

\section{Computation of the local integrals} \label{Appendix C}
In this appendix, we analyze the expressions for \( I_j^{N}(t) \) and \( I_j^{S}(t) \) introduced in eqs.~\eqref{Iintegralsoccupationnumber} and \eqref{Iintegralssqueezing}, which are made tractable by the power-law ansatz for the occupation number and squeezing spectra given in eq.~\eqref{ansatzspectra}. We begin with the contribution from the occupation number.
\begin{equation}
 I_j^N \, = \, \frac{1}{2\pi^2 am} \int_0^{\infty} dk\, \frac{ N_k \, k^{2j}}{ \sqrt{1+\frac{k^2}{a^2m^2}}}.
\end{equation}
Due to condition~\eqref{nonrelativisticcondition}, we expand the denominator as a power series in $k^2/(a^2 m^2)$:
\begin{equation} \label{phaseexpansion}
    \bigg(1+\dfrac{k^2}{a^2 m^2} \bigg)^{\!-\frac{1}{2}} = \sum_{n=0}^{\infty} \frac{(-1)^n (1/2)_n}{n! \, (a m)^{2n}} \, k^{2n},
\end{equation}
where $(a)_n$ denotes the Pochhammer symbol. Substituting this expansion transforms the integral $I_j^N$ into an infinite sum of standard Gaussian moments:
\begin{equation}
    \begin{split}
        I_j^N &= \sum_{n=0}^{\infty} \mathcal{I}^N_{n,j}, \\[3pt]
        \mathcal{I}^N_{n,j} &=
        \frac{(-1)^n (1/2)_n}{n! \, (a m)^{2n}} \,
        \frac{1}{2\pi^2 a m}
        \int_0^{\infty} dk \, N_k \, k^{2j+2n}.
    \end{split}
\end{equation}
Inserting the power-law ansatz \(N_k = N_0 \, (k/k_*)^{n_N} e^{-k^2/k_{\text{UV}}^2}\) and defining \(\beta \equiv 2j+2n+n_N\) and \(\lambda \equiv k_{\text{UV}}^{-2}\), the integral becomes
\begin{equation}
    \mathcal{I}^N_{n,j} =
    \frac{(-1)^n (1/2)_n}{n! \, (a m)^{2n}} \,
    \frac{1}{2\pi^2 a m} \,
    \frac{N_0}{k_*^{n_N}}
    \int_0^{\infty} dk \, k^{\beta} e^{-\lambda k^2}.
\end{equation}
The change of variable $x \equiv \lambda k^2$ rewrite the momentum integral in terms of the Gamma function:
\begin{equation} \label{approximationN}
    \begin{split}
        \mathcal{I}^N_{n,j}
        &= \frac{(-1)^n (1/2)_n}{n! \, (a m)^{2n}} \,
        \frac{1}{4\pi^2 a m} \,
        \frac{N_0}{k_*^{n_N}} \,
        \lambda^{-\frac{\beta+1}{2}}
        \int_0^{\infty} dx \, x^{\frac{\beta-1}{2}} e^{-x} \\[3pt]
        &= \frac{(-1)^n (1/2)_n}{n! \, (a m)^{2n}} \,
        \frac{1}{4\pi^2 a m} \,
        \frac{N_0}{k_*^{n_N}} \,
        k_{\text{UV}}^{2j+2n+1+n_N} \,
        \Gamma\!\left( \frac{n_N + 2j + 2n + 1}{2}\right).
    \end{split}
\end{equation}
Summing over $n$ identifies the series as a hypergeometric function:
\begin{equation}
    \begin{split}
        I_j^N &=
        \frac{1}{4\pi^2 a m} \,
        \frac{N_0}{k_*^{n_N}} \,
        k_{\text{UV}}^{2j+1+n_N} \,
        \Gamma\!\left( \frac{n_N + 2j + 1}{2}\right) \\[3pt]
        &\quad\times\sum_{n=0}^{\infty}
        \frac{ \bigl( \frac{n_N+1+2j}{2} \bigr)_n \, (-1)^n (1/2)_n }{n!}
        \left( \frac{k_{\text{UV}}}{a m} \right)^{2n} \\[3pt]
        &=
        \frac{1}{4\pi^2 a m} \,
        \frac{N_0}{k_*^{n_N}} \,
        k_{\text{UV}}^{2j+1+n_N} \,
        \Gamma\!\left( \frac{n_N + 2j + 1}{2}\right) \,
        {}_2F_0\!\left(
            \frac{n_N+1+2j}{2};\,\frac{1}{2}
            \;;\; -\frac{k_{\text{UV}}^2}{a^2 m^2}
        \right).
    \end{split}
\end{equation}
Although the exact result is expressible in closed form, the physical regime of interest justifies a severe simplification. As seen from \eqref{approximationN}, the $n=1$ term is already suppressed relative to the leading ($n=0$) term by a factor of order $(k_{\text{UV}}/a m)^2 \ll 1$, a direct consequence of the non‑relativistic condition \eqref{nonrelativisticcondition}. Higher‑order terms are further suppressed by additional powers of this small parameter. Hence, to an good approximation we retain only the leading contribution:
\begin{equation}\label{solutionlocalintegralN}
    I_j^N \;\approx\; \mathcal{I}^N_{0,j}
    = \frac{1}{4\pi^2 a m} \,
      \frac{N_0}{k_*^{n_N}} \,
      k_{\text{UV}}^{2j+1+n_N} \,
      \Gamma\!\left( \frac{n_N + 2j + 1}{2}\right).
\end{equation}
For the squeezing local integral, we use a similar approach,
\begin{equation} 
 I_j^S \, = \, \frac{1}{2\pi^2 am} \int_0^{\infty} dk\, \frac{ S_k \, k^{2j}}{ \sqrt{1+\frac{k^2}{a^2m^2}}} \cos\bigg(2m \int^t_{t_{\text{eq}}} d\bar{t} \, \sqrt{1+\frac{k^2}{a^2(\bar{t})m^2}}  \bigg).
 \end{equation}
In the denominator, we adopt the non-relativistic approximation \(\sqrt{1 + \frac{k^2}{a^2 m^2}} \approx 1\), incurring a maximum error of order \(\mathcal{O}\big(k_{\text{UV}}^2/(a m)^2\big)\), which is negligible in the non-relativistic regime, as ensured by condition \eqref{nonrelativisticcondition}. Using the expansion \eqref{phaseexpansion} for the integrand of the phase, the cumulative error in the phase at \(n\)th order, \(\Delta \Phi_n\), is found to be:
\begin{equation}
\Delta \Phi_n \sim \frac{k_{\text{UV}}^{2n}}{a^{2n} m^{2n-1} H_{\text{eq}}}.
\end{equation}
For simplicity, we adopt the first-order approximation for the phase, which is justified when the following condition holds:
\begin{equation} \label{app:phasecondition}
    \frac{k_{\text{UV}}}{a_{\text{eq}}} \ll \sqrt{m H_{\text{eq}}}.
\end{equation}
This condition can be relaxed by including higher orders in the expansion. In the limit \(n \to \infty\), the validity of the approximation requires only the smallness of the relativistic parameter \eqref{nonrelativisticcondition}.
Substituting the expression for the scale factor during matter domination, the phase integral can be solved analytically. The integral we approach is the following:
    \begin{equation}
        \begin{split}
             I_j^S \, = \, \frac{1}{2\pi^2 am} \frac{S_0}{k_*^{n_S}}\int_0^{\infty} dk \, k^{n_S+2j} \, e^{-\frac{k^2}{k_{\text{UV}}^2}} \cos\bigg[\bigg(2m - \frac{3k^2}{a^2 m} \bigg) \, t - 2mt_{\text{eq}} +\frac{3k^2t_{\text{eq}}}{ma^2_{\text{eq}}}\bigg],
        \end{split}
    \end{equation}
We factor the cosine into a product of \(k\)-independent oscillations and \(k\)-dependent oscillations, yielding two distinct terms:
\begin{equation}
        \begin{split}
             I_j^S \,& = \, \frac{1}{2\pi^2 am} \frac{S_0}{k_*^{n_S}} \, \cos\big[2m(t-t_{\text{eq}})\big]\, \int_0^{\infty} dk \, k^{n_S+2j} \, e^{-\frac{k^2}{k_{\text{UV}}^2}} \cos\bigg[\frac{3k^2}{m}\bigg(\frac{t}{a^2} - \frac{t_{\text{eq}}}{a_{\text{eq}}^2} \,\bigg) \bigg] \\
             & + \frac{1}{2\pi^2 am} \frac{S_0}{k_*^{n_S}} \, \sin \big[2m(t-t_{\text{eq}})\big] \, \int_0^{\infty} dk \, k^{n_S+2j} \, e^{-\frac{k^2}{k_{\text{UV}}^2}} \sin \bigg[\frac{3k^2}{m}\bigg(\frac{t}{a^2} - \frac{t_{\text{eq}}}{a_{\text{eq}}^2} \,\bigg) \bigg]\\[5pt]
             & := \, I_{\text{cos}} + I_{\text{sin}}.
        \end{split}
    \end{equation}
We evaluate the first integral. The calculation for the second follows similarly. Calling $\beta \equiv 2j+n_N $, $\lambda \equiv 1/ k_{\text{UV}}^{2}$ and $\gamma \equiv 3/m \times (t/a^2 - t_{\text{eq}}/a^2_{\text{eq}}) $ we obtain the following integral:
\begin{equation}
\begin{split}
    I_{\text{cos}} \, &= \,  \frac{1}{2\pi^2 am} \frac{S_0}{k_*^{n_S}} \, \cos\big[2m(t-t_{\text{eq}})\big] \, \int_0^{\infty} dk \, k^{\beta} \, e^{-\lambda k^2} \, \cos\big(\gamma k^2\big) \\
    &=  \frac{1}{4\pi^2 am} \frac{S_0}{k_*^{n_S}} \, \cos\big[2m(t-t_{\text{eq}})\big] \, \sum_{\pm} \int_0^{\infty} dk \, k^{\beta} \, e^{-\big( \lambda\pm i \gamma \big) k^2}.
\end{split}
\end{equation}
We call $\alpha_{\pm} \equiv \lambda\pm i \gamma $ and bring our integral to the complex plane by introducing the complex variables $z_{\pm}=\alpha_{\pm} k^2$.
\begin{equation}
    \begin{split}
         I_{\text{cos}} \, &= \,  \frac{1}{8\pi^2 am} \frac{S_0}{k_*^{n_S}} \, \cos\big[2m(t-t_{\text{eq}})\big] \sum_{\pm} \alpha_{\pm}^{-\frac{\beta + 1}{2}} \int_{\mathcal{C_{\pm}}} dz \,  z_{\pm}^{\frac{\beta -1}{2}} e^{-z}.
    \end{split}
\end{equation}
The contours $\mathcal{C}_+$ and $\mathcal{C}_-$ are rays from the origin to infinity in the right half-plane, with positive and negative phase, respectively. Each contour can be closed by adding the positive real axis and an arc at infinity. For $n_S > -1 - 2j$ (in our case, $n_S > -5$), the integrand is entire, so the integral over each closed contour vanishes by Cauchy's theorem. Moreover, the contribution from the arc at infinity is zero due to the exponential suppression $e^{-z}$. Hence, the integrals along $\mathcal{C}_+$ and $\mathcal{C}_-$ are equal to the corresponding integrals along the positive real axis. This equivalence allows us to express the result in terms of the Gamma function:
\begin{equation}
    \begin{split}
 I_{\text{cos}} \, &= \,  \frac{1}{8\pi^2 am} \frac{S_0}{k_*^{n_S}} \, \cos\big[2m(t-t_{\text{eq}})\big] \, \int_{0}^{\infty}dx\,  x^{\frac{\beta -1}{2}} e^{-x} \, \sum_{\pm} \alpha_{\pm}^{-\frac{\beta + 1}{2}} \\
 &= \frac{1}{8\pi^2 am} \frac{S_0}{k_*^{n_S}} \, \cos\big[2m(t-t_{\text{eq}})\big] \, \Gamma \bigg(\frac{n_S +2j+1}{2}\bigg) \, \sum_{\pm} \bigg[\frac{1}{k_{\text{UV}}^2} \pm i \frac{3}{m}\bigg(\frac{t}{a^2} - \frac{t_{\text{eq}}}{t_{\text{eq}}^2}\bigg)\bigg]^{-\frac{n_S +2j+1}{2}}\\
 &= \frac{1}{4\pi^2 am} \frac{S_0}{k_*^{n_S}} \, \cos\big[2m(t-t_{\text{eq}})\big] \, \Gamma \bigg(\frac{n_S +2j+1}{2}\bigg) \\
 &  \times
 \bigg[\frac{1}{k_{\text{UV}}^4} + \frac{9}{m^2}\bigg( \frac{t}{a^2} - \frac{t_{\text{eq}}}{a_{\text{eq}}^2}\bigg)^2\bigg]^{-\frac{n_S +2j+1}{4}} \, \cos\bigg\{\frac{n_S +2j+1}{2} \arctan\bigg[\frac{3k_{\text{UV}}^2}{m } \bigg(\frac{t}{a^2} - \frac{t_{\text{eq}}}{t_{\text{eq}}^2}\bigg)\bigg] \bigg\}.
    \end{split}
\end{equation}
Applying the analogous procedure to $I_{\text{sin}}$ and summing the two contributions we obtain the following:
\begin{equation}
    \begin{split}
        I_j^S &=  \frac{1}{4\pi^2 am} \, \frac{S_0}{k_*^{n_S}}\, k_{\text{UV}}^{2j+1+n_S} \, \Gamma \, \bigg(\frac{n_S +2j+1}{2}\bigg)  \bigg[1 + \frac{9k_{\text{UV}}^4}{m^2}\bigg( \frac{t}{a^2} - \frac{t_{\text{eq}}}{a_{\text{eq}}^2}\bigg)^2\bigg]^{-\frac{n_S +2j+1}{4}} \\[2pt]
 &  \quad \times
  \, \cos\bigg\{2m(t-t_{\text{eq}}) -\frac{n_S +2j+1}{2} \arctan\bigg[\frac{3k_{\text{UV}}^2}{m } \bigg(\frac{t}{a^2} - \frac{t_{\text{eq}}}{a_{\text{eq}}^2}\bigg)\bigg] \bigg\}.
    \end{split}
\end{equation}
The quantity 
\begin{equation}
    \frac{3k_{\text{UV}}^2}{m } \bigg(\frac{t}{a^2} - \frac{t_{\text{eq}}}{a_{\text{eq}}^2}\bigg),
\end{equation}
is rendered small by condition \eqref{app:phasecondition}. This suppression allows us to perform an asymptotic expansion in this parameter: we retain the leading-order term in the amplitude, while keeping the phase to linear order due to its cumulative sensitivity. This yields
\begin{equation}
    \begin{split} \label{squeezinglocalintegral}
        I_j^S &=  \frac{1}{4\pi^2 am} \, \frac{S_0}{k_*^{n_S}}\, k_{\text{UV}}^{2j+1+n_S} \, \Gamma \, \bigg(\frac{n_S +2j+1}{2}\bigg)   \\[2pt]
 &  \quad \times
  \, \cos\bigg[\bigg(2m +\frac{3(n_S +2j+1)}{2} \frac{k_{\text{UV}}^2}{ma^2} \bigg)\, t \, - \varphi_{j,k}\bigg] ,
    \end{split}
\end{equation}
where we have defined $\varphi_{j} \equiv 2mt_{\text{eq}}+ \frac{3(n_S +2j+1)}{2} \frac{k_{\text{UV}}^2t_{\text{eq}}}{ma^2_{\text{eq}}} $.

\vskip 1cm

\section{Estimation of non-local integrals} \label{Appendix D}

In this Appendix we estimate the value of the non-local integrals appearing in the right-hand side of the equation governing the propagation of gravitational waves \eqref{xiresonanceequation}. Our estimate is performed in cosmic time \(t\) for convenience. A rigorous numerical treatment would be more naturally formulated in conformal time \(\eta\), but that lies beyond the scope of this estimation. Whenever possible, we denote the magnitude of a vector by \(p \equiv \|\vec{p} \, \|\), and use the notation $\bar{a} \equiv a(\bar{t})$. We begin by evaluating the contributions that depend on the occupation number. The combination of the integrals $\bar{J}^t_{ru}$ summed over the internal sign indices takes the form
\begin{equation}
\begin{split}
    \sum_t \bar{J}^t_{ru}(k^{\mu};\bar{t}) = \frac{i}{\bar{a}^2} \int &\frac{d^3p}{(2\pi)^3} \frac{N_p}{\omega(p)} \,  \bigg\lVert \vec{p}-\frac{\vec{k}}{2}\bigg\rVert^r \, \bigg(\frac{\vec{p} \cdot \vec{k}}{k } -\frac{k }{2}\bigg)^{\! \! u}\\
    & \times \frac{(k^0)^2 - (\omega^2( \vec{p}-\vec{k}) - \omega^2(p))}{\big[ (k^0+i\epsilon)^2- (\omega(\vec{p}-\vec{k}) + \omega(p))^2\big]\, \big[(k^0+i\epsilon)^2- (\omega(\vec{p}-\vec{k}) - \omega(p))^2\big]}.
    \end{split}
\end{equation}
The presence of $N_p$ imposes a Gaussian cutoff on the values of the modulus $ p \lesssim k_{\text{UV}} $. Therefore, we approximate the frequencies that appear in the integral:
\begin{equation}
    \begin{split}
       & \omega(p) = m + \mathcal{O}\bigg(\frac{p^2}{a^2m^2}\bigg) \approx m \\
       &\omega(\vec{p}-\vec{k}) = \omega(k)  + \mathcal{O}\bigg(\frac{p}{am}\bigg) \approx \omega(k),
    \end{split}
\end{equation}
where the error is kept small under the non-relativistic condition \eqref{nonrelativisticcondition}.
Then, we define the define the $k^0$-independent factor
\begin{equation}
    \bar{J}'_{ru}(\vec{k},\bar{t}) := \frac{1}{4\bar{a}^2m^2} \int \frac{d^3p}{(2\pi)^3} N_p  \, \bigg\lVert \vec{p}-\frac{\vec{k}}{2}\bigg\rVert^r \, \bigg(\frac{\vec{p} \cdot \vec{k}}{k } -\frac{k}{2}\bigg)^{\! \! u},
\end{equation}
Switching to spherical coordinates, aligning the $z$-axis with $\vec{k}$, and introducing the coordinate $x:= \cos(\theta)$, the integral reads as follows:
\begin{equation}
    \bar{J}'_{ru}(\vec{k},\bar{t}) = \frac{1}{16 \pi^2 \bar{a}^2 m^2} \int_0^{\infty} \, dp \, N_p\, p^2 \, \int_{-1}^{1} \bigg( p^2 + \frac{k^2}{4} -pkx\bigg)^{\! \! \frac{r}{2}}\bigg( p^2 x^2 + \frac{k^2}{4} -pkx\bigg)^{\! \! \frac{u}{2}}\,.
\end{equation}
Finally, we denote as  $\sigma_N$ the combination of these integrals that appears in the equation of motion of the gravitational wave field, with the addition of a factor that will be useful later on:
\begin{equation}
\begin{split}
    \sigma_N(\vec{k}) &:= m\bar{a}^2\Bigl( \bar{J}'_{40}(\vec{k}, \bar{t}) - 2\bar{J}'_{22}(\vec{k}, \bar{t}) + \bar{J}'_{04}(\vec{k}, \bar{t}) \Bigr)\\[2pt]
    &=  \frac{1}{16 \pi^2  m} \int_0^{\infty} \, dp \, N_p\, p^6 \, \int_{-1}^{1} (1-x)^2 = - \frac{ \bar{a}\, I^N_3}{15 m} \,\\[2pt]
    &=  \frac{1}{30 \pi^2  m} \frac{N_0}{k_*^{n_N}} \,
      k_{\text{UV}}^{7+n_N} \,
      \Gamma\!\left( \frac{n_N +7}{2}\right),
    \end{split}
\end{equation}
which is independent of $\vec{k}$. Having obtained this result, we can move on to the evaluation through contour integration of the integral over the energy component $k^0$:
\begin{equation}
 \begin{split}
   &\frac{i}{4} \int \frac{dk^0}{2\pi} \, e^{-ik^0 \Delta t} \sum_{t=1}^4    \left( \bar{J}^t_{40} -2 \bar{J}^t_{22} + \bar{J}^t_{04} \right)\\
   & = -\frac{1}{2\pi}  \frac{\sigma_N}{\bar{a}^2} \int \, d k^0 \,e^{-ik^0 \Delta t} \frac{(k^0)^2 -(\omega^2(k;\bar{t}) - m^2) }{[(k^0+i\epsilon)^2-(\omega(k;\bar{t})+m)^2][(k^0+i\epsilon)^2-(\omega(k;\bar{t})-m)^2]}\\[5pt]
   &=  \frac{\sigma_N}{2\bar{a}^2\,\omega(k;\bar{t})}  \Theta \big(\Delta t \big) \big\{ \sin\big[(m +\omega(k;\bar{t}) )\Delta t \big] -\sin\big[(m- \omega(k;\bar{t}) )\Delta t \big] \big\} .
 \end{split}
\end{equation}
As is evident from the above expression, the non-local contributions arising from the occupation number split into two distinct oscillatory terms, with angular frequencies \(\omega_1 \equiv m + \omega(k)\) and \(\omega_2 \equiv m - \omega(k)\). In this section we explicitly compute only the former; the latter follow by direct analogy. In estimating this contribution, we refrain from integrating the adiabatic time dependence of the scale factor. The resulting contribution to the equation of motion then reads:
\begin{equation} \label{integral}
\frac{\sigma_N}{a^5(t) \omega(k;t)}\,\int_{t_{\text{eq}}}^{t} dt' \, \, \sin\!\big[\omega_1(t)\Delta t\big] \, \xi^{ij}(\vec{k};t').
\end{equation}
In evaluating these integrals, we approximate $\omega(k;\bar{t}) \approx \omega(k;t)$. This simplification is adopted because retaining the full time dependence of $\omega(k)$ would render an analytic solution intractable. Given the adiabatic evolution of $\omega(k)$, the error introduced by this approximation remains negligible for the purpose of estimating these integrals.
We shift the integration variable to $s:=t-t'$, yielding
\begin{equation} \label{deltatintegralN}
 \frac{\sigma_N}{a^5(t) \omega(k;t)}\,\int_{0}^{t-t_{\text{eq}}} ds  \, \sin\!\big[\omega_1(t)s\big]  \, \xi_{ij}(\vec{k};t-s).
\end{equation}
To evaluate the integral, we introduce an adiabatic ansatz for the rescaled tensor perturbation field, assuming that its phase evolves as a power law in the scale factor with exponent \(\alpha\):
\begin{equation} \label{ansatz}
\begin{split}
\xi_{ij}(\vec{k};t-s) &= f_{ij}(\vec{k};t-s) \exp\!\bigg(-i \Omega_{0}(\vec{k}) \int_{t_{\text{eq}}}^{t-s} d\bar{t}\, a^{\alpha}(\bar{t}) \bigg) \\[4pt]
&= f_{ij}(\vec{k};t-s) \exp\!\Big\{-i \Big[ \mathcal{W}(\vec{k};t-s)(t-s) - \mathcal{W}(\vec{k};t_{\text{eq}})\, t_{\text{eq}} \Big] \Big\},
\end{split}
\end{equation}
where we have defined
\begin{equation} \label{adiabatic phase}
\mathcal{W}(\vec{k};t) := \Omega_{0}(\vec{k}) \, \frac{3a^{\alpha}(t)}{2\alpha + 3}.
\end{equation}
For sub-horizon modes in an unperturbed background, the gravitational wave frequency takes the familiar form $\Omega(k) = k/a$, corresponding to free propagation in an expanding universe. In the presence of matter sources, however, this frequency receives corrections whose precise scaling with the scale factor is not known a priori. By leaving $\alpha$ as a free parameter, our analysis remains sufficiently general to accommodate a variety of possible physical effects arising from the coupling to the ULDM field.
To estimate the resulting integral we approximate the graviton amplitude $f_{ij}(\vec{k};t-s)$ to first order around $s=0$ and the angular frequency of the perturbation field $\mathcal{W}(\vec{k};t-s)\approx \mathcal{W}(\vec{k};t)$ to the leading order. The resulting simplified integral reads as follows:
\begin{equation}\label{newansatztermsN}
    \begin{split}
  &\frac{\sigma_N}{\omega(k;t)\,a^5(t)} \xi_{ij}(\vec{k};t) \big( \mathcal{A}_+ (\vec{k},t) - \mathcal{A}_- (\vec{k},t)\big) \\
  & - \frac{\sigma_N}{\omega(k;t)\,a^5(t)} \partial_t \big( f_{ij}(\vec{k};t)\big) \, e^{-i\mathcal{W}(\vec{k}) (t -   t_{\text{eq}})  } \big( \mathcal{B}_+ (\vec{k},t) - \mathcal{B}_- (\vec{k},t)\big),
    \end{split}
\end{equation}
where we defined the integral expressions:
\begin{equation}
    \begin{split}
         &\mathcal{A}_{\pm}(\vec{k},t) :=\frac{1}{2i} \int_{0}^{t-t_{\text{eq}}} ds \,e^{i(\mathcal{W} \pm \omega_1) s} = -i \frac{t-t_{\text{eq}}}{2} e^{\frac{i}{2}(\mathcal{W}\pm \omega_1) (t-t_{\text{eq}})} j_0 \bigg( \frac{\mathcal{W} \pm \omega_1}{2} (t-t_{\text{eq}})\bigg); \\[6pt]
         &\mathcal{B}_{\pm}(\vec{k},t) :=\frac{1}{2i} \int_{0}^{t-t_{\text{eq}}} ds \, s\, \,e^{i(\mathcal{W}  \pm \omega_1) s}=\frac{(t-t_{\text{eq}})^2}{4} e^{\frac{i}{2}(\mathcal{W}\pm \omega_1) (t-t_{\text{eq}})} \\
         & \qquad \qquad \qquad \qquad \qquad \qquad \qquad \qquad \times \bigg[  j_1 \bigg( \frac{\mathcal{W} \pm \omega_1}{2} (t-t_{\text{eq}}) \bigg) -i j_0 \bigg( \frac{\mathcal{W} \pm \omega_1}{2} (t-t_{\text{eq}})\bigg)\bigg].
    \end{split}
\end{equation}
Here \(j_n(x)\) is the spherical Bessel function of the first kind, and we have omitted the explicit dependencies of \(\mathcal{W}\) and \(\omega_1\) to avoid overburdening the notation.
Analogous expressions hold for the contributions with angular frequency \(\omega_2\). Substituting these results, together with their \(\omega_2\) counterparts, into eq. \eqref{newansatztermsN} yields the complete occupation number non-local contribution to the equation of motion.
\begin{equation}
    \begin{split}
        \xi_{ij}(\vec{k};t) \times \kappa^2  M^2_N(\vec{k},t) \, -  \,  \partial_t \big(f_{ij}(\vec{k};t)\big)\times \kappa^2  \gamma_N(\vec{k},t),
    \end{split}
\end{equation}
where we have defined the non-local contribution to the effective mass
\begin{equation} \label{mass-nonlocal-N}
    \begin{split}
        M^2_N(\vec{k},t) := &\frac{-i \sigma_N (t-t_{\text{eq}})}{4 \omega(k)a^5(t)} e^{\frac{i}{2}\mathcal{W} (t-t_{\text{eq}})}  \\[5pt] & \times \bigg[e^{\frac{i}{2}\omega_1 (t-t_{\text{eq}})} j_0 \bigg( \frac{\mathcal{W} + \omega_1}{2} (t-t_{\text{eq}})\bigg)
         - e^{-\frac{i}{2}\omega_1 (t-t_{\text{eq}})} j_0 \bigg( \frac{\mathcal{W} - \omega_1}{2} (t-t_{\text{eq}})\bigg) \\[5pt]
         & \quad \, - e^{\frac{i}{2}\omega_2 (t-t_{\text{eq}})} j_0 \bigg( \frac{\mathcal{W} + \omega_2}{2} (t-t_{\text{eq}})\bigg) + e^{-\frac{i}{2}\omega_2 (t-t_{\text{eq}})} j_0 \bigg( \frac{\mathcal{W} - \omega_2}{2} (t-t_{\text{eq}})\bigg)\bigg],
    \end{split}
\end{equation}
and to the friction coefficient
\begin{equation} \label{friction-nonlocal-N}
\begin{split}
    \gamma_N(\vec{k},t):= &\frac{\sigma_N (t-t_{\text{eq}})^2}{8 \omega(k)a^5(t)} e^{-\frac{i}{2}\mathcal{W} (t-t_{\text{eq}})} \\[5pt] 
    & \times \bigg[e^{\frac{i}{2}\omega_1 (t-t_{\text{eq}})} g \bigg( \frac{\mathcal{W} + \omega_1}{2} (t-t_{\text{eq}})\bigg) - e^{-\frac{i}{2}\omega_1 (t-t_{\text{eq}})} g \bigg( \frac{\mathcal{W} - \omega_1}{2} (t-t_{\text{eq}})\bigg)\\[5pt] 
    & \quad \, 
        - e^{\frac{i}{2}\omega_2 (t-t_{\text{eq}})} g \bigg( \frac{\mathcal{W} + \omega_2}{2} (t-t_{\text{eq}})\bigg) + e^{-\frac{i}{2}\omega_2 (t-t_{\text{eq}})} g \bigg( \frac{\mathcal{W} - \omega_2}{2} (t-t_{\text{eq}})\bigg)\bigg],
\end{split}
\end{equation}
where, for simplicity, we define
\begin{equation} \label{gapp}
    g(x) := j_1(x) - i j_0(x).
\end{equation}
These estimates remain bounded for all values of \(\vec{k}\). Moreover, the amplitude of the mass term exhibits two distinct behaviors: one for frequencies that satisfy
\begin{equation}
    \mathcal{W}(\vec{k};t) \approx  \omega_i (\vec{k};t), \quad i=1,2,
\end{equation}
and another for all other frequencies.
This amplitude, compared to that of the local terms, is found to be
\begin{equation} \label{Ncases}
   a^2 \frac{M^2_N}{I_2^N} \sim 
    \begin{cases}
        \dfrac{k_{\text{UV}}^2}{a^2 m^2} & \text{for } \mathcal{W} \neq s_1 \omega+ s_2 m,\quad s_1, s_2 \in \{-1;1\}; \\[10pt]
        \dfrac{k_{\text{UV}}^2 (t - t_{\text{eq}})}{a^2 m} & \text{for } \mathcal{W} \approx s_1 \omega+ s_2 m,\quad s_1, s_2 \in \{-1;1\}.
    \end{cases}
\end{equation}
Therefore, under conditions~\eqref{nonrelativisticcondition} and \eqref{phasecondition}, the non-local terms are suppressed relative to the local ones and can be considered as relativistic corrections. If we assume an adiabatic evolution characterized by \(\partial_t f_{ij} \sim H f_{ij}\), we obtain the same suppression for the friction term. \bigskip \\ 
Let us now approach the contributions due to squeezing of the matter field modes. From their definition in eq.~\eqref{nonlocalSintegrals}, the combination of the integrals \(\widetilde{J}^t_{ru}\) summed over the internal sign indices takes the form 
\begin{equation}
\begin{split}
    \sum_t \widetilde{J}^t_{ru}(k^{\mu};\bar{t}) = \frac{i}{\bar{a}^2} \int \frac{d^3p}{(2\pi)^3} \frac{S(p)}{\omega(p)}& \, \bigg\lVert \vec{p}-\frac{\vec{k}}{2}\bigg\rVert^{\,r} \, \left( \frac{\vec{p}\cdot\vec{k}}{\|\vec{k}\|} - \frac{\|\vec{k}\|}{2} \right)^{\! \! u} \\
    & \times \frac{1}{ (k^0+i\epsilon)^2 - \omega^2(\vec{p}-\vec{k}) } \, \cos\!\left( 2 \int^{\bar{t}}_{t_{\mathrm{eq}}} dt' \, \omega(p;t') \right),
\end{split}
\end{equation}
where we have restored the exact adiabatic phase \(\Phi(k;\bar{t}) \equiv \int_{t_{\mathrm{eq}}}^{\bar{t}} dt'\,\omega(k;t')\) in place of the product \(\omega(k;\bar{t})\bar{t}\) to facilitate comparison with the local terms. As in the occupation number case, we approximate at leading order in $p /(am)$ the angular frequencies appearing in the amplitude but keep their leading correction in the oscillatory phase:
\begin{equation}
    \begin{split}
        \cos\bigg(2 \int^{\bar{t}}_{t_{\text{eq}}} dt' \, \omega(p;t')  \bigg)&\approx     \cos\bigg[2m \int^{\bar{t}}_{t_{\text{eq}}} dt' \, \bigg(1 + \frac{k^2}{2 a^2(t') m^2}\bigg) \bigg]\\
       &= \cos\bigg[\bigg(2m - \frac{3k^2}{\bar{a}^2 m} \bigg) \, \bar{t} - 2mt_{\text{eq}} +\frac{3k^2t_{\text{eq}}}{ma^2_{\text{eq}}}\bigg].
    \end{split}
\end{equation}
This prescription mirrors the approximation scheme employed for the local contributions, and is justified under the condition \eqref{phasecondition}, which can be relaxed through the inclusion of higher order terms, as specified in Appendix \ref{Appendix C}. The \(k^0\) and \(\vec{p}\) integrals can then be performed independently.
\begin{equation}
    \begin{split}
        &\int_{-\infty}^{\infty} \frac{dk^0}{2\pi} \frac{e^{-ik^0\Delta t}}{(k^0+i \epsilon)^2 - \omega^2(k)}\, = \, - \frac{\sin(\omega(k)\Delta t)}{\omega(k)} \Theta(\Delta t).
    \end{split}
\end{equation}
For the integral over the momentum, we first define the following expressions:
\begin{equation}
    \widetilde{J}'_{ru}(\vec{k},\bar{t}) := \frac{1}{4a\bar{a}^2 m^2} \int \frac{d^3p}{(2\pi)^3} S_p  \, \bigg\lVert \vec{p}-\frac{\vec{k}}{2}\bigg\rVert^r \, \bigg(\frac{\vec{p} \cdot \vec{k}}{k } -\frac{k }{2}\bigg)^{\! \! u}\cos\bigg[\bigg(2m - \frac{3k^2}{\bar{a}^2m} \bigg) \, \bar{t} - 2mt_{\text{eq}} +\frac{3k^2t_{\text{eq}}}{ma^2_{\text{eq}}}\bigg]. \raisetag{-4mm}
\end{equation}
The angular dependence of these integrals is resolved in a similar way to the case of $\bar{J}'_{ru}$ above. The contribution of these integrals to the equation of motion turns out to be a factor $\sigma_S$ similar to that found for the contribution arising from the occupation spectrum, multiplied by an oscillatory term caused by the squeezing of the ULDM field:
\begin{equation}
\begin{split}
    &m\bar{a}^2\Bigl(\widetilde{J}'_{40}(\vec{k}, \bar{t}) - 2\widetilde{J}'_{22}(\vec{k}, \bar{t}) + \widetilde{J}'_{04}(\vec{k}, \bar{t}) \Bigr)\\[2pt]
    &=  \frac{1}{16 \pi^2  m} \int_0^{\infty} \, dp \, S_p\, p^6\, \cos\bigg[\bigg(2m - \frac{3k^2}{\bar{a}^2 m} \bigg) \, \bar{t} - 2mt_{\text{eq}} +\frac{3k^2t_{\text{eq}}}{ma^2_{\text{eq}}}\bigg] \, \int_{-1}^{1} (1-x)^2 = \frac{ 2\bar{a} I^S_3}{15 m} \,\\[2pt]
    &=   \frac{1}{30 \pi^2  m} \frac{S_0}{k_*^{n_S}} \,
      k_{\text{UV}}^{7+n_S} \,
      \Gamma\!\left( \frac{n_S +7}{2}\right) \cos\big[2m_S(\bar{t})  \bar{t} - \varphi_3\big] = \sigma_S \cos\big[2m_S(\bar{t})  \bar{t}- \varphi_3\big] ,
    \end{split}
\end{equation}
where $\varphi_j$ is the same phase shift that appears in \eqref{squeezinglocalintegral}, and we have introduced the following quantities that depend on the squeezing parameters:
\begin{equation}
    \begin{split}
        &\sigma_S :=  \frac{1}{30 \pi^2  m} \frac{S_0}{k_*^{n_S}} \,
      k_{\text{UV}}^{7+n_S} \,
      \Gamma\!\left( \frac{n_S +7}{2}\right);\\[5pt]
      &m_S(t)= \, m+ \frac{3 (n_S+7)}{4} \frac{k_{\text{UV}}^2}{m\bar{a}^2}.
    \end{split}
\end{equation}
Therefore, neglecting the adiabatic time dependence of the scale factor, the contribution to the equation of motion for $\xi_{ij}(\vec{k};t)$ coming from the non-local squeezing terms takes the following form:
\begin{equation} \label{integral2}
   \frac{\sigma_S}{a^5(t)\omega(k)}\,\int_{0}^{t- t_{\text{eq}} } du   \,\cos\big[2m_S(t-u/2)  \big(t-u/2\big) - \varphi_3\big]  \sin\!\big(\omega(k) u\big) \, \xi_{ij}(\vec{k};t-u),
\end{equation}
where, as for the occupation number case, we do not integrate the adiabatic time dependence of \(\omega(k)\), treating it as constant over the relevant timescales to maintain analytic tractability.
For the purpose of estimating the integral, we adopt the ansatz~\eqref{ansatz} for the perturbation field.  We expand the amplitude $f_{ij}(\vec{k};t-s)$ to the first order around $s=0$, and approximate the pulsation of the perturbation field $\mathcal{W}(\vec{k};t-s)\approx \mathcal{W}(\vec{k};t)$ and the squeezing mass $m_S(t-s/2)\approx m_S(t)$ to the leading order. The contribution to the equation of motion of gravitational waves is proportional to the following expression
\begin{equation}
    \begin{split}\label{newansatztermsS}
       & \frac{\sigma_S}{2\omega(k)a^5(t)} \xi_{ij}(\vec{k};t) \big[ e^{-i[2m_S(t)\, t -\varphi_3]} \big( \mathcal{A}_{++} - \mathcal{A}_{-+} \big) + e^{i[2m_S(t)\, t -\varphi_3]} \big( \mathcal{A}_{+-} - \mathcal{A}_{--} \big)\big]\\[4pt]
       & - \frac{\sigma_S}{2\omega(k)a^5(t)} \partial_t \big( f_{ij}(\vec{k};t)\big) \, e^{-i \mathcal{W}(\vec{k}) (t - t_{\text{eq}})  } \\[3pt]
       & \qquad \qquad \qquad \qquad \times \big[ e^{-i[2m_S(t)\, t -\varphi_3]} \big( \mathcal{B}_{++} - \mathcal{B}_{-+} \big) + e^{i[2m_S(t)\, t -\varphi_3]} \big( \mathcal{B}_{+-} - \mathcal{B}_{--} \big)\big].\\[2pt]
    \end{split}
\end{equation}
Here, the integrals $\mathcal{A}_{s_1s_2}, \mathcal{B}_{s_1 s_2}$ can be expressed in terms of exponential functions
\begin{equation}
    \begin{split}
         &\mathcal{A}_{s_1 s_2}(\vec{k},t) :=\frac{1}{2i} \int_{0}^{t-t_{\text{eq}}} du \,e^{i \omega_{s_1 s_2} u} = -\frac{i (t-t_{\text{eq}})}{2} e^{\frac{i}{2} \omega_{s_1 s_2} (t-t_{\text{eq}}) } j_0\bigg( \frac{\omega_{s_1 s_2}}{2}(t-t_{\text{eq}})\bigg) ;\\[6pt]
         &\mathcal{B}_{s_1 s_2}(\vec{k},t) :=\frac{1}{2i} \int_{0}^{t-t_{\text{eq}}} du \, u\,e^{i\omega_{s_1 s_2} u}=  \frac{(t-t_{\text{eq}})^2}{4} e^{\frac i2 \omega_{s_1 s_2} (t-t_{\text{eq}})} g\bigg( \frac{ \omega_{s_1 s_2}}{2}(t-t_{\text{eq}})\bigg) ,
    \end{split}
\end{equation}
where the function $g(z)$ has been defined in~\eqref{gapp}. We introduced a simplified notation for the typical angular frequencies governing this term:
\begin{equation}
    \omega_{s_1 s_2}(\vec{k};t) := \mathcal{W}(\vec{k};t) + s_1 \omega(k;t) + s_2 m_S(t).
\end{equation}
To conclude, we plug in these results in eq. \eqref{newansatztermsS}, yielding the following expression:
\begin{equation}
    \begin{split}
        \xi_{ij}(\vec{k};t) \times \kappa^2 M^2_S(\vec{k},t)  \, -  \,  \partial_t \big( f_{ij}(\vec{k};t)\big)\times \kappa^2 \gamma_S(\vec{k},t).
    \end{split}
\end{equation}
Here, we have defined the squeezing non-local contribution to the effective mass
\begin{equation} \label{mass-nonlocal-S}
    \begin{split}
        M^2_S(\vec{k},t)& := \frac{-i \sigma_S (t-t_{\text{eq}}) }{4 \omega(k)a^5(t)}  e^{\frac{i}{2} \mathcal{W} (t-t_{\text{eq}})}\\[4pt]
         &  \times  \bigg\{ e^{-i[2m_S\, t -\varphi_3]} e^{\frac i2 m_S (t-t_{\text{eq}})} \bigg[ e^{\frac i2 \omega (t-t_{\text{eq}})} j_0 \bigg(\frac{\omega_{++}}{2} (t-t_{\text{eq}}) \bigg) \\[2pt]
         & \qquad \qquad \qquad \qquad \qquad \qquad -e^{-\frac i2 \omega (t-t_{\text{eq}})} j_0 \bigg(\frac{\omega_{-+}}{2} (t-t_{\text{eq}}) \bigg)  \bigg]\\
        &\quad +  e^{i[2m_S\, t -\varphi_3]} e^{-\frac i2 m_S (t-t_{\text{eq}})} \bigg[ e^{\frac i2 \omega (t-t_{\text{eq}})} j_0 \bigg(\frac{\omega_{+-}}{2} (t-t_{\text{eq}}) \bigg)\\[2pt]
         & \qquad \qquad \qquad \qquad \qquad \qquad -e^{-\frac i2 \omega (t-t_{\text{eq}})} j_0 \bigg(\frac{\omega_{--}}{2} (t-t_{\text{eq}}) \bigg)  \bigg] \bigg\},
    \end{split}
\end{equation}
and to the friction factor
\begin{equation} \label{friction-nonlocal-S}
\begin{split} 
    \gamma_S(\vec{k},t) &:= \frac{\sigma_S (t-t_{\text{eq}})^2 }{8 \omega(k)a^5(t)}  e^{-\frac{i}{2} \mathcal{W} (t-t_{\text{eq}})}\\[4pt]
         &  \times  \bigg\{ e^{-i[2m_S\, t -\varphi_3]} e^{\frac i2 m_S (t-t_{\text{eq}})} \bigg[ e^{\frac i2 \omega (t-t_{\text{eq}})} g \bigg(\frac{\omega_{++}}{2} (t-t_{\text{eq}}) \bigg)\\[2pt]
         & \qquad \qquad \qquad \qquad \qquad \qquad -e^{-\frac i2 \omega (t-t_{\text{eq}})} g \bigg(\frac{\omega_{-+}}{2} (t-t_{\text{eq}}) \bigg)  \bigg]\\
        &\quad +  e^{i[2m_S\, t -\varphi_3]} e^{-\frac i2 m_S (t-t_{\text{eq}})} \bigg[ e^{\frac i2 \omega (t-t_{\text{eq}})} g \bigg(\frac{\omega_{+-}}{2} (t-t_{\text{eq}}) \bigg)\\[2pt]
         & \qquad \qquad \qquad \qquad \qquad \qquad -e^{-\frac i2 \omega (t-t_{\text{eq}})} g \bigg(\frac{\omega_{--}}{2} (t-t_{\text{eq}}) \bigg)  \bigg] \bigg\}.
\end{split}
\end{equation}
Similarly to the previous case, the amplitude of the mass term exhibits two distinct behaviors: one for frequencies that satisfy
\begin{equation}
    \mathcal{W}(\vec{k};t) \approx  s_1 \omega(k;t) + s_2 m_S(t), \qquad s_1, s_2 \in \{-1, 1\},
\end{equation}
and another for all other frequencies. Assuming condition~\eqref{phasecondition} holds, the mass term is suppressed relative to the local squeezing contribution in both cases and can be treated as a relativistic correction:
\begin{equation} \label{Scases}
   a^2 \frac{M^2_S}{I_2^S} \sim 
    \begin{cases}
        \dfrac{k_{\text{UV}}^2}{a^2 m^2} & \text{for } \mathcal{W} \neq s_1 \omega + s_2 m_S,\quad s_1, s_2 \in \{-1, 1\}; \\[10pt]
        \dfrac{k_{\text{UV}}^2 (t - t_{\text{eq}})}{a^2 m} & \text{for } \mathcal{W} \approx s_1 \omega + s_2 m_S,\quad s_1, s_2 \in \{-1, 1\}.
    \end{cases}
\end{equation}
If we assume an adiabatic evolution characterized by \(\partial_t f_{ij} \sim H f_{ij}\), we obtain the same suppression for the friction term.

\bibliographystyle{JHEP}

\end{fmffile}
\end{document}